\documentclass[aps,prc,superscriptaddress,showpacs,nofootinbib,floatfix,twocolumn]{revtex4}

\usepackage{graphicx}	
\usepackage[intlimits]{amsmath}
\usepackage{amssymb}
\usepackage[hypertex]{hyperref}
\usepackage{xspace}
\usepackage{epsfig}

\def\pt{$p_T$}

\def\gev2{GeV/$c^2$}

\def\mev2{MeV/$c^2$}

\def\ncoll{$N_{\rm coll}$}

\begin{document}

\title{Detailed measurement of the $e^+e^-$ pair continuum in $p+p$
and Au~+~Au collisions at $\sqrt{s_{NN}}$=200~GeV and implications for
direct photon production}

\newcommand{\abilene}{Abilene Christian University, Abilene, TX 79699, USA}
\newcommand{\banaras}{Department of Physics, Banaras Hindu University, Varanasi 221005, India}
\newcommand{\bnlcoll}{Collider-Accelerator Department, Brookhaven National Laboratory, Upton, NY 11973-5000, USA}
\newcommand{\bnlphys}{Brookhaven National Laboratory, Upton, NY 11973-5000, USA}
\newcommand{\caucr}{University of California - Riverside, Riverside, CA 92521, USA}
\newcommand{\charlesczech}{Charles University, Ovocn\'{y} trh 5, Praha 1, 116 36, Prague, Czech Republic}
\newcommand{\ciae}{China Institute of Atomic Energy (CIAE), Beijing, People's Republic of China}
\newcommand{\cns}{Center for Nuclear Study, Graduate School of Science, University of Tokyo, 7-3-1 Hongo, Bunkyo, Tokyo 113-0033, Japan}
\newcommand{\colorado}{University of Colorado, Boulder, CO 80309, USA}
\newcommand{\columbia}{Columbia University, New York, NY 10027 and Nevis Laboratories, Irvington, NY 10533, USA}
\newcommand{\czechtech}{Czech Technical University, Zikova 4, 166 36 Prague 6, Czech Republic}
\newcommand{\dapnia}{Dapnia, CEA Saclay, F-91191, Gif-sur-Yvette, France}
\newcommand{\debrecen}{Debrecen University, H-4010 Debrecen, Egyetem t{\'e}r 1, Hungary}
\newcommand{\elte}{ELTE, E{\"o}tv{\"o}s Lor{\'a}nd University, H - 1117 Budapest, P{\'a}zm{\'a}ny P. s. 1/A, Hungary}
\newcommand{\fit}{Florida Institute of Technology, Melbourne, FL 32901, USA}
\newcommand{\fsu}{Florida State University, Tallahassee, FL 32306, USA}
\newcommand{\gsu}{Georgia State University, Atlanta, GA 30303, USA}
\newcommand{\hiroshima}{Hiroshima University, Kagamiyama, Higashi-Hiroshima 739-8526, Japan}
\newcommand{\ihepprot}{IHEP Protvino, State Research Center of Russian Federation, Institute for High Energy Physics, Protvino, 142281, Russia}
\newcommand{\illuiuc}{University of Illinois at Urbana-Champaign, Urbana, IL 61801, USA}
\newcommand{\instpasczech}{Institute of Physics, Academy of Sciences of the Czech Republic, Na Slovance 2, 182 21 Prague 8, Czech Republic}
\newcommand{\isu}{Iowa State University, Ames, IA 50011, USA}
\newcommand{\jinrdubna}{Joint Institute for Nuclear Research, 141980 Dubna, Moscow Region, Russia}
\newcommand{\kaeri}{KAERI, Cyclotron Application Laboratory, Seoul, Korea}
\newcommand{\kek}{KEK, High Energy Accelerator Research Organization, Tsukuba, Ibaraki 305-0801, Japan}
\newcommand{\kfki}{KFKI Research Institute for Particle and Nuclear Physics of the Hungarian Academy of Sciences (MTA KFKI RMKI), H-1525 Budapest 114, POBox 49, Budapest, Hungary}
\newcommand{\korea}{Korea University, Seoul, 136-701, Korea}
\newcommand{\kurchatov}{Russian Research Center ``Kurchatov Institute", Moscow, Russia}
\newcommand{\kyoto}{Kyoto University, Kyoto 606-8502, Japan}
\newcommand{\labllr}{Laboratoire Leprince-Ringuet, Ecole Polytechnique, CNRS-IN2P3, Route de Saclay, F-91128, Palaiseau, France}
\newcommand{\lawllnl}{Lawrence Livermore National Laboratory, Livermore, CA 94550, USA}
\newcommand{\losalamos}{Los Alamos National Laboratory, Los Alamos, NM 87545, USA}
\newcommand{\lpc}{LPC, Universit{\'e} Blaise Pascal, CNRS-IN2P3, Clermont-Fd, 63177 Aubiere Cedex, France}
\newcommand{\lund}{Department of Physics, Lund University, Box 118, SE-221 00 Lund, Sweden}
\newcommand{\muenster}{Institut f\"ur Kernphysik, University of Muenster, D-48149 Muenster, Germany}
\newcommand{\myongji}{Myongji University, Yongin, Kyonggido 449-728, Korea}
\newcommand{\nagasaki}{Nagasaki Institute of Applied Science, Nagasaki-shi, Nagasaki 851-0193, Japan}
\newcommand{\newmex}{University of New Mexico, Albuquerque, NM 87131, USA}
\newcommand{\nmsu}{New Mexico State University, Las Cruces, NM 88003, USA}
\newcommand{\ornl}{Oak Ridge National Laboratory, Oak Ridge, TN 37831, USA}
\newcommand{\orsay}{IPN-Orsay, Universit{\'e} Paris Sud, CNRS-IN2P3, BP1, F-91406, Orsay, France}
\newcommand{\peking}{Peking University, Beijing, People's Republic of China}
\newcommand{\pnpi}{PNPI, Petersburg Nuclear Physics Institute, Gatchina, Leningrad region, 188300, Russia}
\newcommand{\riken}{RIKEN Nishina Center for Accelerator-Based Science, Wako, Saitama 351-0198, JAPAN}
\newcommand{\rikjrbrc}{RIKEN BNL Research Center, Brookhaven National Laboratory, Upton, NY 11973-5000, USA}
\newcommand{\rikkyo}{Physics Department, Rikkyo University, 3-34-1 Nishi-Ikebukuro, Toshima, Tokyo 171-8501, Japan}
\newcommand{\saispbstu}{Saint Petersburg State Polytechnic University, St. Petersburg, Russia}
\newcommand{\saopaulo}{Universidade de S{\~a}o Paulo, Instituto de F\'{\i}sica, Caixa Postal 66318, S{\~a}o Paulo CEP05315-970, Brazil}
\newcommand{\seoulnat}{System Electronics Laboratory, Seoul National University, Seoul, Korea}
\newcommand{\stonybrkc}{Chemistry Department, Stony Brook University, Stony Brook, SUNY, NY 11794-3400, USA}
\newcommand{\stonycrkp}{Department of Physics and Astronomy, Stony Brook University, SUNY, Stony Brook, NY 11794, USA}
\newcommand{\subatech}{SUBATECH (Ecole des Mines de Nantes, CNRS-IN2P3, Universit{\'e} de Nantes) BP 20722 - 44307, Nantes, France}
\newcommand{\tenn}{University of Tennessee, Knoxville, TN 37996, USA}
\newcommand{\titech}{Department of Physics, Tokyo Institute of Technology, Oh-okayama, Meguro, Tokyo 152-8551, Japan}
\newcommand{\tsukuba}{Institute of Physics, University of Tsukuba, Tsukuba, Ibaraki 305, Japan}
\newcommand{\vandy}{Vanderbilt University, Nashville, TN 37235, USA}
\newcommand{\waseda}{Waseda University, Advanced Research Institute for Science and Engineering, 17 Kikui-cho, Shinjuku-ku, Tokyo 162-0044, Japan}
\newcommand{\weizmann}{Weizmann Institute, Rehovot 76100, Israel}
\newcommand{\yonsei}{Yonsei University, IPAP, Seoul 120-749, Korea}
\affiliation{\abilene}
\affiliation{\banaras}
\affiliation{\bnlcoll}
\affiliation{\bnlphys}
\affiliation{\caucr}
\affiliation{\charlesczech}
\affiliation{\ciae}
\affiliation{\cns}
\affiliation{\colorado}
\affiliation{\columbia}
\affiliation{\czechtech}
\affiliation{\dapnia}
\affiliation{\debrecen}
\affiliation{\elte}
\affiliation{\fit}
\affiliation{\fsu}
\affiliation{\gsu}
\affiliation{\hiroshima}
\affiliation{\ihepprot}
\affiliation{\illuiuc}
\affiliation{\instpasczech}
\affiliation{\isu}
\affiliation{\jinrdubna}
\affiliation{\kaeri}
\affiliation{\kek}
\affiliation{\kfki}
\affiliation{\korea}
\affiliation{\kurchatov}
\affiliation{\kyoto}
\affiliation{\labllr}
\affiliation{\lawllnl}
\affiliation{\losalamos}
\affiliation{\lpc}
\affiliation{\lund}
\affiliation{\muenster}
\affiliation{\myongji}
\affiliation{\nagasaki}
\affiliation{\newmex}
\affiliation{\nmsu}
\affiliation{\ornl}
\affiliation{\orsay}
\affiliation{\peking}
\affiliation{\pnpi}
\affiliation{\riken}
\affiliation{\rikjrbrc}
\affiliation{\rikkyo}
\affiliation{\saispbstu}
\affiliation{\saopaulo}
\affiliation{\seoulnat}
\affiliation{\stonybrkc}
\affiliation{\stonycrkp}
\affiliation{\subatech}
\affiliation{\tenn}
\affiliation{\titech}
\affiliation{\tsukuba}
\affiliation{\vandy}
\affiliation{\waseda}
\affiliation{\weizmann}
\affiliation{\yonsei}
\author{A.~Adare} \affiliation{\colorado}
\author{S.~Afanasiev} \affiliation{\jinrdubna}
\author{C.~Aidala} \affiliation{\columbia}
\author{N.N.~Ajitanand} \affiliation{\stonybrkc}
\author{Y.~Akiba} \affiliation{\riken} \affiliation{\rikjrbrc}
\author{H.~Al-Bataineh} \affiliation{\nmsu}
\author{J.~Alexander} \affiliation{\stonybrkc}
\author{A.~Al-Jamel} \affiliation{\nmsu}
\author{K.~Aoki} \affiliation{\kyoto} \affiliation{\riken}
\author{L.~Aphecetche} \affiliation{\subatech}
% ...
\collaboration{PHENIX Collaboration} \noaffiliation

\date{\today}

\begin{abstract}


PHENIX has measured the $e^+e^-$ pair continuum in $\sqrt{s_{NN}}$=200
~GeV Au~+~Au and $p+p$ collisions over a wide range of mass and
transverse momenta.  The $e^+e^-$ yield is compared to the 
expectations from hadronic sources, based on PHENIX measurements.  In 
the intermediate-mass region, between the masses of the $\phi$ and the 
$J/\psi$ meson, the yield is consistent with expectations from 
correlated $c\bar{c}$ production, although other mechanisms are not 
ruled out.  In the low mass region, below the $\phi$, the $p+p$ 
inclusive mass spectrum is well described by known contributions from 
light meson decays.  In contrast, the Au~+~Au minimum bias inclusive 
mass spectrum in this region shows an enhancement by a factor of 
$4.7\pm0.4^{\rm stat}\pm1.5^{\rm syst}\pm0.9^{\rm model}$.  At low 
mass ($m_{ee} <$ 0.3~GeV/$c^2$) and high $p_T$ (1 $<p_T<$5~GeV/$c$) an 
enhanced $e^+e^-$ pair yield is observed that is consistent with 
production of virtual direct photons.  This excess is used to infer the 
yield of real direct photons.  In central Au~+~Au collisions, the 
excess of the direct photon yield over the $p+p$ is exponential in 
$p_T$, with inverse slope 
$T = 221 \pm 19^{\rm stat} \pm 19^{\rm syst}$~MeV.  
Hydrodynamical models with initial temperatures ranging 
from $T_{\rm init} \simeq$ 300--600 MeV at times of 0.6--0.15 fm/$c$ 
after the collision are in qualitative agreement with the direct 
photon data in Au~+~Au.  For low $p_T<1$~GeV/$c$ the low-mass region 
shows a further significant enhancement that increases with centrality 
and has an inverse slope of $T \simeq 100$ MeV.  Theoretical models 
under predict the low-mass, low-$p_T$ enhancement.

\end{abstract}
\pacs{25.75.Dw}
\maketitle


\section{INTRODUCTION} \label{sec:intro}
Experimental results from the Relativistic Heavy Ion Collider (RHIC)
have established that in Au~+~Au collisions at $\sqrt{s_{NN}}$~=~200~GeV
matter is created with very high energy density~\cite{whitepaper}, as indicated by the
large energy produced transverse to the beam direction~\cite{ppg019},
as well as by the large energy loss of light
~\cite{ppg003,ppg051} and heavy quarks~\cite{ppg056,ppg066}, and is thermalized rapidly, as
indicated by the large elliptic flow of these
partons~\cite{ppg022,ppg062,ppg066,ppg073}.  Such a high density
thermalized medium is expected to emit thermal radiation~\cite{Stankus}
in the form of direct photons and dileptons.

Electron-positron pairs, or dileptons in general, are excellent tools
for studying collisions of heavy ions at ultra-relativistic energies.
Since they are not affected by the strong interaction, and therefore
can escape from the dense medium without final state interaction,
dilepton spectra can probe the whole time evolution and dynamics of
the collision.  Dileptons can also be used to study the properties of
low-mass vector mesons $\rho$, $\omega$, and $\phi$ in the
medium, since their lifetime is shorter or similar ($\phi$) to that
of the medium.  Their mass and width inside the dense medium can be directly
measured through their dilepton decay channels, and thereby one can
study the effect of chiral symmetry restoration on these
mesons.  Furthermore, production of photons can be measured through
their conversion to dileptons.

\begin{figure}[htbp]
\begin{center} 
\includegraphics[width=1.0\linewidth]{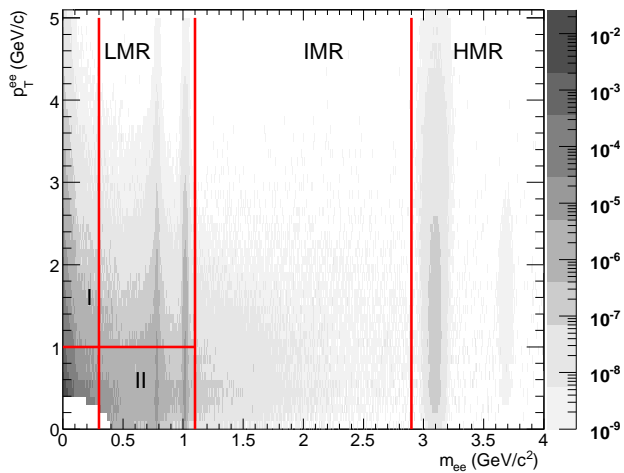}
\caption{(Color online) Dilepton spectrum as a function of mass and transverse
momentum from a simulation of hadron decays.  The high-mass
region (HMR; $m_{ee}>3.2$~GeV/$c^2$) goes from $J/\psi$ mass and above, the
low-mass region (LMR; $m_{ee}<1.2$~GeV/$c^2$) from the $\phi$ mass and
below, and the intermediate mass region (IMR; $1.2<m_{ee}<2.9$~GeV/$c^2$)
between them.  In the LMR at low-$p_T$ (II), dilepton production is
expected to be dominated by the hadronic gas phase.  Part of the LMR,
where $p_T\gg m_{ll}$, specifically $m_{ee}<0.3$~GeV/$c^2$ and $p_T$
$>$ 1~GeV/$c$ (I), is the quasi-real virtual photon region.  The $z$ axis shows the dilepton yield from the hadron decays according to the color scheme plotted on the right.  
\label{fig:cartoon}} \end{center}
\end{figure}

As schematically shown in Fig.~\ref{fig:cartoon}, the dilepton spectra
can be classified into the high-mass region ({\bf HMR}; $m>3.2$~GeV/$c^2$)
from $J/\psi$ mass and above, the low-mass region ({\bf LMR};
$m<1.2$~GeV/$c^2$) from the $\phi$ mass and below, which is further
subdivided in {\bf LMR~I} and {\bf LMR~II} as described below, and the
intermediate mass region ({\bf IMR}; $1.2<m<2.9$~GeV/$c^2$) between them.

\begin{description}
\item[{\bf In the HMR}]
hard scattering on partons in colliding nuclei produces dileptons 
through the Drell-Yan process ($\bar{q}q \rightarrow
l^+l^-$) and correlated semi-leptonic decays of heavy quark pairs ($b
\bar{b} \rightarrow l^+l^-$,$c \bar{c} \rightarrow l^+l^-$).  
Dileptons from these hard processes are expected to dominate
in the HMR since their mass spectra are harder than that
from other possible sources.
Thus dileptons in the HMR probe the initial
stage of the collision~\cite{rapp_shuryak}.
Charmonia ($J/\psi, \psi'$) and Upsilons are in this mass region and
deconfinement~\cite{masui_satz} and recombination~\cite{pbm, coalescence} 
effects can be studied from their yields.  Little
contribution from thermal radiation is expected in the HMR at RHIC
energies~\cite{rapp_shuryak}.

\item[{\bf In the IMR}]
theoretical models predict that dileptons from the thermalized
deconfined phase, the quark gluon plasma (QGP), are the dominant
source of dileptons~\cite{Rapp1, Kaempfer, Shuryak}.  The
measurement of thermal dileptons from QGP can be used to determine the
initial temperature of the matter.  Here a competing source of
dileptons is the semi-leptonic decay of $c$ and $\bar{c}$, correlated
through flavor conservation.  The continuum yield in this mass region
is sensitive to the energy loss of charm quarks in the medium.

\item[{\bf In the LMR}] dilepton production is expected to be dominated by
in-medium decay of $\rho$ mesons in the hadronic gas
phase.~\cite{rapp_RHIC,dusling_RHIC,cassing_RHIC}.  The $\rho$ has a
strong coupling to the $\pi\pi$ channel, and its lifetime (1.3 fm/$c$)
is much shorter than the expected lifetime of the hadronic gas.  The
shape and the yield of the mass spectrum can test predicted in-medium
modifications of the properties (the mass and the width) of $\rho$
mesons due to chiral symmetry restoration~\cite{brown0}.  
Dileptons can also arise from other hadronic sources.  
These dilepton sources compete with a large contribution of
$e^+e^-$ pairs from Dalitz decays of pseudoscalar mesons ($\pi^0,
\eta, \eta'$) and decays of vector mesons ($\rho, \omega,
\phi$).

\item[{\bf In the LMR~I}] (marked with I in Fig.~\ref{fig:cartoon}) is 
the quasi-real virtual photon region, where the $p_T$ of the dilepton is
much greater than its mass ($p_T\gg m_{ll}$).   Any source
of real photons must also emit virtual photons which convert to
low-mass $e^+e^-$ pairs.  These low-mass pairs are produced by a
higher order QED correction to the real photon emission process, and their yield
is related to that of real photons.  Thus $e^+e^-$ pairs in this region
provide an alternative method for measuring direct photons.
The measurement of the direct photon yield using
low-mass lepton pairs was first used at the CERN ISR~\cite{Cobb:1978gj}.
UA1 observed that the low-mass dimuon cross section was consistent with 
the so-called ``internal conversion'' of direct photons~\cite{Albajar:1988iq}.

\end{description}

The discovery of a large enhancement of the dilepton yield in the LMR
in ion-ion collisions by HELIOS/3~\cite{HEL} and CERES~\cite{CER1} at
the CERN SPS has triggered a broad theoretical investigation of
modifications of properties of hadrons in a dense medium and of how
these modifications relate to chiral symmetry restoration
~\cite{rapp0, cassing0, brown0}.  These studies advanced with the availability of more
precise data from NA60 and CERES~\cite{NA60_rho,CER2} and HADES~\cite{HADES}.  Most
theoretical studies suggest that in-medium modifications of the $\rho$
meson, with its short lifetime and its strong coupling to the $\pi\pi$
channel, are primarily responsible for the enhancement.  

\begin{description}

\item[{\bf In the LMR~II}] (marked with II in Fig.~\ref{fig:cartoon}) 
the CERES
data show that the enhancement increases significantly faster than
linearly with charged particle density and is concentrated at very low
pair-$p_T$~\cite{CER2}.  This behavior is consistent with the
interpretation that the excess is due to annihilation
processes.  NA60 recently confirmed the excess of dileptons in the LMR
in In+In collisions at 158 $\mathrm A \cdot \mathrm{~GeV}$ with a high statistics dimuon
measurement~\cite{NA60_rho}.  NA60 also observed that the inverse slope
parameter $T_{\rm eff}$ of the pair-$p_T$ spectra rises with dimuon mass in
the LMR~\cite{NA60_pt}.

\end {description}

An enhanced yield was also observed in the IMR by HELIOS/3
~\cite{HEL,HELIOS3}, NA38/50~\cite{NA38,NA50}, and NA60
~\cite{NA60_rho,NA60_pt}.  NA60 data suggest that the enhancement
cannot be attributed to decays of $D$ mesons but may result from
prompt production, as expected for thermal radiation
~\cite{NA60_therm}.  
Furthermore NA60 measures the inverse slope parameter of pair-$p_T$
spectra in the IMR around 190 MeV, independent of mass and
significantly lower than those found at masses below 1~GeV/$c^2$
~\cite{NA60_therm, NA60_pt}.

The PHENIX experiment at the Relativistic Heavy Ion Collider (RHIC)
has measured the dilepton continuum in a new energy regime,
$\sqrt{s_{NN}}$~=~200~GeV, for $p+p$ and Au~+~Au collisions.  In this
article we present results from Au~+~Au collisions taken in 2004 and
$p+p$ collisions taken in 2005.  We show the results in the LMR and in
the IMR as well as the result of direct photon measurement from the
analysis of quasi-real virtual photons.  The main physics results in
the LMR and IMR in Au~+~Au and $p+p$ have been reported in~\cite{ppg075}
and~\cite{ppg085}, respectively, and the results of the virtual photon
analysis have been reported in~\cite{ppg086}.  New results on the
centrality and $p_T$ dependence of the $e^+e^-$ pairs in the LMR are
presented in this paper.

This article is organized as follows.
Sec.~\ref{sec:detector}
describes the PHENIX detector system related to the analysis.  
Sec.~\ref{sec:analysis} presents the
analysis details including the systematic uncertainties.
Sec.~\ref{sec:cocktail} describes the methods used to calculate the pair
yield expected from hadronic decays.
Sec.~\ref{sec:results} shows the $e^+e^-$ results as a function of $m_{ee}$ and $p_T$,
which are then discussed in Sec.~\ref{sec:discussion} and compared to
available theoretical predictions.
Finally Sec.~\ref{sec:summary} provides a conclusion.

\section{PHENIX DETECTOR}\label{sec:detector}
\begin{figure}[htbp]
\begin{center}
\includegraphics[width=1.0\linewidth]{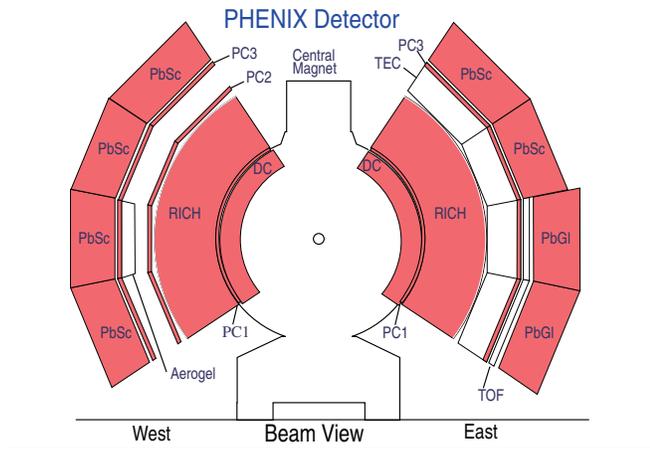}
\caption{(Color online) 
Beam view (at $z=0$) of the PHENIX central arm detector in
Run-4 Au~+~Au and Run-5 $p+p$.  The detectors used in the present
analysis are the drift chamber (DC) and the multi-wire proportional
pad chamber (PC) for charged particle tracking, the ring-imaging
\v{C}erenkov counter (RICH) for electron identification, and the
electromagnetic calorimeter (lead-scintillator (PbSc) and lead-glass  (PbGl))
for energy measurement.  \label{fig:Phenix_2004}} \end{center}
\end{figure}

A detailed description of the complete PHENIX
detector system can be found elsewhere
~\cite{Adcox:2003zm,Adler:2001a,Aronson:2003a,Adcox:2003a,Aizawa:2003a,Aphecetche:2003a,Allen:2003a}.
Here we describe the parts of the detector system that are used in
this analysis, namely, two global detectors and two central arm
spectrometers.  The global detectors are the beam-beam counters (BBC)
and the zero-degree calorimeters (ZDC).  Each central arm covers the
pseudorapidity range $|\eta| < 0.35$ and an azimuthal angle of
$\pi/2$, and includes a drift chamber (DC) and multi-wire proportional
pad chambers (PC) for the charged particle tracking, a ring-imaging
\v{C}erenkov counter (RICH) for electron identification, and an
electromagnetic calorimeter (EMCal) for energy measurement and further
electron identification.  Figure~\ref{fig:Phenix_2004} shows a beam
view of the PHENIX detector.

\subsection{Global Detectors}
The BBC and the ZDC measure the start time and the collision vertex
position $z$-vertex along the beam axis and are used to determine the
centrality of the collision~\cite{Allen:2003a}.  They also provide 
first level trigger information.

The BBC consists of two sets of 64 \v{C}erenkov counter modules,
located $\pm 1.44$ m from the nominal interaction point along the beam
line and measure the number of charged particles in the pseudorapidity
region $3.1<|\eta|<3.9$.  They provide the start-time of the collision
with a resolution of 20 ps, which gives the $z$-vertex position with
a resolution of $\sim$~2~cm in $p+p$ collisions.  
For Au~+~Au central collisions we achieve a resolution of $\sim 0.6$~cm.

The ZDC consists of two hadronic calorimeters, located $\sim 18$ m
from the interaction point, that measure neutral energy emitted within
$\sim$~2 milliradians of the beam direction.  This energy is carried
by neutrons produced either by Coulomb dissociation of the beam
particle or by evaporation from beam spectators.  The energy
resolution of the ZDC is $\delta E/E \sim 218\%/\sqrt{E~\rm{(~GeV)}}$
~\cite{Adler:2001a}.

The centrality of Au~+~Au collisions is determined by the correlation
between the BBC charge sum and the ZDC total energy~\cite{ppg001}.

\subsection{Central Magnet}
The Central Magnet (CM) is an axial field magnet energized by two
pairs of concentric coils, roughly in a Helmholtz configuration, which
can be run separately, together, or in opposition so that the momentum
and the charge of a particle can be determined by its bending
curvature~\cite{Aronson:2003a}.  In the mode in which both coils are
running in the same direction, the single particle momentum resolution
is better than 1\% between 0.2 and 1~GeV/$c$.
During the Au~+~Au measurement in Run-4 and the $p+p$ measurement in
Run-5, the field component parallel to the beam axis had an
approximately Gaussian dependence on the radial distance from the beam
axis, dropping from $0.9$ T at the center to $0.096$ T ($0.048$ T) at
the inner (outer) radius of the DCs.  The total field integral is $\int
B\cdot dl=1.15$ Tm.

\subsection{Tracking Detectors}
The drift chambers (DCs) and the pad chambers (PCs)~\cite{Adcox:2003a}
in the central arms measure charged particle trajectories in the
azimuthal direction to determine the transverse momentum ($p_T$) of
each particle.  The DC provides the most precise measurement of
particle trajectories in the plane perpendicular to the collision
axis.  The first layer of PC provides the most precise measurement of
the track space point along the collision axis.  Additional layers
principally supply pattern recognition support.

The DCs are located in the radial region $2.02<r<2.46$~m.
Each DC volume consists of 20 sectors, each of which covers
$4.5$ degrees in azimuth and $|\eta|<0.35$.  Each sector has six types
of wire modules stacked radially: X1, U1, V1, X2, U2 and V2.  Each
module is further divided into 4 drift cells in the $\phi$ direction.
A plane of sense wires is at the center of each drift cell, with 2 to
2.5 cm drift space on either side.  The X wires run parallel to the
beam axis and measure the particle trajectory in the $r$-$\phi$
plane.  The U and V wires have a stereo angle of about 6.0 degrees
relative to the X wires in order to measure the $z$-coordinate of the
track.  Each X- and U, V-stereo cell contains 12 and 4 sense wires,
respectively.  The single X wire position resolution is $\sim
150~\mu$m.  The intrinsic tracking efficiency of the X modules
is greater than $99~\%$.

The pad chambers (PC) are multi-wire proportional chambers that form
three separate layers.  They determine
space points along the straight line particle trajectories outside the
magnetic field.  The first PC layer (PC1) is located between the DC
and the RICH, the second layer (PC2) is placed behind RICH (west arm
only) and the third layer (PC3) is located in front of the EMCal.  PC1
and the DC, along with the $z-$vertex position measured by the BBC, are
used in the global track reconstruction to determine the polar angle
of each charged track.  The position resolution is $\pm 1.7$~mm for
PC1 along the wire ($z$-direction).

Helium bags were installed between the beam pipe and the DCs to reduce
the conversion material prior to the first tracking layer to
$\sim$~0.4\% of a radiation length.  The material budget is known with
an uncertainty of $\sim$~5\%.

\subsection{Ring-imaging \v{C}erenkov Counter (RICH)}
The RICH is a threshold gas \v{C}erenkov detector and is the primary
detector to identify electrons in PHENIX~\cite{Aizawa:2003a}.  It is
located in the radial region $2.5<r<4.1$~m, just outside PC1.  Each arm
contains spherical mirror panels (0.53\% of a radiation length) which
focus \v{C}erenkov light onto two arrays of $80(\phi) \times 16(z) =
1280$ PMTs.  The PMTs are located outside the acceptance, on either
side of the RICH entrance window, and are shielded to allow operation
in a magnetic field up to 100 Gauss.  The \v{C}erenkov radiator gas,
CO$_2$ at atmospheric pressure, has $n=1.000410$ ($\gamma=35$) which
corresponds to a momentum threshold of 20~MeV/$c$ for an electron and
4.65~GeV/$c$ for a pion.

The average number of hit PMTs per electron track is about 5, and the
average number of photo-electrons detected is about 10.  
Simulation studies show that pion rejection by the RICH alone,
for isolated tracks, is limited by the production rate of
``co-linear'' delta electrons to one part in $10^4$.  However in high
multiplicity collisions pion tracks may be mistaken for electrons via
overlap of their trajectory with a true electron's ring.  This effect
worsens the RICH-alone pion rejection to roughly one part in $10^3$ for
central Au~+~Au collisions and requires additional cuts in the offline
analysis as described below.

\subsection{Electromagnetic Calorimeter (EMCal)}
The EMCal~\cite{Aphecetche:2003a} provides a measurement of the
energies and the spatial positions of photons and electrons.  Each arm
consists of four rectangular sectors in $\phi$: the two bottom sectors of the
east arm are lead-glass (PbGl) calorimeters; whereas, the remaining
sectors are lead-scintillator (PbSc) calorimeters.  The radial
distance from the $z$ axis is $510$~cm for the PbSc and $550$~cm for
the PbGl.

The PbSc is a Shashlik-type sampling calorimeter made of alternating
tiles of lead and scintillator.
It consists of 
$10.5\times 10.5\times 37$ cm$^3$ ($18.2~X_0$) 
modules, constructed of alternating layers of $1.5$ mm
thick lead, reflecting paper, and $4$ mm thick polystyrene-based
scintillator.  Each module is divided into four equal 
towers, from which the light is collected separately by scintillating
fibers.  Each PbSc sector consists of $36 (\phi) \times 72 (z) = 2592$
towers.  The nominal energy resolution is $\delta E/E \sim 4.5\%
\oplus 8.3\%/\sqrt{E~\rm{(~GeV)}}$.

The PbGl is a \v{C}erenkov counter that measures the light
emitted by the particles in an electromagnetic shower and collected by
one PMT at the back end.  Each PbGl sector consists
of 4608 $4.0\times 4.0~\times 40.0$~cm$^3$ ($16~X_0$) modules made of lead-glass crystals.  
The PbGl has a nominal energy resolution of $\delta E/E \sim
4.3\% \oplus 7.7\%/\sqrt{E~\rm{(~GeV)}}$.

\section{ANALYSIS}\label{sec:analysis}
In this Section we present all steps of the data analysis.  We start by
introducing the data set, the event selection for $p+p$ 
and Au~+~Au 
and the centrality definition for Au~+~Au collisions 
(\ref{sub:dataset}).  We present the single electron analysis,
including track reconstruction (\ref{sub:tracking}) and electron
identification (\ref{sub:eID}).  We present the details of the pair
analysis (\ref{sub:pair}) including pair cuts and photon rejection
(\ref{subsub:photons}), combinatorial and correlated
(\ref{subsub:correlated}) background subtraction.  An alternative
method to subtract combinatorial and correlated background together is
described in~\ref{subsub:likesign}.  
Section~\ref{subsub:rawmass} presents the final raw mass spectrum.  In
Section~\ref{subsub:rawmass_conversion} we present the mass spectrum
obtained from the analysis of runs with increased conversion material,
a technique employed to estimate the systematic uncertainty on the
background subtraction.  Next we describe the efficiency
(\ref{sub:effi}) and acceptance (\ref{sub:acc}) corrections, trigger
efficiency (\ref{sub:ert}) (for $p+p$ collisions) and occupancy
correction (for Au~+~Au collisions) (\ref{sub:embedding}).  Finally we
describe the calculation of the associated systematic uncertainties
involved in the analysis (\ref{sub:systematic}).

\subsection{Data Sets and Event Selection}\label{sub:dataset}
The data for $p+p$ collisions at $\sqrt{s}$~=~200~GeV were collected
during the polarized $p+p$ run in 2005.  For this analysis two data
sets were used: a reference sample of events selected with a minimum
bias interaction trigger (Min. Bias) and a data set selected with a single
electron trigger (ERT: EMCal and RICH Trigger).  The Min. Bias trigger 
for $p+p$ requires at least one hit in both the North and South BBC 
detectors in coincidence with the beam bunch crossing and a $z$-vertex 
position (determined online by the BBCs) within 38~cm:
\begin{eqnarray}
{\rm Min.~Bias}{\equiv}({\rm BBCN}{\ge1})\times({\rm BBCS}{\ge}1){\times}(|z|<38~cm).
\end{eqnarray}
The Min. Bias trigger cross section is measured to be $\sigma_{\rm
BBC}=23.0\pm$2.2~mb or~54.5~$\pm$~5\% of the total inelastic $p+p$
cross section at this center of mass energy
$\sigma^{p+p}_{\rm inel}=42\pm3$~mb~\cite{ppg065}.
Data collected without requiring the BBC trigger show that the BBC
fires on 79\% of events with particles in the central arm acceptance.
We assume that regardless of the electron $p_{\rm T}$ or electron
pair $p_{\rm T}$ and invariant mass, the BBC always fired with the same
probability of 79\%.  Therefore, in the $p+p$ data the yield is divided
by 0.79/0.545 to account for the fraction of tracks (0.79) and
inelastic $p+p$ collisions (0.545) missed by the Min. Bias trigger.

The ERT trigger requires a minimum energy deposit of 0.4~GeV in a
tile of 2$\times$2 EMCal towers matched to a hit in the RICH, in coincidence
with the Min. Bias trigger.  In the active area the ERT trigger has a very
high efficiency for electrons; it reaches approximately 50\% at $\sim$
0.5~GeV/$c$ and saturates at $\sim$ 1~GeV/$c$ close to 100\% (for the
EMCal) and close to 90\% (for the RICH).

After applying a $z$-vertex cut $|z|<25$~cm, and
discarding any run with unusual beam or detector
conditions, the total integrated luminosities were 65.6 nb$^{-1}$ and
2.49 pb$^{-1}$ for the Min. Bias and ERT data sets, respectively.

The data for Au~+~Au collisions at $\sqrt{s_{NN}}$~=~200~GeV were
collected during the run in 2004.  
Collisions were triggered using beam-beam counters (BBC).  
The Min. Bias trigger requires at least 2 hits
in each of the BBCs and $|z|<38$~cm: 
\begin{eqnarray}
{\rm Min. Bias} &\equiv& (\rm{BBCN} \ge 2) \times (\rm{BBCS} \ge 2) \times (|z|<38~cm).
\end{eqnarray}
The offline Min. Bias trigger also requires one hit in one of the ZDCs.  The
same $z$-vertex cut $|z|<25$~cm as in the $p+p$ data is
applied offline.  This corresponds to $92^{+2.5}_{-3.0}$\% of the
Au~+~Au inelastic cross section.

The centrality is determined for each Au~+~Au collision by the correlation in
the measurement of the BBC charge and ZDC energy~\cite{ppg001}.  Using
simulations based on a Glauber model calculation~\cite{ppg019} the
average number of participants $N_{\rm part}$ and the average number of
binary collisions $N_{\rm coll}$ associated with each centrality bin are
determined.  Table~\ref{tab:evt} summarizes the average $N_{\rm part}$ and
$N_{\rm coll}$ and the corresponding systematic uncertainties in each
centrality class used in the analysis.

We analyzed a sample of 8$\times 10^8$ minimum bias events, divided
into five centrality classes  (0-10\%, 10-20\%, 20-40\%, 40-60\%, and
60-92\%) for which the number of events is summarized in Table~\ref{tab:evt}.

\begin{table*}[t]
\caption{$N_{\rm part}$, $N_{\rm coll}$ for
Au~+~Au collisions at $\sqrt{s_{NN}}$~=~200~GeV with the corresponding uncertainties derived from a Glauber
calculation~\cite{ppg019} and the number of events and pairs for each
centrality class.  Note that the uncertainties are correlated~\cite{ppg002}.
\label{tab:evt}}
\begin{ruledtabular}   \begin{tabular}{ccccccc}
&   Centrality class    & $\langle N_{part} \rangle$ (syst)
& $\langle$  \ncoll $\rangle$ (syst) & $N_{\rm events}$ & Signal Pairs & \\
\\
\hline
&  ~0-10 \%   & 325.2 (3.3)    & 955.4 (93.6)      &  $8.6\times 10^7$  & $9.2\times 10^4$  &  \\
&   10-20 \%   & 234.6 (4.7)    & 602.6 (59.3)      &  $8.6\times 10^7$  & $6.6\times 10^4$  &  \\
&   20-40 \%   & 140.4 (4.9)    & 296.8 (31.1)      
&  $1.7\times 10^8$  & $8.1\times 10^4$\\
&   40-60 \%   & 59.95 (3.6)    & 90.70 (11.8)      &  $1.7\times 10^8$  & $3.3\times 10^4$  &  \\
&   60-92 \%   & 14.50 (2.5)    & 14.50 (4.00)      &  $2.9\times 10^8$  & $1.1\times 10^4$  &  \\
&  ~0-92 \%   & 109.1 (4.1)    & 257.8 (25.4)      &  $8.1\times 10^8$  & $28.3\times 10^4$ &  \\
&  ~$p+p$ (Min. Bias)  & 2              & 1               &  $1.5\times 10^9$  &  $1.4\times 10^4$  &  \\
&  ~$p+p$ (ERT) & 2              & 1               &  $2.7\times 10^8$  & $22.8 \times 10^4$ & \\
\end{tabular} \end{ruledtabular}
\end{table*}

\subsection{Track Reconstruction}\label{sub:tracking}
The PHENIX tracking system reconstructs charged particles with
momentum above 0.2~GeV/$c$ with a momentum resolution of
$\sigma_{p_T}/p_T=0.7\%\oplus 1\% \cdot p_T$ for $p_T$ in~GeV/$c$.  A
track is reconstructed by 2 sets of at least 4 hits in the X1 and X2
plane separated by 20~cm in radial direction, i.e.  in the main bend
plane of the central magnet, using a Hough transform performed over
all possible hit combinations.  The UV1 and UV2 wires provide up to 6
measurements in the $z$ direction, which are associated with the
three-dimensional space point provided by PC1.  After the pattern
recognition and track reconstruction by the Hough transform technique,
the initial momentum vector of the track at the $z$-vertex is
calculated.  Each reconstructed track is then associated with hit
information from the outer detectors (PC2, PC3, RICH, and EMCal).

The transverse momentum ($p_T$) is determined by measuring the angle
$\alpha$ between the reconstructed particle trajectory and a line that
connects the $z$-vertex point to the particle trajectory at a
reference radius $R$=220~cm.  The angle $\alpha$ is approximately
proportional to charge/$p_T$.  Note that this procedure assumes tracks
originate from the vertex.  As a result, tracks which originate off
vertex are reconstructed with an incorrect momentum.
Conversion pairs are reconstructed with
invariant mass $m_{ee}>0$ and contaminate the spectrum up to
$m_{ee}\sim$0.3~GeV/$c^2$ (see Section~\ref{subsub:photons}).

Because charged particles are deflected in 
the azimuthal direction by the magnetic field, the single-track acceptance depends on the 
momentum and charge of the particle, and also on the radial location of 
the detector component (DC and RICH).  The acceptance for a track with 
charge $q$, transverse momentum $p_T$ and azimuthal emission angle 
$\phi$ can be described by the logical AND of these conditions:
\begin{eqnarray} \label{eq:acc}
\phi_{\rm min} \leq \phi+ q \frac{k_{\rm DC} }{p_T} \leq
\phi_{\rm max} \nonumber \\
\phi_{\rm min} \leq \phi+ q \frac{k_{\rm RICH}}{p_T} \leq \phi_{\rm max}
\end{eqnarray}
where $k_{\rm DC}$ and $k_{\rm RICH}$ represent the effective azimuthal bend 
to DC and RICH ($k_{\rm DC}=0.206$ rad~$\cdot$~GeV/$c$ and $k_{\rm RICH}=0.309$ rad~$\cdot$~GeV/$c$).  One 
arm covers the region from $\phi_{\rm min}=\frac{-3}{16}\pi$ to $\phi_{\rm 
max}=\frac{5}{16}\pi$, the other arm from $\phi_{\rm min}=\frac{11}{16}\pi$ to 
$\phi_{\rm max}=\frac{19}{16}\pi$.  

Results in Sec.~\ref{sec:results} will show the dilepton invariant
mass spectrum ``in the PHENIX acceptance,'' where the data will be
compared to the expectations filtered according to this simple
parameterization of the acceptance.

\subsection{Electron Identification}\label{sub:eID}
Electrons in the range 0.2 $<p_T<$ 20~GeV/$c$ are identified by hits
in the Ring Imaging \v{C}erenkov detector (RICH) and by matching the
momentum with the energy measured in the electromagnetic calorimeter
(EMCal)~\cite{ppg066}.  Specifically we consider the following
variables for electron identification (eID), as summarized in Table
~\ref{tab:RGeid}.

\begin{description}
\item[\textsf{track quality}]
A bit pattern representing the reconstruction quality of the track.  If the
track is reconstructed by both of the X1 and X2 sections of the DC and
is uniquely associated with hits in U or V stereo wires, the value of
{\tt quality} is 63 (in case a unique PC1 hit is found too) or 31 (in case
the PC1 hit is found but ambiguous).  If there are no UV hits found,
but a PC1 hit is, {\tt quality} is 51.

\item[\textsf{EMCal match ($\sigma_{\Delta\phi}$)}]
Displacement in $\phi$ between the position of the associated EMCal
cluster and the projection of the track onto the EMCal.  The quantity
is measured in units of momentum-dependent resolution.  For
example, $\Delta\phi < 2$ means that the position of the associated
EMCal cluster in $\phi$ is within 2$\sigma$ of the projected
track position.  The particle hit position of an EMCal cluster is
particle-species dependent due to different shower shapes.  Here the
parameterization has been optimized for electrons.

\item[\textsf{EMCal match ($\sigma_{\Delta z}$)}]
Analogous to the previous variable, for the $z$ coordinate.

\item[\textsf{n0}]
Number of hit RICH PMTs in an annular region with an inner radius of
3.4~cm and outer radius of 8.4~cm around the track projection on the
RICH.  The expected radius of a \v{C}erenkov ring emitted by an electron is
5.9~cm.

\item[\textsf{chi2/npe0}]
A $\chi^2$-like shape variable of the RICH ring associated with the
track divided by the number of photo-electrons
measured in a given ring (\textsf{npe0}).

\item[\textsf{RICH match}]
The displacement of the RICH ring center
from the projected track position.  Units are cm.

\item[\textsf{E/p} or \textsf{dep}]
A variable quantifying  energy-momentum matching.  This variable is calculated as
\textsf{dep} $= (E/p - 1)/\sigma_{E/p}$, where $E$ is the energy
measured by EMCal; $p$ is the momentum of the track; and $\sigma_{E/p}$ is the
standard deviation of the Gaussian-like $E/p$ distribution calculated
for electrons.

\end{description}

\begin{table}
\caption{Electron ID cuts used in the Au~+~Au and $p+p$ analyses.}
\label{tab:RGeid}
\begin{ruledtabular} \begin{tabular}{ccccc}
& eID cuts & for Au~+~Au & for $p+p$ & \\
\hline
& \textsf{track quality} = & 63 $||$ 31 $||$ 51 & 63 $||$ 31 $||$ 51 & \\
& $\sqrt{\sigma_{\Delta\phi}^2 + \sigma_{\Delta z}^2} <$ & 3.0 & 5.0 & \\
& \textsf{n0} $\geq$ &  2 & 1 & \\
& \textsf{chi2/npe0} $<$ & 10.0 & 15.0 & \\
& \textsf{RICH match} $<$ & 5.0 & 10 & \\
& \textsf{dep} $>$ & -2.0 & - & \\
& $\frac{E}{p} >$ & - & 0.5 & \\
\end{tabular} \end{ruledtabular} 
\end{table}

\begin{figure}[!ht]
\includegraphics[width=1.0\linewidth]{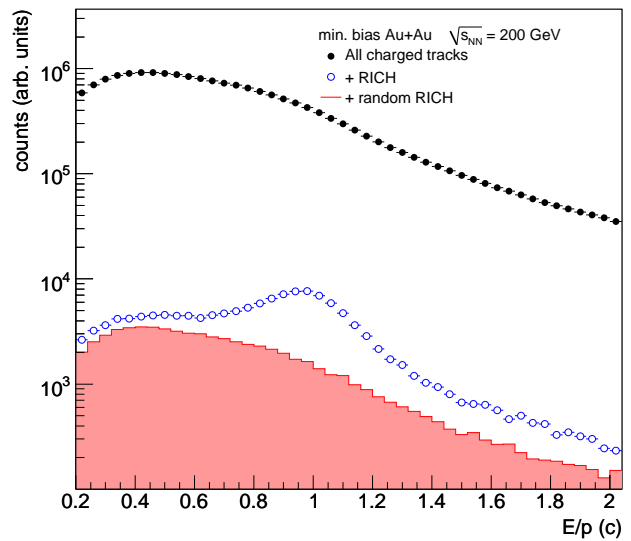}
\caption{\label{fig:eop} (Color online) 
$E/p$ distribution in minimum bias Au~+~Au
for all charged tracks and for tracks after applying all the RICH
cuts in Table~\ref{tab:RGeid} except the \textsf{E/p} or \textsf{dep}.  The
contribution from randomly associated hadrons is shown by the filled
histogram.}
\end{figure}
Figure~\ref{fig:eop} shows\footnote{Data tables for this and other
data plots are available at
https://www.phenix.bnl.gov/WWW/p/info/ppg/088/datatables.}  the $E/p$
distribution for all charged tracks and for electron candidates,
i.e.  tracks which fulfill all the RICH eID cuts except the
\textsf{E/p} or \textsf{dep}.  While the distribution of all charged
tracks shows no clear electron peak, requiring the eID cuts greatly
improves the signal-to-background ratio.  However, there still remains
some background underneath the peak even below $p_T < 4.9$~GeV/$c$,
the \v{C}erenkov threshold for pions.  This background, due to random
coincidences between hadron tracks and hits in the RICH, is estimated
by swapping the north and south sides of the RICH in software, and
reconstructing the track matching to the RICH once again.  This
contamination ranges from~2\% in $p+p$ to 30\% in the most central
Au~+~Au collisions.  It is $\sim$24\% for Min. Bias Au~+~Au collisions.  This
pion contamination in Au~+~Au contributes to the
large combinatorial background (see Sec.~\ref{subsub:combinatorial}).

\subsection{Backgrounds}\label{sub:pair}
The source of any particular electron or positron in an event is
unknown; therefore, all electrons and positrons are combined into
a foreground of pairs, like-sign $N_{++},N_{--}$ and unlike-sign $N_{+-}$.  
This results in a large combinatorial background which must be removed.  
In the following we will use the notation for the foreground 
$N_{\pm\pm}=N_{++}+N_{--}$ and for the background 
$B_{\pm\pm}=B_{++}+B_{--}$.  The analysis steps to achieve this 
are outlined here and presented in detail in the subsections below.  
We can distinguish our background sources in two types:

Type I background (\ref{subsub:type1bg}) consists of two classes of
fake combinations that can be identified on a pair-by-pair basis:
\begin{itemize}
\item 
Overlapping pairs are fake electron pairs that arise from overlapping
hits in the detectors, mostly in the RICH.

\item 
Photon conversions are fully reconstructed pairs originating from
photon conversions in the detector material are removed by a cut on
the orientation of the pairs in the magnetic field.
\end{itemize}

Type II background (\ref{subsub:type2bg}) consists of all those pairs
that cannot be identified on a pair-by-pair basis and are therefore
removed statistically:
\begin{itemize}
\item
Combinatorial background $B^{\rm comb}$ arises from all the combinations
where the origin of the two electrons is totally uncorrelated.  

\item
Correlated background $B^{\rm corr}$ occurs if there are two
$e^+e^-$ pairs in the final state of a meson or when two hadrons,
either within the same jet or in back-to-back jets, decay into
electron pairs.  
\end{itemize}

Since accurate background subtraction is essential for this analysis,
we have developed two independent methods to subtract the type II
background.  In the first method we calculate the shapes of
combinatorial and correlated background with event mixing or
simulations and use the yield of the like-sign spectra for the
normalization.  In the second method (\ref{subsub:likesign}) we use
measured the real like-sign distributions, corrected for the
acceptance difference, without making any assumption about the mass
dependence nor about the decomposition of the background into correlated and
uncorrelated component.

\subsubsection{Type I Background} \label{subsub:type1bg}

\begin{figure}[!ht]
\includegraphics[width=1.0\linewidth]{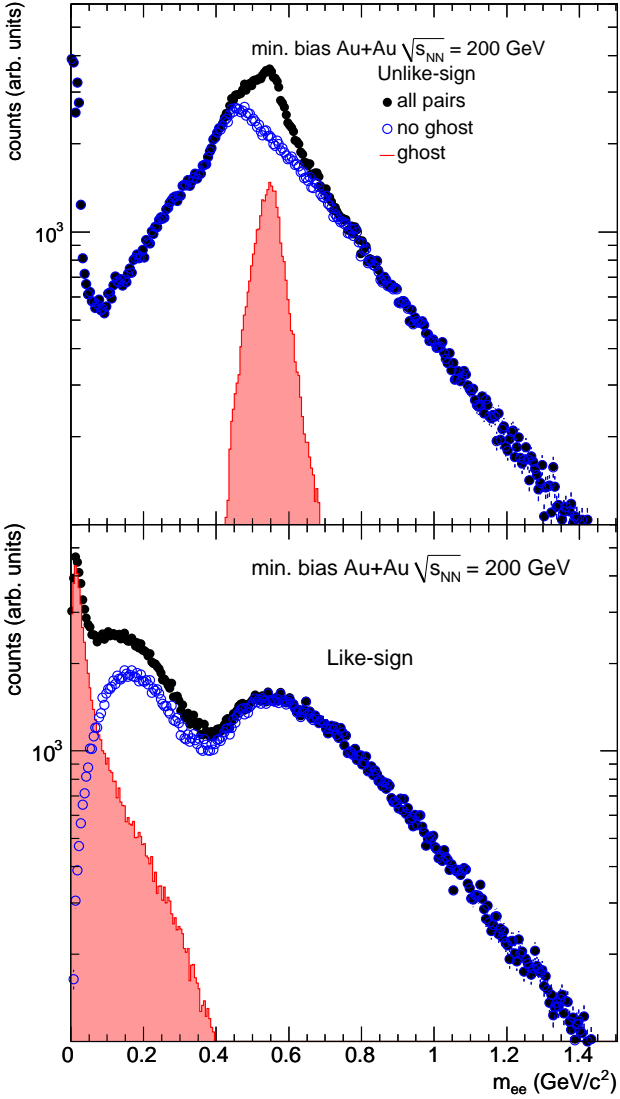}
\caption{\label{fig:ghosts} (Color online) 
Invariant mass distribution for all like-
and unlike-sign pairs.  Overlapping pairs are shown
separately.  Also shown are the distributions after the overlapping pairs are removed.}
\end{figure}

{\bf Overlapping Pairs} \label{subsub:ghosts}
Fake electron pairs can be created if two particles are in close
proximity in any of the detectors.  These correlations within an event
are particularly noticeable when two tracks share the same RICH
ring.  This issue can be illustrated with a simplified model of the
RICH based on spherical mirror optics.  In this case, tracks that are
parallel to each other while passing through the RICH radiator,
i.e.  after they have been bent in the magnetic field, share the same
search region for \v{C}erenkov light in the RICH PMT plane.  Therefore
an overlapping pair is created whenever a track after the field is parallel
to a true electron.  These overlapping pairs typically have a small opening
angle and are therefore reconstructed with small invariant mass.
Because like-sign pairs are bent in the same direction, in contrast to
unlike-sign pairs which are bent in opposite directions, like-sign
overlapping pairs have smaller mass than unlike-sign overlapping pairs.
Figure~\ref{fig:ghosts} shows the invariant mass distribution for all
like- and unlike-sign pairs, the distributions for overlapping pairs, and the
distributions after removing overlapping pairs.

Overlapping pairs are eliminated by applying a cut on the physical
proximity of every pair projected to every detector.  The cut value on every
detector is determined by the corresponding double hit resolution.
In the RICH the cut is applied at 36 cm, which is roughly twice the
predicted maximum diameter ($\sim$~16.8~cm) of a RICH ring.  In the
Pad Chambers the cut is applied at $\Delta z \le 0.5$ cm and
$\Delta\phi\le$ 20 mrad.  In the EMCal the cut is applied to a 3$\times$3
tower region around the hit.  These cuts remove a fraction of real
pairs which varies from 4\% in the most central to 2\% in the most
peripheral collisions, estimated using mixed events.  These cuts
remove 10\% more real like-sign than unlike-sign pairs.  The ratio of
like- to unlike-sign pairs lost was determined with an uncertainty of
50\% by varying the cut values chosen.  While the efficiency of the
pair cut depends on the centrality, the ratio of like- to unlike-sign
pairs lost was found to be independent of centrality.

Because these overlapping pairs are rare, whenever we encounter one, we
remove the entire event.  
This results in a loss of $\sim$ 0.08\% of all events.
In mixed events the same cuts are applied and whenever an event is
discarded due a pair cut, another event is generated with the same
electron multiplicity and in the same centrality, $z$-vertex and reaction
plane class.

{\bf Photon Conversions} \label{subsub:photons}
Since the tracking algorithm assumes that all tracks originate from the
collision vertex, pairs from photons that convert off-vertex are
reconstructed with an artificial opening angle, which leads to an
invariant mass that increases with the radius at which the conversion
occurs.

Conversion pairs have no intrinsic opening angle (i.e.  their opening
angle is exactly zero at the point of creation), they are bent only
in the azimuthal direction by the magnetic field, which is parallel to
the beam axis $\vec{z}$.  We can define unit vectors $\hat{u}$ in the 
direction of the pair momentum and $\hat{v}$ perpendicular to the plane 
defined by the pair
\begin{eqnarray}
\hat{u} &=& \frac{\vec{p}_+ + \vec{p}_-}{|\vec{p}_+ + \vec{p}_-|}
\\
\hat{v} &=& \hat{p}_+ \times \hat{p}_- 
\end{eqnarray}
where 
$\hat{p}_{\pm} = \vec{p}_{\pm} / |\vec{p}_{\pm}|$.
We can define the orientation of the actual opening angle as
\begin{eqnarray}
\hat{w} &=& \hat{u} \times \hat{v} 
\end{eqnarray}
We can also define the expected orientation of the opening angle for
conversion pairs
\begin{eqnarray}
\hat{w_c} &=& \hat{u} \times \hat{z} 
\end{eqnarray}
Finally we can define 
$\phi_V$ as the angle between these two vectors
\begin{eqnarray}
\cos \phi_V = \hat{w} \cdot \hat{w_c} 
\end{eqnarray}
For pairs originating from photon conversions $\phi_V$ is zero.  (By
consistently ordering positive and negative tracks within the pair we
avoid $\phi_V = \pi$ as a solution for photon conversions.)  In
contrast, $e^+e^-$ pairs from hadron decays, as well as combinatorial
pairs, have no preferred orientation.
\begin{figure}[t]
\includegraphics[width=1.0\linewidth]{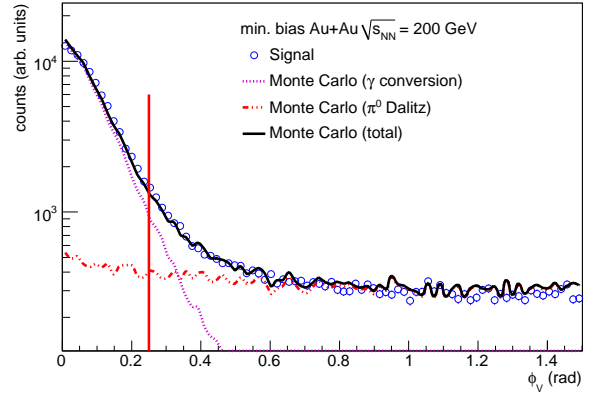} 
\caption{\label{fig:phiV}  (Color online) 
Comparison of the $\phi_V$ angle distributions in
Monte Carlo and data in the mass range 10 $<m_{ee}<$ 30
MeV/c$^2$.  The vertical red line indicates the cut value used to
eliminate photon conversions in this mass bin.}
\end{figure}
Figure~\ref{fig:phiV} presents a comparison of the $\phi_V$ angle
distributions of real data and Monte Carlo simulated data in a mass
bin, 10 $<m_{ee}<$ 30 MeV/c$^2$, dominated by photon conversion in the
beam-pipe.  In the simulations we can distinguish which $e^+e^-$ pairs
originate from a $\pi^0$ Dalitz decay (dotted-dashed line) and which
originates from photon conversion (dotted line).  The simulations show
that the distribution for all unlike-sign pairs originating from
photon conversion is strongly peaked at $\phi_V=0$.  In contrast,
simulated pairs originating from $\pi^0$ Dalitz decays have no
strongly preferred orientation.  The sum of the Monte Carlo data can
be compared to the signal, which in this mass region contains
conversion photons, as well as $\pi^0$'s.  Figure~\ref{fig:phiV} shows
that the agreement between the signal and the simulation is good.
The width of the $\phi_V$ peak for photon conversions increases with
the path length of the $e^+e^-$ pair in the magnetic field, where
residual field in the polar direction as well as multiple scattering cause
the pair to lose its perfect alignment perpendicular to the $\vec{z}$
axis.

\begin{figure}[t]
\includegraphics[width=1.0\linewidth]{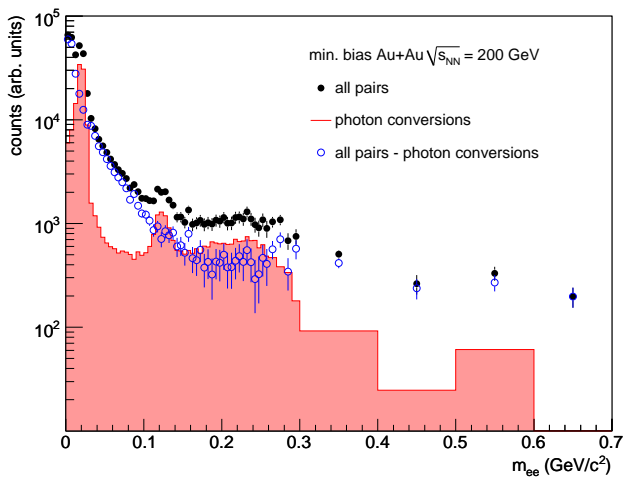}
\caption{\label{fig:gamma_spectrum} (Color online) 
Invariant mass spectrum of all
the unlike-sign pairs after subtraction of combinatorial pairs in
Min. Bias Au~+~Au collisions at $\sqrt{s_{NN}}$~=~200~GeV.  The
filled histogram shows the pairs removed by the $\phi_V$ angle cut.}
\end{figure}

The contribution from conversion pairs as a function of the
(mis-reconstructed) invariant mass is shown in
Figure~\ref{fig:gamma_spectrum} (filled histogram).  As the
(mis-reconstructed) mass is essentially proportional to the radius
where the conversion happens, the mass spectrum of those pairs allows
a ``tomography'' of the material in the spectrometer.  The peaks
correspond to the conversions in the beam-pipe material (r~=~4~cm, or
$m_{ee}$~=~20~MeV/$c^2$) and detector support structures (r~=~25~cm,
or $m_{ee}$~=~125~MeV/$c^2$).  Conversions in the He bag generates
pairs with $m_{ee} \lesssim 0.3$~GeV/$c^2$.  For this value of mass the
corresponding radius would be the entrance window of the DC.  At this
point though the electrons do not bend anymore because the region is
field-free, and are therefore removed with a $p_T$ cut
($p_T<$20~GeV/$c$) on the single electrons.
The $\phi_V$ resolution improves for increasing conversion radius
because electrons are less modified from their original
direction by multiple scattering or the residual polar field.
The cut on $\phi_V$
(indicated in Fig.~\ref{fig:phiV} by the line at $\phi_V>0.25$ for
$m_{ee}<30$~MeV/$c^2$) is reduced for larger masses due to the 
improved resolution of $\phi_V$ at larger radii.
By varying the cut values on real pairs we estimate that our $\phi_V$ cut
removes more than 98\% of the conversion pairs with a mass-dependent
efficiency of more than 90\%.  The uncertainty of 
on the final $e^+e^-$ pair signal is 6\%.

\subsubsection{Type II Background} \label{subsub:type2bg}
After removing type I background, the unlike-sign foreground spectrum
$N_{+-}$ measures the physical signal plus background, while the
like-sign spectra $N_{++}, N_{--}$ measure only background.  We have
developed two methods to measure and subtract the unlike-sign
background

\begin{itemize}
\item
One solution is to use a mixed-event technique, which combines tracks
from different events.  With this method the background has much larger
statistics than the foreground.  The accuracy in the determination of
the shape of the background is tested by comparing the like-sign
distribution in real and mixed events.  We find good
agreement between real and mixed-events like-sign spectra in some
regions of the ($m_{ee}, p_T$) plane, while in others they clearly
deviate.  This indicates that not all the type II background is of
uncorrelated origin, but there are also some correlated pairs in the
background (\ref{subsub:correlated}) that one needs to
separately account for.

\item
Another solution is to still use the measured like-sign spectra and
correct them for the different acceptance (\ref{subsub:likesign}).
This solution does not require any assumption on the decomposition of
combinatorial and correlated since the like-sign spectra measures all
the backgrounds simultaneously.  In experiments with equal acceptance
for electrons and positrons, the background can be measured directly
through the geometric mean of the like-sign pair distributions
$2\sqrt{N_{++}N_{--}}$.  With this method the background has similar
statistics as the foreground.  In PHENIX however the mass and $p_T$
dependence of the background is different for the two charge
combinations (see Equation~\ref{eq:acc} and Fig.~\ref{fig:ghosts}).
\end{itemize}

We have used both these methods, which are
compared in Section~\ref{subsub:sys_bg}, to assign a systematic
uncertainty to the background subtraction.  The first method is used
for the final analysis.

\begin{figure}[!ht]
\includegraphics[width=1.0\linewidth]{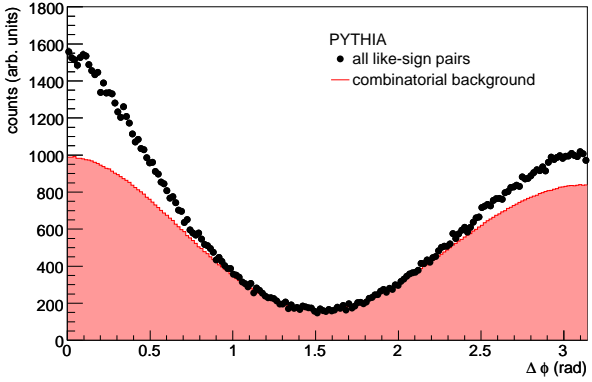}
\caption{\label{fig:delta_phi} (Color online) 
$\Delta\phi$ distribution of like-sign
pairs for real and mixed events calculated 
by {\sc pythia}~\protect\cite{pythiamb}.}
\end{figure}

The structure of the type II background was studied using minimum bias 
events generated with {\sc pythia}~\cite{pythiamb} with the branching ratio
of the $\pi^0$ Dalitz decay set to 100\% to enhance the sample of
$e^+e^-$ pairs per event.  All the electrons are filtered through the
PHENIX acceptance.  From {\sc pythia} events we make real and mixed-event
like- and unlike-sign distributions as in the real data.  Now we
analyze like-sign pairs, that contain only background.  The
$\Delta\phi$ distribution shown in Fig.~\ref{fig:delta_phi} compares
the difference of azimuthal emission angle of the two electrons (or
positrons) in real and mixed {\sc pythia} events.  If the two shapes were
identical, we would conclude that all the background come
from uncorrelated sources.  However the shapes clearly deviate at
$\Delta\phi\sim0$ and $\Delta\phi\sim\pi$.  This indicates the presence
of combinations that arise from the \emph{same} jet ($\Delta\phi\sim0$) or
\emph{back-to-back} ($\Delta\phi\sim\pi$) jets (see Section
~\ref{subsub:correlated}).  
In addition, correlations can occur if there
are two $e^+e^-$ pairs in the final state of a meson, e.g.  double
Dalitz decays, Dalitz decays followed by a conversion of the decay
photon or two-photon decays followed by conversion of both photons.

Therefore, we decompose the type II background into two components: a
combinatorial background made of uncorrelated pairs and a background
of correlated pairs.  The distributions for combinatorial and
correlated background are determined with methods that are explained
below (\ref{subsub:combinatorial}) for like and unlike-sign
spectra.  The like-sign background distributions are fit to the
measured like-sign spectra (that contain only background).  The same
fit parameters are then applied to the unlike-sign background.  The
normalized unlike-sign background is finally subtracted from the
foreground of all pairs $N_{+-}$ to obtain the signal.

{\bf Combinatorial Background} \label{subsub:combinatorial}
The \emph{combinatorial background} $B^{\rm comb}$ is determined with a
mixed-event technique, which combines tracks from different events
with similar centrality, $z$-vertex and reaction plane.  In the $p+p$
data, where we use a triggered data set, the mixed-event pairs are
constructed from the Min. Bias data set requiring that at least one of the
two partners has fulfilled the ERT trigger condition.  Since the
tracks are from different events, this technique reproduces the
uncorrelated background by definition.  
This technique also allows computation of background spectra with
negligible statistical errors.

We compare like-sign spectra in real and mixed-event data to locate
a region in the ($m_{ee}, p_T$) plane where their
shapes agree.  In this region we normalize the combinatorial background spectra
$B_{\pm\pm}^{\rm comb}$ to the measured like-sign pairs $N_{\pm\pm}$.
We define:
\begin{eqnarray} \label{eq:normalization}
A_{+} = \frac {\int_{N.R.} N_{++}(m_{ee},p_T)} {\int_{N.R.} B_{++}(m_{ee},p_T)} dm_{ee}dp_T\nonumber \\
A_{-} = \frac {\int_{N.R.} N_{--}(m_{ee},p_T)} {\int_{N.R.} B_{--}(m_{ee},p_T)}  dm_{ee}dp_T
\end{eqnarray}
where $N.R.$ is the chosen normalization region.
Then we calculate the integral of the normalized like-sign background
over the full phase-space:
\begin{eqnarray} \label{eq:normalization2}
{\cal B}_{++} = 
\int_{0}^{\infty} A_{+} \cdot B_{++}(m_{ee},p_T)  dm_{ee}dp_T
\nonumber \\
{\cal B}_{--} = 
\int_{0}^{\infty} A_{-} \cdot B_{--}(m_{ee},p_T)  dm_{ee}dp_T
\end{eqnarray}
The unlike-sign background is then normalized such that its yield
equals the geometric mean of the like-sign pairs
$2\sqrt{\cal{B}_{++}\cal{B}_{--}}$:
\begin{eqnarray} \label{eq:normalization3}
B_{+-} (m_{ee},p_T) = \frac {2\sqrt{\cal{B}_{++}\cal{B}_{--}}} {\int_{0}^{\infty} B_{+-}(m_{ee},p_T) dm_{ee}dp_T}
B_{+-} (m_{ee},p_T)
\end{eqnarray}
Appendix~\ref{app:2sqrt} shows that as long as electrons and
positrons are produced in pairs the absolute normalization of the
unlike-sign background is given by the geometric mean of the observed
positive and negative like-sign pairs $2\sqrt{N_{++}N_{--}}$, without
any further assumption about efficiencies, acceptances or probability
distribution functions for the pair.  Using $\cal{B}_{++}$ and
$\cal{B}_{--}$ instead of directly taking $N_{++}$ and $N_{--}$ simply
avoids counting correlated pairs, which will be measured and normalized
separately (\ref{subsub:correlated}).  

The systematic uncertainty of the normalization is therefore
determined by the statistical accuracy of the measured like-sign yield
in the region chosen for the normalization.  

{\bf Combinatorial Background in $p+p$ data} \\
In our {\sc pythia}~\cite{pythiamb} studies we found 
(Fig.~\ref{fig:delta_phi}) that at
$\Delta\phi\sim 0$ and $\Delta\phi\sim\pi$ the real events background
deviates from the shape of uncorrelated sources, while at
$\Delta\phi\sim\pi/2$ it looks consistent with the shape from mixed
events.  Considering that $m_{ee}^2 = 2 p_1 p_2 (1-\cos
\Delta\omega)$, with $\Delta\omega$ being the opening angle, this
condition corresponds to a region in the ($m_{ee}, p_T$) plane where
$m_{ee} \sim p_T$.

\begin{figure}[!ht]
\includegraphics[width=1.0\linewidth]{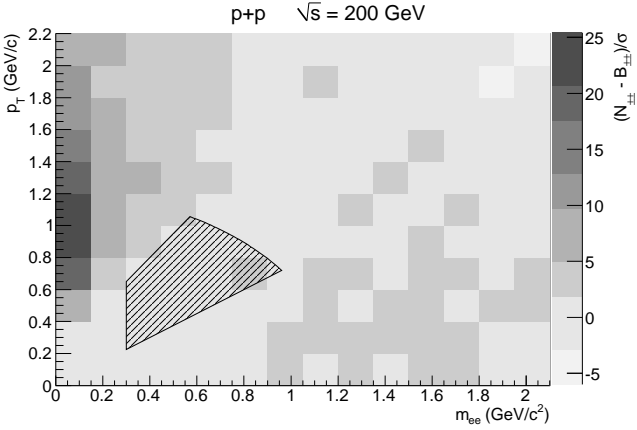}
\caption{\label{fig:fgbg_like_pp} Difference between real and
mixed-events like-sign distributions divided by its standard deviation
$(N_{\pm\pm} - B_{\pm\pm}^{\rm comb}) / \sigma_{(N_{\pm\pm} -
B_{\pm\pm}^{\rm comb})}$.  The background $B_{\pm\pm}^{\rm comb}$ is
normalized to the foreground $N_{\pm\pm}$ in the normalization region
shown by the dashed area.  }
\end{figure}

We define this region empirically by a set of equations:
\begin{eqnarray} \label{eq:norm_region}
m_{ee}>0.3 \rm{~GeV}/c^2 \nonumber \\
m_T<1.2 \rm{~GeV}/c^2 \nonumber \\
p_T/c-1.5 m_{ee} \leq 0.2 \rm{~GeV}/c^2 \nonumber \\ 
p_T/c-0.75 m_{ee} \geq 0 \rm{~GeV}/c^2
\end{eqnarray}
where $m_T = \sqrt{p^2_T + m^2_{ee}}$ is the transverse mass of the
pair.  This region is shown in Fig.~\ref{fig:fgbg_like_pp} by the
dashed area.

Figure~\ref{fig:fgbg_like_pp} shows the difference between the
like-sign distributions in real and mixed-events, as a function of
$m_{ee}$ and $p_T$.  The background $B_{\pm\pm}^{\rm comb}$ is
normalized to the foreground $N_{\pm\pm}$ in the normalization region
(from Equation~\ref{eq:norm_region}).  The absolute normalization of the
unlike-sign combinatorial background is determined with an uncertainty
given by the statistical error of the measured like-sign spectra in
this region of 3\%.  The difference between real and mixed $N_{\pm\pm}
- B_{\pm\pm}^{\rm comb}$ is divided by its standard deviation.  
Figure~\ref{fig:fgbg_like_pp} shows that in this region the 
background does not deviate from the foreground by more than 
2~$\times~\sigma$.  The stability of the results has been checked 
by varying the normalization region, and the difference 
is included in the systematic uncertainty.

\begin{figure}[!ht]
\includegraphics[width=1.0\linewidth]{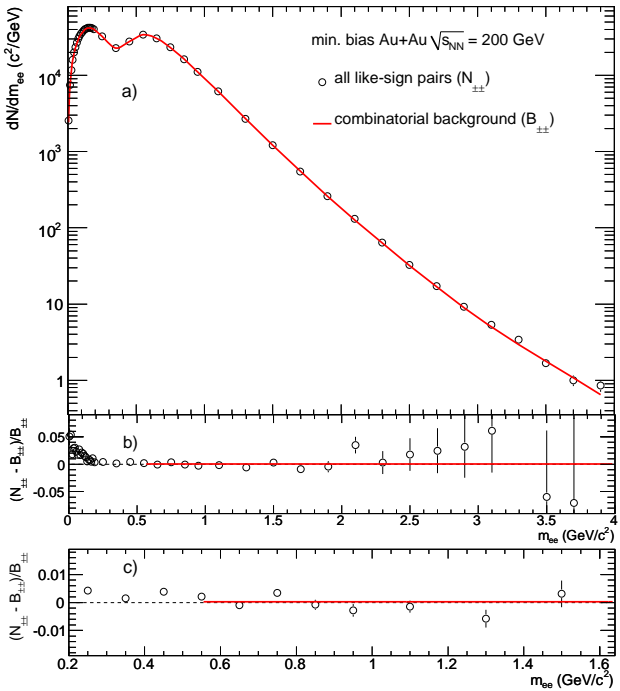}
\caption{\label{fig:like} (Color online) 
(a) like-sign distribution
for real $N_{\pm\pm}$ and mixed events $B_{\pm\pm}^{\rm comb}$.
(b) and (c) Ratio of
$(N_{\pm\pm}-B_{\pm\pm}^{\rm comb})/B_{\pm\pm}^{\rm comb}$, with two different scales.}
\end{figure}

{\bf Combinatorial Background in Au~+~Au data} \\
Figure~\ref{fig:like} shows a comparison between the like-sign
distribution from real and mixed events.
\begin{figure}[!ht]
\includegraphics[width=1.0\linewidth]{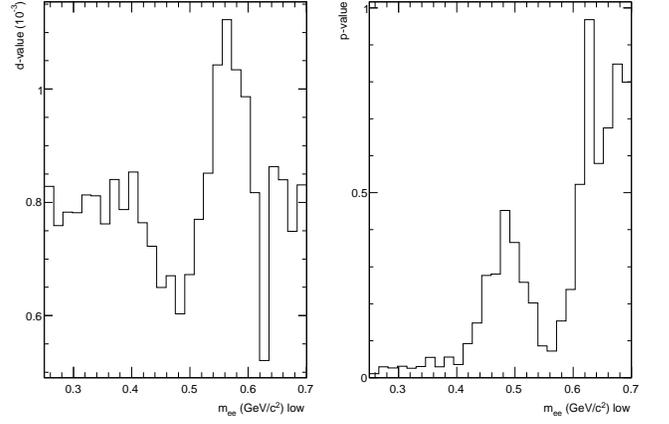}
\caption{\label{fig:like_test} 
(left) Maximum cumulative fractional
distance $d$ of the two like-sign distributions.  The distance is
calculated from a lower endpoint $m_{\rm low}$ to infinity.  The test
gives a maximum deviation of 0.1\%.  (right) Corresponding
Kolmogorov-Smirnov $p$-value as a function of the lower endpoint
$m_{\rm low}$ , i.e.  for $m_{ee} > m_{\rm low}$.  The $p$-value increases for
$m_{\rm low}>$~0.55~GeV/$c^2$, reaching values of $\sim$ 90\%, confirming
the hypothesis of compatibility of the two distributions in the
region chosen for the normalization.}
\end{figure}
The comparison shows that the mixing technique reproduces the mass
dependence within the statistical accuracy of the data not only in the
normalization region for the $p+p$ data, but also for all masses above
0.55~GeV/$c^2$.  In general the larger combinatorial background
produced in the Au~+~Au environment reduces the capability to
distinguish between different background shapes.  Also the agreement
at high mass, which was not observed in $p+p$ (see
Fig.~\ref{fig:fgbg_like_pp}), can be qualitatively explained by the
suppression of away-side jets observed in Au~+~Au~\cite{ppg083}.

\begin{table*}[tbp]
\caption{Fit parameters for the mass dependence comparison of real
and mixed-events like-sign pairs for different centrality and $p_T$
bins.  The second column reports the results of
$(N_{\pm\pm}-B_{\pm\pm}^{\rm comb})/B_{\pm\pm}^{\rm comb}$
distributions to a constant and the third is $\chi^2$ value divided
by the number of degrees of freedom.  The fourth and fifth columns
report the result of a $\chi^2$ statistical test and the
corresponding $p$-value for $m_{ee}>0.55$~GeV/$c^2$.  The last one
gives the maximum deviation of the $N_{\pm\pm}$ and $B_{\pm\pm}^{\rm
comb}$ distribution in a Kolmogorov-Smirnov
test.~\label{tab:like_cent}}
\begin{ruledtabular}  \begin{tabular}{cccccccc} 
& Centrality & $p_0$ & $\chi^2$/NDF & $\chi^2$ test & $p$-value &max dev.\\ 
\hline
& 0-10\%  &$ 6.3  \pm 8.8 \times 10^{-4}$& 30.2/19 & 1.05 & 0.25 & 0.0014  & \\ 
& 10-20\% &$ -9.4 \pm 1.4 \times 10^{-4}$& 18.6/19 & 0.97 & 0.61 & 0.0018  & \\
& 20-40\% &$ -2.4 \pm 1.8 \times 10^{-3}$& 18.7/19 & 1.02 & 0.40 & 0.0034  & \\
& 40-60\% &$ -8.5 \pm 4.9 \times 10^{-3}$& 21.9/19 & 1.65 & 0.02 & 0.0071  & \\ 
& 60-92\% &$ -1.8 \pm 1.6 \times 10^{-2}$& 21.5/14 & 1.51 & 0.04 & 0.0321  & \\ \hline
& 00-92\% &$ 2.6  \pm 6.3 \times 10^{-4}$& 27.6/19 & 0.92 & 0.83 & 0.0010  & \\ 
& $p_T<$~1~GeV/$c$ &$ 9.2 \pm  5.1 \times 10^{-4}$  & 18.9/18 & 0.95 & 0.73 & 0.0011 & \\ 
& 1$<p_T<$~2~GeV/$c$ &$ -3.4 \pm 1.6 \times 10^{-3}$& 27.9/18 & 0.91 & 0.84 & 0.0029 & \\ 
& $p_T>$~2~GeV/$c$ &$ -9.6 \pm 5.4 \times 10^{-3}$  & 15.2/18 & 0.97 & 0.63 & 0.0038 & \\ 
\end{tabular} \end{ruledtabular}  
\end{table*}

A small signal from correlated background remains at low-masses (see
Section~\ref{subsub:correlated}).  To quantitatively compare the mass
dependence of the data to the mixed events we calculate the ratio
$(N_{\pm\pm}-B_{\pm\pm}^{\rm comb})/B_{\pm\pm}^{\rm comb}$ shown in
the bottom panels of Fig.~\ref{fig:like} and fit it with a constant
above the $\eta$ mass (0.55~GeV/$c^2$).  The result is $(- 2.59 \pm
6.33)\times 10^{-4}$ with $\chi^2$/NDF = 27.6/19.  Panel (b) shows  
the entire mass range and allows to distinguish the signal from
correlated background at very low mass.  Panel (c) shows a zoom in 
the region where we fit.

Two statistical tests are performed to test the hypothesis that the
two distributions ($N_{\pm\pm}$ and $B_{\pm\pm}^{\rm comb}$) represent
identical distributions: the Pearson $\chi^{2}$ test and the
Kolmogorov-Smirnov test.

The Pearson test statistic $\chi^2$ resulting from the comparison of
real and mixed events for $m_{ee}>0.55$~GeV/$c^2$ returns a
$p$-value greater than 0.83.  

The Kolmogorov-Smirnov test is used to decide if a sample comes from a
population with a specific distribution, either comparing one
fluctuating distribution (the test) to a truth hypothesis (reference),
or querying two fluctuating distributions as to whether they have a
common truth origin.  This latter one was done in our case since we do
not have a truth reference distribution without fluctuations.  The
test is based on the maximum cumulative difference between the
distribution under test and a specified reference distribution.  The
left panel of Fig.~\ref{fig:like_test} shows the maximum cumulative
difference $d$ of the two like-sign distributions $N_{\pm\pm}$ and
$B_{\pm\pm}^{\rm comb}$ for $m_{ee} > m_{\rm low}$, where $m_{\rm low}$ is the
lower endpoint chosen between 0.25 and 0.7~GeV/$c^2$.  The
Kolmogorov-Smirnov test gives a maximum deviation of 0.1\% which is
small compared to the uncertainty of the absolute normalization of the
mixed-event background.  The corresponding Kolmogorov-Smirnov $p$-value
is also shown in the right panel of Fig.~\ref{fig:like_test} as a
function of mass.  It is small below 0.55~GeV/$c^2$, where there is
some contribution of the correlated background.  However, for
$m_{\rm low}>$~0.55~GeV/$c^2$, the $p$-value is $\sim$90\%, therefore
confirming that the hypothesis of compatibility of the two
distributions for $m_{\rm low}>$~0.55~GeV/$c^2$ is valid for any commonly
used significance level.
\begin{figure}[!ht]
\includegraphics[width=1.0\linewidth]{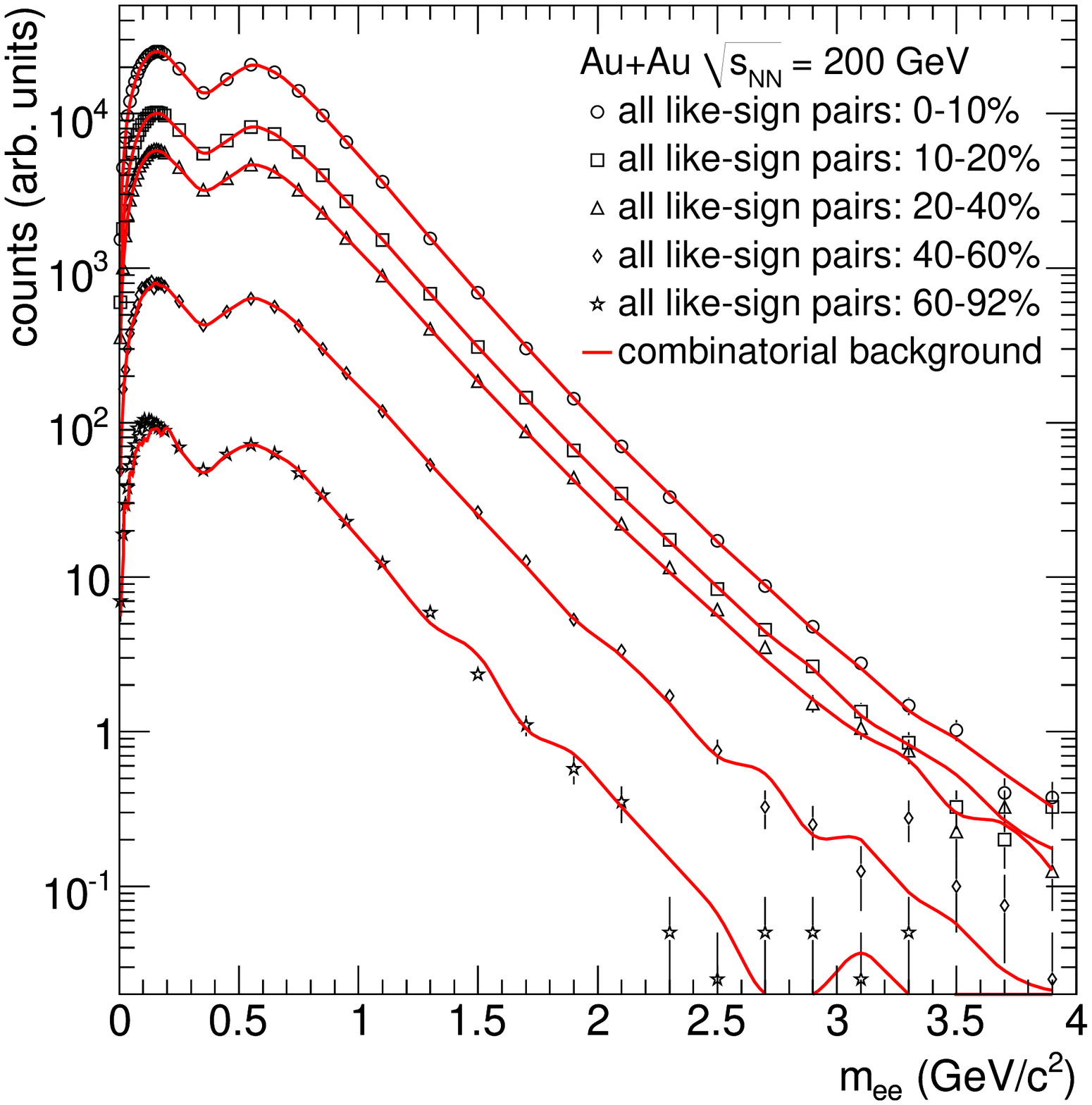}
\includegraphics[width=1.0\linewidth]{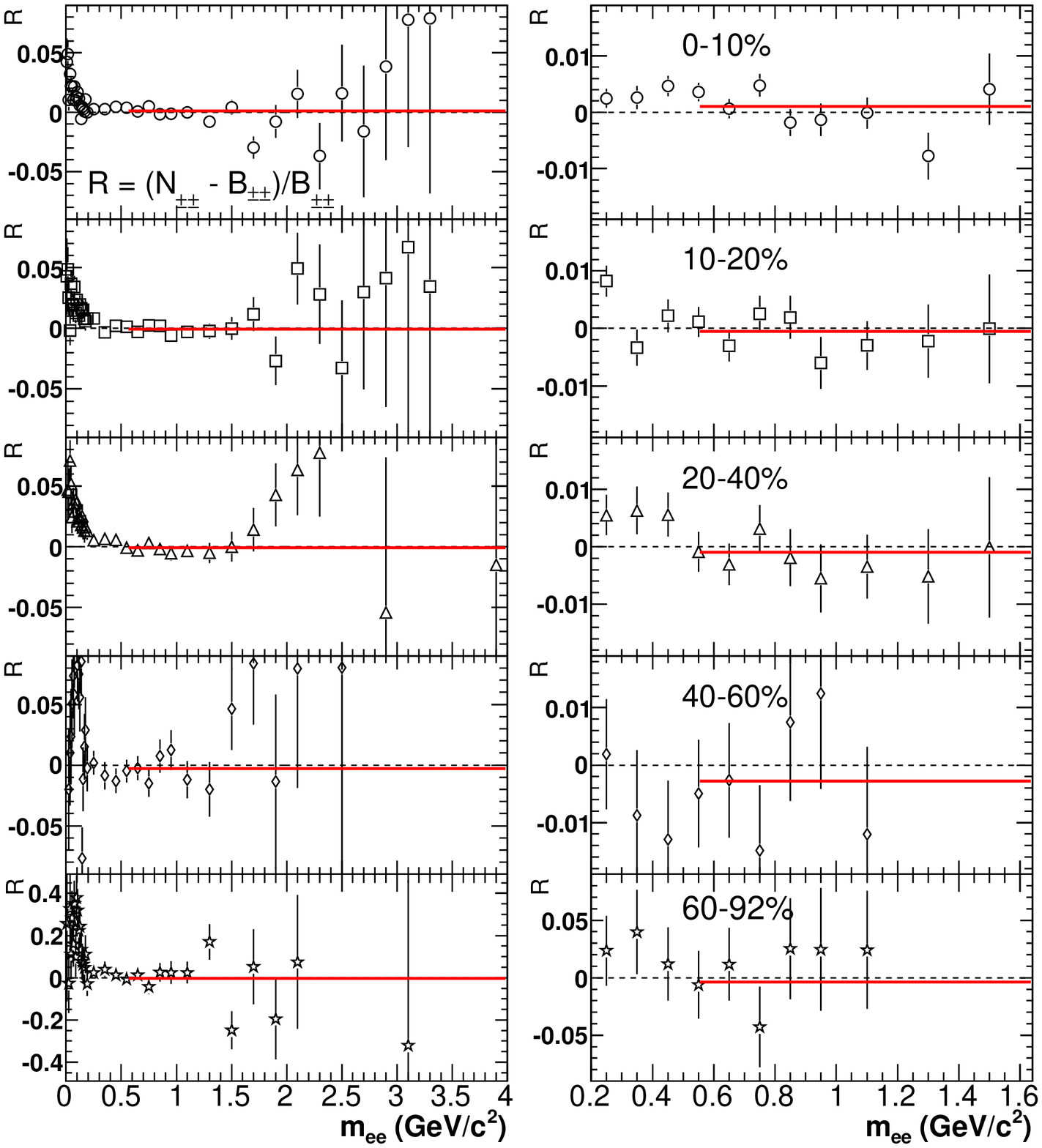} 
\caption{ \label{fig:like_cent} (Color online)  
Like-sign distribution for real $N_{\pm\pm}$
and mixed events $B_{\pm\pm}^{\rm comb}$ for different centrality
data sets.  The bottom panels show the ratios
$(N_{\pm\pm}-B_{\pm\pm}^{\rm comb})/B_{\pm\pm}^{\rm comb}$ with
different scales.  The left ones show all the mass range and allow to
identify the correlated background at low masses.  The right ones
focus on mass region where we fit.}
\end{figure}

Figure~\ref{fig:like_cent} shows the like-sign mass distribution for
real and mixed events in the different centrality bins used in the
analysis.  The bottom panels show the ratio
$(N_{\pm\pm}-B_{\pm\pm}^{\rm comb})/B_{\pm\pm}^{\rm comb}$ which are
fit to a constant for $m_{ee}>0.55$~GeV/$c^2$.  The fit results
for all centralities are reported in Table~\ref{tab:like_cent}
together with the results of the statistical tests described above.
The results reported in Table~\ref{tab:like_cent} demonstrate that the
agreement between real and mixed-event like-sign mass spectra
demonstrated for minimum bias collisions also holds for all centrality
classes.

\begin{figure}[!ht]
\includegraphics[width=1.0\linewidth]{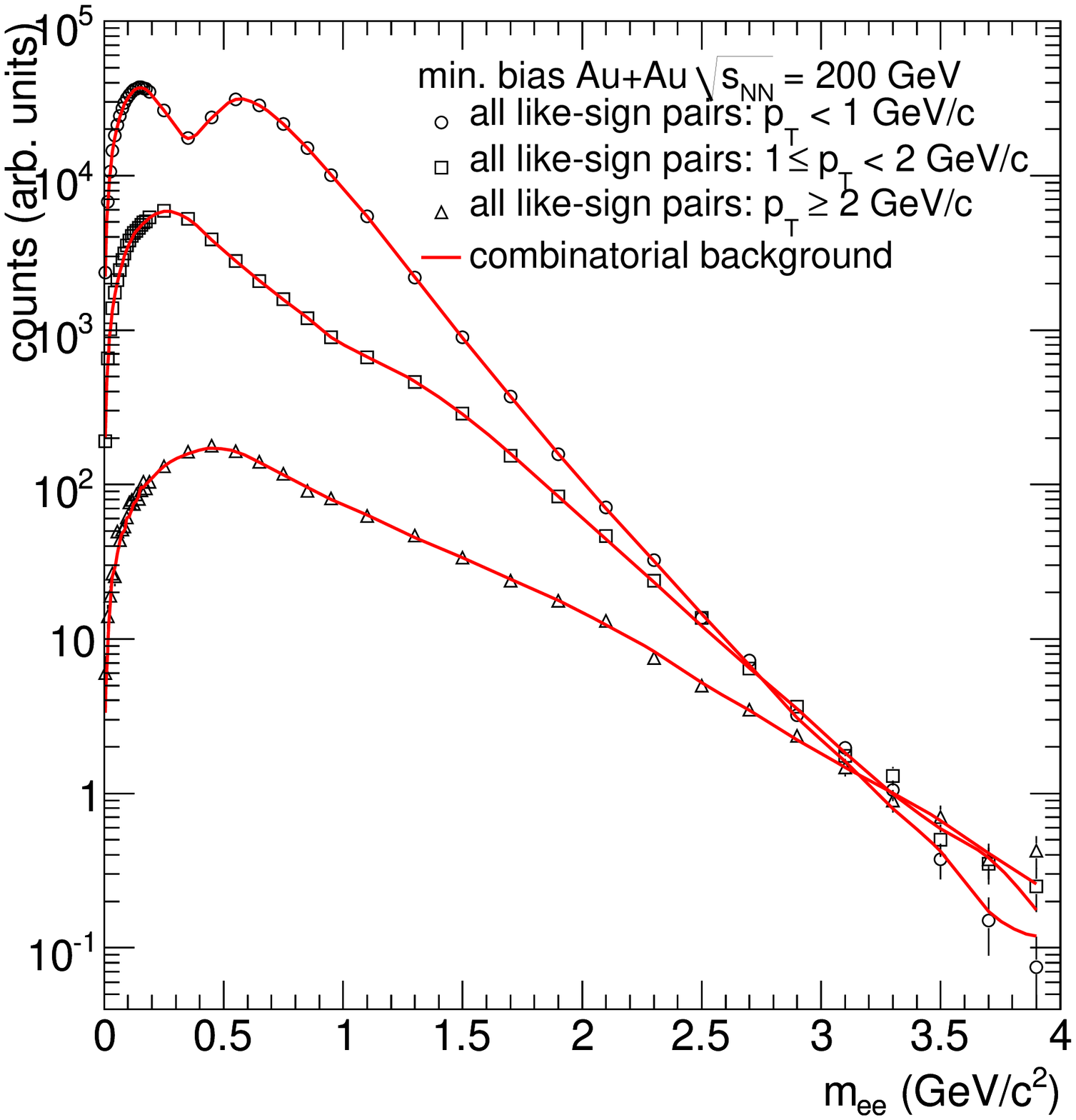}
\includegraphics[width=1.0\linewidth]{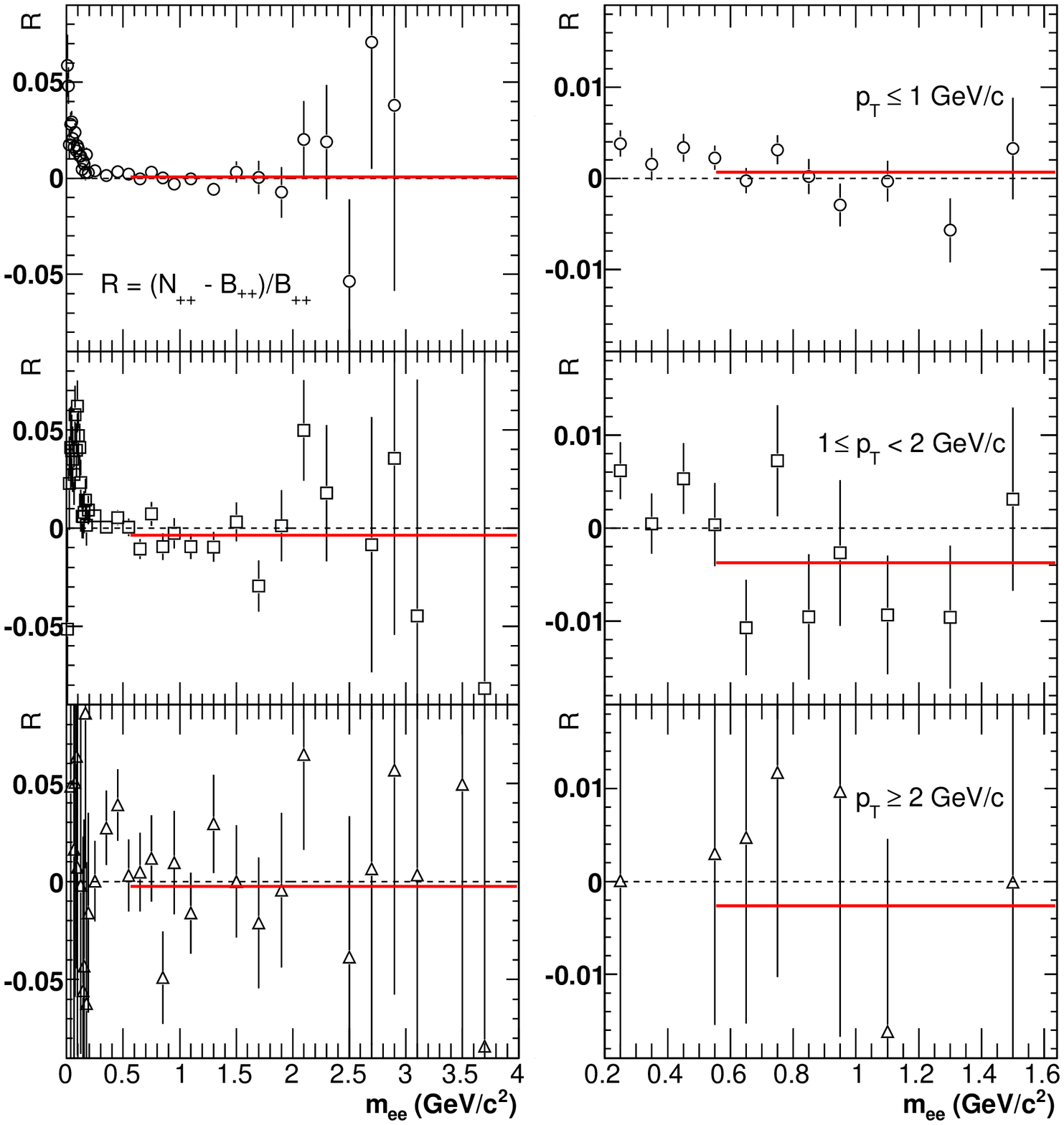}
\caption{ \label{fig:like_pt} (Color online) 
Like-sign distribution for real and
mixed events for different $p_T$ bins.  The bottom panels show the
ratio of $(N_{\pm\pm}-B_{\pm\pm}^{\rm comb})/B_{\pm\pm}^{\rm comb}$
with different scales.  The left ones show all the mass range and
allow to identify the correlated background at low masses.  The right
ones focus on mass region where we fit.}
\end{figure}

Figure~\ref{fig:like_pt} shows the like-sign mass distribution for
real and mixed events in different $p_T$ bins.  The ratio
$(N_{\pm\pm}-B_{\pm\pm}^{\rm comb})/B_{\pm\pm}^{\rm comb}$ shows a
good agreement between real and mixed-event distribution for all
bins.  This demonstrates that, within the statistical error of the
foreground, there is no deviation from uncorrelated combinatorial
behavior for masses above 0.55~GeV/$c^2$ in any $p_T$ range.

The absolute normalization of the unlike-sign combinatorial background
is determined with an uncertainty given by the statistical error of
the measured like-sign spectra for $m_{ee}>0.7$~GeV/$c^2$ of 0.12\%.
The mass interval ($m_{ee}>0.7$~GeV/$c^2$) is sufficiently large to
achieve the desired statistical accuracy and is conservatively chosen
to exclude any possible region which may be contaminated by the
correlated background.  The results are stable when varying the lower
end of the normalization region between 0.55 and 0.7~GeV/$c^2$, and the
difference is included in the systematic uncertainty.

Because the pair cuts remove more like-sign than unlike-sign pairs,
(see Section~\ref{subsub:ghosts}) the normalization factor is
corrected by this asymmetry, which is estimated with mixed events to
be $1.004 \pm 0.002$, independent of centrality.

For the various centrality bins and minimum bias collisions, the 0.2\%
uncertainty on the event rejection is added in quadrature to the
uncertainty on the normalization, which is determined by the
statistics of the like-sign pairs.  Since the ratios shown in
Figs.~\ref{fig:like},~\ref{fig:like_cent}, and \ref{fig:like_pt} 
show no systematic
deviation in shape between the like-sign real- and mixed-event
distribution for  $m_{ee}>0.55$~GeV/$c^2$, the uncertainty due to the
shape is negligible compared to the uncertainty on the normalization.
The total uncertainty on the combinatorial background is given in
Table~\ref{tab:errBG}.

\begin{table}[htbp]
\caption{Systematic uncertainty on the combinatorial background for
$p+p$ and various centrality classes of Au~+~Au collisions.~\label{tab:errBG}}
\begin{ruledtabular}  \begin{tabular}{cccc}
&   Centrality   &   $\delta B^{\rm comb}/B^{\rm comb}$ (\%) & \\
\hline
&   0-92\%       &  0.25  & \\
&   0-10\%       &  0.25  & \\
&   10-20\%      &  0.30  & \\
&   20-40\%      &  0.35  & \\
&   40-60\%      &  0.85  & \\
&   60-92\%      &  2.50  & \\
&   $p+p$        &  3.00  & \\
\end{tabular} \end{ruledtabular}  
\end{table}

\begin{figure*}[t]
\includegraphics[width=0.45\linewidth]{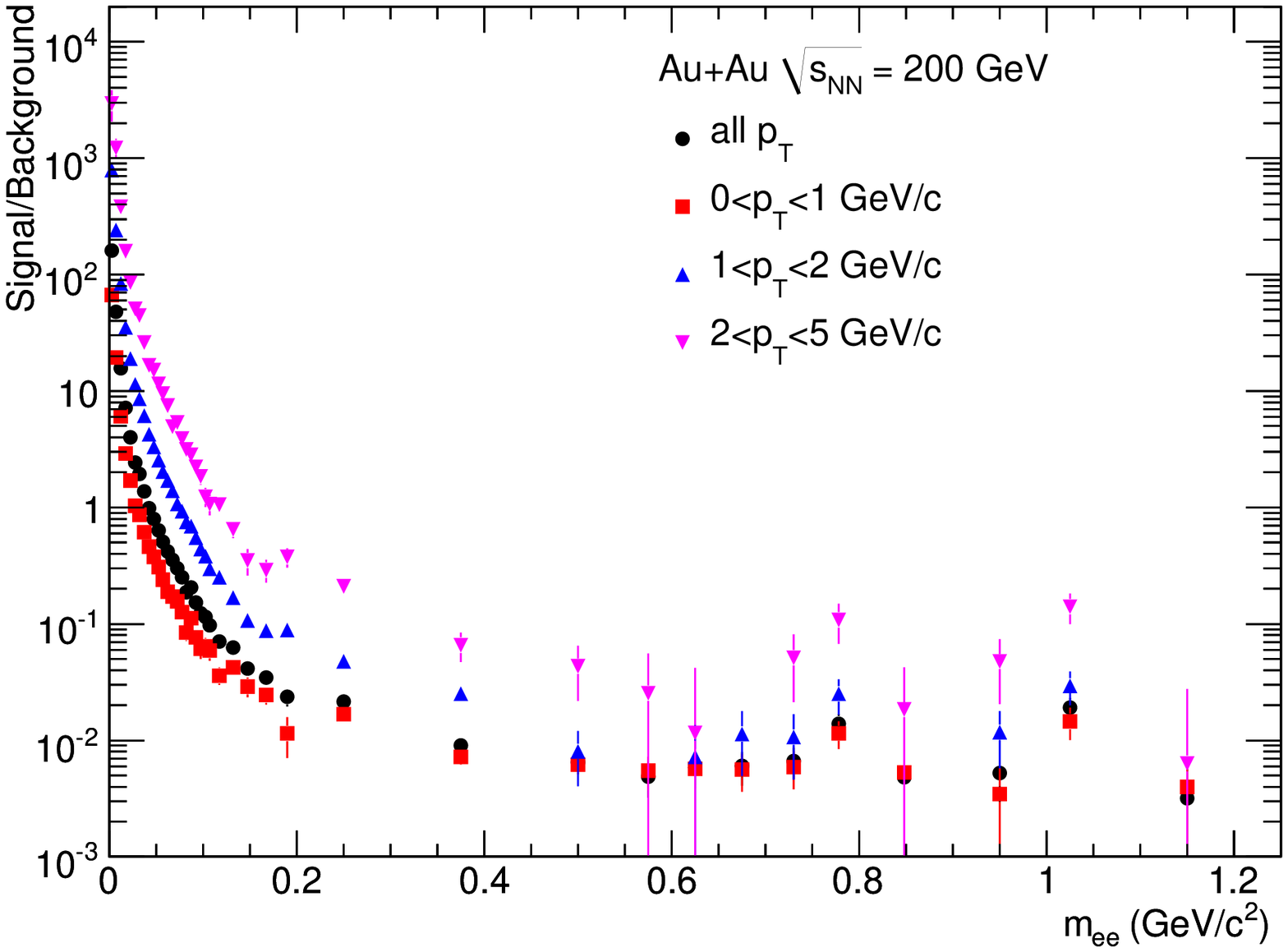}
\includegraphics[width=0.45\linewidth]{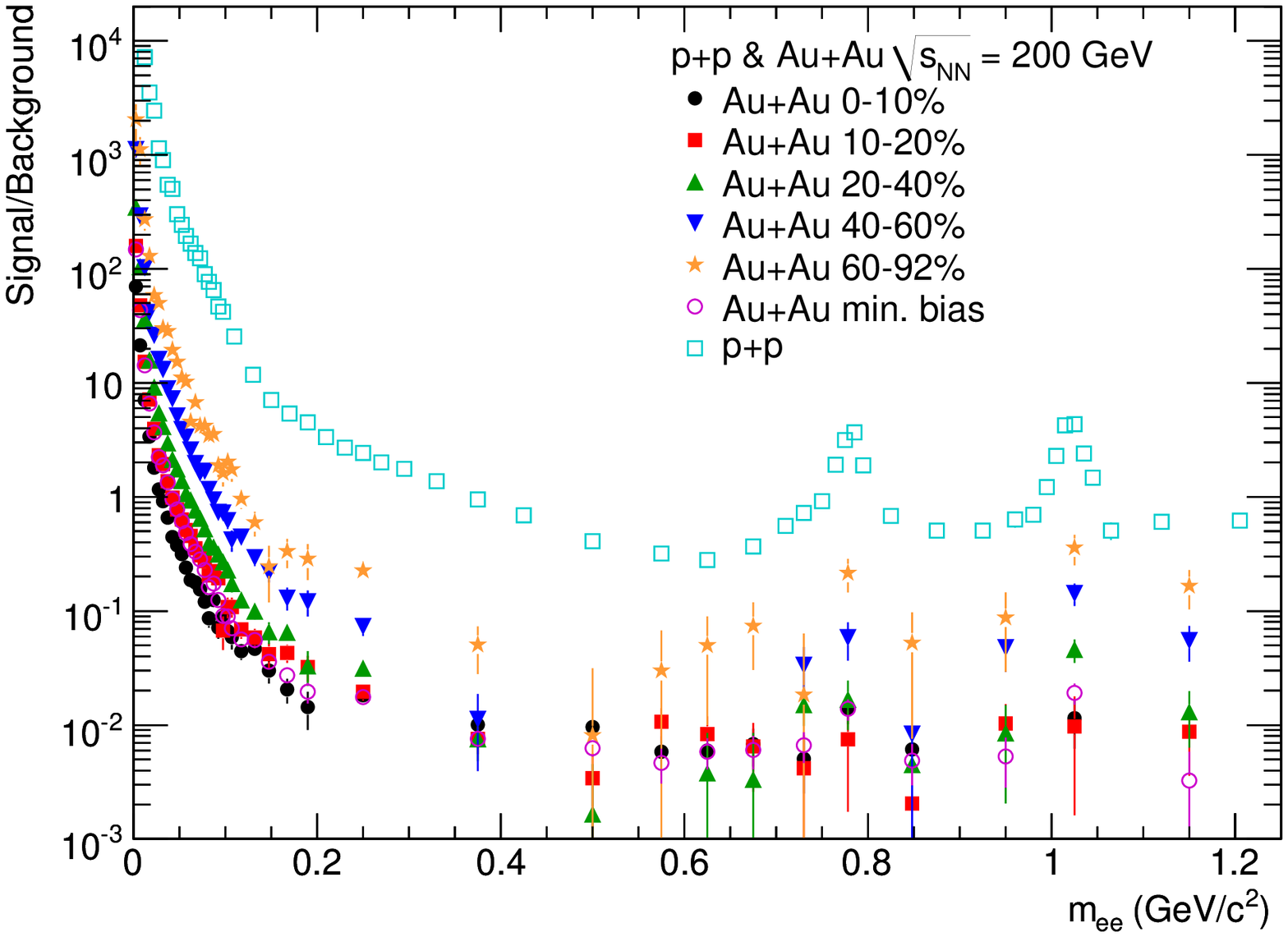}
\caption{
\label{fig:sigbg} (Color online) 
Ratio of signal to background for different
ranges of pair-$p_T$ (left) and centrality classes (right).}
\end{figure*}

This translates into a systematic uncertainty $\delta S$ on the signal
$S$ of $\delta S/S = (\delta B^{\rm comb}/B^{\rm comb}) / (S/B^{\rm comb})$.
$\delta B^{\rm comb}/B^{\rm comb}$ for minimum bias collisions
and all centrality bins are listed in Table~\ref{tab:errBG}.
Figure~\ref{fig:sigbg} shows $S/B^{\rm comb}$ for different 
ranges of pair-$p_T$ and centrality.

{\bf Correlated Background} \label{subsub:correlated}
After subtracting the combinatorial background, the remaining pair
distributions, like- and unlike-sign, are considered correlated pairs, where
the like-sign distribution only contains correlated background pairs
while the unlike-sign also contains the signal.
\begin{eqnarray}
N_{\pm\pm} - B_{\pm\pm}^{\rm comb} = B_{\pm\pm}^{\rm corr} \\
N_{+-} - B_{+-}^{\rm comb} = B_{+-}^{\rm corr} + S_{+-}
\end{eqnarray}
The \emph{correlated background} $B^{\rm corr}$ arises from two sources.

The first source is ``cross pairs''.  They occur if there are two
$e^+e^-$ pairs in the final state of a meson, e.g.  double Dalitz
decays, Dalitz decays followed by a conversion of the decay photon or
two-photon decays followed by conversion of both photons.  Besides the
real unlike-sign signal, this leads to like- and unlike-sign cross
pairs.  
While all mesons in
principle produce cross pairs, all contributions above the $\eta$ mass
(0.55~GeV/$c^2$) can be safely neglected.  Like- and unlike-sign cross
pairs were simulated using our hadron decay generator including the
PHENIX acceptance (Equation~\ref{eq:acc}).  Because their rate is
proportional to the Dalitz decay probability only, the ratio of cross
pairs to pions is independent of centrality.\\

The second source is ``jet pairs''.  They are produced by two
independent hadron decays yielding electron pairs, either within the
same jet or in back-to-back jets.  Jet pairs were simulated using the
minimum bias events generated with {\sc pythia}~\cite{pythiamb} as 
described above.  As
noted previously, correlated pairs from the same jet typically have
small mass and large $p_T$ while those from back-to-back jets have
large mass and smaller $p_T$.

Correlated background pairs equally populate like- and unlike-sign
combinations.  Since the like-sign spectrum measures only the
background, we can determine the cross and jet pair yields by
simultaneously fitting simulated cross and jet pair distributions to
the measured correlated like-sign pairs, after subtraction of
combinatorial background.  The resulting normalization factors, one for
cross-, one for jet-pairs, are then applied to the unlike-sign
correlated background.\\
\\

{\bf Correlated Background in $p+p$ data} \\
\begin{figure*}
\includegraphics[width=0.45\linewidth]{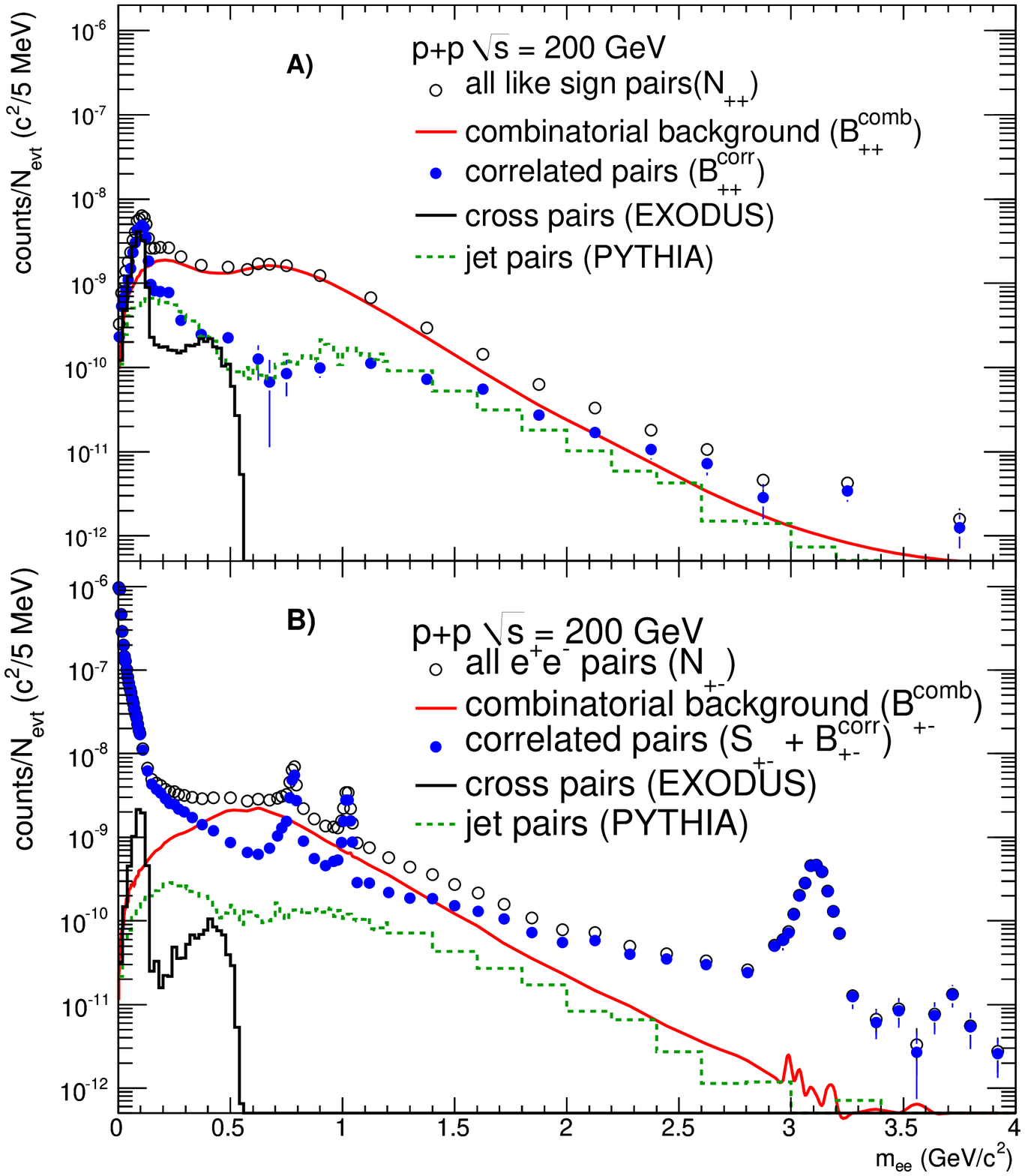}
\includegraphics[width=0.45\linewidth]{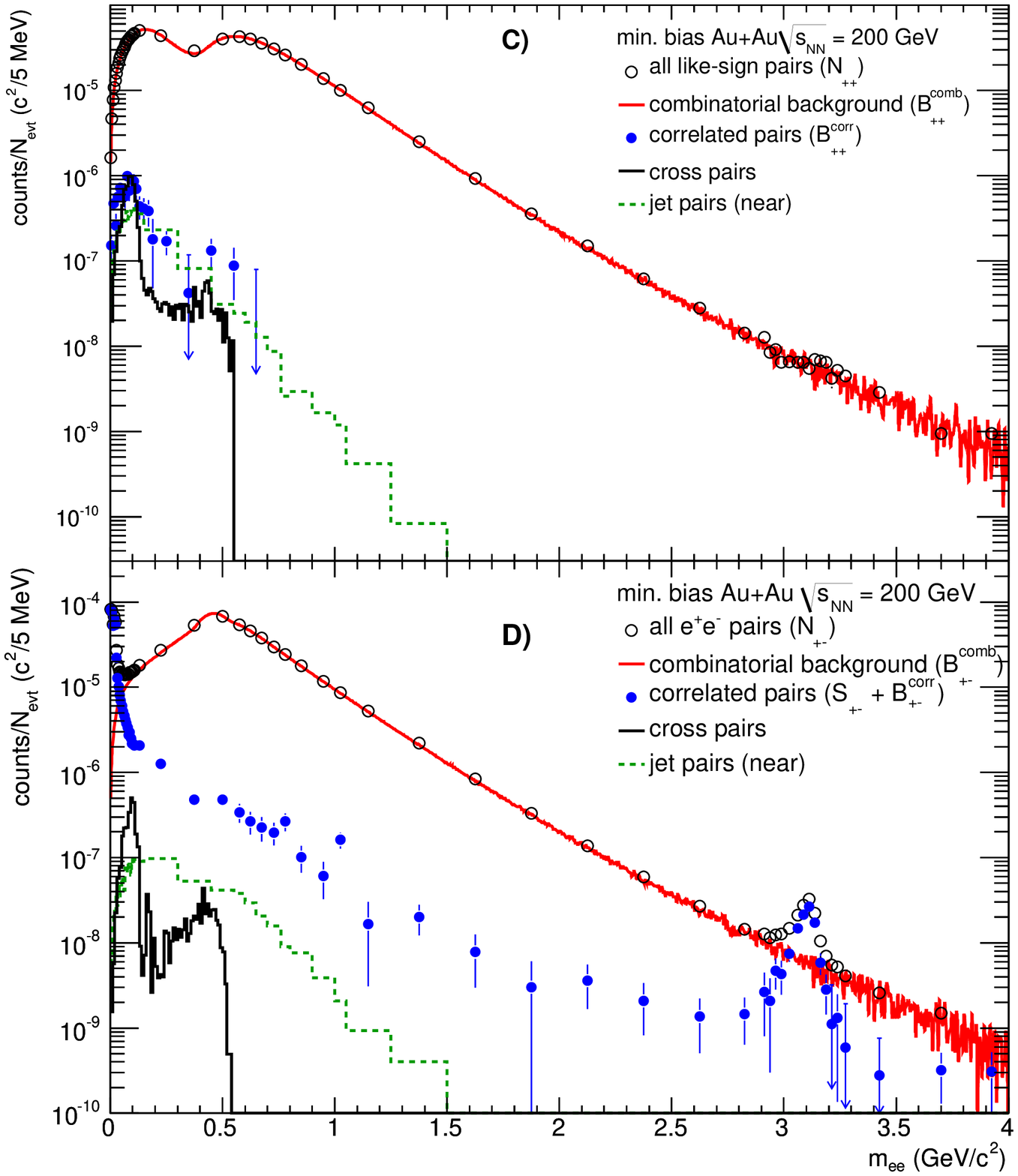}
\caption {\label{fig:rawspectra} (Color online) 
Raw dielectron spectra in $p+p$
(left) and Au~+~Au (right) collisions.  The top panels show like-sign
pairs $N_{\pm\pm}$ as measured in the experiment, the combinatorial
background from mixed-events $B_{\pm\pm}^{\rm comb}$, the correlated
pair background $B_{\pm\pm}^{\rm corr}$ obtained by subtracting the
combinatorial background, and the individual contributions from cross
and jet pairs to the correlated background (see text).  The bottom
panels show the same distributions for unlike-sign pairs.  The
correlated like-sign background $B_{\pm\pm}^{\rm corr}$, is normalized
to the measured like-sign pairs remaining after subtracting the
combinatorial background $N_{\pm\pm}-B_{\pm\pm}^{\rm comb}$, and the
same factors are applied to the unlike-sign distribution $B_{+-}^{\rm
corr}$.  }
\end{figure*}
Panel a) of Fig.~\ref{fig:rawspectra} shows
like-sign pair distributions: $N_{\pm\pm}$, $B^{\rm comb}_{\pm\pm}$ and
$B^{\rm corr}_{\pm\pm} = N_{\pm\pm} - B{\pm\pm}$.  Panel b) shows
analogous distributions for unlike-sign pairs: $N_{+-}$, $B_{+-}$ and
$B^{\rm corr}_{+-} = N_{+-} - B_{+-}$.  

The like-sign correlated background distribution is fit to
\begin{equation}
B_{\pm\pm}^{\rm corr} = A \cdot cross + B \cdot jet
\end{equation}
The resulting normalization factors, $A$ and $B$, are then applied to
the unlike-sign correlated background.  The unlike-sign signal results
from subtracting the correlated background from all correlated
unlike-sign pairs.\\

{\bf Correlated Background in Au~+~Au data} \\
In Au~+~Au data the like-sign mixed-event distribution reproduces the
mass dependence of the real-event distribution not only in region of
Equation~\ref{eq:norm_region}, but for all masses above 0.55~GeV/$c^2$ for
every centrality, as shown in Fig.~\ref{fig:like_cent}, and at every
$p_T$, as shown in Fig.~\ref{fig:like_pt}.  This means that there is no
room for correlated background for $m_{ee}>0.55$~GeV/$c^2$.  Here the
contribution that typically arises from back-to-back jets in the Au~+~Au
data is indeed expected to be different than $p+p$ because of the
observed jet modifications~\cite{ppg083}.

We therefore separate the jet distribution into ``near-side''
($jet_{near}$: $\Delta\phi<\pi/2$) and ``away-side'' contributions
($jet_{away}$: $\Delta\phi>\pi/2$) and we fit the like-sign correlated background
distribution to the sum of
\begin{equation}
B_{\pm\pm}^{\rm corr} = A \cdot cross + B \cdot jet_{near} + C \cdot jet_{away}
\end{equation}

Panels c) and d) of Fig.~\ref{fig:rawspectra} show the like- and
unlike-sign pair distributions, the normalized mixed-event background
and the distributions after subtraction.  We note that the like-sign
yield is well described by the sum of combinatorial and correlated
background, and that the contribution from ``away-side-jet pairs'' is
consistent with zero, i.e.  $C=0$, as the mixed-event distribution was
normalized to the real data in the IMR: the ``away-side-jet pairs''
are therefore not shown.  The unlike-sign signal ($S_{+-}$) is
obtained by subtracting from the distribution of all pairs the
mixed-event combinatorial background ($B_{+-}^{\rm comb}$) and the correlated
background ($B_{+-}^{\rm corr}$) normalized with the factors $A$,$B$
and $C$ measured in the like-sign spectrum.

\begin{figure*}
\includegraphics[width=0.45\linewidth]{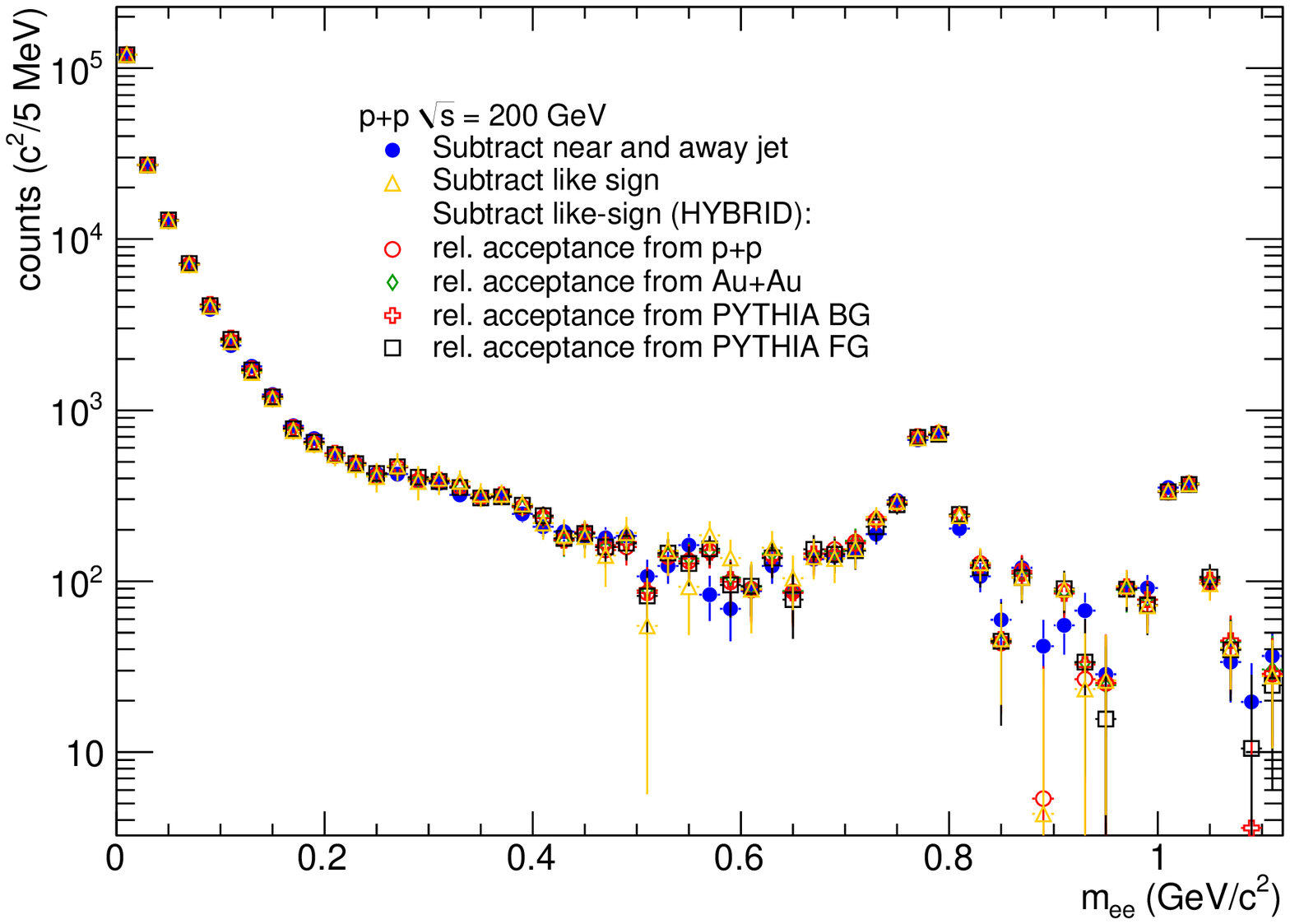}
\includegraphics[width=0.45\linewidth]{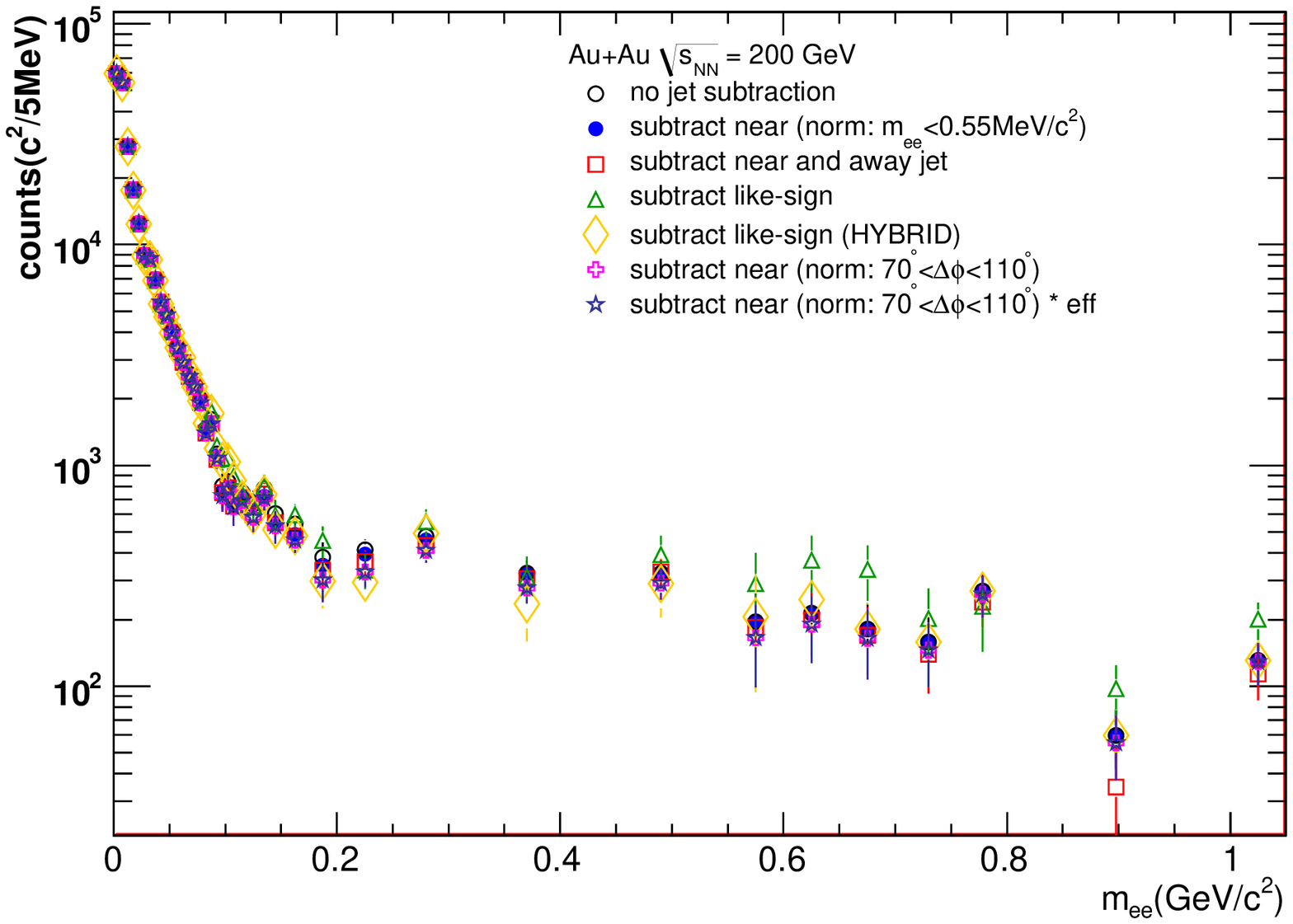}
\includegraphics[width=0.45\linewidth]{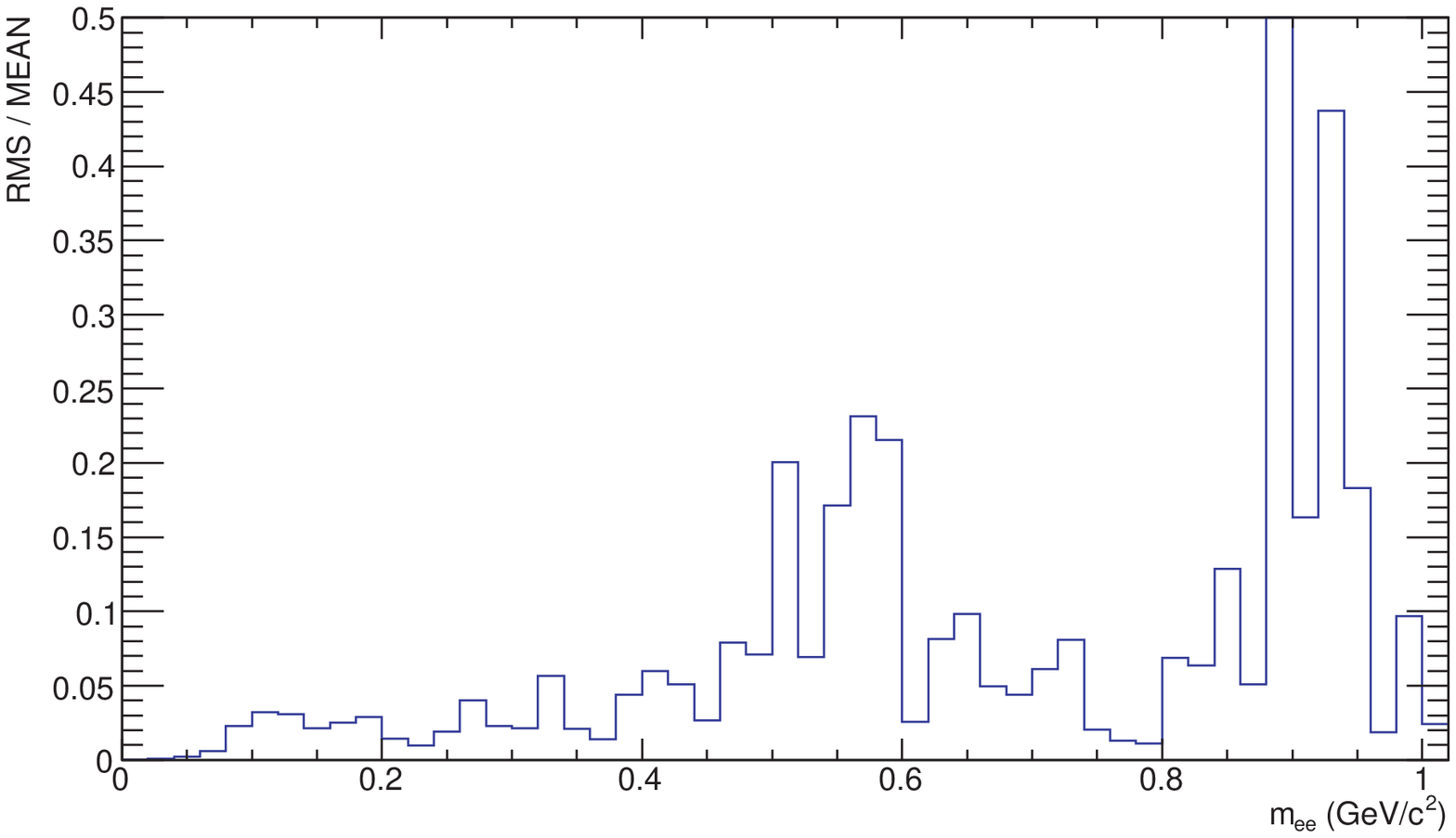}
\includegraphics[width=0.45\linewidth]{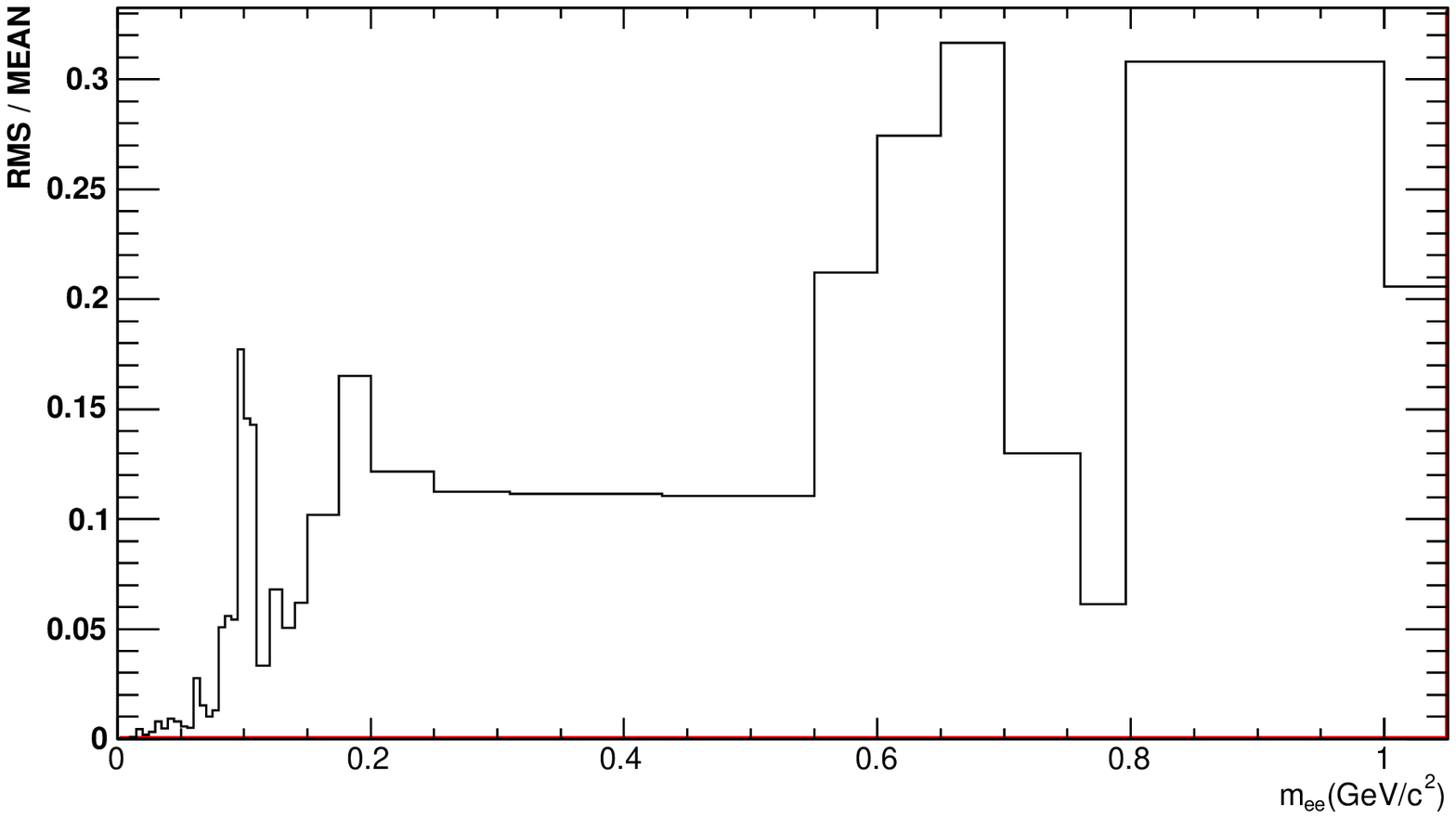}
\caption {\label{fig:corrBG} (Color online) 
Unlike-sign dielectron spectra obtained by
subtracting the background (combinatorial and correlated) with the
methods explained in the text for $p+p$ (left) and Au~+~Au (right).  The
RMS/MEAN of all the spectra shown in the bottom panels 
allows to assign a systematic uncertainty on the signal due to
correlated background subtraction.}
\end{figure*}

\subsubsection{Like-sign Subtraction Method} \label{subsub:likesign} 
The event-mixing technique, along with the determination of correlated
backgrounds, used to estimate the background contribution to the
measured dilepton mass spectra, was developed in order to get around
the problems introduced in a traditional like-sign background
calculation by the asymmetric PHENIX acceptance.  With a like-sign
calculation we need to make only the assumption that the mass
dependence of the correlated background is symmetric for like- and
unlike-sign pairs.  We do not need to make any assumption about the decomposition
of the background into ``cross'', ``jet-near-side'' and
``jet-away-side'' pairs.  The like-sign distribution measures only the
background.  Correlated and uncorrelated backgrounds are charge
symmetric, i.e.  they yield the same number of like-and unlike-sign
pairs, with the same distribution.  The \emph{measured} distributions
however are different because of the acceptance.

The acceptance difference between like- and unlike-sign pairs can be
measured with the ratio of mixed-event distributions,
i.e.  $\frac{B_{+-}^{\rm comb}}{2\sqrt{B_{++}^{\rm comb}\cdot B_{--}^{\rm comb}}}$ where $B_{+-}^{\rm comb},
B_{++}^{\rm comb}, B_{--}^{\rm comb}$ are the number of $e^+e^-, e^+e^+,
e^-e^-$ pairs in a given mass and $p_T$ bin.
The like-sign distribution can therefore be
acceptance-corrected and then subtracted from the unlike-sign pairs.  
\begin{equation}
S_{+-}=N_{+-} - 2\sqrt{N_{++}N_{--}} \cdot
\frac{B_{+-}^{\rm comb}}{2\sqrt{B_{++}^{\rm comb}\cdot B_{--}^{\rm comb}}} 
\end{equation}
This technique however measures the background distribution with similar
statistical error as the foreground, therefore the resulting error on
the signal is much larger than when using the mixed events technique.

One can also use a ``hybrid'' method where first the mixed events are
normalized in the region where real and mixed-event distributions
agree (region in Equation~\ref{eq:norm_region} for $p+p$ and
$m_{ee}>0.7$~GeV/$c^2$ for Au~+~Au).  Then the correlated background is
obtained by correcting the subtracted like-sign distribution by the
acceptance difference $\frac{B_{+-}^{\rm comb}}{2\sqrt{B_{++}^{\rm comb}\cdot
B_{--}^{\rm comb}}}$.

\subsubsection{Systematic Uncertainty of the Background Subtraction} \label{subsub:sys_bg}
The background subtraction is the most critical aspect of this
analysis, therefore it is crucial to assign a proper systematic
uncertainty to it.

In Section~\ref{subsub:combinatorial} we reported the systematic
uncertainty on the normalization of the combinatorial background 
determined by the statistical accuracy of the measured like-sign
yield (see Table~\ref{tab:errBG}).  The uncertainty on the shape of
combinatorial background is everywhere negligible compared to the
uncertainty on the normalization.

To evaluate the systematic uncertainty due to the correlated pairs we
compare the results obtained with the two subtraction methods.
It is important to note that the contribution of the correlated
background is everywhere small compared to the signal.  Therefore any
uncertainty in its estimate
would result in a small uncertainty on the signal.

In the first subtraction method the correlated background 
$B_{+-}^{\rm corr}$ was calculated using Monte Carlo simulations.
We have estimated
potential spectral shape modifications of $B_{+-}^{\rm corr}$ in
several ways, shown in Fig.~\ref{fig:corrBG}.  In Au~+~Au
\begin{itemize}
\item we have not subtracted any jet background (this would clearly
result in an upper limit);
\item we have subtracted the near side normalizing it for
$m_{ee}>0.55$~GeV/$c^2$;
\item we have subtracted both near
and away-side using for the away-side the same normalization as for
the near-side (this would clearly result in a lower limit);
\item we have subtracted the near side normalizing it for
$70^{\circ}<\Delta\phi<110^{\circ}$;
\item we have subtracted the near side normalizing it for
$70^{\circ}<\Delta\phi<110^{\circ}$ and we have corrected the shape
for the eID efficiency.
\end{itemize}

In the second subtraction method, both in $p+p$ and Au~+~Au we use the like-sign corrected for
the different acceptance, or a hybrid method which uses the
mixed-event distribution for the combinatorial background and the
like-sign distribution corrected for the different acceptance for the
correlated background only.  

Since the acceptance for pairs is a
function of mass and $p_T$, we have checked that for different
$e^+e^-$ pair sources, which span reasonable variations in mass and
$p_T$ shapes of the $e^+e^-$ pairs, the relative acceptance is
unchanged.  For this purpose in $p+p$ we have calculated the relative
acceptance:

\begin{itemize}
\item using data from $p+p$ run only;
\item using data from Au~+~Au run only;
\item using data from {\sc pythia} real events;
\item using data from {\sc pythia} mixed events.
\end{itemize}

We added the results from these two methods in Fig.~\ref{fig:corrBG}
as well.  The agreement within the different methods (see RMS/MEAN in
the bottom panel) allows to assign a systematic uncertainty on the
signal.

\subsection{Raw Mass Spectrum} \label{subsub:rawmass}
Figure~\ref{fig:mass} shows the mass distribution of $e^+e^-$ pairs,
the normalized mixed-event background ({\it{B}}), and the signal yield
({\it{S}}) obtained by subtracting the mixed-event background and the
correlated background (cross and jet pairs) for Min. Bias Au~+~Au
collisions.  The right panel shows the signal-to-background ratio
($S/B^{\rm comb}$).  The systematic uncertainties (boxes) reflect the
uncertainty from the background subtraction, which is given by
$\delta_S/S~=~0.25\%~\cdot~B^{\rm comb}/S$, added in quadrature to the
uncertainty due to the correlated background subtraction,
conservatively assumed to
be around $10\%S$ below 0.55~GeV/$c^2$.  Despite the small $S/B^{\rm comb}$
ratio, an $e^+e^-$ pair continuum is visible up to 4.5~GeV/$c^2$.  Due
to the limited statistical precision in Au~+~Au binning was chosen
such that the bins near the $\phi$ and $\omega$ meson are centered at
the nominal meson mass.  The bin width was chosen to correspond
approximately to twice the mass resolution observed in $p+p$
collisions.  In this way the $\phi$ and $\omega$ mesons can be resolved
from the continuum.

\subsection{Runs with Increased Conversion Material} \label{subsub:rawmass_conversion}

In order to check the background subtraction, a subset of data
(5$\times 10^7$ events) was collected with a brass sheet of 1.68\%
radiation length ($X_0$) wrapped around the beam pipe to increase the
number of photon conversions.  We make an estimate on the uncertainty
in the radiation length (5\%) by comparing photon conversions in data
and simulation~\cite{ppg066}.  Because the additional conversion leads
to an increased electron multiplicity, in this data set the
combinatorial and correlated background ($B$) contribution is larger
by a factor of $\sim$2.5.

If there is a systematic bias in background normalization, and the
yield of background is off by a small fraction $f$, it can lead to
significant difference between the ``apparent'' signal and the
``true'' signal $S^{\rm true}$.  The difference should be larger in the
converter run since signal-to-background is smaller.  Thus the
``apparent'' signal $S_{\rm C}$ in the converter run should become larger
than the ``apparent'' signal $S_{\rm NC}$ in normal run without the
converter.  The relation between $S^{\rm true}$, $S_{\rm C}$, $S_{\rm NC}$ can be
written as:
\begin{eqnarray}
S_{\rm NC} &=& N_{\rm NC} - (1-f) \cdot B \nonumber \\
&=& S^{\rm true} + f B \label{eq:s_nc} 
\\ 
S_{\rm C} &=& N_{\rm C} - 2.5 \cdot (1-f) \cdot B
\nonumber \\
&=& S^{\rm true} + 2.5 f B
\end{eqnarray}
where $N_{\rm C}$ ($N_{\rm NC}$) is the foreground of all $e^+e^-$
pairs in converter (non-converter) runs.
If we divide Equation~\ref{eq:s_nc} by $B$, we obtain:
\begin{eqnarray}
\frac{S}{B} =
\frac{S_{\rm NC}}{B} =
\frac{S^{\rm true} + f B}{B} \nonumber \\
S^{\rm true} + f B = (S/B) B
\end{eqnarray}
Then the ratio between the apparent signal in converter and non-converter
runs is given by:
\begin{eqnarray}
\frac{S_{\rm C}}{S_{\rm NC}} 
&=& \frac{S^{\rm true} + 2.5 f B} {S^{\rm true} + f B}
\nonumber \\
&=& \frac{S^{\rm true} + f B + 1.5  f  B} {S^{\rm true} + f  B}
\nonumber \\
&=& \frac{(S/B) B + 1.5 f B}
{(S/B) B}
\nonumber \\
&=& \frac{S/B + 1.5 f}{S/B}
\nonumber \\
&=& 1 + 1.5 f  \frac{B}{S}
\end{eqnarray}
Therefore
\begin{eqnarray}
f = \left( \frac{S_{\rm C}}{S_{\rm NC}} - 1 \right) 
\frac{S}{B} 
\frac{1}{1.5}
\end{eqnarray}

\begin{figure*}[t]
\includegraphics[width=0.625\linewidth]{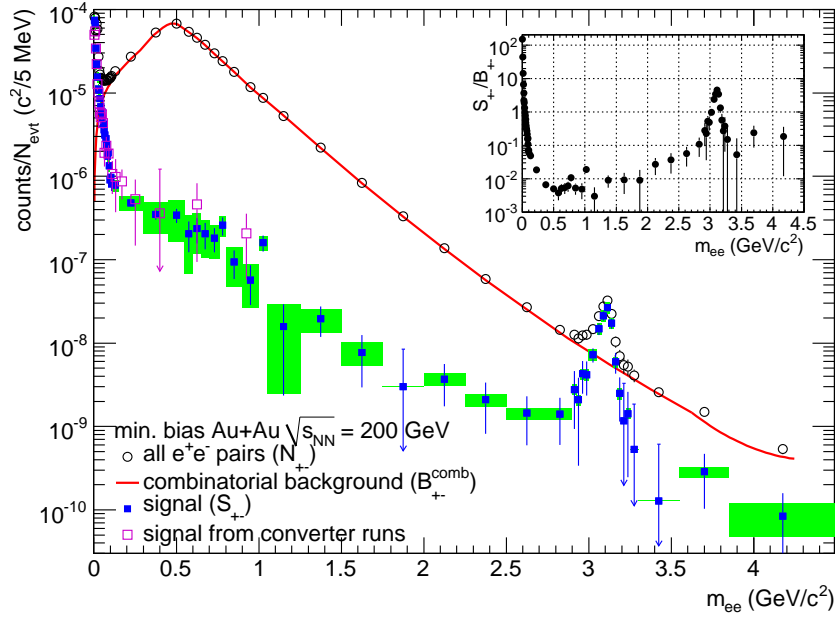}
\caption{\label{fig:mass} (Color online) 
Uncorrected mass spectra of all $e^+e^-$
pairs, mixed-event background ($B^{\rm comb}$) and signal ($S$) in
minimum bias Au~+~Au collisions.  Statistical (bars) and systematic
(boxes) uncertainties are shown separately.  The signal from the runs
with additional converter material is shown with statistical errors
only.  The two spectra are normalized by the number of events.  The
insert shows the $S/B^{\rm comb}$ ratio.  The mass range covered by
each data point is given by horizontal bars.  }
\end{figure*}

This means we can use the agreement between converter and non-converter runs
$S_{\rm C}/S_{\rm NC}$, 
shown in Fig.~\ref{fig:mass}, and the signal-to-background ratio
$S/B$ also shown in the insert of Fig.~\ref{fig:mass}, to constrain a
potential bias $f$ in the background normalization.

In $0.3<m_{ee}<0.75$~MeV/$c^2$
\begin{eqnarray}
\frac{S_{\rm C}}{S_{\rm NC}} = 1.29 \pm 0.92 \nonumber
\end{eqnarray}
and 
\begin{eqnarray}
\frac{S}{B} = (7.47 \pm 0.55) 10^{-3} \nonumber
\end{eqnarray}
constrain $f = (0.14 \pm 0.46) 10^{-2}$.
If we extend the mass range to lower values
$0.04<m_{ee}<0.75$~MeV/$c^2$,
\begin{eqnarray}
\frac{S_{\rm C}}{S_{\rm NC}} = 1.08 \pm 0.28 \nonumber
\end{eqnarray}
and
\begin{eqnarray}
\frac{S}{B} = (2.75 \pm 0.05) 10^{-2}, \nonumber
\end{eqnarray}
the constraint becomes $f = (0.15 \pm 0.51) 10^{-2}$.

The agreement between the converter run and the normal run confirms
that the systematic bias in the background normalization is small
(0.15$\pm$0.51\%).  The result is consistent with our estimate of the
systematic uncertainty in the background normalization (0.25\%).  It
should be noted that the converter run provides an independent test on
the background normalization.

\subsection{Efficiency Correction}\label{sub:effi}
The $e^+e^-$ mass spectra are corrected for the total pair reconstruction
efficiency $\epsilon^{\rm total}_{\rm pair}$
to give the $e^+e^-$ pair yield inside the PHENIX aperture as defined
by Equation~\ref{eq:acc}:
\begin{equation}
\frac{dN}{dm_{ee}}=
\frac{1}{N_{\rm evt}}
\frac{N_{ee}}{\Delta m_{ee}}
\frac{1}{\epsilon^{\rm total}_{\rm pair}} 
\end{equation}

The $p_T$ spectra are further corrected for the pair geometric acceptance
($\epsilon^{\rm geo}_{\rm pair}$) to give the $e^+e^-$--pair yield
in one unit of rapidity:
\begin{equation}
\frac{1}{2\pi p_T^{ee}} \frac{d^2N}{dp_T^{ee}dy^{ee}}=
\frac{1}{2\pi p_T^{ee}} 
\frac{1}{N_{\rm evt}}
\frac{N_{ee}}{\Delta p_T^{ee} \Delta y}
\frac{1}{\epsilon^{\rm total}_{\rm pair}} 
\frac{1}{\epsilon^{\rm geo}_{\rm pair}} 
\end{equation}

The $p+p$ data are further corrected by the factor
$\frac{\epsilon_{\rm BBC}}{\epsilon_{bias}}$ where $\epsilon_{\rm BBC}=54.5\pm
5$\% is the BBC efficiency and $\epsilon_{bias}=79\pm 2$\% is the
BBC trigger bias, described above (Section~\ref{sub:dataset}).

The total pair reconstruction efficiency
$\epsilon^{\rm total}_{\rm pair}$ 
depends on the single electron efficiency for reconstruction and eID 
($\epsilon^{\rm eID}_{\rm single}$), the efficiency from the detector
live area ($\epsilon^{\rm live}_{\rm single}$), the
occupancy efficiency ($\epsilon^{\rm occ}_{\rm single}$) (for Au~+~Au).
The occupancy efficiency ($\epsilon^{\rm occ}_{\rm single}$) is
different for each Au~+~Au centrality class and is described in
Section~\ref{sub:embedding}.  The $p+p$ data are also corrected for the
ERT trigger efficiency, described in Section~\ref{sub:ert}.  In addition,
the effect of the pair cuts ($\epsilon^{\phi_V}_{\rm pair}$ and
$\epsilon^{\rm ghost}_{\rm pair}$) is taken into account:
\begin{eqnarray}
\epsilon^{\rm total}_{\rm pair}  = \epsilon^{\rm eID}_{\rm pair}  \cdot
\epsilon^{\rm live}_{\rm pair} \cdot \epsilon^{\phi_V}_{\rm pair} \cdot
\epsilon^{\rm ghost}_{\rm pair} \nonumber \\
\cdot \epsilon^{\rm occ}_{\rm pair} \;\;(\rm{for}\; \rm{Au~+~Au})
\nonumber \\
\cdot \epsilon^{\rm ERT}_{\rm pair} \;\;(\rm{for}\; p+p)
\end{eqnarray}
These efficiencies depend on the eID cuts used to determine the
electron sample, therefore they do factorize only on the condition
that the electron sample used to calculate them is the same for all
them.

The pair-eID and reconstruction efficiency $\epsilon^{\rm eID}_{\rm pair}$ and pair-live
efficiency $\epsilon^{\rm live}_{\rm pair}$, as well as the ERT
efficiency $\epsilon^{\rm ERT}_{\rm pair}$ for $p+p$ data
are derived, as a function of mass and
pair-$p_T$, from the corresponding \emph{\rm single} electron efficiency
($\epsilon^{\rm eID}_{\rm single}$, $\epsilon^{\rm live}_{\rm single}$
and $\epsilon^{\rm ERT}_{\rm single}$ respectively) using
the pair kinematic properties implemented in our hadron decay
generator, as explained below.  

The $\epsilon^{\rm eID}_{\rm single}$ is the fraction of signal loss due to
track reconstruction and eID cuts within the detector active area.  It
depends only on the momentum of the track.  The shape is very similar in
$p+p$ and Au~+~Au, but the scale is different since different cut values
are used.  It is calculated as follows:
\begin{eqnarray}
{\epsilon}^{\rm eID}_{\rm single} (p_T^e)
= \frac{dN_e^{\rm out}/dp_T^e}{dN_e^{\rm in}/dp_T^e},
\end{eqnarray}
The $dN_e^{\rm in}/dp_T^e$ is the $p_T^e$ distribution
of the input electron
yield that falls into the \emph{real} PHENIX acceptance, which
includes the boundary described by Equation~\ref{eq:acc} and the active
areas of the detector.  $dN_e^{\rm out}/dp_T^e$
is the $p_T^{e}$ distribution of the output electron yield in the same
acceptance after passing all the eID cuts.  

The input distribution comes from a simulation of 450M 
$\pi^0$'s flat in phase space ($0 < p_T < 25$~GeV/$c$,
$|y| < 0.5$, and $0< \phi <2\pi$) with the branching ratio of the
$\pi^0$ Dalitz decay set to 100\% to enhance the sample of $e^+e^-$ pairs
per event.  These events were processed by the full GEANT
simulation program of the PHENIX detector~\cite{geant} that includes
the details of the detector response.  The output simulation data files
were processed by the event reconstruction chain of PHENIX.  Standard
eID cuts are applied to the output.  The
reconstructed $p_T^{e}$ of each output electron is weighted 
according to $p_T^{in}$ with a realistic exponential $p_T$ weight.

\begin{figure*}[t]
\includegraphics[width=1.0\linewidth]{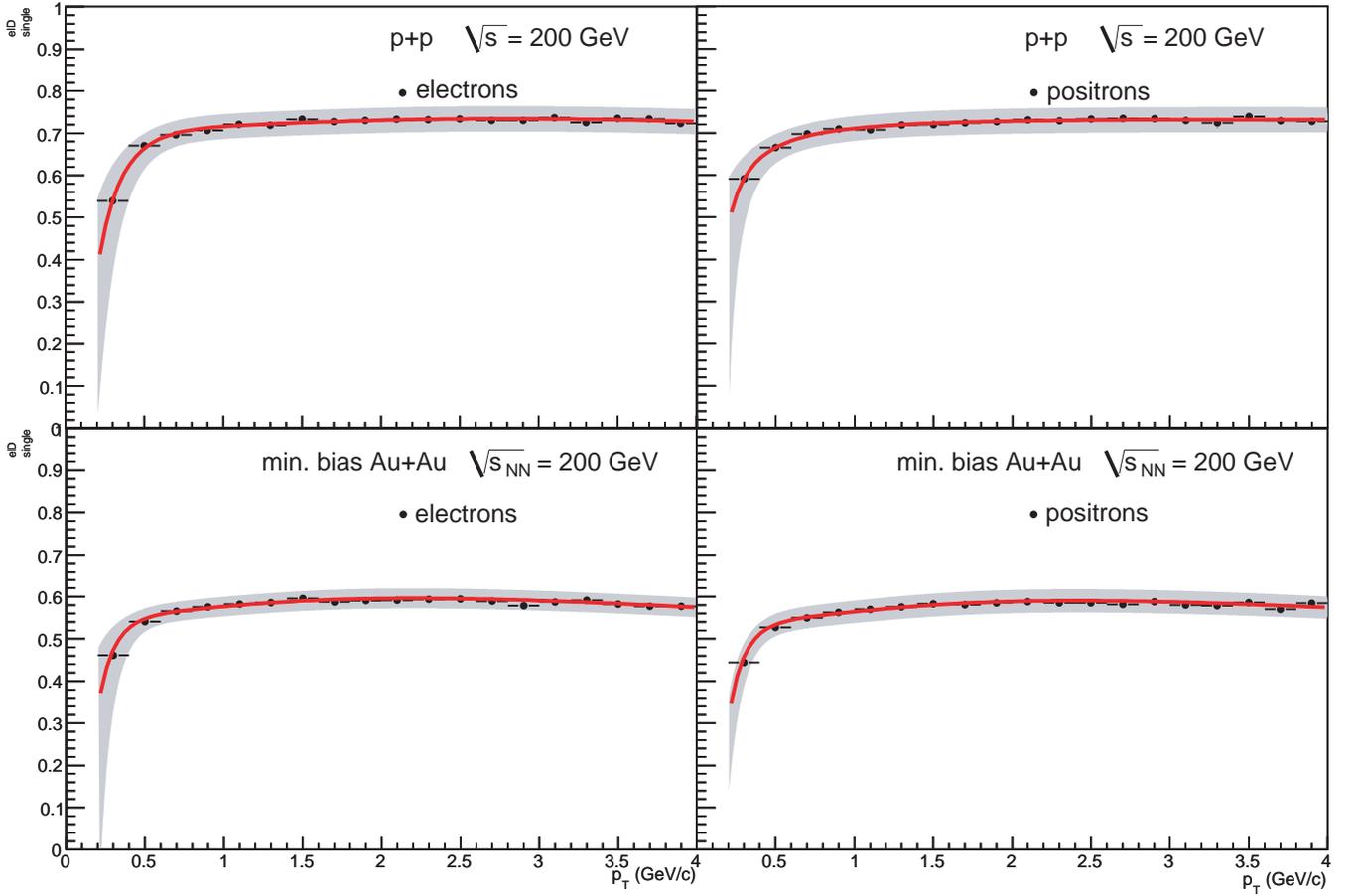}
\caption{ \label{fig:single_effi} (Color online) 
Single electron (positron)
efficiency as a function of $p_T^{e}$ for electrons (left) and positrons
(right) in $p+p$ (top) and Au~+~Au (bottom) collisions.}
\end{figure*}

\begin{figure*}[t]
\includegraphics[width=0.45\linewidth]{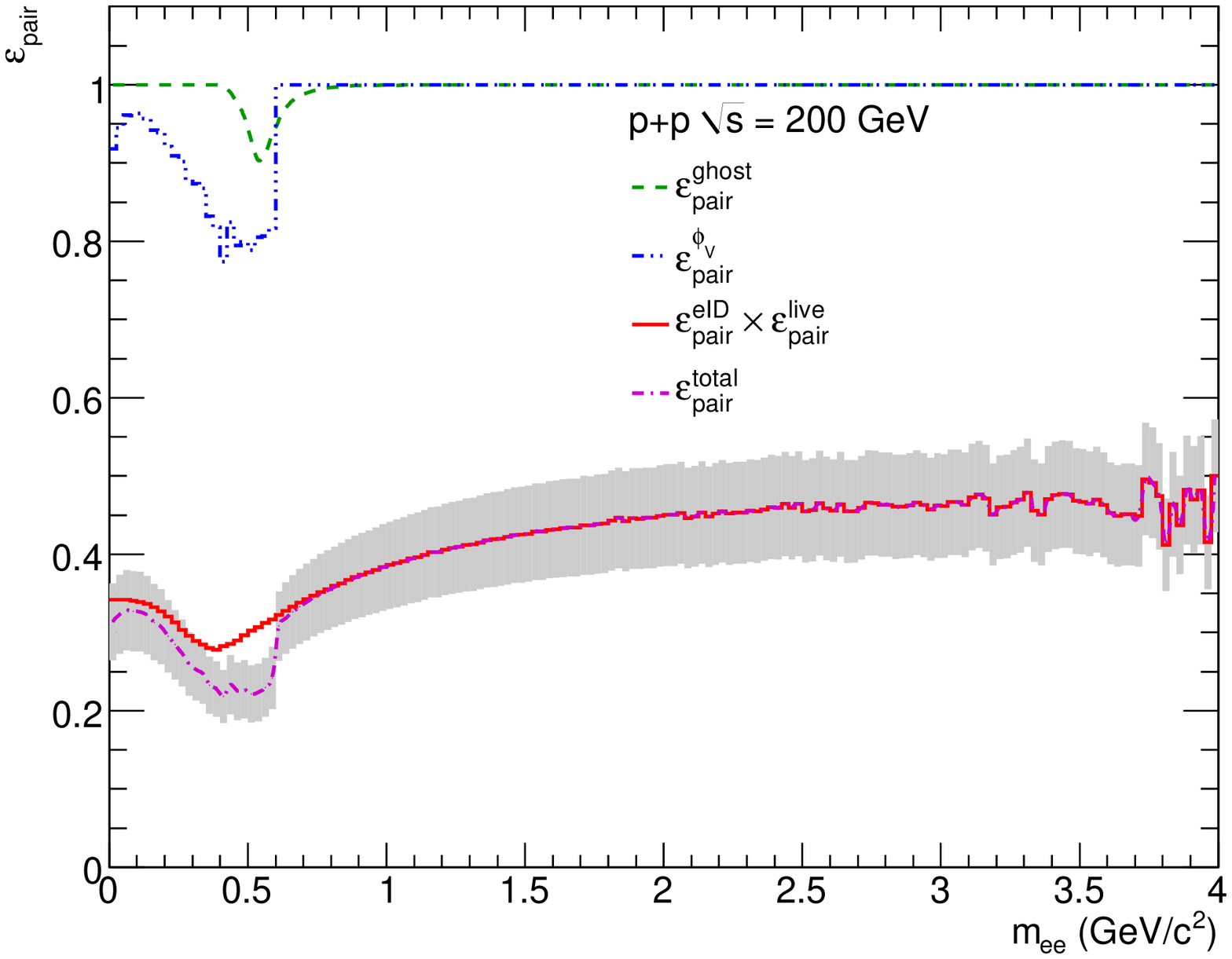}
\includegraphics[width=0.45\linewidth]{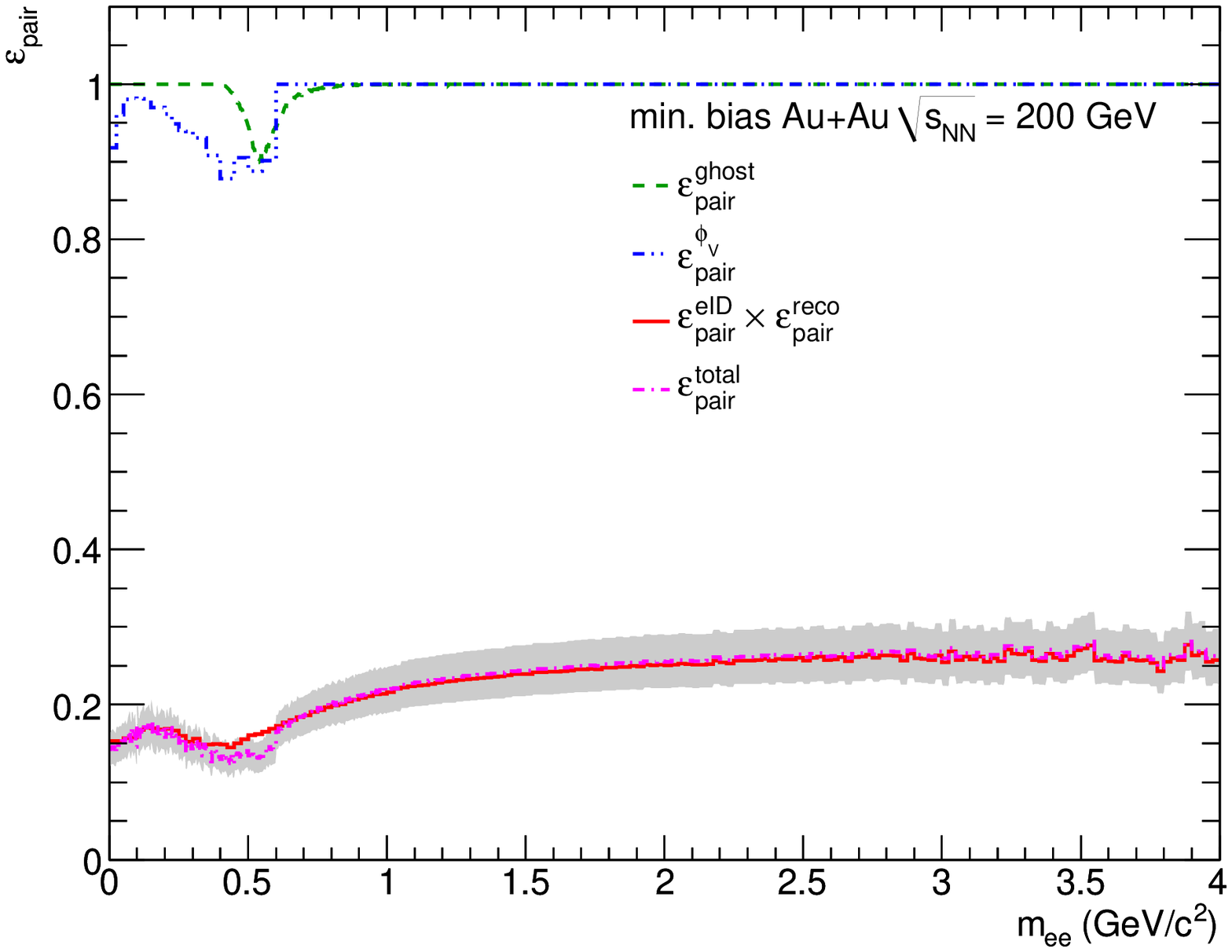}
\caption{  \label{fig:pair_effi} (Color online)
Different components of the total pair efficiency as a function of pair mass, obtained with the
procedure described in the text, for the $p+p$ and the Au~+~Au collisions.}  
\end{figure*}

Figure~\ref{fig:single_effi} shows the single electron (positron)
efficiency as a function of $p_T^{e}$ for $p+p$ and
Au~+~Au.  The different scales corresponding to the two data sets are due
to more stringent cuts applied in the Au~+~Au data set.  The band around
the curve shows only the $p_T^{e}$ dependence of the
systematic uncertainty corresponding to a shift of $\pm$ 0.1~GeV/$c$
of the efficiency curve.  This in turns leads to a distortion of the
$e^+e^-$ mass shape shown by the band in Fig.~\ref{fig:pair_effi}.  
The total uncertainties on the pair reconstruction, including the range of applicability, are reported in Table~\ref{tab:errors}.

The $\epsilon^{\rm live}_{\rm single}$ is the fraction of signal loss due
to inactive areas of the detector.  The active areas of the detector
are parameterized as a function of the particle momenta and azimuth
using real data.  There are small differences between $p+p$ and Au~+~Au.

In the {\sc exodus} cocktail of hadron decays we have implemented the
parameterization of the single electron (positron) efficiency (shown
as a curve in Fig.~\ref{fig:single_effi}).  
The single electron (positron) efficiency was applied as a
weight to each track.  The pair will therefore get a weight given by
the product of the electron and the positron weight.  This weight, a
function of mass and pair-$p_T$, represents
$\epsilon^{\rm eID}_{\rm pair}$.  To calculate
$\epsilon^{\rm live}_{\rm pair}$, a fiducial cut corresponding to the
detector active areas is also implemented in {\sc exodus} and a pair is
rejected if at least one track falls out of the active areas.  We
therefore determine the product 
$\epsilon^{\rm eID}_{\rm pair} \times \epsilon^{\rm live}_{\rm pair}$
double differentially as a function of mass and pair-$p_T$.

Figure~\ref{fig:pair_effi} shows $\epsilon^{\rm eID}_{\rm pair} \times
\epsilon^{\rm live}_{\rm pair}$ as a function of mass.  At high masses
the efficiency is constant with a value nearly equal to the square of
the single electron efficiency.  At low mass the pair efficiency
results from the convolution of the single electron reconstruction
efficiency, which drops toward low momentum, and the geometric
acceptance, which effectively truncates the single electron $p_T$
distribution ($p_T^{e} > 0.2$~GeV/$c$).  For pairs with $0.4 \le
m_{ee} \le 0.8$~GeV/$c^2$ the efficiency drops as a consequence of the
drop at low-$p_T$ of the single electron efficiency.  However, for
$m_{ee}\le 0.4$~GeV/$c^2$, the lower limit on single electron $p_T$
results in a larger average momentum; and the pair-efficiency
consequently increases.
The band around the 
$\epsilon^{\rm eID}_{\rm pair} \times \epsilon^{\rm live}_{\rm pair}$
curve in Fig.~\ref{fig:pair_effi} again shows only the possible distortion
of the mass distribution, due to the uncertainty shown in
Fig.~\ref{fig:single_effi}.  It reaches a maximum of $~$8\% around
$m_{ee}=0.4$~GeV/$c^2$.  

The $\epsilon^{\rm ghost}_{\rm pair}$ represents the loss of
real pairs which accidentally fulfill the detector overlap
criteria.  It is determined by the corresponding loss in mixed events as:
\begin{eqnarray}
{\epsilon}^{\rm ghost}_{\rm pair}
= \frac{ dB1_{ee}^2/(dm_{ee}dp_T)} { dB2_{ee}^2/(dm_{ee}dp_T) }
\end{eqnarray}
where $dB1$ ($dB2$) are the mixed-event unlike-sign pair (mass, $p_T$)
distribution with (without) applying the overlap pair cuts.

The $\epsilon^{\phi_V}_{\rm pair}$ represents the loss of real pairs
which are accidentally oriented like conversion pairs in the magnetic
field.  At low masses it is calculated using the $\pi^0$ GEANT
simulations described above.  Above the pion mass, we use our hadron
decay generator {\sc exodus} where we have implemented an empirical smearing
for the detector resolution in the determination of the magnitude and
direction of the momentum vector.  

\begin{figure}[!ht]
\includegraphics[width=1.0\linewidth]{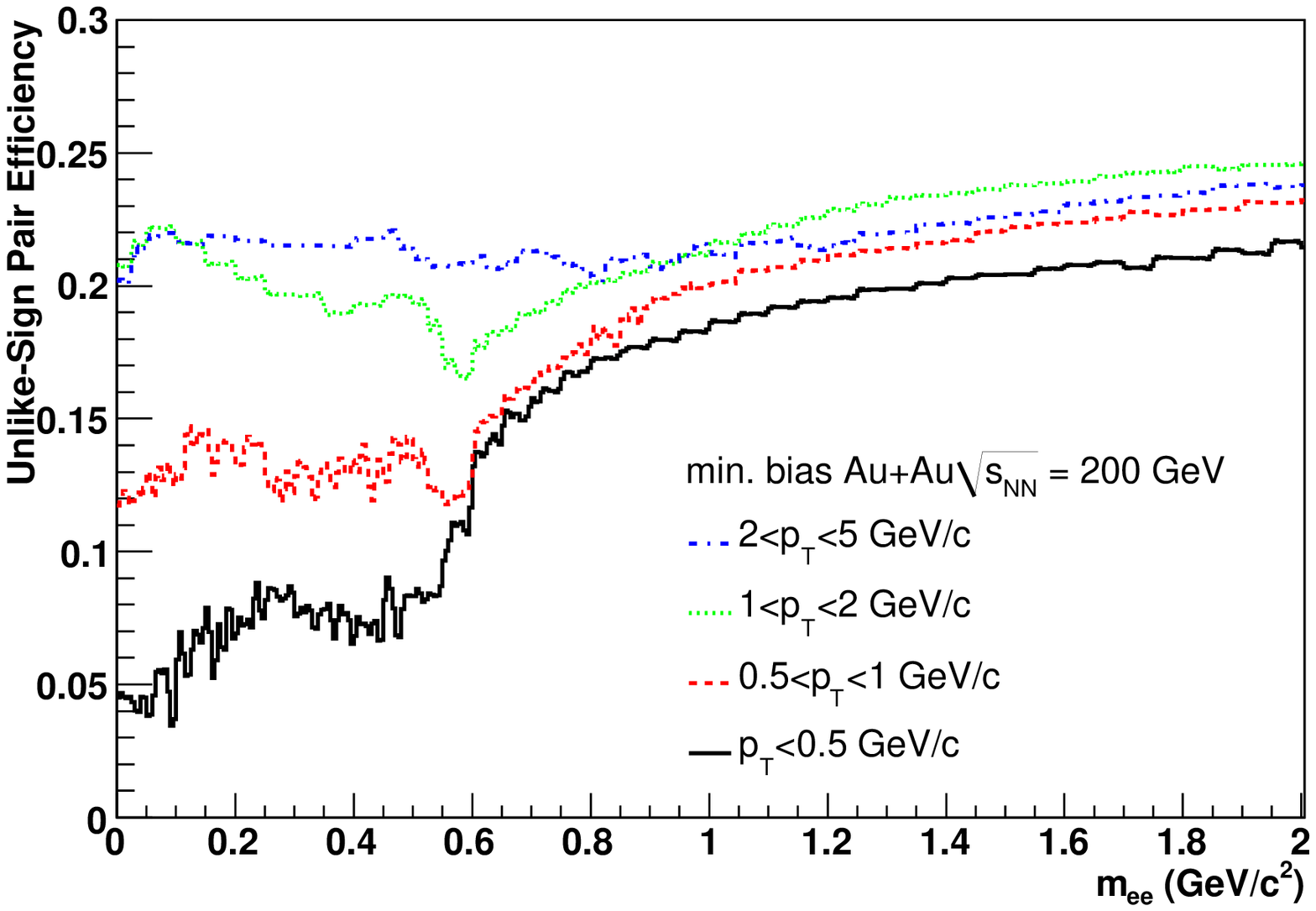}
\caption{ \label{fig:pair_effi_pt} (Color online)
$\epsilon^{\rm eID}_{\rm pair} \times \epsilon^{\rm live}_{\rm pair}$
as a function of invariant mass for different ranges of
pair-$p_T$ for Au~+~Au data set.}
\end{figure}

Figure~\ref{fig:pair_effi_pt} shows the total pair-efficiency as a 
function of invariant mass for different ranges of pair-$p_T$ for 
Au~+~Au collisions.

The systematic uncertainty on $\epsilon^{\rm eID}_{\rm pair} \times
\epsilon^{\rm live}_{\rm pair}$ is given by twice the uncertainty on
the single-particle efficiency, determined by varying the cut values.  In
addition, uncertainties in the active areas have been determined by
varying the active areas in the simulations.

The systematic uncertainties depend somewhat on $p_T$ and mass, they
are largest at low $p_T$ and low mass where they reach 14.4\% for
$p+p$ and 13.4\% for Au~+~Au.  We use these errors as an estimate for all
momenta and masses.  Since the pair-efficiency was derived using
realistic kinematics, the uncertainty arises mostly from low-$p_T$
tracks and neglecting the mass and pair-$p_T$ dependence is a
conservative approach.

The $\epsilon^{\rm ghost}_{\rm pair}$ and $\epsilon^{\phi_V}_{\rm pair}$ 
affect only the LMR.  Therefore, the corresponding systematic
uncertainties have been independently evaluated by varying the cut
values.  We estimate an uncertainty of 5\% due to the ghost cut and
6\% due to the photon conversion cut on the final $e^+e^-$ yield and
we apply it to the LMR by varying the cut value in a suitable range.

The (mass, $p_T$) distributions from every individual centrality bin
has been corrected using the same 2D efficiency correction function
calculated for Min. Bias.  This procedure avoids an assumption for the input
kinematic distribution.  However, since the individual centrality bins
have limited statistics compared to the Min. Bias sample, this procedure may
suffer statistical fluctuations.  Alternatively we use an effective
efficiency correction, as a function of mass only, obtained by
weighting the 2-dimensional (mass, $p_T$) corrections with a realistic
$p_T$ distribution provided by the minimum bias data set.  This curve
is given by the product of all the curves shown in
Fig.~\ref{fig:pair_effi}.  We take the difference of 10\% between the
1D and the 2D corrected result as an additional systematic uncertainty
for the efficiency correction of the different centrality data sets
(see Fig.~\ref{fig:ratio}).

We verified this efficiency using a GEANT simulation of~1 million
e+e- pairs.  Roughly half were generated flat in mass
(0--4\,~GeV/$c^2$), $p_T$ (0--4\,~GeV/$c$), azimuthal angle (0--$2\pi$),
and rapidity ($|y|<0.5$).  The other half were generated with a
probability inversely proportional to $p_T$ in order to enhance the
statistics in the low-mass and low-$p_T$ region, where the efficiency
varies most.  Only pairs with both an electron and a positron in the
ideal acceptance (given by Equation~\ref{eq:acc}) are processed by GEANT
and reconstructed with the same analysis chain.  The efficiency is
determined double differentially in $p_T$ and mass of the $e^+e^-$
pair.  This second method gives consistent results, but is limited by
MC statistics.  It was used as cross-check in the final analysis.

\subsection{Acceptance Correction}\label{sub:acc}
\begin{figure*}[t]
\includegraphics[width=1.0\linewidth]{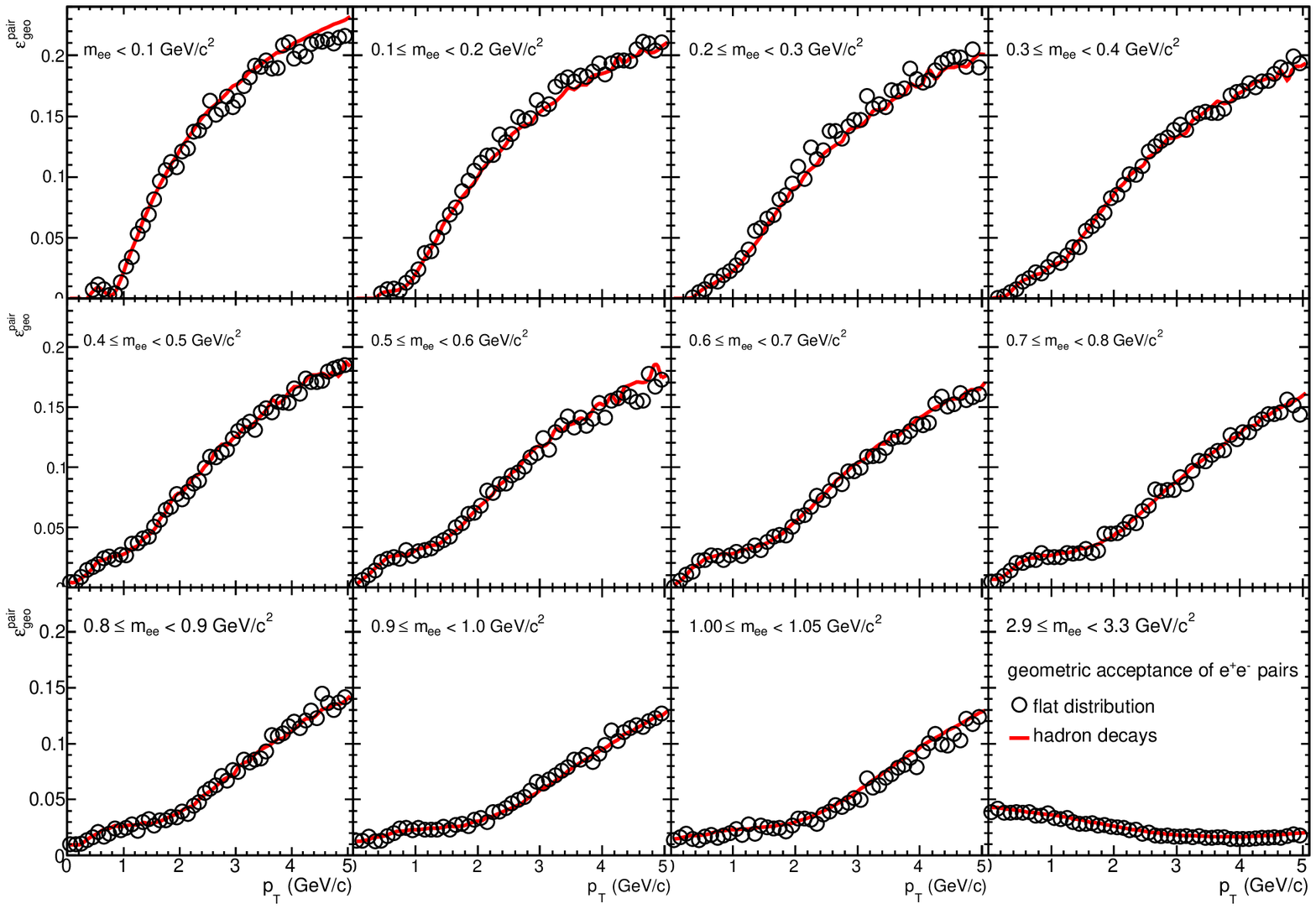}
\includegraphics[width=1.0\linewidth]{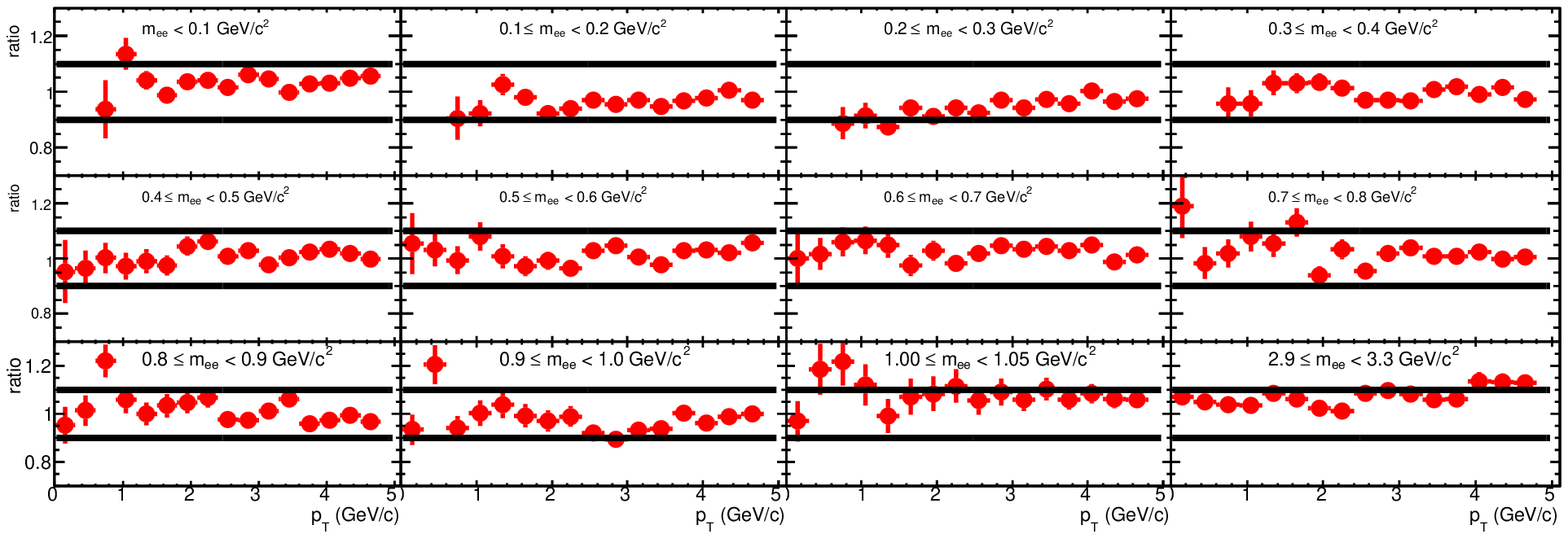}
\caption {\label{fig:acceptance} (Color online)
$e^+e^-$-pair acceptance as a function
of pair-$p_T$ for different mass ranges for the hadronic cocktail
(line) and a flat distribution of $e^+e^-$ pairs (empty circles) and ratio of the two
(bottom figure).  The lines mark the limits of the systematic uncertainty.}
\end{figure*}
In addition to the efficiency corrections, the $p_T$ spectra are also
corrected for the detector geometric acceptance
$\epsilon^{\rm geo}_{\rm pair}$, to give the $e^+e^-$--pair yield over the
full azimuth in one unit of rapidity.  $\epsilon^{\rm geo}_{\rm pair}$
accounts for the fraction of pairs produced in one unit of rapidity
over the full azimuthal range that are lost because either one or both
particles from the pair miss the PHENIX detector
\begin{equation} \label{eq:e_geo}
\epsilon^{\rm geo}_{\rm pair} = 
\frac{\frac{d^{2}N_{ee}}{dp_{T}^{ee}dm_{ee}} (\rm{in~PHENIX~acceptance}) }
{\frac{d^{2}N_{ee}}{dp_{T}^{ee}dm_{ee}} (|y_{ee}|<0.5) }
\end{equation}
$\epsilon^{\rm geo}_{\rm pair}$ has been calculated using a Monte Carlo
simulation of Dalitz decays of pseudoscalar mesons ($\pi^0, \eta,
\eta'$) and direct decays of vector mesons ($\rho, \omega, \phi$,
J/$\psi$, $\psi'$).  For all mesons the rapidity distribution is
assumed to be flat around mid-rapidity.  This assumption is well
justified as PHENIX measures in $|y_{ee}|<0.35$, where the natural
distribution is flat.  The acceptance is therefore 
uniform in $|y_{ee}|<0.35$ and we do not assign a systematic
uncertainty to it.  The $p_T$ distributions are taken from
PHENIX measurements (see Section~\ref{sec:cocktail}), and meson
polarizations are taken from~\cite{PDG}.
The acceptance correction is performed double differentially in
pair-mass and $p_T$ with 0.005~GeV/$c^2$ bins in mass and 0.1~GeV/$c$
bins in $p_T$.  

The systematic uncertainty due to the $p_T$ parameterization and the
polarization of the Dalitz pairs is studied with a simulation of
unpolarized pairs with a flat distribution in mass and pair-$p_T$.  In
Fig.~\ref{fig:acceptance} we compare the acceptance for the full
hadronic cocktail and the simulation of unpolarized pairs as a
function of pair-$p_T$ in 0.1~GeV/$c^2$ wide mass bins.  The cocktail
consists of the sum of polarized Dalitz decays and unpolarized vector
mesons, while the flat simulations are always unpolarized.  The shape
of the acceptance is very similar, and the relative normalization
agrees within 5\% in the lowest mass bins, and better for higher bins.
Based upon this comparison, illustrated by the ratio in the bottom
figure, we assign an upper limit of 10\% (marked by the lines in the
figure) for the systematic uncertainty of $\epsilon^{\rm geo}_{\rm
pair}$.

Since it arises from the independent fragmentation of two charm
quarks, the contribution of charmed meson decays has a different
acceptance.  This component has been simulated with 
{\sc pythia}~\cite{pythia} and
normalized according to the cross section measured in~\cite{ppg065}
scaled by $N_{\rm coll}$.  However, due to the observed modifications of
charm quarks in the medium~\cite{ppg066}, the acceptance could
potentially be different than what is simulated by {\sc pythia}.  For
$m_{ee}<0.5$~GeV/$c^2$ the charm contribution is negligible.
For $0.5<m_{ee}<1$~GeV/$c^2$ a systematic uncertainty of 5\% due to the
uncertainty of the charm cross section ($\sigma_{c\bar{c}}=N_{coll}
\times 567 \pm 57^{\rm stat} \pm 224^{\rm syst}\ \mu$b
~\cite{ppg065, ppg066}) has been added in
quadrature to the other systematic uncertainties on $\epsilon^{\rm geo}_{\rm pair}$.

\subsection{Trigger Efficiency ($p+p$)}\label{sub:ert}
\begin{figure}[!ht]
\includegraphics[width=1.0\linewidth]{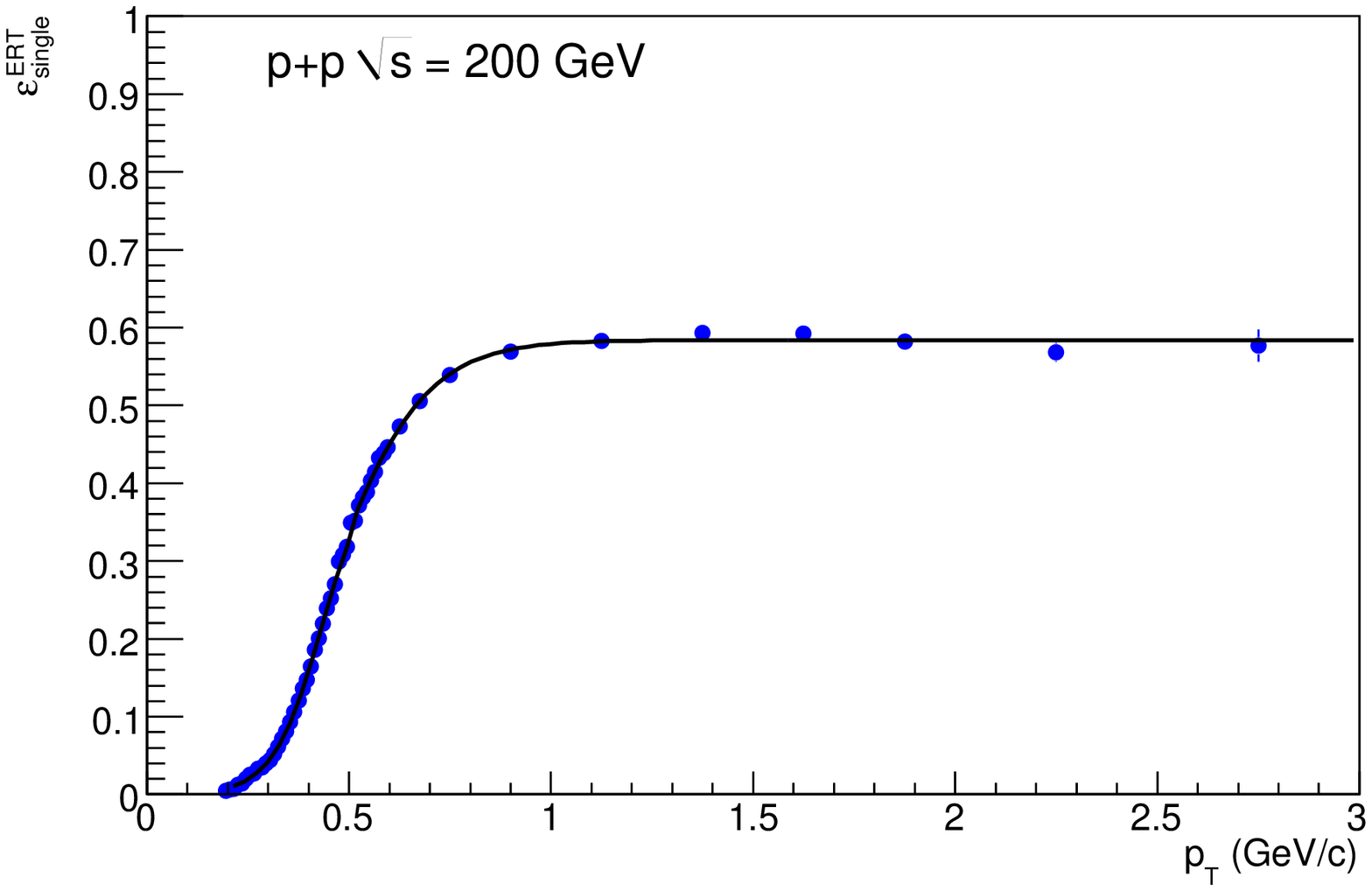}
\caption{ \label{fig:trig_eff} (Color online)
Trigger efficiency for single
electrons $\epsilon^{\rm ERT}_{\rm single}$ as a function of $p_T$ of
the ERT trigger in $p+p$ collisions determined from the Min. Bias data set.}
\end{figure}

The efficiency of the ERT trigger ($\epsilon^{\rm ERT}_{\rm pair}$) in $p+p$ collisions, as a function of pair
mass and $p_T$, is determined with a fast Monte Carlo simulation of pairs
which uses a parameterization of the single electron ERT efficiency $\epsilon^{\rm ERT}_{\rm single}$.
The single electron ERT efficiency is determined using the Min. Bias data
set.  The online level-1 trigger decision is recorded in the Min. Bias 
data set even though it is not used to select the event.  We require that 
the trigger tile that fires the ERT trigger is used by the
electron candidate selected by the offline analysis.  The ratio of
triggered electrons relative to all electrons candidates gives the trigger
efficiency as function of the single electron $p_T$ and is shown in
Fig.~\ref{fig:trig_eff}.  
The trigger efficiency at the plateau is $\approx$ 60\%, consistent
with the fraction of the active trigger tiles.  The trigger part and
the offline part of the RICH and EMCAL read-out are handled in 
separate electronics chains, and the trigger part has more noisy tiles
which are masked out.  This results in less active areas in the
trigger.  The efficiency is fit to the sum of two Fermi functions
\begin{equation}\label{eq:fermi}
f(p_T) = \frac{\epsilon_0 \cdot \theta(p_T-0.5)}{e^{-(p_T - p_0)/k} + 1} +
\frac{\epsilon_0' \cdot \theta(p_T+0.5)}{e^{-(p_T - p_0')/k'} + 1}
\end{equation}
where $\epsilon_0$ ($\epsilon_0'$), $p_0$ ($p_0'$) and $k$ ($k'$) are
free fit parameters and $\theta$ is the usual Heaviside theta function.

\begin{figure}[!ht]
\includegraphics[width=1.0\linewidth]{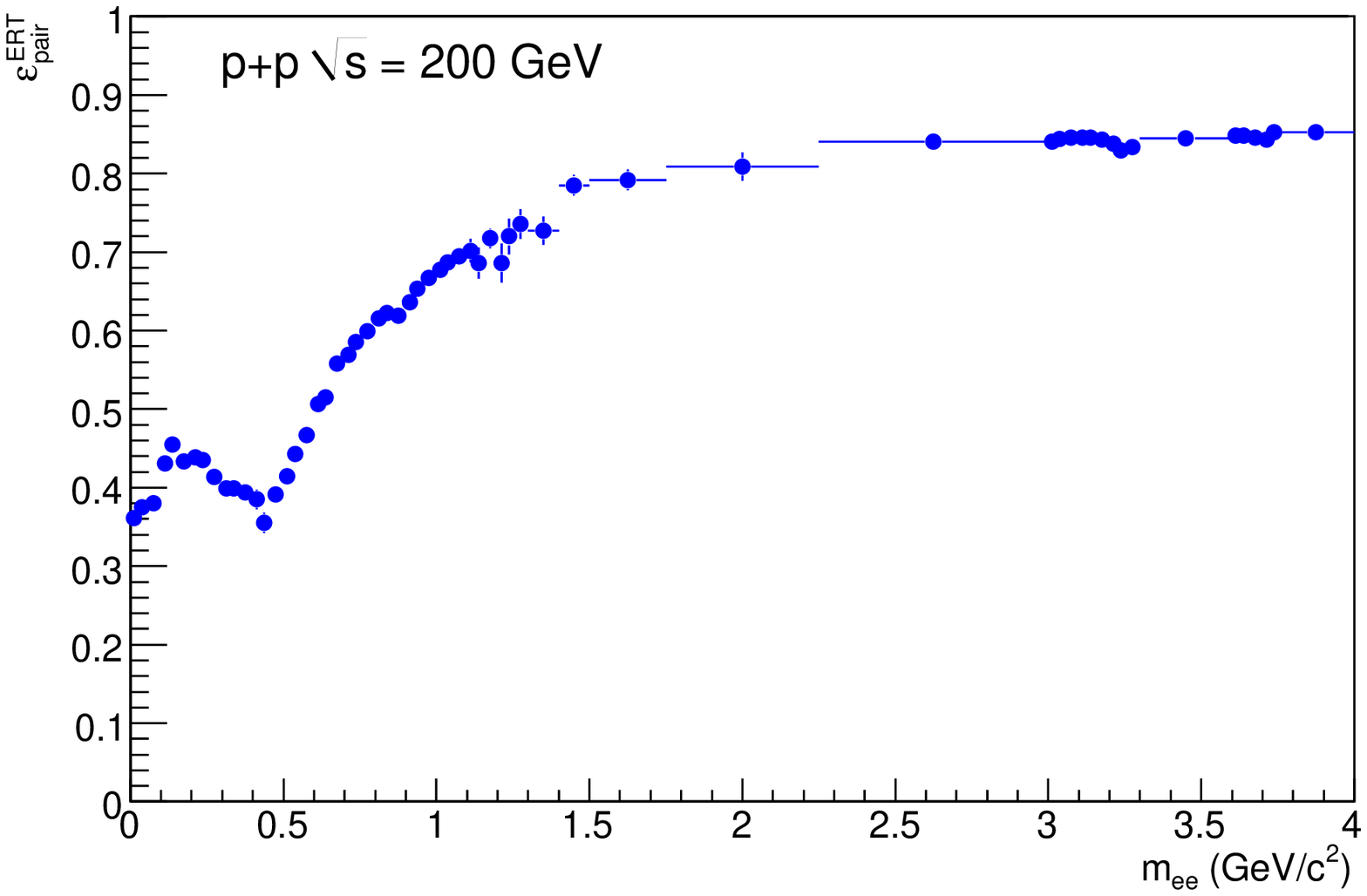}
\caption{  \label{fig:trig_eff_pair} (Color online)
Trigger efficiency $\epsilon^{\rm ERT}_{\rm pair}$ for $e^+e^-$-pairs as a function of $e^+e^-$-pair invariant
mass.}
\end{figure}
This parameterization is used in a fast Monte Carlo simulation of
hadron decays into $e^+e^-$.  Each electron fires the trigger with a
probability given by the trigger efficiency.  We require that both
electrons are within the PHENIX acceptance and at least one of them
fires the trigger.  The mass and $p_T$ distribution of those pairs is
compared to the one obtained with no trigger requirement.  The ratio is
the pair trigger efficiency, and is shown in
Fig.~\ref{fig:trig_eff_pair} as a function of $e^+e^-$ invariant mass.
The structures in Fig.~\ref{fig:trig_eff_pair} result from the turn-on
curve of the trigger threshold convoluted with the acceptance of the
detector.  

The systematic uncertainty has been studied by varying the
parameterization of the single electron ERT efficiency, as well as
varying the eID cuts which define the reference sample.
The effect of changing the eID cuts leads to a larger
uncertainty at low-$p_T$ and low-masses.  We therefore assign a
systematic uncertainty of 20\% for $m_T <$ 1~GeV/$c^2$ and 5\% elsewhere.
Finally, the systematic uncertainty includes potential shape distortion
to the efficiency correction due to the variation of active
trigger tiles during the data collection.

\begin{figure}[!ht]
\includegraphics[width=1.0\linewidth]{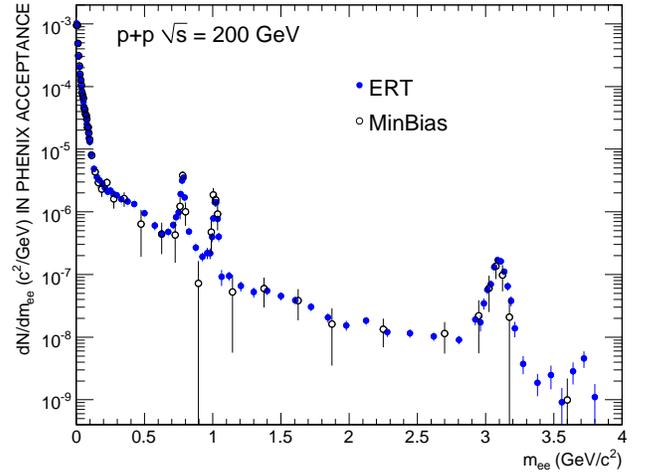}
\caption{  \label{fig:ertmb} (Color online)
Invariant mass spectra for $p+p$ data with the Min. Bias (hollow) and ERT (solid) data sets.  The
agreement between the two data sets is excellent.}
\end{figure}

Fig.~\ref{fig:ertmb} compares the invariant mass spectra for the
$p+p$ data obtained with the Min. Bias and the ERT data sets.  The ERT data set has
been corrected by ERT trigger efficiency
$\epsilon^{\rm ERT}_{\rm pair}$ and the total pair reconstruction
efficiency $\epsilon^{\rm total}_{\rm pair}$, while the
Min. Bias is corrected only for the $\epsilon^{\rm total}_{\rm pair}$.  
This comparison of the two data sets confirms that the Min. Bias and 
the ERT agree within their respective statistical errors.

\subsection{Occupancy Correction (Au~+~Au)}\label{sub:embedding}
In Au~+~Au collisions there is an additional efficiency loss of particle
detection due to the presence of other particles nearby.  To study
this effect single electrons and positrons are simulated through the
GEANT simulator of PHENIX and then embedded into data files containing
detector hits from real Au~+~Au events.  Next, these new files
containing the embedded $e^\pm$ are run through the entire
reconstruction software.  As the particle density reduces the
efficiency but does not introduce additional $p_T$ dependence, the
occupancy correction can be factored out.  Since all the detectors
used in the analysis are located after the pair has been opened by the
magnetic field, the pair embedding efficiency in each centrality bin
is defined as the square of the single electron embedding efficiency
\begin{eqnarray*}
\epsilon^{\textrm{occ}}_{\rm pair} &=&
(\epsilon^{\textrm{occ}}_{\rm single})^2 \\
&=& \left(\frac{\# \ \textrm{reconstructed} \ e^\pm
\ \textrm{from embedded data}}{\# \ \textrm{reconstructed} \ e^\pm \
\textrm{from single track data}}\right)^2
\end{eqnarray*}
where a reconstructed particle from embedded data is required to have most of its
detector hits associated with hits from the simulated particle.

\begin{table}[htbp]
\caption{Embedding efficiency for different centrality classes
of Au~+~Au collisions.~\label{tab:embedding}}
\begin{ruledtabular}    \begin{tabular}{ccccc}
&   Centrality   &  $\epsilon^{\textrm{occ}}_{\rm single}$ &
$\epsilon^{\textrm{occ}}_{\rm pair}$ & \\
\hline
&   0-10\%       &  0.86 & 0.74 & \\
&   10-20\%      &  0.91 & 0.83 & \\
&   20-40\%      &  0.93 & 0.87 & \\
&   40-60\%      &  0.97 & 0.95 & \\
&   60-92\%      &  0.99 & 0.98 & \\
\end{tabular} \end{ruledtabular}  
\end{table}

Table~\ref{tab:embedding} displays the embedding efficiencies for the
centrality classes used in the analysis.  For the minimum bias results
we have weighted the occupancy correction by the fraction of pairs in
each centrality class.  Since most of the yield is concentrated in the
most central classes, the resulting pair efficiency is 0.81 instead of
the square of the minimum bias value $0.96^2=0.92$.

A second, data-driven method was developed to determine
$\epsilon^{\textrm{occ}}_{\rm pair}$.  This method uses the conversion
pairs created at the beam-pipe to select a track sample of pure $e^+$
or $e^-$ from real data.  The conversion pairs originating at the
beam-pipe result in a clear invariant mass peak around $m_{ee}\sim
20$~MeV/$c^2$ (see Fig.~\ref{fig:gamma_spectrum}).  We assign tight eID
cuts to only one track of a pair and measure the efficiency of the
other track of the same pair without any eID cut applied.  Differences
between the two methods are accounted for in a 3\% systematic
uncertainty \cite{ppg077}.

\subsection{Systematic Uncertainty} \label{sub:systematic}
The systematic uncertainties are summarized in Table~\ref{tab:errors}.
The uncertainties in Table~\ref{tab:errors} are categorized by type:
\begin{itemize}
\item A: point-to-point uncertainty uncorrelated between mass or $p_T$ bins,
\item B: correlated uncertainty, all points move in the same direction but
not by the same factor,
\item C: an overall normalization uncertainty in which all points move
by the same factor independent of mass or $p_T$.
\end{itemize}
The uncertainty on pair reconstruction efficiency includes eID cuts,
geometric acceptance, and run-by-run fluctuations.  The uncertainty
due to the conversion rejection cut and the overlap cuts as well as
the uncertainty due to the ERT and minimum bias trigger efficiency
(for $p+p$) and occupancy (for Au~+~Au) are also listed.
\begin{table*}[th]
\caption{\label{tab:errors}
Systematic uncertainties on the dilepton yield and mass range of
applicability.}
\begin{ruledtabular}  \begin{tabular}{ccccc}
Syst.  Err.  component       & $p+p$      & Au~+~Au  & Mass Range & Type\\
\hline
pair reconstruction        & 14.4\% &  13.4\%  & 0--4~GeV/$c^2$ & B\\
conversion rejection       & 6\%    &   6\%    & 0--0.6~GeV/$c^2$ & B\\
pair cuts                  & 5\%    &   5\%    & 0.4--0.6~GeV/$c^2$ & B\\
occupancy efficiency       & -      &   3\%    & 0--4~GeV/$c^2$ & C\\
BBC and trigger bias       & 11.3\% &   -      & 0--4~GeV/$c^2$ & C\\
ERT efficiency             & 5\% (20\% for $m_T<$1~GeV/$c^2$)    &   -
& 0--4~GeV/$c^2$ & B\\
combinatorial background   & 3\%$\cdot B/S$&   0.25\%$\cdot B/S$   &
0--4~GeV/$c^2$ & B\\
correlated background & mass-dependent (Fig.~\ref{fig:corrBG})  &
mass-dependent (Fig.~\ref{fig:corrBG})  &0--4~GeV/$c^2$ & B\\
centrality                 & -   &  10\%    & 0--4~GeV/$c^2$ &B\\
acceptance correction      & 10\%&  10\%    & 0--4~GeV/$c^2$ & B\\
charm acceptance           & 5\% &   5\%    & $>$0.5~GeV/$c^2$ & B\\
\end{tabular} \end{ruledtabular}  
\end{table*}
These uncertainties are included in the final systematic uncertainty on the invariant $e^+e^-$--pair 
yield.  The uncertainties deriving from reconstruction and occupancy do not have a
strong $p_T$ dependence, so we keep the assigned values for every
$p_T$ bin.  The pair cuts and the conversion rejection are localized in mass
($m<0.6$~GeV/$c^2$) and are rather $p_T$-independent.  The uncertainty on the
combinatorial background is the largest contribution to the systematic
uncertainty in Au~+~Au and the value is estimated to be 0.25\%
$\cdot B/S$.  $S/B$ rises with $p_T$,
therefore the uncertainty on the combinatorial background has been
propagated separately for each $p_T$ bin.  The uncertainty on the
correlated background is approximately 2-3\% in $p+p$ and 10\% in
Au~+~Au for $m_{ee}<0.6$~GeV/$c^2$ and $p_T$-independent.  However it
increases in $p+p$ towards high masses.  In the $p+p$ data the
uncertainty on the ERT trigger efficiency, as well as on the BBC and
trigger bias is included.  For individual centrality bins, we add
10\% uncertainty arising from the $p_T$ dependence of the efficiency
correction: this was obtained from the difference between the
1D-corrected mass spectra and the 2D-corrected mass spectra.  For the
$p_T$ spectra, which are further corrected by the geometric acceptance,
we added 10\% uncertainty from the acceptance correction and 5\% for
$m_{ee}>0.5$~GeV/$c^2$ due to the charm contribution.
Most of these uncertainties are mass-$p_T$ correlated, i.e.  all points
move in the same direction, but not by the same factor.  Only the BBC
and trigger bias (in $p+p$) and the occupancy (in Au~+~Au) are
normalization uncertainties in which all points move by the same
factor independent of mass and $p_T$.  Since those uncertainties are
small compared to the total uncertainty, they are included in the
total uncertainty, without plotting it separately.

\section{COCKTAIL OF HADRONIC SOURCES} \label{sec:cocktail}
In this Section we describe the methods used to calculate the pair yield
expected from hadronic decays which will be compared to the
experimental data.

\begin{figure*}
\includegraphics[width=0.45\linewidth]{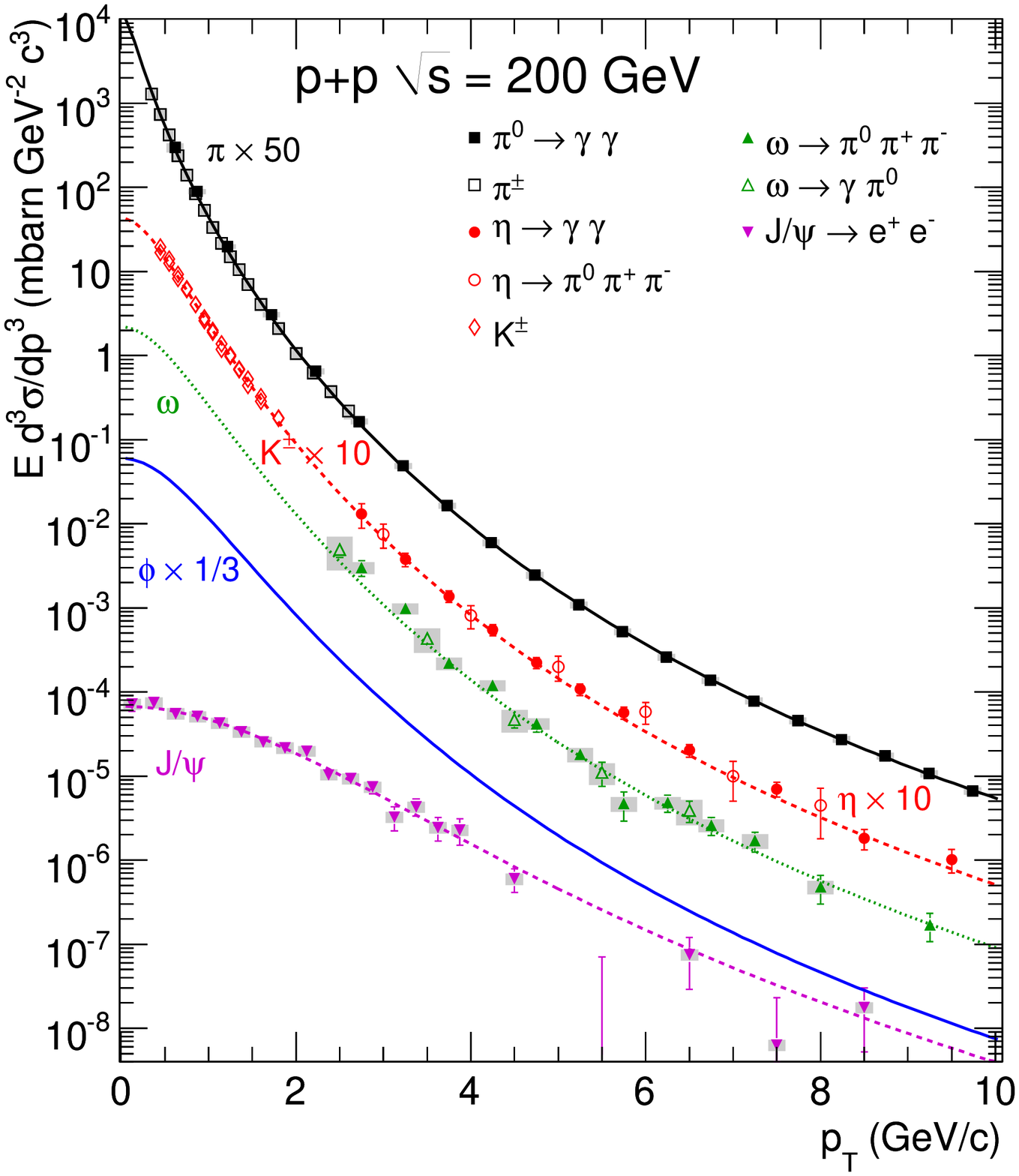}
\includegraphics[width=0.45\linewidth]{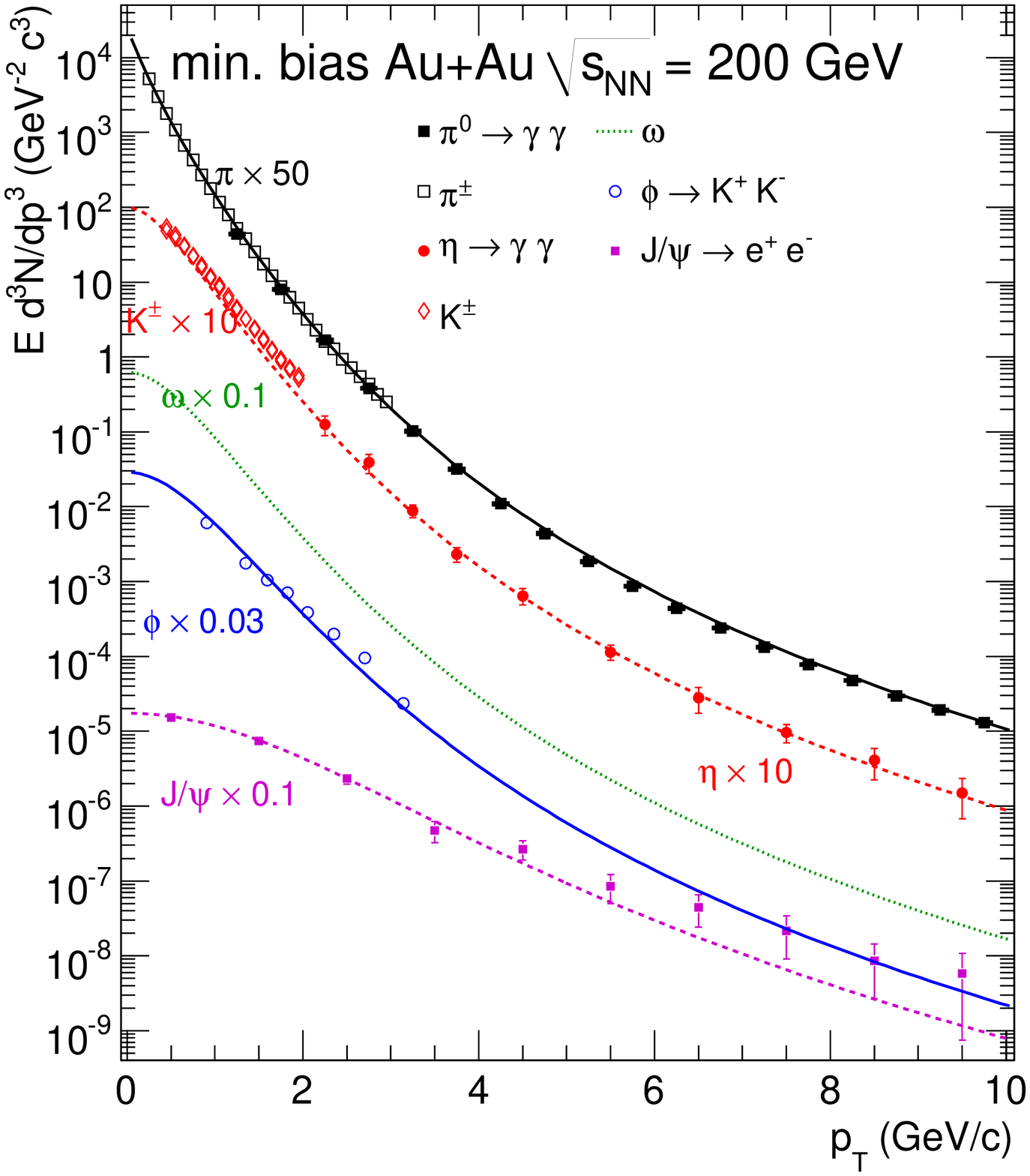}
\caption {\label{fig:mtfits} (Color online)
Compilation of meson production cross sections in $p+p$ (left) and
Au~+~Au (right) collisions at $\sqrt{s_{NN}}$~=~200~GeV.  Shown for $p+p$ are data
for neutral~\cite{pi0} and charged pions~\cite{pikp}, $\eta$
~\cite{eta}, kaons~\cite{pikp}, $\omega$~\cite{omega}, 
$\phi$~\cite{phi}, and $J/\psi$~\cite{jpsi}.  Shown for Au~+~Au are data
for neutral~\cite{AUpi0} and charged pions~\cite{AUpikp}, $\eta$
~\cite{eta}, kaons~\cite{AUpikp}, $\omega$,
$\phi$~\cite{AUphi}, and $J/\psi$~\cite{AUjpsi}.  The data are
compared to the parameterization based on $m_T$ scaling used in our
hadron decay generator.}
\end{figure*}

We model the $e^+e^-$ pair contributions from hadron decays using the
decay generator {\sc exodus}.  {\sc exodus} is a phenomenological event generator
that allows to simulate the phase-space distribution of all relevant
sources of electrons and electron pairs and the decay of these
sources.  Also it allows to include the filtering for the geometrical
acceptance and the detector resolution.  The relevant primary mesons
that involve electrons in the final state are $\pi^0$, $\eta$,
$\eta'$, $\rho$, $\omega$, $\phi$, $J/\psi$ and $\psi'$.

We assume that all hadrons have a constant rapidity density in the
range $\left|\Delta\eta\right|\le 0.35$ and a uniform distribution in
azimuthal angle.  Transverse momentum distributions are largely based
on measurements in PHENIX.  The key input is the rapidity density
$dN/dy$ of neutral pions, which we determine from a fit to PHENIX data
on charged and neutral pions~\cite{pi0, pikp, AUpi0, AUpikp} with a
modified Hagedorn function, given by:
\begin{equation}
\label{eq:hagedorn}
E\frac{d^3\sigma}{dp^3} = A ( e^{-(a p_T + b p_T^2)} + p_T/p_0)^{-n}
\end{equation}
Fit parameters and $dN/dy$ for $p+p$ and Au~+~Au are given in Table
~\ref{tab:exo_pion}.
\begin{table}[th]
\caption{\label{tab:exo_pion}
Fit parameters from the modified Hagedorn function (Equation~\ref{eq:hagedorn})
for $p+p$ and Au~+~Au pion spectra ($\pi^0$ and $\pi^{\pm}$) and the corresponding rapidity density $dN/dy$.
}
\begin{ruledtabular}  \begin{tabular}{ccccc}
& Parameter & $p+p$ & Au~+~Au & \\
\hline
& $dN/dy$                       & 1.06$\pm$0.11   & 95.7$\pm$6.9   & \\
& $A$[mb~GeV$^{-2}c^3$]         & 377$\pm$60      & 504.5$\pm$10   & \\
& $a$[(GeV/$c$)$^{-1}$]         & 0.356$\pm$0.014 & 0.52$\pm$0.007 & \\
& $b$[(GeV/$c$)$^{-2}$]         & 0.068$\pm$0.019 & 0.16$\pm$0.010 & \\
& $p_{0}$[GeV/$c$]            & 0.7$\pm$0.02    & 0.7$\pm$0.005  & \\
& $n$	                    & 8.25$\pm$0.04   & 8.27$\pm$0.02    & \\
\end{tabular} \end{ruledtabular}  
\end{table}

For all other mesons we assume $m_T$ scaling, replacing $p_T$ by 
$\sqrt{m^2-m_{\pi}^2+(p_T/c)^2}$, where $m$ is the mass of the meson 
and we fit a normalization factor to PHENIX data~\cite{pi0, pikp, eta, 
omega, phi, etaprime, jpsi, AUpi0, AUpikp, AUomega, AUphi, AUjpsi}, 
where available.  Figure~\ref{fig:mtfits} shows the excellent agreement 
with published PHENIX data.  The $\eta$ meson is measured only at 
higher $p_T$; however, in the $p+p$ collisions the fit is in good 
agreement with the $p_T$ distribution of kaons, which have similar 
mass (see discussion below).

In order to extract the meson yield we integrate the fits over all
$p_T$.  Results, systematic uncertainties, and references to data are
given in Table~\ref{tab:cocktail} and the ratio of the integrated
yields $meson/\pi^0$ are compared for $p+p$ and Au~+~Au data.  For the
$\rho$ meson we assume $\sigma_\rho/\sigma_\omega=1.15\pm0.15$,
consistent with values found in jet fragmentation~\cite{PDG}.  The
$\eta'$ yield is scaled to be consistent with jet fragmentation
$\sigma_{\eta'}/\sigma_\eta=0.15\pm0.15$~\cite{PDG}.  The $\psi'$ is
adjusted to be $\sigma_{\psi'}/\sigma_{J/\psi}=0.14\pm0.03$
~\cite{psiprime} in agreement with PHENIX measurements~\cite{psiprimePP}.  
\begin{table*}[th]
\caption{\label{tab:cocktail} Hadron rapidity densities used in our
hadron decay generator.  For the $\omega$ and $\phi$, data from
this analysis were used together with data from the quoted
references.}
\begin{ruledtabular}  \begin{tabular}{ccccc}
& $\frac{dN}{dy}\rvert_{y=0}$  
	       & relative uncertainty.  
	       & meson/$\pi^0$ 
	       & data used \\
\hline \\
&& $p+p$ && \\
$\pi^0$        & $1.065\pm0.11$     &   10\%        & 1.0 &  PHENIX~\cite{pi0},~\cite{pikp}\\
$\eta$         & $(1.1\pm0.3) \times 10^{-1}$      &   30\%      & $1.032\times 10^{-1}$ &  PHENIX~\cite{eta} \\
$\rho$         & $(8.9\pm2.5) \times 10^{-2}$    &   28\%        & $8.34\times 10^{-2}$ &  jet fragmentation~\cite{PDG} \\
$\omega$       & $(7.8\pm1.8) \times 10^{-2}$    &   23\%        & $7.32\times 10^{-2}$ &  PHENIX~\cite{omega} \\
$\phi$         & $(9.0\pm2.0) \times 10^{-3}$    &   24\%        & $8.4\times 10^{-3}$ &  PHENIX~\cite{phi}   \\
$\eta'$        & $(1.3\pm0.5) \times 10^{-2}$    &   40\%        & $1.27\times 10^{-2}$ &  PHENIX~\cite{etaprime} \\
$J/\psi$       & $(1.77\pm0.27) \times 10^{-5}$ & 15\% & $1.66\times 10^{-5}$ & PHENIX~\cite{jpsi} \\
$\psi'$        & $(2.5\pm0.7) \times 10^{-6}$   &   27\% & $2.3\times 10^{-6}$ & PHENIX~\cite{psiprime},~\cite{psiprimePP} \\
\\
&& Au~+~Au && \\
$\pi^0$        & $(9.572\pm0.95) \times 10$   &   10\%        & 1.0     &  PHENIX~\cite{AUpi0},~\cite{AUpikp} \\
$\eta$         & $(1.077\pm0.32) \times 10$    &   30\%        &
$1.12\times 10^{-1}$  & PHENIX~\cite{eta} \\
$\rho$         & $8.60\pm2.8$    &   33\%        & $8.98\times 10^{-2}$ & jet fragmentation~\cite{PDG} \\
$\omega$       & $9.88\pm3.0$    &   30\%        & $1.03\times 10^{-1}$  & PHENIX~\cite{AUomega} \\
$\phi$         & $2.05\pm0.6$    &   30\%        & $2.14\times 10^{-2}$  & PHENIX~\cite{AUphi} \\
$\eta'$        & $2.05\pm0.2$    &  100\%        & $2.15\times 10^{-2}$ & PHENIX~\cite{etaprime}, and~\cite{PDG} \\
$J/\psi$       & $(1.79\pm0.26) \times 10^{-3}$  & 15\%  & $1.82\times 10^{-5}$ & PHENIX~\cite{AUjpsi} \\
$\psi'$        & $(2.6\pm0.7 ) \times 10^{-4}$   & 27\%  & $2.70\times 10^{-6}$& PHENIX~\cite{psiprime}, and~\cite{psiprimePP}\\

\end{tabular} \end{ruledtabular}  
\end{table*}

For the $\eta$, $\omega$, $\phi$, and $J/\psi$, and also $\eta'$, and
$\psi'$ (in $p+p$) the quoted uncertainties include those on the data
as well as those using different shapes of the $p_T$ distributions to
extrapolate to zero $p_T$.  Specifically we have fit the functional
form given in Equation~\ref{eq:hagedorn} with all parameters free and also
with an exponential distribution in $m_T$.  For the $\rho$, which is
not measured in $p+p$ nor in Au~+~Au, and $\eta'$ and $\psi'$, which are
not measured in Au~+~Au, the uncertainty in the table represent the
quadrature sum of the uncertainty of the cross section and the
uncertainty relative to other mesons.

All the mesons shown in Fig.~\ref{fig:mtfits} can be described by the
$m_T$-scaling parameterization of the pion spectrum.  The fact that
the $\eta$'s and the kaons follow the same $m_T$-scaling prediction over
all $p_T$ appears to be due to the fact that the masses of the
particles are almost the same.  In Au~+~Au however $\eta$ and kaons do
not follow the same $m_T$-scaling prediction.  At high $p_T$, where we
measure $\eta$'s, we see that they are suppressed as much as pions and
the trend of $R_{\rm AA}$ for these two mesons looks identical.  However we have
observed a different trend (i.e.  a smaller suppression) for strange
particles and $\eta$ has a strangeness content too
~\cite{milov,naglis}.  Therefore, since the $\eta$ cannot be measured at
low-$p_T$, we take as systematic uncertainty in the low-$p_T$ region
(and consequently on the extrapolated $dN/dy$), the difference of the
two spectra ($\sim$ 30\%) in the low-$p_T$ range.  We note that this is
a conservative estimate of the systematic uncertainty.  Statistical
models which reproduce well the particle spectra and ratios measured
at RHIC~\cite{pbm,bec} calculate a $dN/dy$ for the $\eta$ which is well
within the assigned systematic uncertainty.  At high-$p_T$ we assign a
smaller systematic uncertainty of 7\% (17\%) for $p+p$ (Au~+~Au)
collisions arising from the asymptotic value of the $\eta/\pi^0$ ratio
of 0.48 $\pm$ 0.03 (0.08) based on PHENIX measurement~\cite{eta}.

Once the meson yields and $p_T$ spectra are known the dilepton spectrum
is given by decay kinematics and branching ratios, which are
implemented in our decay generator {\sc exodus} following earlier work
published in~\cite{ppg065}.  The branching ratios are taken from
the compilation of particle properties in~\cite{PDG}.  For the Dalitz
decays $\pi^0$, $\eta$, $\eta'\rightarrow e^+e^-\gamma$ and the decay
$\omega\rightarrow e^+e^-\pi^0$ we use the Kroll-Wada expression
~\cite{kroll-wada} with electromagnetic transition form factors
measured by the Lepton-G collaboration~\cite{lepton-g,landsberg}.  For
the decays of the vector mesons $\rho$, $\omega$, $\phi\rightarrow
e^+e^-$ we use the expression derived by Gounaris and Sakurai
~\cite{gounaris-sakurai}, extending it to 2~GeV/$c^2$, slightly beyond its
validity range.  For the $J/\psi$ and $\psi'\rightarrow e^+e^-$ we use
the same expression discussed in~\cite{jpsi} modified to include radiative corrections.
All vector mesons are assumed to be unpolarized.  For the Dalitz decays,
where the third body is a photon, the angular distribution is sampled
according to $1+\cos^2\theta_{CS}$, where $\theta_{CS}$
is the polar angle of the electrons in the Collins-Soper frame.

The resulting systematic uncertainties on the mass spectrum depend on
mass, and range from 10 to 30\%.  They result primarily from the uncertainty on
the measured pion yield and on the meson-to-pion ratios.  
The uncertainty from the measured
electromagnetic transition form factors, in particular for the
$\omega \rightarrow e^+e^- \pi^0$ decay, is also included but
contributes significantly only in the range around 0.5 to 0.6~GeV/$c^2$.
The uncertainty from polarization is negligible but also included.

\section{RESULTS} \label{sec:results}
This Section presents the results for the $p+p$ and the Au~+~Au
analyses.  The $p+p$ data provide a good baseline for understanding
the results of the Au~+~Au analysis presented in this Section.  This
Section is organized as follows.  Each subsection concentrates on a
different region of the ($m_{ee}, p_T$) phase-space.  For each region
we will present the results for the $p+p$ and for the Au~+~Au data.  In
Section~\ref{sub:mass} we will show the inclusive mass spectrum for
$p+p$ and minimum bias Au~+~Au collisions and we will compare it with
the cocktail.  In Section~\ref{sub:charm} we discuss the results in
the IMR by comparing the data with the charm expectations from
($N_{\rm coll}$-scaled) {\sc pythia}~\cite{pythia}.  In 
Section~\ref{sub:LMR_enh} we
present the yields in the LMR and discuss their centrality dependence.
Section~\ref{sub:mass_pt} discusses the $p_T$ dependence of the mass
spectra.  Part of the LMR, denoted with LMR~I, where the $p_T$ of the
$e^+e^-$ pair is much larger than its mass, is the region of
quasi-real virtual photons: we present the results in this region for
$p+p$ and Au~+~Au data in Section~\ref{sub:photon}.
Section~\ref{sub:pt_lowmass} discusses what we can learn from the
measurement of direct photons (in LMR~I) about the yields in the
$p_T$-inclusive LMR.  Finally Section~\ref{sub:ptspectra} presents the
$p_T$ spectra for different mass bins for $p+p$ and Min. Bias Au~+~Au 
data and compares them with the expectations from cocktail plus charm plus
direct photons.

\subsection{$p_T$-inclusive Mass Spectra} \label{sub:mass}
\begin{figure}[tb]
\includegraphics[width=1.0\linewidth]{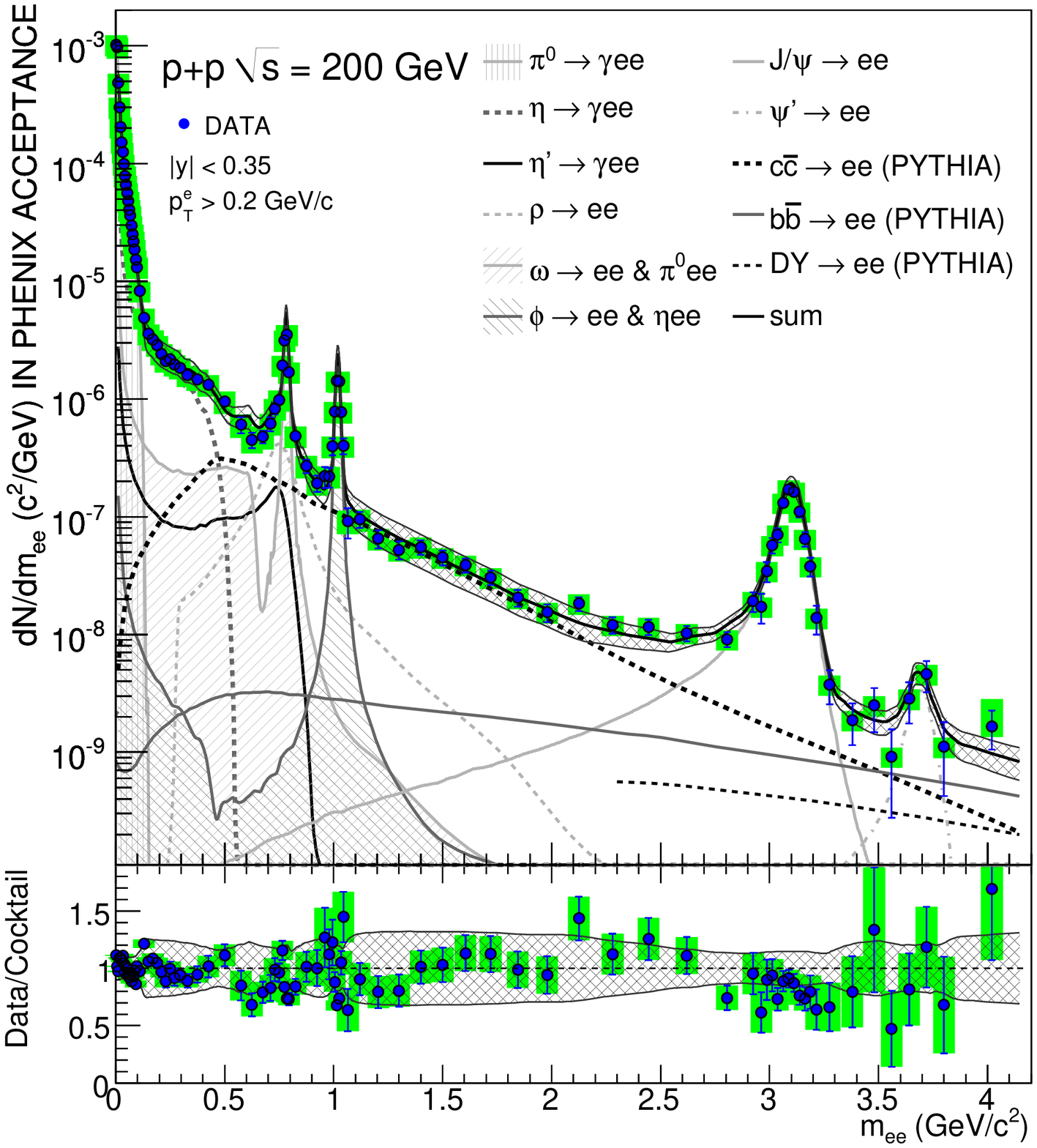}
\caption { \label{fig:pp_cock} (Color online)
Inclusive mass spectrum of $e^+e^-$
pairs in the PHENIX acceptance in $p+p$ collisions compared to the
expectations from the decays of light hadrons and correlated decays
of charm, bottom, and Drell-Yan.  The contribution from hadron decays
is independently normalized based on meson measurements in PHENIX.
The bottom panel shows the ratio of data to the cocktail of known
sources.  The systematic uncertainties of the data are shown as
boxes, while the uncertainty on the cocktail is shown as band around
1.}
\end{figure}
Figure~\ref{fig:pp_cock} compares the yield of $e^+e^-$ pairs in the PHENIX
acceptance in $p+p$ data to the expected yield from the contributions
of the cocktail and shows the various sources of the cocktail (hadron 
decays, charm, bottom and Drell-Yan pairs).  
The $p+p$ data are very well described by the
expectation from the hadronic cocktail and heavy flavor decays for the
entire mass range within the uncertainty of the data and the cocktail.

\begin{figure}[htb]
\includegraphics[width=1.0\linewidth]{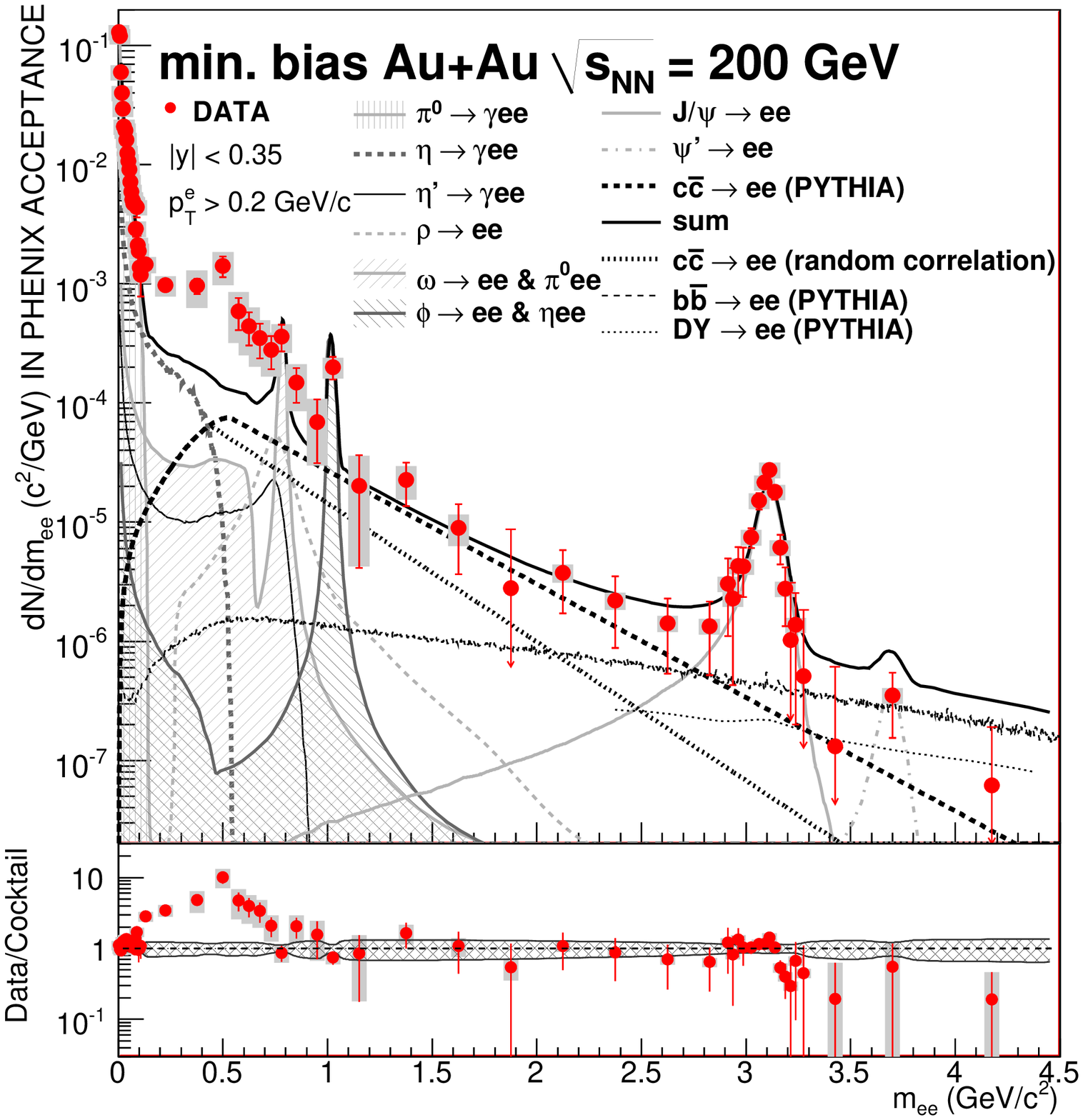}
\caption { \label{fig:AuAu_cock} (Color online)
Inclusive mass spectrum of $e^+e^-$ pairs in the PHENIX acceptance in
minimum-bias Au~+~Au compared to expectations from the decays
of light hadrons and correlated decays of charm, bottom and Drell-Yan.
The charm contribution expected if the dynamic correlation of
$c$ and $\bar{c}$ is removed is shown separately.  Statistical
(bars) and systematic (boxes) uncertainties are shown separately.
The contribution from hadron decays is
independently normalized based on meson measurements in PHENIX.  The
bottom panel shows the ratio of data to the cocktail of known
sources.  The systematic uncertainties of the data are shown as boxes,
while the uncertainty on the cocktail is shown as band around 1.  }
\end{figure}
\begin{figure}[tb]
\includegraphics[width=1.0\linewidth]{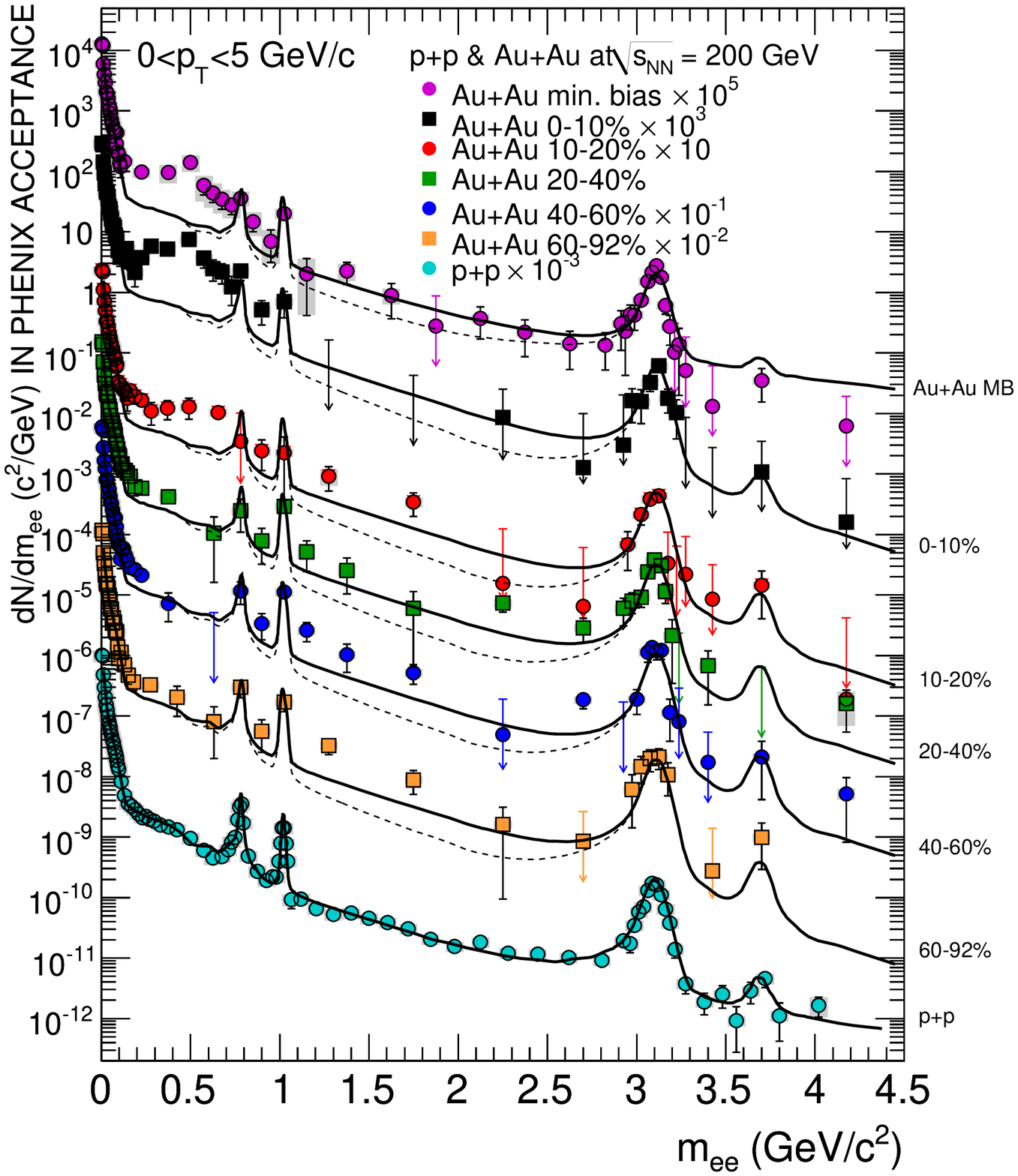}
\caption{\label{fig:mass_cent} (Color online)
Invariant mass spectrum of $e^+e^-$
pairs inclusive in $p_T$ compared to expectations from the model of
hadron decays for $p+p$ and for different Au~+~Au centrality
classes.  The charmed meson decay contribution based on 
{\sc pythia}~\protect\cite{pythia} is
included in the sum of sources (solid black line).  The uncorrelated
charm is shown with the dotted line.  Statistical (bars) and
systematic (boxes) uncertainties are shown separately.  The
systematic uncertainty on the expected hadronic sources is not shown:
it ranges from $\sim$10\% in the $\pi^0$ region to $\sim$30\% in the
region of the vector mesons.  The uncertainty on the charm cross
section, which dominates the IMR, is $\sim$30\% both in $p+p$ and in
Au~+~Au collisions.}
\end{figure}

Figure~\ref{fig:AuAu_cock} compares the $e^+e^-$ yield in the PHENIX
acceptance in minimum bias Au~+~Au collisions to the expected yield from
the contributions of various sources.  The cocktail sources are the
same as in the $p+p$ data, but tuned separately to the Au~+~Au
measurements.  The data below the pion mass, where $\pi^0$ Dalitz
decays dominate, are well described by the cocktail.  The vector meson
yields in the cocktail are partly based on the $e^+e^-$ pair data (see
Sec.~\ref{sec:cocktail}) and consequently agree well with the data.
In particular the $J/\psi$ yield in the cocktail is exclusively based
on the $e^+e^-$ measurement~\cite{AUjpsi} 
(shown in Fig.\ref{fig:mtfits} and therefore the data in
Fig.~\ref{fig:AuAu_cock} are expected to agree with the cocktail at
the $J/\psi$ peak.  $J/\psi$ suppression would be observed if the
cocktail would instead use the $N_{\rm coll}$-scaled $p+p$
measurement~\cite{jpsi}.  In the IMR, besides the charm, bottom and
Drell Yan calculated with {\sc pythia}~\cite{pythia}, which are 
the same as in the
$p+p$ data, scaled by $N_{\rm coll}$, there is another curve drawn for
the charm.  This will be described in Section~\ref{sub:charm}.

Figure~\ref{fig:mass_cent} shows the mass spectra for $p+p$ (bottom), 
for minimum bias Au~+~Au (top) and for five different centrality 
classes in Au~+~Au.  The data are compared to the sum of hadronic 
cocktail and charmed meson decays.  The charm cross section, measured 
in $p+p$ $\sigma_{c\bar{c}}=567 \pm 57^{\rm stat} 
\pm 224^{\rm syst}\ \mu$b~\cite{ppg065} 
has been scaled by $N_{\rm coll}$ (given in Table~\ref{tab:evt}).  
For each centrality class, the data and the 
cocktail are absolutely normalized.  
Each data set is compared with two corresponding cocktail lines, 
shown in solid and dotted curves.  
The difference between the cocktails is due to uncertainty 
in the $c\bar{c}$ contribution (see discussion below).

Unlike the $p+p$ mass spectrum, the Au~+~Au mass spectra show
enhancement above the cocktail, in particular for the LMR (0.15 -- 0.75
~GeV/$c^2$).  There is little enhancement for peripheral (60-92\%)
data, but very strong enhancement for two most central classes (0-10\%
and 10-20\%).  The enhancement increases rapidly with increasing
centrality.

In order to quantitatively describe this enhancement, more information
is needed about other components that can potentially contribute to
the LMR, namely the open heavy flavor and internal conversion of real
direct photons.  We discuss them in the next sections.

\subsection{Open Heavy Flavor Contribution} \label{sub:charm}
The dilepton yield in the IMR is dominated by semi-leptonic decays of
charm hadrons correlated through flavor conservation.  Small
contributions also arise from bottom hadrons and Drell-Yan.  For $p+p$
data we determine the heavy flavor contribution by subtracting the
hadronic cocktail from the dilepton data.  We integrate the subtracted
yield in the IMR, extrapolate to zero $e^+e^-$ pair mass to get the
entire cross section, correct for geometric acceptance, and convert to
a production cross section using known branching ratios of
semi-leptonic decays~\cite{PDG}.  Details of the analysis of the charm
cross section are reported in~\cite{ppg085}.  

We find a rapidity density of $c\bar{c}$ pairs at mid rapidity:
\begin{equation}
\frac{d\sigma_{c\bar{c}}}{dy}\rvert_{y=0} = 118.1 \pm 
8.4^{\rm stat} \pm 30.7~{\rm syst} \pm 39.5^{\rm model} {\rm\mu b}.  \nonumber
\end{equation}
This corresponds to a total charm cross section of $\sigma_{c\bar{c}}
= 544 \pm 39^{\rm stat} \pm 142^{\rm syst} \pm 200^{\rm model} \mu$b,
consistent with previous measurement of single electrons by PHENIX
($\sigma_{c\bar{c}}=567 \pm 57^{\rm stat} \pm 224^{\rm syst}\ \mu$b)
~\cite{ppg065} and with a fixed-order-plus-next-to-leading-log (FONLL)
pQCD calculation ($\sigma_{c\bar{c}} = 256^{+400}_{-146} \mu$b)
\cite{cacciari}.

\begin{figure}[!ht]
\includegraphics[width=1.0\linewidth]{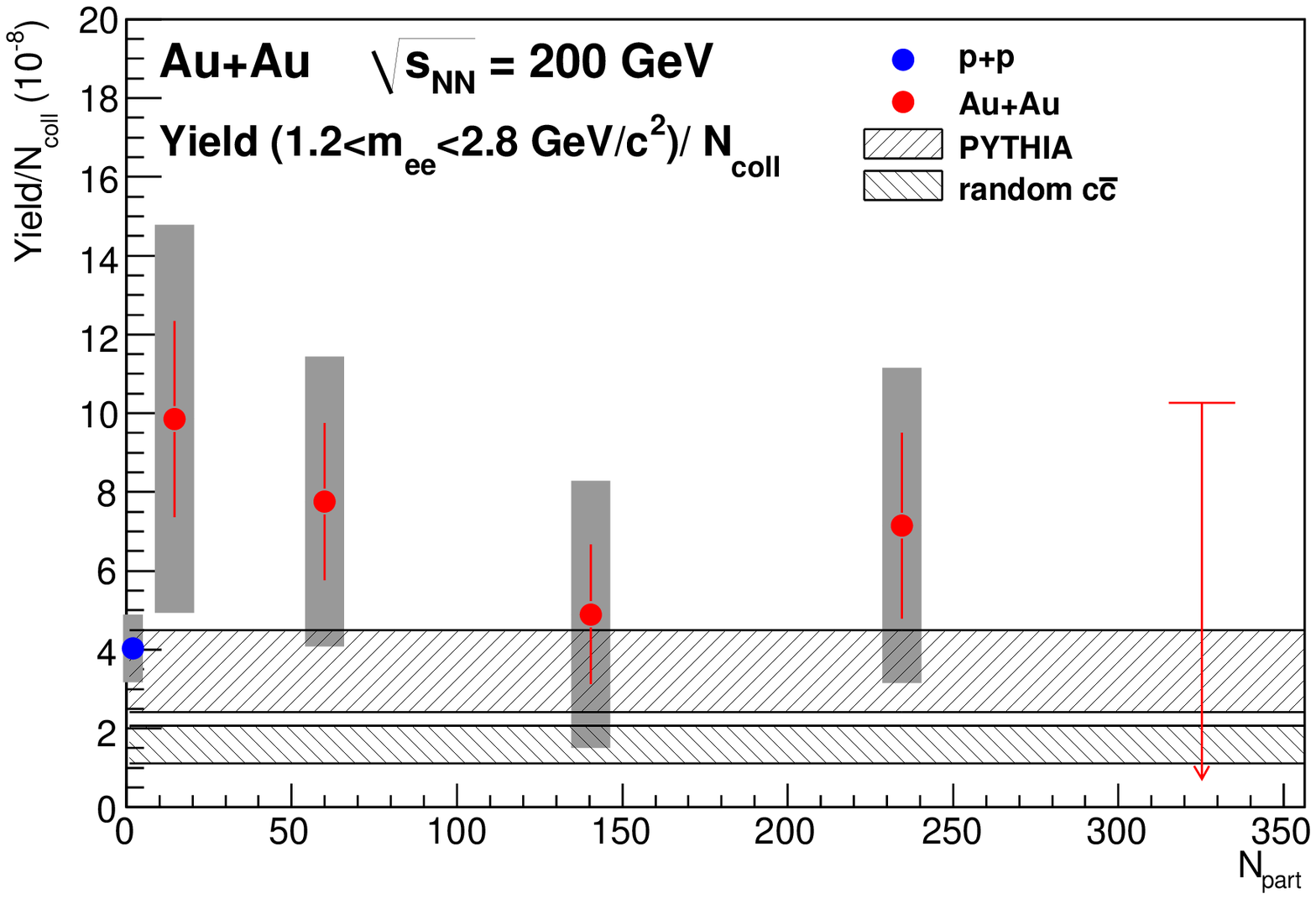}
\caption{  \label{fig:ratio2} (Color online)
Dielectron yield per binary collision in the mass range 1.2 to 
2.8~GeV/$c^2$ as a function of $N_{\rm part}$.  Statistical and systematic 
uncertainties are shown separately.  Also shown are two bands corresponding to 
different estimates of the contribution from charmed meson decays.  The 
width of the bands reflects the uncertainty of the charm cross section
only.  }
\end{figure}

In Au~+~Au the dynamic correlation of $c$ and $\bar{c}$, which is
essential to determine the mass spectral shape, could be modified
compared to $p+p$ collisions.  The observed suppression and the
elliptic flow of non-photonic electrons indicates that charm quarks
interact with the medium~\cite{ppg066}, which should change the
correlations between the produced $c\bar{c}$ pairs.  We also note that
the $p_T$ distribution for electrons generated 
by {\sc pythia}~\cite{pythia}
is softer than the spectrum measured in $p+p$ data but coincides with
that observed in minimum bias Au~+~Au collisions.  Thus we compare our
Au~+~Au data to two extreme scenarios that bracket the charm
contribution:

(i) The correlation is unchanged by the medium and equals what is
known from $p+p$ collisions.  In this case we can use the same {\sc pythia}
calculation scaled to match the cross section measured in $p+p$ and
scale it by the mean number of binary $N+N$ collisions (as given
in Table~\ref{tab:evt}).

(ii) The $c\bar{c}$ dynamical correlation is washed out by medium
interactions, i.e.  the direction of $c$ and $\bar{c}$ quarks are
uncorrelated.  We sample from the heavy flavor single-electron $p_T$
spectra, choose the angle randomly and keep the overall cross section
fixed to the experimental data~\cite{ppg066}.  Because the average 
opening angle of uncorrelated pairs is smaller than the one resulting
from the back-to-back correlation predicted by {\sc pythia}, the mass
spectral shape of uncorrelated pairs is much softer than the one
calculated by {\sc pythia}.

The charm contribution determined by case (i) is shown as the upper
dashed curve in Fig.~\ref{fig:AuAu_cock} or the upper solid curves in
Fig.~\ref{fig:mass_cent}.  The charm contribution determined by case
(ii) is shown as the dotted curve in Fig.~\ref{fig:AuAu_cock} or the
dashed curves in Fig.~\ref{fig:mass_cent}.  
In both cases the total yield of charm is normalized
to the value measured by PHENIX.

In the Min. Bias Au~+~Au data set, the IMR seems to be well described
by the continuum calculation based on case (i).  This is somewhat
surprising, since single electron distributions from charm show
substantial medium modifications~\cite{ppg066}.  Thus, it is hard to
understand how the dynamical correlation at production of the $c\bar{c}$
remains unaffected by the medium.  Case (ii) leads to a much softer
mass spectrum, as shown by the dotted curve in
Fig.~\ref{fig:AuAu_cock}).  This would leave significant room for 
other
contributions, e.g.  thermal radiation.  

We have integrated the yield in the mass region 1.2 to 2.8~GeV/$c^2$
and normalized to the number of binary collisions $N_{\rm coll}$
(Fig.~\ref{fig:ratio2}).  The systematic uncertainty due to $N_{\rm coll}$
(as indicated in Table~\ref{tab:evt}) has been included in the overall
systematic uncertainty.  Within uncertainties $N_{\rm coll}$ scaling is observed
for the production of non-photonic electrons, i.e., for those
electrons arising from decays of heavy-flavor hadrons
~\cite{ppg066}.  The normalized yield shows no significant centrality
dependence and is consistent within systematic uncertainties with the
expectation based on $N_{\rm coll}$-scaled {\sc pythia}, with the cross section
measurement of~\cite{ppg065} (case (i)).  However the scaling with
$N_{\rm coll}$ may be a mere coincidence resulting from two balancing
effects: the energy loss of charm, which increases with $N_{\rm part}$,
would lead to a softer mass distribution and therefore less yield in
the IMR (case (ii)), while a thermal contribution could increase
faster than linearly with $N_{\rm part}$ resulting in more yield in the
IMR.  Such a coincidence may have been observed at the SPS
~\cite{NA50}, where a prompt component has now been suggested by NA60
~\cite{NA60_therm}.

\subsection{Low-Mass Excess in Au~+~Au Data} \label{sub:LMR_enh}
\begin{figure}[!ht]
\includegraphics[width=1.0\linewidth]{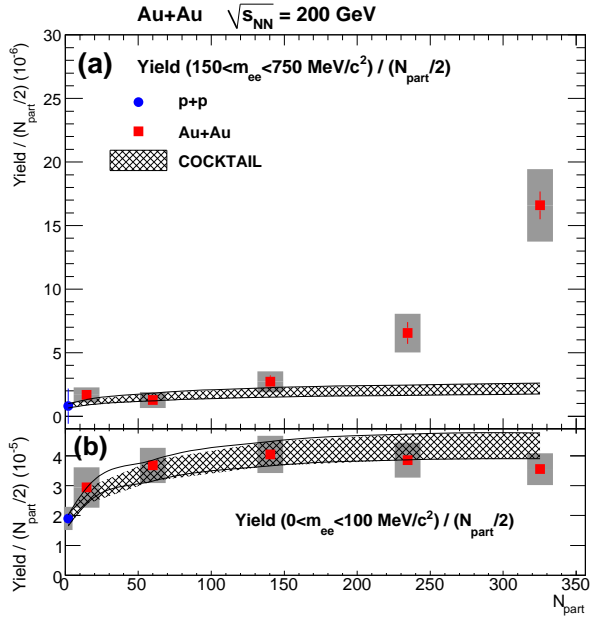}
\caption{\label{fig:ratio} (Color online)
Dielectron yield per participating
nucleon pair ($N_{\rm part}/2$) as function of $N_{\rm part}$ for
two different mass ranges (a: $0.15<m_{ee}<0.75$~GeV/$c^2$, b:
$0<m_{ee}<0.1$~GeV/$c^2$) compared to the expected yield from the
hadron decay model.  The two lines give the systematic uncertainty of
the yield from cocktail and charmed hadron decays.  For the data
statistical and systematic uncertainties are shown separately.}
\end{figure}

Figure~\ref{fig:mass_cent} shows that
the low-mass enhancement is concentrated in the first two centrality
classes, i.e.  0-10\% and 10-20\%.  For more peripheral collisions the
enhancement diminishes.  Only some small excess is visible for 20-40\%
and 60-92\% while no deviation is observed in 40-60\% with respect to the cocktail
beyond systematic uncertainties.

To quantify the centrality dependence of the enhancement, we have
integrated the yield in two mass windows: below 0.1~GeV/$c^2$, and 0.15
to 0.75~GeV/$c^2$.  
Since the cocktail yield in these regions arises mostly from hadrons
(more than 90\% from $\pi^0$ below 0.1~GeV/$c^2$ and more than 99\%
from the sum of light hadrons in $0.15<m_{ee}<0.75$~GeV/$c^2$) we
compare the measured yield to the rate of pion production.  Pions were
found to scale approximately with $N_{\rm part}$~\cite{white},
therefore we compare the measured yield in data to $N_{\rm part}$.

The top panel of Fig.~\ref{fig:ratio} shows the centrality dependence
of the yield in the mass region 0.15--0.75~GeV/$c^2$ divided by the
number of participating nucleon pairs ($N_{\rm part}/2$).  The
systematic uncertainty due to $N_{\rm part}$ (as indicated in
Table~\ref{tab:evt}) has been included in the overall systematic
uncertainty.  For comparison the yield below 0.1~GeV/$c^2$, which is
dominated by low-$p_T$ pion decays, is shown in the lower
panel.  

For both mass intervals the yield is compared to the yield calculated 
from the hadron cocktail.  Two solid curves on each panel show the 
upper and lower limit of the expected yield from the cocktail.  The 
cocktail uncertainty includes the uncertainty in the charm 
contribution discussed in the previous section.  In the lower mass 
range the yield agrees with expectations, and is proportional to the 
pion yield (bottom panel of Fig.~\ref{fig:ratio}).  In contrast, in 
the range from 0.15 to 0.75~GeV/$c^2$ the observed yield rises 
significantly above expectations.

The enhancement factor, defined as the ratio between the measured
yield and the expected yield for $0.15 < m_{ee} < 0.75$~GeV/$c^2$, is
$4.7\pm0.4^{\rm stat}\pm1.5^{\rm syst}\pm0.9^{\rm model}$, for 
Min. Bias data.  The first error is the statistical error, the 
second the systematic uncertainty of the data, and the last error is 
an estimate of the uncertainty in the cocktail, i.e.  the expected 
yield from hadronic sources.  For the various centrality bins the 
enhancement factor is reported in Table~\ref{tab:enhancement}.

\begin{table}[htbp]
\caption{The enhancement factor, defined as the ratio
between the measured yield and the expected yield for
0.15~$<m_{ee}<$~0.75~GeV/$c^2$, for different centrality
bins.  The meaning of the errors is defined in the
text.~\label{tab:enhancement}} 
\begin{ruledtabular}  \begin{tabular}{cccc} 
&  Centrality & Enhancement ($\pm$stat $\pm$syst $\pm$model) & \\ 
\hline 
& 00-10 \% & 7.6 $\pm$ 0.5 $\pm$ 1.5   $\pm$ 1.5 &  \\
& 10-20 \% & 3.2 $\pm$ 0.4 $\pm$ 0.1   $\pm$ 0.6 &  \\
& 20-40 \% & 1.4 $\pm$ 1.3 $\pm$ 0.02  $\pm$ 0.3 &  \\
& 40-60 \% & 0.8 $\pm$ 0.3 $\pm$ 0.03  $\pm$ 0.2 &  \\
& 60-92 \% & 1.5 $\pm$ 0.3 $\pm$ 0.001 $\pm$ 0.3 &  \\
& Min. Bias       & 4.7 $\pm$ 0.4 $\pm$ 1.5   $\pm$ 0.9 &  \\
\end{tabular} \end{ruledtabular}  
\end{table}

The increase is qualitatively consistent with the conjecture that an
in-medium enhancement of the dielectron continuum yield arises from
scattering processes like $\pi\pi$ or $q\bar{q}$ annihilation.  In
this case the enhancement would scale proportional to $N^2_{\rm
part}$, different from the hadronic cocktail which scales proportional
to $N_{\rm part}$.

\subsection{$p_T$ Dependence of the Mass Spectra}\label{sub:mass_pt}

\begin{figure*}
\includegraphics[width=0.45\linewidth]{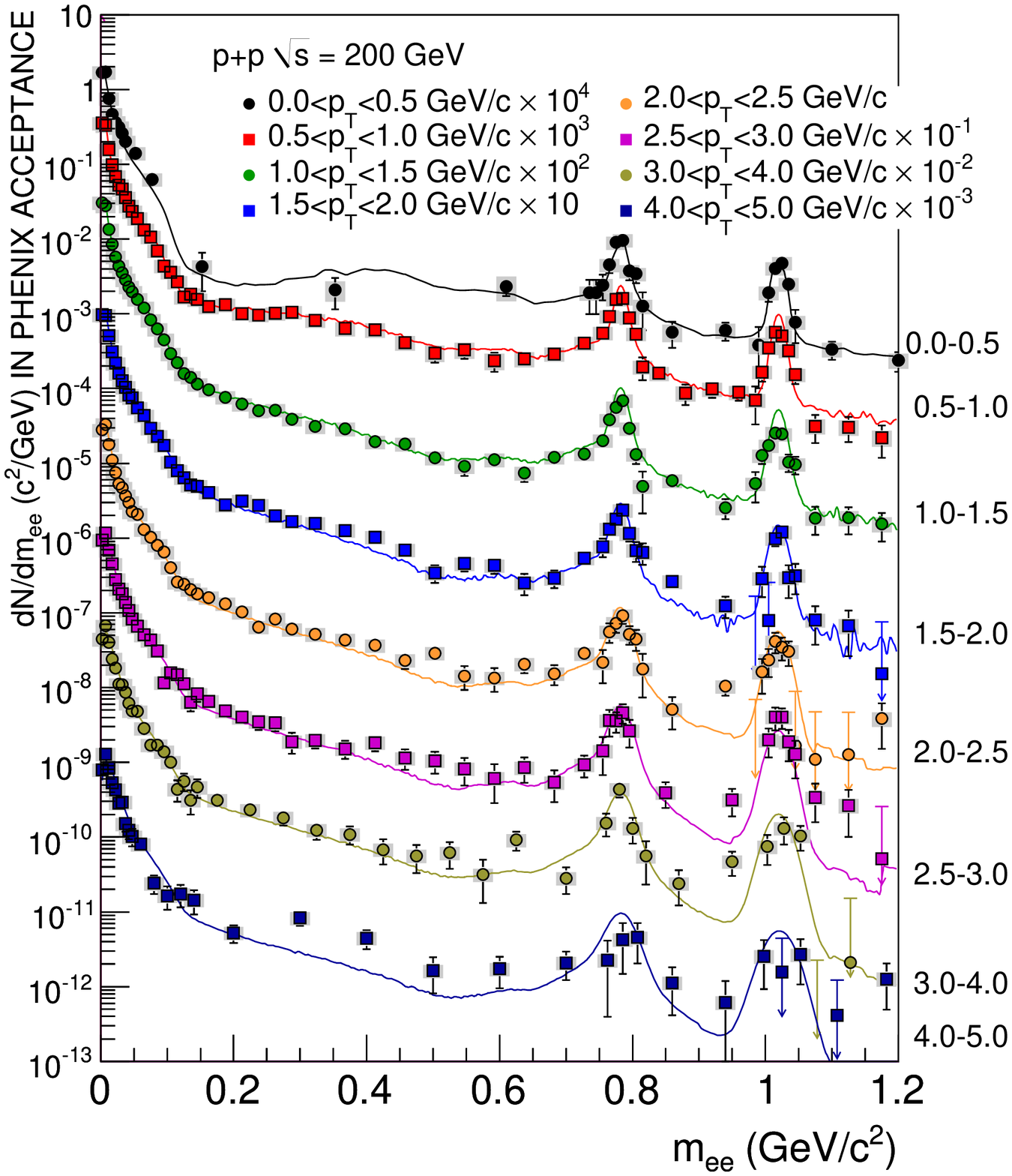}
\includegraphics[width=0.45\linewidth]{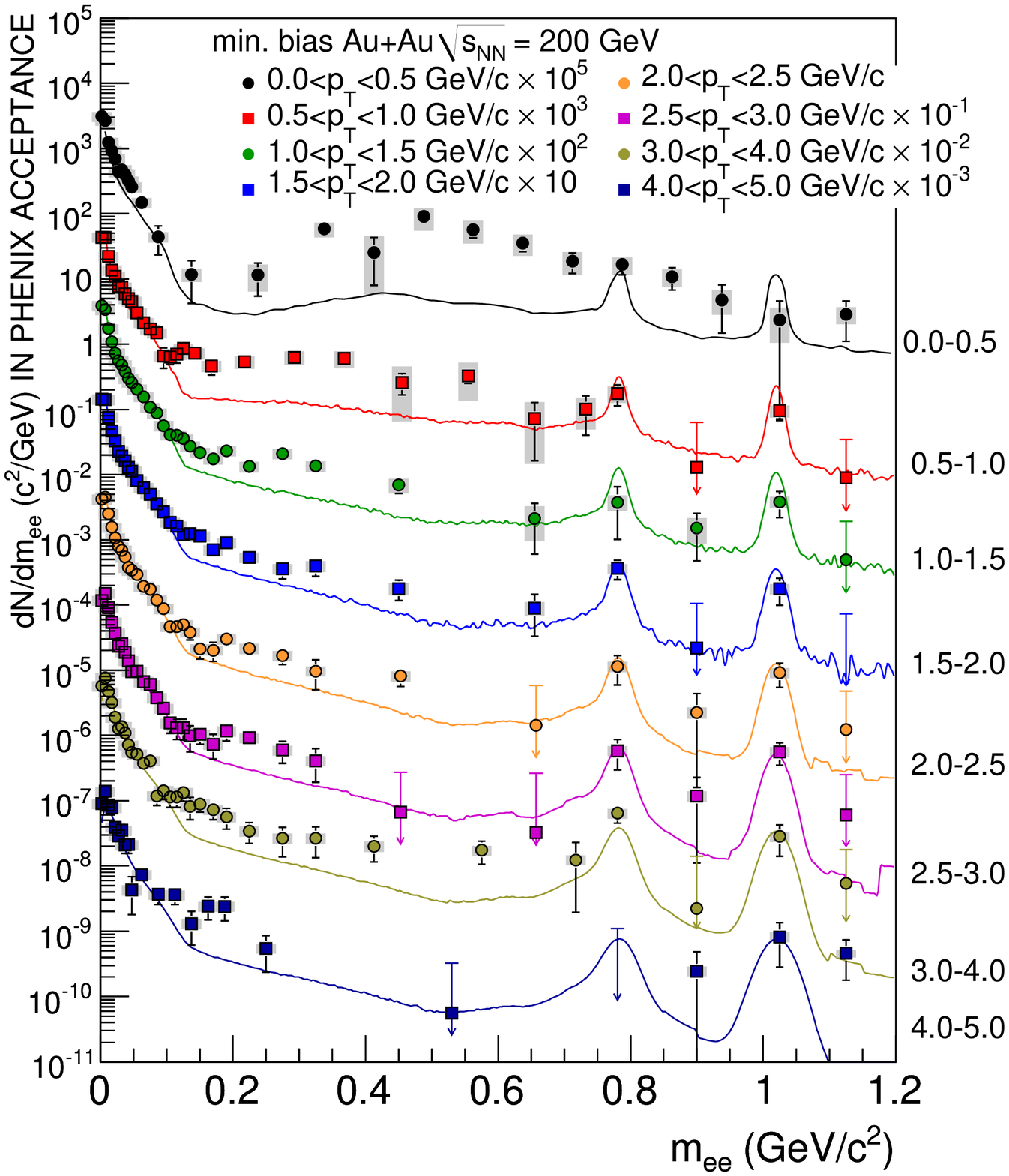}
\caption{\label{fig:mass_pp_au} (Color online)
$e^+e^-$ pair invariant mass distributions in
$p+p$ (left) and minimum bias Au~+~Au collisions (right).  The $p_T$ ranges are
shown in the legend.  The solid curves represent the cocktail of
hadronic sources (see Section~\ref{sec:cocktail}) and include
contribution from charm calculated by {\sc pythia} using the cross section
from~\cite{ppg065} scaled by $N_{\rm coll}$.   }
\end{figure*}

Figure~\ref{fig:mass_pp_au} compares $e^+e^-$ invariant mass spectra
measured in $p+p$ and in Min. Bias Au~+~Au collisions to the corresponding
expectations from the cocktail of hadron decays and open charm, in
different ranges of $p_T$.  Data and cocktail are absolutely
normalized.

The ``knee'' in the mass spectra beginning at $m_{ee} \simeq
0.1$~GeV/$c^2$ corresponds to the cut-off of the $\pi^0$ Dalitz
decays.  
For $m_{ee}>m_{\pi}$, the $\eta$ Dalitz decay ($\eta \rightarrow
e^+e^-\gamma$) is the dominant hadronic source of $e^+e^-$ pairs,
followed by the $\omega$ Dalitz decay ($\omega \rightarrow
e^+e^-\pi^0$), with small contributions from other sources such as
$\eta'$ and $\phi$.  
The detector acceptance and resolution effects broaden the low-mass
Dalitz peak and smear the ``knee'' of the $\pi^0$ Dalitz contribution.
These detector effects are included in the cocktail calculation.

The $p+p$ data are consistent with expectations from the cocktail over
the full mass range in the low $p_T$ bin.  In the highest $p_T$ bins
however the data are enhanced with respect to the cocktail.  The
deviations are however small in contrast to the Au~+~Au data which show
a large enhancement in the LMR above $m_\pi^0$, which is concentrated
at low-$p_T$.  For $p_T >$ 1.0~GeV/$c$ the enhancement becomes smaller
than at lower $p_T$ but it is still larger than the one observed in
the $p+p$ data.

In the introduction, we classified the low mass region into LMR~I (low
mass, high $p_T$) and LMR~II (low mass, low $p_T$)
(see Fig.~\ref{fig:cartoon}).  The behavior of the low mass excess in
Au~+~Au shown in Fig.~\ref{fig:mass_pp_au} is different for LMR~I and
LMR~II.  In LMR~I, the excess has a similar shape to the cocktail and
the level of the excess with respect to the cocktail is approximately
constant.  In this region, a contribution from internal conversion of
virtual direct photons is expected.  In LMR~II, the excess increases with
increasing mass and decreasing $p_T$.

In the following we first analyze the data in the low mass
high $p_T$ region (LMR~I), which allows the measurement of direct
photons; then we study how much they contribute to the inclusive
enhancement, dominated by the yield in the LMR~II.  

\subsection{Measurement of Direct Photons}\label{sub:photon}

\begin{figure}[tb]
\includegraphics[width=1.0\linewidth]{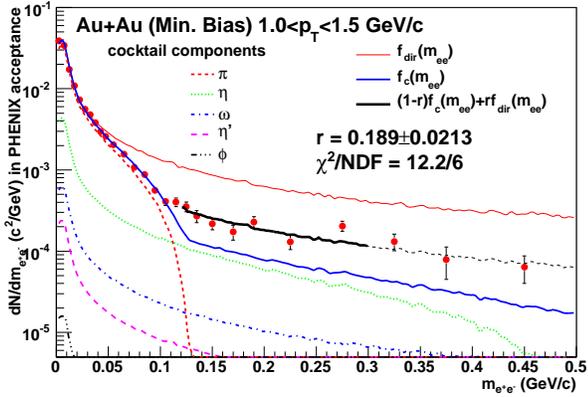}
\caption{\label{fig:fit} (Color online)
Electron pair mass distribution for Au~+~Au (Min. Bias) events
for $1.0<p_T<1.5$ GeV/$c$.  The two-component fit is explained
in the text.  The fit range is $0.12<m_{ee}<0.3$~GeV/$c^2$.
The dashed (black) curve at greater $m_{ee}$ shows
$f(m_{ee})$ outside of the fit range.
}
\end{figure}


In LMR~I (Fig.~\ref{fig:cartoon}), where the $p_T$ of the $e^+e^-$
pair is much greater than its mass ($m_{ee} \ll p_T$), the yield of
the virtual photons is approximately the same as that of
real photons.  Therefore, in this quasi-real virtual photon region, 
the production of real direct photons can be deduced from measurements 
of $e^+e^-$ pairs.  Theoretical details are given in
Appendix~\ref{app:ICA_theory}.  In this Section, we determine the real
direct photon cross section for $1 < p_T < 5$~GeV/$c$ from the data
shown in Fig.~\ref{fig:mass_pp_au}.  We use the mass range
$0.1<m_{ee}<0.3$~GeV/$c^2$.

The relation between real photon production and the associated $e^+e^-$ production can
be written as (Equation~\ref{eq:ICA})
\begin{eqnarray}
\frac{d^2N_{ee}}{dm_{ee}dp_T} = \frac{2\alpha}{3\pi}\frac{1}{m_{ee}} L(m_{ee}) S(m_{ee},p_T) 
\frac{dN_{\gamma}}{dp_T}, \label{eq:Conversion}\\
L(m_{ee})=\sqrt{1-\frac{4m_e^2}{m_{ee}^2}}\Bigl( 1+\frac{2m_e^2}{m_{ee}^2} \Bigr).  
\end{eqnarray}
Here $\alpha$ is the fine structure constant, $m_{ee}$ is the mass of $e^+e^-$ pair, $m_e$ is the electron mass, 
and $S(m_{ee},p_T)$ is a process-dependent factor that accounts for differences between
real and virtual photon production, such as form factors, phase space, and
the spectral function.
Equation~\ref{eq:Conversion} holds for any process emitting
real photons, in particular direct or thermal emission.
For high $p_T$ ($p_T \gg m_{ee}$) 
the process dependence becomes negligible and the factor $S(m_{ee},p_T)$ becomes 1
as $m_{ee} \rightarrow 0$ or $m_{ee}/p_T \rightarrow 0$.
For $m_{ee} \gg m_e$, the factor $L(m_{ee})$ also becomes very close to unity.
Thus the relation simplifies to
\begin{eqnarray}
\frac{d^2N_{ee}}{dm_{ee}dp_T} \simeq \frac{2\alpha}{3\pi}\frac{1}{m_{ee}} 
\frac{dN_{\gamma}}{dp_T}.  \label{eq:simple}
\end{eqnarray}
Here the mass distribution of electron pairs for a given
$p_T$ bin takes on a very characteristic $1/m_{ee}$ shape.
If there is real direct photon production in a given $p_T$ bin, there
should be a corresponding electron pair contribution that behaves like $1/m_{ee}$ in the
same $p_T$ bin.  Therefore, the real direct photon production can
be determined from the yield of the excess electron pairs.

For Dalitz decays, the $1/m_{ee}$ behavior is truncated by the kinematic
limit and $S(m_{ee})$ becomes zero for $m_{ee} > m_h$,
where $m_h$ is the mass of the hadron.
The functional form of $S(m_{ee})$
for Dalitz decays is given in Appendix~\ref{app:ICA_theory}.
In contrast, 
the factor $S(m_{ee},q)$ for the direct photon process is unity for
$p_T \gg m_{ee}$.
We exploit this difference to separate the direct photon signal
from the hadronic background.  Since 80\% of the hadronic photons are
from $\pi^0$ Dalitz decays,
the signal to background (S/B) ratio for the direct photon
signal improves by a factor of five
for $m_{ee}> m_{\pi^0}\approx$~0.135~GeV/$c^2$, thereby allowing
a real direct photon
signal that is 10\% of the yield of hadronic decay photons
to be observed as a 50\% excess of $e^+e^-$ pairs
for this mass range.

Figure~\ref{fig:mass_pp_au} shows a visible excess above the $\pi^0$
cutoff for all $p_T$ bins of the Au~+~Au data.  For
$p_T > 1$~GeV/$c$, the excess is almost a constant factor above
the cocktail.  
As we examine later
the mass distribution for
$p_T > 1$~GeV/$c$ is consistent with the
$1/m_{ee}$ shape expected for the electron pairs
from internal conversion of virtual direct photons.

\begin{figure}[!ht]
\includegraphics[width=1.0\linewidth]{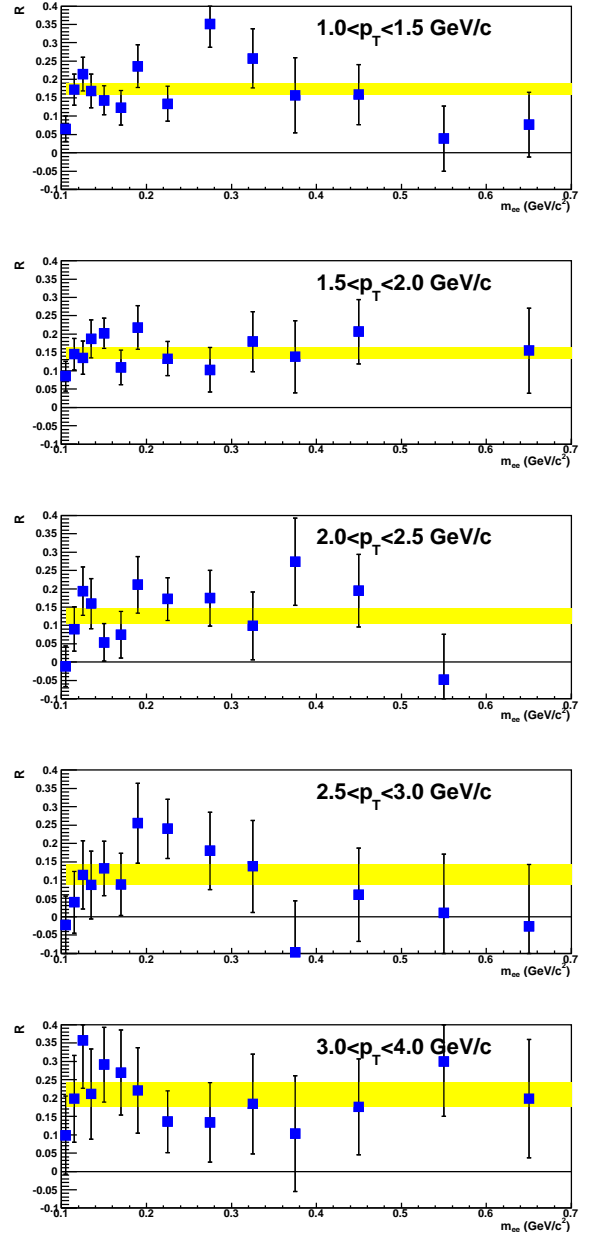}
\caption{\label{fig:real_virtual_ratio_MB} (Color online)
Ratio $R = ({\rm data - cocktail})/f_{\rm dir}(m_{ee})$ of
electron pairs for different $p_T$ bins in Min. Bias Au~+~Au
collisions.  The $p_T$ range of each panel is indicated in the
figure.
The yellow band in each panel shows $\pm 1 \sigma$ range of a constant
fit value to the data points.
}
\end{figure}

In the following we assume the excess for $p_T>1$~GeV/$c$ and
$m_{ee} < 0.3$~GeV/$c^2$ is entirely due to internal conversion of
virtual direct photons and deduce the real direct photon yield from the
$e^+e^-$ pair yield using Equation~\ref{eq:Conversion}.  
We demonstrate the validity of this assumption later.  Although the data are
consistent with $1/m_{ee}$ over a wider mass range ($m_{ee} \sim 0.7$
~GeV/$c^2$), as can be seen in Fig.~\ref{fig:real_virtual_ratio_MB},
we limit our analysis for $0.1 < m_{ee} <
0.3$~GeV/$c^2$.  We do so in order (i) to ensure the condition $m_{ee} \ll p_T$
for the lowest $p_T$ bin ($1.0<p_T<1.5$~GeV/$c$), (ii) to
keep the correction factor $S(M,q)$ close to unity, and (iii) to minimize
uncertainty due to $c\bar{c}$.  In this kinematic range, the
contribution of $c\bar{c} \rightarrow e^+e^-$, estimated by {\sc pythia}, is
less than 5\% of the excess.

In order to quantify the excess, we fit a two-component
function,
\begin{equation}
f(m_{ee};r) = (1-r)f_{c}(m_{ee}) + r f_{\rm dir}(m_{ee}) \label{eq:ICAfit}
\end{equation}
to the mass distribution.
Here $f_{c}(m_{ee})$ is the shape of the cocktail
mass distribution (shown in Fig.~\ref{fig:mass_pp_au}),
$f_{\rm dir}(m_{ee})$ is the expected
shape of the virtual direct photon internal conversion mass distribution, and
$r$ is the only fit parameter.
In the low mass region used for the fit, 
the functional form of $f_{c}(m_{ee})$ is the sum of
Dalitz decay mass distributions of hadrons
(Equation~\ref{eq:Dalitz1} -- Equation~\ref{eq:DalitzPS})
filtered through the PHENIX acceptance and smeared by
the detector effects.  It is calculated by a
Monte Carlo simulation which takes into account detector
effects such as finite mass resolution.  The functional
form of $f_{\rm dir}(m_{ee})$ corresponds to
Equation~\ref{eq:Conversion} with $S(m_{ee})=1$.  It is
also filtered through the PHENIX acceptance and smeared by
detector effects.

Both $f_{c}(m_{ee})$ and $f_{\rm dir}(m_{ee})$ are
separately normalized to the data for $m_{ee}<30$~MeV/$c^2$.
In this mass region $S(m_{ee})$ of $\pi^0$ Dalitz decays is
very close to unity.  Thus the functional shapes of
$f_{c}$ and $f_{\rm dir}$ are essentially identical and equal to
$L(m_{ee})/m_{ee}$ smeared by the detector effects.
This means that the fit function $f(m_{ee};r)$ in this
mass range is independent of the fit parameter $r$ as
$(1-r)L(m_{ee})/m_{ee} + rL(m_{ee})/m_{ee} = L(m_{ee})/m_{ee}$.
Thus this normalization ensures that the yield of fit function $f(m_{ee};r)$
is always normalized to that of the data for $m_{ee}<30$ MeV/$c^2$.
The parameter $r$ can be interpreted as the direct photon
fraction of the inclusive photon yield.

This fitting method has the advantage of canceling most
of the systematic uncertainties of the cocktail
normalization relative to the data.
The PHENIX acceptance for
electron pairs with $m_{ee}<0.3$~GeV/$c^2$ and
$p_T>1$~GeV/$c$ is almost constant, and its shape
can be calculated accurately as a function of mass.
Many systematic effects, such as electron identification
efficiency, detector dead area, etc.  can influence
the absolute value of the acceptance but not its
shape.

For each $p_T$ bin, $f(m_{ee})$ is fit to the data for several
mass ranges with $r$ the only fit parameter.  Figure~\ref{fig:fit} shows
$f_{\rm dir}(m_{ee})$ and $f_{c}(m_{ee})$ together with the fit result for
Au~+~Au Min. Bias data for $1.0<p_T<1.5$~GeV/$c$ and 
the cocktail components.  The dashed curve
shows $f(m_{ee})$ extended outside of the fit range.  Although the mass
region is not used in the fit, the fit function describes the data
for $m_{ee} > 0.3$~GeV/$c^2$.

The fit shown in Fig.~\ref{fig:fit} has $\chi^2/$NDF$ = 12.2/6$.
The somewhat large $\chi^2$ values is due to the large contribution
from the lowest mass bins, where statistical errors are small and
systematic errors due to the detector resolution are significant.
The $\chi^2$ value is calculated from the statistical errors only.
The results for the fit range $0.12<m_{ee}<0.3$~GeV/$c^2$
are summarized in Table~\ref{tab:ICAfit_summary}.  For
$p_T>$~1.5~GeV/$c^2$ the fit gives good $\chi^2/$NDF,
demonstrating that the shape of the excess is consistent with
$1/m_{ee}$ as expected for internal conversion.
\begin{table}[b]
\caption{Summary of the fits to Equation~\ref{eq:ICAfit} in the range $0.12<m_{ee}<0.3$~GeV/$c^2$.}
\begin{ruledtabular} \begin{tabular}{lcc}
$p_T$(GeV/$c$) & $r$ &$\chi^2/$NDF\\
\hline
1.0-1.5& $0.189\pm0.021$ & 12.2/6\\
1.5-2.0& $0.165\pm0.022$ & 4.6/6\\
2.0-2.5& $0.146\pm0.029$ & 6.6/6\\
2.5-3.0& $0.165\pm0.040$ & 3.3/6\\
3.0-4.0& $0.224\pm0.048$ & 3.7/6\\
4.0-5.0& $0.206\pm0.093$ & 4.2/3\\
\end{tabular} \end{ruledtabular}
\label{tab:ICAfit_summary}
\end{table}

To evaluate the systematic uncertainty due to the mass
range used for the fit, the fit was repeated for three
mass ranges: $0.08<m_{ee}<0.3$~GeV/$c^2$, 
$0.1<m_{ee}<0.3$~GeV/$c^2$, and
$0.12<m_{ee}<0.3$~GeV/$c^2$.  The value of $r$ is
taken as the average of the results for these three
fit ranges.

The sources of systematic uncertainty on the fit include (1)
the fit range, (2) the mass
spectrum of the data, and (3) the cocktail.
The sources of the systematic uncertainty on the 
mass spectrum relative to the cocktail include 
(1) uncertainties on the correlated background due to jet pairs ($\simeq 2$\% for $p+p$),
(2) uncertainties in the acceptance and efficiency
in the mass range of the signal ($0.1<m_{ee}<0.3$~GeV/$c^2$)
relative to the mass range ($m_{ee}<30$~MeV/$c^2$) used for the
normalization ($\simeq 1$\% for acceptance and
$\simeq 1$\% for efficiency), and
(3) uncertainty in the mixed-event normalization ($0.25$\%/$(S/B)$),
where $S/B$ is the signal-to-background ratio in
$0.1<m_{ee}<0.3$~GeV/$c^2$.  The uncertainty (2) is small
since uncertainties in the absolute normalization cancel
when the cocktail is normalized to the data in the low-mass
peak ($m_{ee}<30$~MeV/$c^2$).
The largest source of uncertainty is the particle composition
in the hadronic cocktail, namely the $\eta/\pi^0$ ratio.
This corresponds to a $\simeq 7$\% ($\simeq 17$\%) uncertainty in
the $p+p$ (Au~+~Au) cocktail for $0.1 < m_{ee} < 0.3$~GeV/$c^2$.
All systematic uncertainties are added in quadrature to obtain
the total systematic uncertainty.

Since the $\eta/\pi^0$ ratio is the largest source of
uncertainty, we also studied fits with a three component
function,
$f_{3}(m_{ee})=(1-r-r_{\eta})f_{c}(m_{ee})+r f_{\rm dir}(m_{ee}) + r_{\eta} f_{\eta}(m_{ee})$,
with a constraint on $r_{\eta}$ such that $\eta/\pi^0 = 0.48\pm0.03 (0.08)$
for $p+p$ (Au~+~Au)~\cite{eta}.  These alternative fits give consistent results
for $r$ within statistical uncertainties.

So far we have assumed that the excess in $p_T >$ 1~GeV/$c$
and $0.1 < m_{ee} < 0.3$~GeV/$c^2$ is entirely due to internal
conversion of virtual direct photons.  This means that we assume
$S(m_{ee},q)$ for excess virtual photons to be unity.  In the
following, we examine the validity of the assumption from the data.

As shown in Appendix~\ref{app:ICA_theory},
the shape of the virtual photon spectrum as a function of mass can be
obtained from the electron pair yield as
\begin{eqnarray}
q_0\frac{dN_\gamma^*}{d^3q} = \frac{3\pi}{2\alpha} m_{ee} \times q_0\frac{dn_{ll}}{d^3qdm_{ee}}.
\end{eqnarray}

Since the shape of $f_{\rm dir}(m_{ee})$ is $1/m_{ee}$ smeared by the
detector effects, a fit of $R = ({\rm data - cocktail})/f_{\rm dir}(m_{ee})$
to a constant can be used to test that
the excess has the shape expected for internal conversion of direct
photons.  Note that the detector effects
in the numerator and denominator cancel in the ratio.

Furthermore, since $f_{\rm dir}(m_{ee})$ is
normalized to the data for $m_{ee}<30$ MeV/$c^2$, $R$ can be interpreted as the ratio
of the virtual photon yield to the inclusive real photon yield: 
\begin{eqnarray}
R(m,p_T) &\simeq& \frac{dN_{\gamma^*}^{\rm excess}(m,p_T)}{dp_T}/
\frac{dN^{\rm incl}_\gamma(p_T)}{dp_T} \\
&=& S(m,p_T) dN^{direct}_\gamma(p_T)/dN^{\rm incl}_\gamma(p_T).
\end{eqnarray}

Figure~\ref{fig:real_virtual_ratio_MB} shows $R$ as a function of
$m_{ee}$ for $0.1<m_{ee}<0.7$~GeV/$c^2$.  The ratio cannot be measured
for $m_{ee}<0.1$~GeV/$c^2$ because in this mass region the signal is
masked by large background from $\pi^0$ Dalitz decays.  The
distributions are consistent with a constant for the five 
$p_{\rm T}$ bins.
For the highest $p_T$ bin
($p_T>4$~GeV/$c$) the shape of the virtual photon mass
spectrum is not well constrained due to limited statistics.  However,
it is reasonable to expect that the same constant behavior continues
for higher $p_T$.
The $\pm 1 \sigma$ band of a constant value fit are shown in each panel.  
Table~\ref{tab:R_const_fit}
summarizes the results of the fits.  For all $p_T$ bins, the constant value
fit gives a good $\chi^2/$NDF value.  This demonstrate that the data 
are consistent with a constant $S(m_{ee})$ for these $p_T$ bins.

As discussed in Appendix~\ref{app:ICA_theory}, the ratio is expected
to be a smooth function of $m_{ee}$.  
Hadronic and partonic direct photon contributions to $S(m_{ee},q)$ are
expected to be nearly constant in this range.  $q\bar{q}$ annihilation
can make a contribution proportional to $m_{ee}^2$, but should be much
smaller than these two components for 
$0.1<m_{ee}<0.3$~GeV/$c^2$~\cite{rapp_RHIC,dusling_RHIC,cassing_RHIC}.  
There is no sign of a
component that scales with $m_{ee}^2$ for $m_{ee}<0.3$~GeV/$c^2$,
suggesting that the $q\bar{q}$ contribution is indeed small.
Figure~\ref{fig:real_virtual_ratio_MB} illustrates that $S(M,q)$ is
indeed constant, and supports the use of $r$ obtained from our fit as
the direct photon fraction in the inclusive photon spectrum.

The absence of any increase in $R(m_{ee})$ for $m_{ee} >
0.5$~GeV/$c^2$ is somewhat surprising.  
A constant $R$ as function of $m_{ee}$ implies that the $S(m_{ee})$
is also a constant.  If the excess electron pairs
are thermal pairs from the medium, we expect an increasing
contribution from $q\bar{q}$ annihilation or from the tail of the
(possibly modified) $\rho$ resonance, which leads to increase of
$S(m_{ee})$ in higher mass.  The data show no indication of
such increase.  For thermal radiation,
$S(m_{ee})$ is the space time average of a product of
electromagnetic spectral function and the Boltzmann factor,
see Equation~\ref{eq:S}.
We note that the Boltzmann factor
$f^B = 1/(e^{E/T}-1)$ can cause significant suppression of
$S(m_{ee},q)$ for $m_{ee} > 0.5$~GeV/$c^2$.  Thus
the nearly constant behavior of $R(m_{ee})$ for $m_{ee}>0.3$
~GeV/$c$ should not be interpreted as absence of any contribution other
than internal conversion of virtual direct photons at large $m_{ee}$.

\begin{table}[b]
\caption{Summary of a constant fit to the ratio data
shown in Fig.~\ref{fig:real_virtual_ratio_MB}.  The fit range 
is $0.11<m_{ee}<0.7$~GeV/$c^2$.}
\begin{ruledtabular} \begin{tabular}{lcc}
$p_T$(GeV/$c$) & $\langle R \rangle$ &$\chi^2/$NDF\\
\hline
1.0 - 1.5 & 0.173 $\pm$ 0.015 & 16.9/12\\
1.5 - 2.0 & 0.149 $\pm$ 0.016 & 14.8/12\\
2.0 - 2.5 & 0.125 $\pm$ 0.020 & 14.7/12\\
2.5 - 3.0 & 0.115 $\pm$ 0.028 & 9.1/12\\
3.0 - 4.0 & 0.210 $\pm$ 0.033 & 4.4/12\\
\end{tabular} \end{ruledtabular}
\label{tab:R_const_fit}
\end{table}

\begin{figure}[tb]
\includegraphics[width=1.0\linewidth]{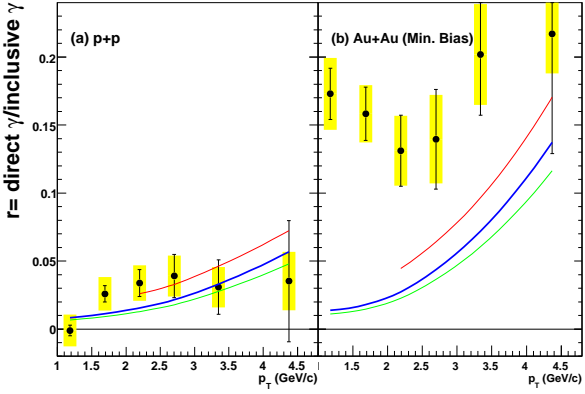}
\caption{\label{fig:dir_ratio} (Color online)
The fraction of the direct photon component as
a function of $p_T$.
The error bars and the error band represent the statistical and
systematic uncertainties, respectively.  The curves are
from a NLO pQCD calculation (see text).
}
\end{figure}

Figures~\ref{fig:dir_ratio} (a) and (b) show the fraction 
$r=\frac{direct\;
\gamma}{inclusive\;\gamma}$ of the direct photon component determined
by the two-component fit (see Equation~\ref{eq:ICAfit}) in $p+p$ and Au~+~Au
collisions, respectively.
The curves represent the expectations from a next-to-leading-order
perturbative QCD (NLO pQCD) calculation~\cite{vogelsang}.  For $p+p$, the curves show
the ratio $d\sigma^{NLO}_{\gamma}(p_{\rm
T})/d\sigma^{\rm incl}_{\gamma}(p_T)$, 
and $d\sigma^{\rm incl}_{\gamma}(p_T)$ is the $p+p$ inclusive photon cross
section (obtained from the data as described later).  For Au~+~Au, the
curves represent $T_{\rm AA}d\sigma^{NLO}_{\gamma}(p_{\rm
T})/dN^{\rm incl}_{\gamma}(p_T)$, where $T_{\rm AA}$ is the Glauber
nuclear overlap function and $dN^{\rm incl}_{\gamma}(p_T)$ is the
Au~+~Au inclusive photon yield.  
The three curves corresponding (from top to bottom)
to the theoretical scales set to $\mu$ = 0.5~$p_T$, $p_T$, and
2~$p_T$, respectively, show the scale dependence of the
calculations.  While the fraction $r$ is consistent with the NLO pQCD
calculation in $p+p$, it is larger than the calculation in Au~+~Au for
$p_T <4.5$~GeV/$c$.

\begin{figure}[tb]
\includegraphics[width=1.0\linewidth]{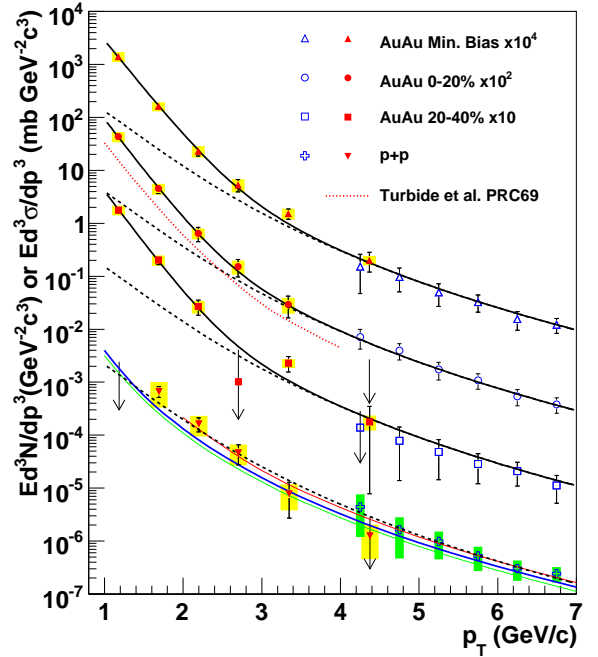}
\caption{\label{fig:photon_spectra} (Color online)
Invariant cross section ($p+p$) and invariant yield (Au~+~Au) of
direct photons as a function of $p_T$.  The filled points are from
this analysis and open points are from~\cite{ppg042,ppg060}.
The three curves on the $p+p$ data represent NLO pQCD
calculations, and the dashed curves show a modified power-law
fit to the $p+p$ data, scaled by $T_{\rm AA}$.  
The dashed (black)
curves are exponential plus the $T_{\rm AA}$ scaled
$p+p$ fit.   The dotted (red) curve near the 0--20\% centrality 
data is a theory calculation~\cite{Turbide:2003si}.
}
\end{figure}

The direct photon fraction $r$ in Fig.~\ref{fig:dir_ratio} is
converted to the direct photon yield using $dN_{\gamma}^{\rm dir}(p_{\rm
T})= r \times dN_{\gamma}^{\rm incl}(p_T)$.
Here we determine the
inclusive photon yield for each $p_T$ bin from the yield of low
mass $e^+e^-$ pairs in the range $m_{ee}<30$~MeV/$c^2$ using the following
method.  
The differential yield of electron pairs is related to 
that of photons by Equation~\ref{eq:Conversion}.
The process dependent factor $S(m_{ee},q)$ is unity within a few
percent for any source of photon for $m_{ee}<30$ MeV/$c^2$.
Thus the measured yield of electron pairs ($N_{ee}^{\rm data}$) in
$m_{ee}<30$ MeV/$c^2$ for a given $p_T$ bin is proportional
to that of inclusive photons in the same $p_T$ bin.
\begin{eqnarray}
N_{ee}^{\rm data}(p_T) = \epsilon^{\rm acc} K \frac{dN_{\gamma}^{\rm incl}}{dp_T},\\
K = \int^{30 \rm{MeV}/c^2}_0 \frac{2\alpha}{3\pi}\frac{L(m_{ee})}{m_{ee}} dm_{ee}
\end{eqnarray}
Here $\epsilon^{\rm acc}$ represents the acceptance of PHENIX.
The same relation holds for the cocktail calculation of photon and electron pairs.
\begin{equation}
N_{ee}^{\rm cocktail}(p_T) = \epsilon^{\rm acc} K \frac{dN_{\gamma}^{\rm cocktail}}{dp_T}\;.
\end{equation}
Here $N_{ee}^{\rm cocktail}$ is the yield of electron pairs for
$m_{ee}<30$ MeV/$c^2$ in the hadronic cocktail calculation, and
$dN_{\gamma}^{\rm cocktail}/dp_T$ is the yield of photons in the
same calculation.  Thus we have
\begin{equation}
\frac{dN_{\gamma}^{\rm incl}}{dp_T}=\frac{N_{ee}^{\rm data}}{N_{ee}^{\rm cocktail}} \times
\frac{dN_{\gamma}^{\rm cocktail}}{dp_T}
\end{equation}
The systematic uncertainty in the inclusive photon spectra equals the systematic uncertainty
in $N_{ee}^{\rm data}$, which is summarized in Table~\ref{tab:errors}.  The total systematic
uncertainty in $N_{ee}$ is approximately 20\%.

In Fig.~\ref{fig:photon_spectra} the direct photon spectra thus obtained are compared
with the direct photon data from~\cite{ppg042,ppg060} and NLO pQCD
calculations.
The systematic uncertainty of the inclusive photon yield is
added in quadrature with the systematic uncertainties of the data.
The $p+p$ data are shown as an invariant cross section
using $d\sigma = \sigma^{\rm inel}_{pp} dN$ using $\sigma^{\rm inel}_{pp} = 42$~mb.

The direct photon data of this analysis
are obtained from the yield of $e^+e^-$ pairs using Equation~\ref{eq:Conversion}
under the assumption $S(m_{ee},q)=1$
for $0.1<m_{ee}<0.3$~GeV/$c^2$.  Although we have shown that our data are consistent
with this assumption, what we actually measure is the yield of $e^+e^-$ pairs 
in this mass range.  For completeness, we give the
relation between the direct photon yield deduced by the analysis and
the electron pair yield that is actually measured.
The relation between the real photon yield
and the $e^+e^-$ pair yield in this mass range is given by Equation~\ref{eq:g_ee_ratio}.
Thus the yield of the excess $e^+e^-$ pairs for $0.1<m_{ee}<0.3$~GeV/$c^2$
can be obtained by multiplying the photon yield by a factor of 
$\frac{2\alpha}{3\pi} \log{\frac{300}{100}} = 1.7 \times 10^{-3}$.

The pQCD calculation is consistent with the $p+p$ data within the
theoretical uncertainties for $p_T>2$~GeV/$c$.  A similarly good
agreement is observed for $\pi^0$~\cite{Adler:2003pb}.
The $p+p$ data can be well described by a modified power-law
function ($A_{pp} (1+p_T^2/b)^{-n}$) as shown by the dashed curve
in Fig.~\ref{fig:photon_spectra}.  
The Au~+~Au data are above
the $p+p$ fit curve scaled by $T_{\rm AA}$
for $p_T<2.5$~GeV/$c$, indicating that the direct photon yield
in the low-$p_T$ range increases faster than the binary-scaled
$p+p$ cross section.

\begin{table}[b]
\caption{Summary of the fits to the Au~+~Au data with
the exponential plus the modified power-law function ($A e^{-p_{\rm
T}/T} + B (1+p_T^2/b)^{-n}$) as explained in the text.  The
first and second errors are statistical and systematic, respectively.}
\begin{ruledtabular}
\begin{tabular}{lccc}
centrality & $dN/dy$($p_T>1$~GeV/$c$) &$T$(MeV)     &$\chi^2/$NDF\\
\hline
0-20\% & $1.50\pm0.23\pm0.35$ & $221\pm19\pm19$ & 4.7/4\\
20-40\%& $0.65\pm0.08\pm0.15$ & $217\pm18\pm16$ & 5.0/4\\
Min. Bias    & $0.49\pm0.05\pm0.11$ & $233\pm14\pm19$ & 3.2/4\\
\end{tabular}
\end{ruledtabular}
\label{tab:summary}
\end{table}

We fit an exponential plus the $T_{\rm AA}$-scaled $p+p$ fit function
($A e^{-p_T/T} + T_{\rm AA} \times A_{pp} (1+p_T^2/b)^{-n}$) to the Au~+~Au data.
The only free parameters in the fit are 
$A$ and the inverse slope $T$ of the exponential term.
The systematic uncertainties in $T$ are estimated
by changing the $p+p$ fit component and the Au~+~Au data points
within the systematic uncertainties.
The results of the fits are summarized in Table~\ref{tab:summary},
where $A$ is converted to $dN/dy$ for $p_T>1$~GeV/$c$.
For central collisions, $T=221\pm19 (stat.)\pm19 (syst.)$MeV.
If an unmodified power-law function ($\propto p_T^{-n}$) is used to fit
the $p+p$ spectrum, we find $n = 5.40 \pm 0.15$, and
$T = 240 \pm 21$ MeV.

\subsection{\pt~Dependence of Low Mass Excess}\label{sub:pt_lowmass}

\begin{figure}[!ht]
\includegraphics[width=1.0\linewidth]{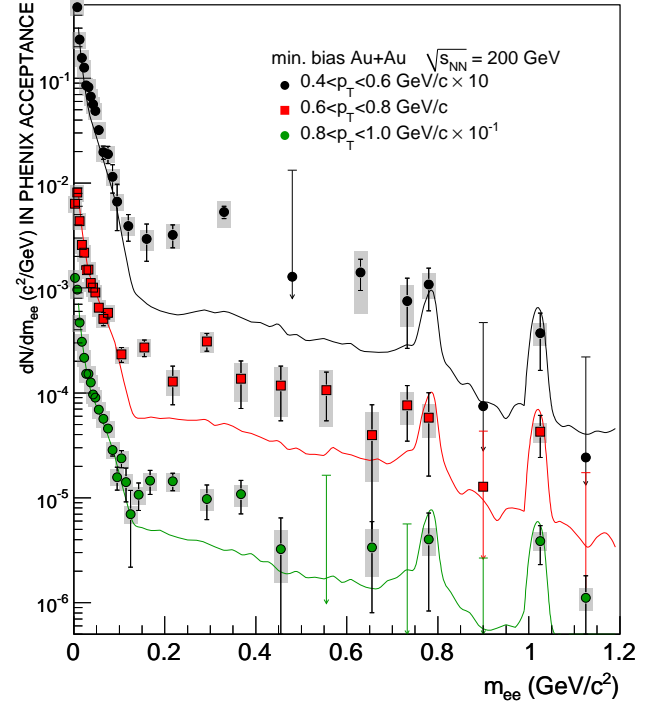}
\caption{ \label{fig:mass_au_lowpt} (Color online) 
The $e^+e^-$ pair invariant mass
distributions in minimum bias Au~+~Au collisions for the low-$p_T$ range.
The solid curves represent the cocktail of hadronic sources (see
Section~\ref{sec:cocktail}) and include contribution from charm
calculated by {\sc pythia} using the cross section from~\cite{ppg065}
scaled by $N_{\rm coll}$.}
\end{figure}

The shape of the enhancement in Au~+~Au data in LMR~II (low mass low $p_T$)
is quite different 
from that in LMR~I (low mass high $p_T$), where it behaves like 
$1/m_{ee}$ and is
consistent with internal conversion of direct photons.
In LMR~II, the enhancement is larger, as seen in the
two lowest $p_T$ bins of Fig.~\ref{fig:mass_pp_au}.  For
these bins, no excess is observed in the $p+p$ data.  In the lowest
$p_T$ bin the enhancement in the Au~+~Au data is approximately a factor
of five above the expectations from the cocktail.  The
data are significantly above the cocktail up to $m_{ee}=1$~GeV/$c^2$,
reaching their maximum around $m_{ee}\simeq 0.4$~GeV/$c^2$.

Figure~\ref{fig:mass_au_lowpt} shows the mass distribution in three
$p_T$ bins (0.4-0.6, 0.6-0.8, and 0.8-1.0~GeV/$c$) in the
LMR and a possible transition from $1/m_{ee}$ behavior at
higher $p_T$ (LMR~I) to much larger enhancement at lower
$p_T$ (LMR~II).  For the highest $p_T$ bin (0.8-1.0
~GeV/$c$) the excess is approximately a constant factor
above the cocktail.  This means that the mass spectrum is
still close to $1/m_{ee}$ expected for internal conversion.
The large enhancement seems to
appear for the next $p_T$ (0.6-0.8~GeV/$c$) bin.  For the
lowest $p_T$ bin the shape appears to be different from the
$1/m_{ee}$ behavior.  

\begin{figure}[!ht]
\includegraphics[width=1.0\linewidth]{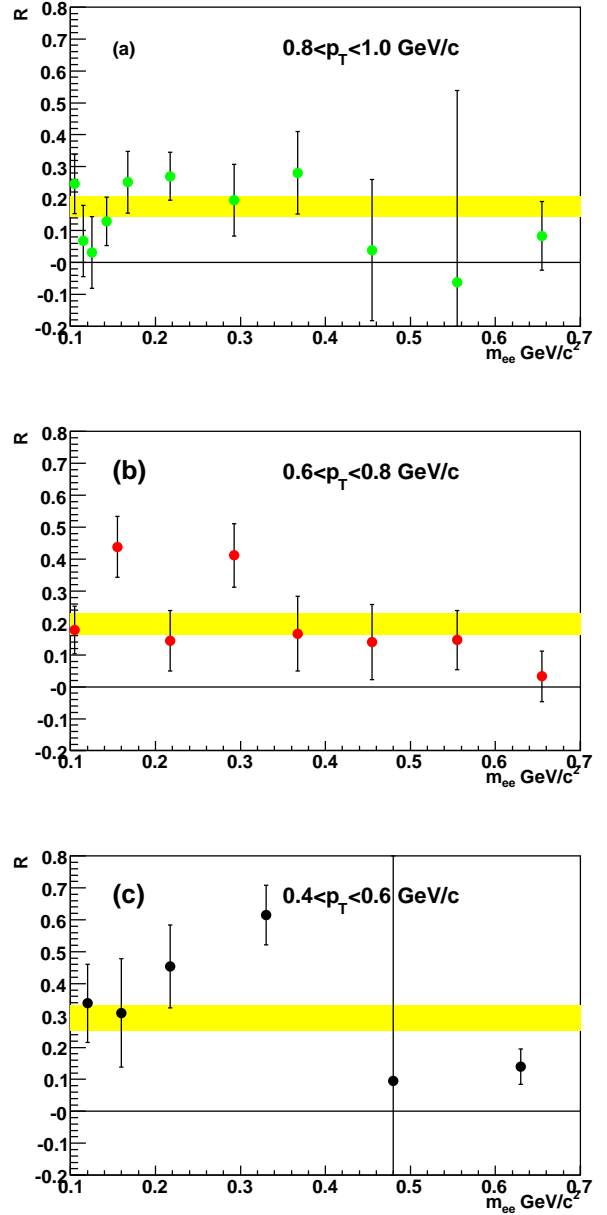}
\caption{ \label{fig:lowpt_ratio} (Color online)
Ratio of $R=({\rm data -
cocktail})/f_{\rm dir}(m_{ee})$ for different $p_T$ bins ((a):
$0.8 < p_T < 1.0$~GeV/$c$ , (b): $0.6< p_T < 0.8$
~GeV/$c$, (c): $0.4< p_T < 0.6$~GeV/$c$) in
minimum bias Au~+~Au collisions.
The yellow band in each panel shows $\pm 1 \sigma$ band of
a constant fit value to the data points.
}
\end{figure}

Figure~\ref{fig:lowpt_ratio} shows
$R$=(data-cocktail)/$f_{\rm dir}(m_{ee})$ for the three
low $p_T$ bins.  These ratios are proportional to the $S(m_{ee})$
factor, and a constant $S(m_{ee})$ leads to a constant ratio
$R$ as a function of mass.
While in Fig.~\ref{fig:lowpt_ratio}(a)
$R$ is still consistent with a constant as a
function of mass, Figure~\ref{fig:lowpt_ratio}(b) suggests that there is
an enhancement for $0.1<m_{ee}<0.4$, although the statistical
error is too large to be conclusive.  Figure~\ref{fig:lowpt_ratio}(c)
suggests a large and broad enhancement around $m_{ee} \simeq
0.4$~GeV/$c^2$.

We test whether the $R$ distributions in Fig.~\ref{fig:lowpt_ratio} are
consistent with a constant.  For each $p_T$ bin, we fit a constant to
the data.  The results of the fits are shown as the horizontal band in
each panel and are summarized in Table~\ref{tab:lowpt_fit}.
For $0.8<p_T<1.0$~GeV/$c$ the fit gives good $\chi^2/$NDF.  Thus the 
data are
consistent with the expected $1/m_{ee}$ behavior.
The next $p_T$ bin, $0.6<p_T<0.8$~GeV/$c$, gives marginally
satisfactory $\chi^2/$NDF.  For $0.4<p_T<0.6$~GeV/$c$ 
the $\chi^2/$NDF is large and the data are statistically inconsistent 
with a constant, suggesting that the electromagnetic spectral function 
is modified at low $p_T$.  
However, due to the large uncertainty of the point at
$m_{ee} \simeq 0.4$~GeV/$c^2$ the shape cannot be well determined.

The value $\langle R \rangle$ obtained from the constant fit
corresponds to the direct photon fraction $r$.
The fit value for $0.8<p_T<1.0$~GeV/$c$,
$\langle R \rangle =0.177\pm0.032$, is consistent with the values of
$r$ for higher $p_T$ shown in Fig.~\ref{fig:dir_ratio}.  
If we extrapolate the
$p_T$ spectrum of direct photons deduced from the previous section to lower 
$p_T$, the expected direct photon fraction for $p_T<1$~GeV/$c$ is 
$\simeq 0.17$ or less, since the spectrum of decay photons is steeper.
The $R(m)$ values for $0.4<p_T<0.6$~GeV/$c$ are larger than this expectation 
for $m_{ee} < 0.4$~GeV/$c^2$,
suggesting that the enhancement in the low $p_T$ region is larger than that
expected from internal conversion of direct photons.

In principle, the distribution of $R$ shown in Fig.~\ref{fig:lowpt_ratio}
can be extrapolated to $m_{ee}=0$ to obtain the fraction of real direct 
photons, even if the distribution of $R$ is not flat.  
However, due to large uncertainties in $R$ for $p_T < 0.8$~GeV/$c$
arising from the combination of multiple dilepton sources, we 
cannot reliably extrapolate the virtual photon yield to $m_{ee}=0$
to determine the real direct photon yield for these two $p_T$ bins.

\begin{table}[b]
\caption{Summary of a constant fit to the ratio data
shown in Fig.~\ref{fig:lowpt_ratio}.  The fit range 
is $0.1<m_{ee}<0.7$~GeV/$c^2$.}
\begin{ruledtabular} \begin{tabular}{lcc}
$p_T$(GeV/$c$) & $\langle R \rangle$ &$\chi^2/$NDF\\
\hline
0.8-1.0& $0.177\pm0.032$ & 7.7/10\\
0.6-0.8& $0.198\pm0.033$ & 16.3/7\\
0.4-0.6& $0.293\pm0.040$ & 21.3/5\\
\end{tabular} \end{ruledtabular}
\label{tab:lowpt_fit}
\end{table}

\subsection{\pt~Spectra for Different Mass bins}\label{sub:ptspectra}

\begin{figure*}[t]
\includegraphics[width=1.0\linewidth]{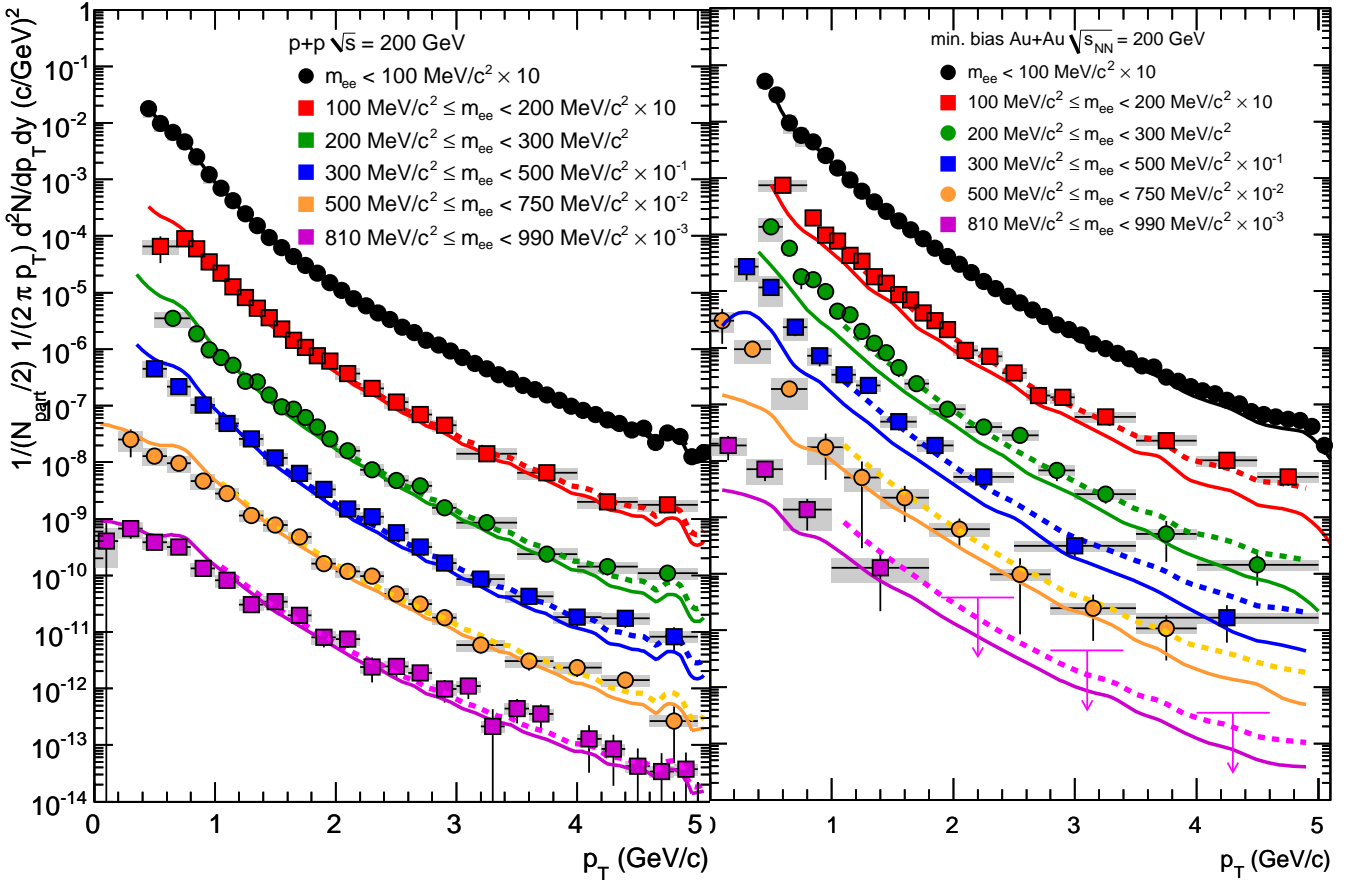} 
\caption{ \label{fig:pt_dir} (Color online)
$p_T$ spectra of $e^+e^-$ pairs in $p+p$ (left)
and Au~+~Au (right) collisions for different mass bins,
which are fully acceptance corrected.  Au~+~Au spectra
are divided by $N_{\rm part}/2$.  The solid curves show the
expectations from the sum of the hadronic decay cocktail and the
contribution from charmed mesons.  The dashed curves show the sum of
the cocktail and charmed meson contributions plus the contribution
from direct photons calculated by converting the photon yield from
Figure~\ref{fig:photon_spectra} to the $e^+e^-$ pair yield using
Eqs.~\ref{eq:Conversion}~and~\ref{eq:gCompton}.  }
\end{figure*}

Figure~\ref{fig:pt_dir} shows the transverse momentum spectra of
dileptons in different
mass windows for Au~+~Au and $p+p$ data:
\begin{equation}
\frac{1}{2\pi p_T}\frac{dN_{ee}}{dp_Tdy} = \int_{m_1}^{m_2} \frac{1}{2\pi p_T} \frac{d^3N}{dp_T dy dm_{ee}} dm_{ee}
\end{equation}
where $m_1$ and $m_2$ are the lower and upper limits of the different
mass slices.  In the low-mass slices ($m_{ee} <$~0.4~GeV/$c^2$) the
spectra are truncated at low pair-$p_T$ due to the
single-track acceptance $p_T > 0.2$~GeV/$c$.  The
pair-$p_T$ cutoff is mass dependent.  The Au~+~Au spectra have
been divided by $N_{\rm part}/2$ in order to ease the comparison with the
corresponding spectra in $p+p$.  The systematic uncertainty due to $N_{\rm part}$
($\sim$10\%) has not been included.  In order to avoid the influence of
$e^+e^-$ decays of narrow vector mesons, the mass regions around the
$\omega$ meson ($0.78 \pm 0.030$~GeV/$c^2$) and the $\phi$ meson ($0.1020
\pm 0.030$~GeV/$c^2$) are excluded.
The solid curves in Fig.~\ref{fig:pt_dir} represent the expectations
from the hadronic cocktail, which includes also the charm decay
contributions.  The charm contribution is calculated with {\sc pythia}.  Using
the random $c\bar{c}$ correlation makes no difference for
$m_{ee}<0.3$~GeV/$c^2$ since the charm contribution is negligible at
low masses.  This difference increases for the higher mass bins,
leading to spectra lower by $\sim$ 10\%, 20\% and 30\% for the mass
bins $0.3<m_{ee}<0.5$~GeV/$c^2$, $0.5<m_{ee}<0.75$~GeV/$c^2$,
$0.81<m_{ee}<0.99$~GeV/$c^2$ respectively, when the random $c\bar{c}$
correlation is used.

First we concentrate on the comparison between data and the sum of
cocktail and charm.  In the low-$p_T$ region ($p_T<$~1~GeV/$c$) all the
$p+p$ spectra are consistent with the expectations from the cocktail
alone for every mass window.  In the high-$p_T$ region however the
$p+p$ data show a small excess above the cocktail.  The Au~+~Au data are
in agreement with the cocktail in the mass region
$m_{ee}<0.1$~GeV/$c^2$.  In higher mass bins the Au~+~Au data show a
large excess both at low- and at high-$p_T$.

As discussed in section~\ref{sub:photon}, we have extracted the direct
photon yield from the dileptons spectrum in the mass range of $0.1 <
m_{ee} < 0.3$~GeV/$c^2$.  The excess in this mass range is consistent
with internal conversion of direct photons.  As shown in
Figure~\ref{fig:real_virtual_ratio_MB} the direct photon component, which
appears as a constant $R$, extends to the
$m_{ee}>0.3$~GeV/$c^2$.  Therefore, there should be sizable
contribution from direct photons in the dilepton spectra
for $m_{ee}>0.3$~GeV/$c^2$.
The relation between real direct photons and
virtual photons is presented in Appendix~\ref{app:ICA_theory}.
Here we use a constant factor $S(m_{ee},q)=1$ to extend
the direct photon component to higher mass ($m_{ee}>0.3$~GeV/$c^2$).
The dashed curves in Fig.~\ref{fig:pt_dir} show the sum of the cocktail, charm and
direct photon contributions to the dilepton spectra for 
$p_T > 1$~GeV/$c$.

The dashed curves describe the data well for all mass bins both in the
Au~+~Au and the $p+p$ data.  This indicates that the excess above the
cocktail and charm at high $p_T$ ($p_T>1$~GeV/$c$)
is consistent with the contribution from direct photons.  It is surprising
that the agreement holds even for $m_{ee}>0.5$~GeV/$c^2$, where
significant modifications of the spectral function may be expected due
to the presence of the vector mesons.  However the data have large
statistical errors for $m_{ee}>0.5$~GeV/$c^2$ and additional
enhancement over the direct photon contribution is not excluded.  The
data at high $p_T$ are also consistent with the cocktail alone for $m_{ee}>0.5$
~GeV/$c^2$.

In the Au~+~Au data, the enhancement over the cocktail is approximately
a constant factor for $p_T>1$~GeV/$c$.  It grows towards
low-$p_T$.  All the Au~+~Au $p_T$ spectra for every
mass bin above 0.3~GeV/$c^2$ seem to indicate that the enhancement
with respect to the cocktail below 1~GeV/$c$ is significantly larger
than above 1~GeV/$c$.  For $p_T>1$~GeV/$c$, the data has
a slope similar to the cocktail, as shown by solid curves.  For
$p_T < 1 $~GeV/$c$, the slope of the data is much steeper
than the cocktail.

In order to study this change of the slope in the Au~+~Au data more
quantitatively, we subtract the cocktail plus charm from the data and
examine the shape of the excess.  The $p_T$ spectra are
combined in the mass range $0.3<m_{ee}<0.75$~GeV/$c^2$.  In this mass
range, the low-$p_T$ cutoff which artificially truncates the
$p_T$ spectra at lower mass is avoided.  The combined data
also have increased statistical significance.

\begin{figure}[!ht]
\includegraphics[width=1.0\linewidth]{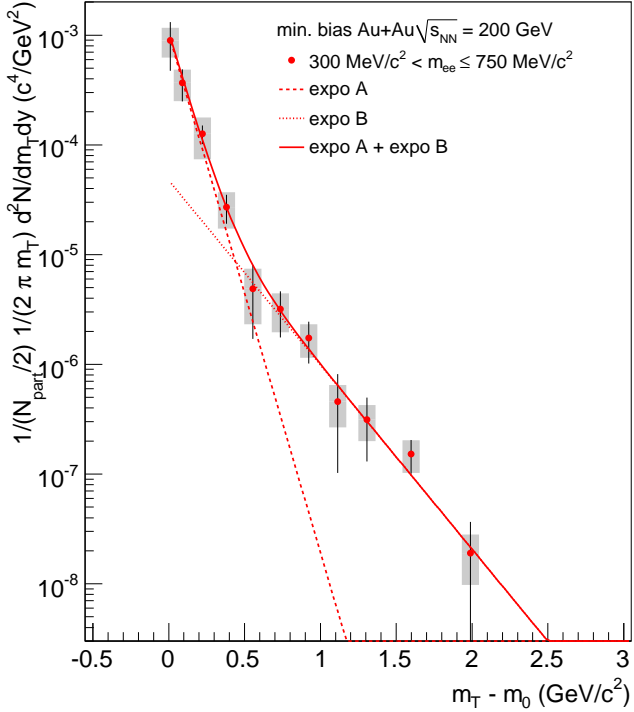}
\caption{ \label{fig:mt_2expo} (Color online)
The $m_T-m_0$
spectrum for the mass range $0.3<m_{ee}<0.75$~GeV/$c^2$ after subtracting
contributions from cocktail and charm.  The spectrum is fully acceptance 
corrected.  The systematic error band includes the difference in charm yields 
in this mass range.  The spectrum is fit to the sum of two exponential 
functions which are also shown separately as the dashed and dotted lines.  
The solid line is the sum.}
\end{figure}

Figure~\ref{fig:mt_2expo} shows the pair $m_T-m_0$ spectrum in
Au~+~Au for the pair mass range $0.3<m_{ee}<0.75$~GeV/$c^2$.  Here
$m_T = \sqrt{p^2_T + m^2_0}$ is the transverse mass of
the pair and $m_0$ is the mean value of $dN/dm_{ee}$ in the given mass
range ($0.3<m_{ee}<0.75$~GeV/$c^2$ in this case).  We plot the data as
function of $m_T - m_0$ since invariant differential cross
sections of hadrons in $p+p$, $p+A$, and $A+A$ collisions are
generally well described by exponential functions in
$m_T$.  Thus the change in the slope can be seen more clearly
in the $m_T$ spectrum.  The $m_T$ spectrum shows a
clear change in the slope around 1~GeV/$c^2$.  The slope below
$m_T-m_0<1$~GeV/$c^2$ is much steeper than that above 1
~GeV/$c^2$.  In order to characterize the change of the slope in the
two $m_T$ regions, we fit the $m_T$ spectrum with
the sum of two exponentials:
\begin{equation} \label{eq:2expo_fit}
\frac{d^2N}{2\pi m_T dm_T dy} = 
A_1 \cdot \exp{-\frac{m_T}{T_1}} + A_2 \cdot \exp{-\frac{m_T}{T_2}}
\end{equation}
where $A_1$ and $A_2$ are the normalization parameters, and $T_1$ and
$T_2$ are the inverse slope parameters.

The result of the fit is shown in Fig.~\ref{fig:mt_2expo}.  
The upper solid curve shows the fit function
and the dashed and dotted lines are the two exponential components.
The fit gives $T_1 = 92.0 \pm 11.4^{\rm stat} \pm 8.4^{\rm syst} $~MeV 
and $T_2 = 258.4 \pm 37.3^{\rm stat} \pm 9.6^{\rm syst}$~MeV with 
$\chi^2/$NDF$=4.00/9$.  The two-exponential fit describes the data 
well.  We note that the value of $T_2$, 
is somewhat higher than, but consistent with, the inverse slope of 
the exponential component of the direct photon spectrum, 
$T = 221 \pm 19^{\rm stat} \pm 19^{\rm syst}$ MeV, obtained in the 
previous section.

\begin{figure}[!ht]
\includegraphics[width=1.0\linewidth]{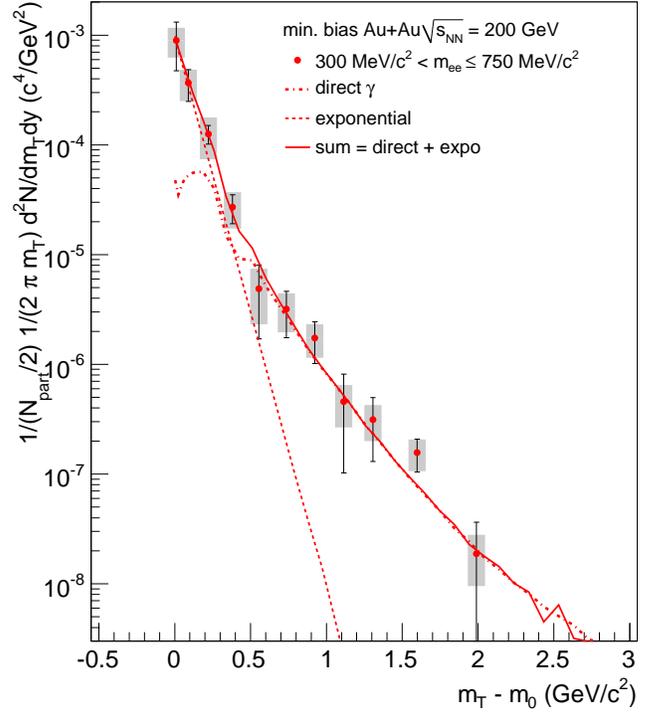}
\caption{ \label{fig:mt} (Color online)
The $m_T-m_0$ spectrum
for the mass range $0.3<m_{ee}<0.75$~GeV/$c^2$ after subtracting the
contributions from cocktail and charm.  
The spectrum is fully acceptance corrected.  
The systematic error band
includes the difference in charm yields in this mass range.  The
spectrum is fit to the sum of direct photons and an exponential
function.  The dashed-dotted line is the direct photon contribution.  The
exponential fit to the low-$m_T$ enhancement is also shown
(dashed line).  The sum is shown with the thick solid line.}
\end{figure}

In Fig.~\ref{fig:mt} the same Au~+~Au $p_T$ spectrum
is fit with the sum of direct photons and an
exponential function in $m_T$.
The exponential function is to characterize the low $p_T$
component.
The direct photon component is obtained from the
direct photon spectrum in Fig.~\ref{fig:photon_spectra},
and extended to the larger mass region assuming $S(m,q)=1$.
We then convert the photon yield to the $e^+e^-$ pair yield using
Equation~\ref{eq:Conversion}.
Thus the direct photon component is fixed and the only 
free parameters of the fits are the normalization and the inverse slope $T$
of the exponential component.

The $p_T$ spectrum and the individual components of the fit function,
i.e.  the direct photon component, and the
exponential component are shown in Fig.~\ref{fig:mt}.  The spectrum is well
reproduced by the fit and $\chi^2/$NDF$ = 16.6/11$.  The systematic
uncertainty accounts for the uncertainty on the data and the
uncertainty ($\sim$20\%) of the cocktail normalization.  From the fit,
we extract a value of 
$T = 86.5 \pm 12.7^{\rm stat}$~+11.0-28.4$^{\rm syst}$~MeV.
The yield of the low-$p_T$ exponential extracted from
the fit contributes more than 50\% of the yield of the spectrum.

Both the two exponentials fit shown in Fig.~\ref{fig:mt_2expo} and the
exponential + direct photon fit shown in Fig.~\ref{fig:mt} show that
there is a low inverse slope component with $T\simeq 100$ MeV for
$m_T-m_0 < 0.6$~GeV/$c^2$.  In order to further study the mass
dependence of the inverse slope, we calculated the local slopes of the
invariant pair $m_T$ spectra obtained from the $p_T$ spectra shown in Fig.~\ref{fig:pt_dir}.  For
all the mass bins the cocktail and charm are subtracted from the data
and the average inverse slope $\langle T(m_0)\rangle$ of the excess at mass $m_0$
has been numerically calculated as
\begin{eqnarray}
\langle T(m_0)\rangle=\frac{\Sigma (m_T^i - m_0) f(m_T^i) \Delta m_T^i}
{\Sigma f(m_T^i) \Delta m_T^i},
\end{eqnarray}
where
\begin{eqnarray}
f(m_T) &=& \frac{1}{2\pi m_T}\frac{d^2N}{dm_Tdy}\\
&=& \frac{1}{2\pi p_T}\frac{d^2N}{dp_Tdy}
\end{eqnarray}
is the invariant spectrum of the electron pairs after cocktail subtraction
shown in Fig.~\ref{fig:pt_dir}.

\begin{figure}[!ht]
\includegraphics[width=1.0\linewidth]{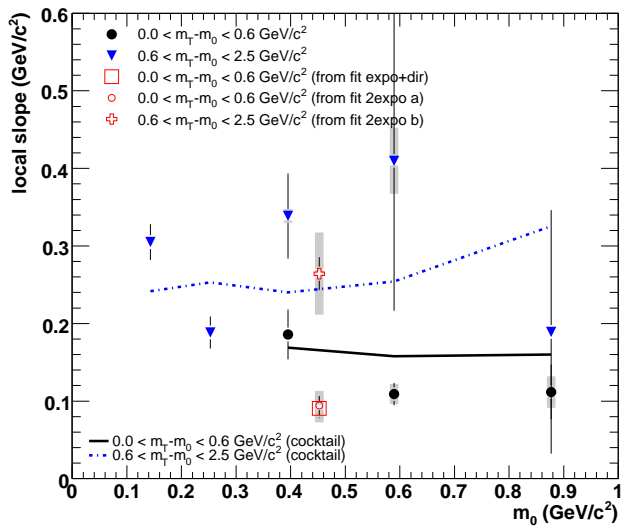} \caption{
\label{fig:meanmt} (Color online)
Local inverse slope of the
$m_T$ spectra of electron pairs, after subtracting the cocktail and
the charm contribution, for different mass bins.  The local slope is
calculated in different mass ranges, $0 < m_T - m_0 < 0.6$~GeV/$c^2$
and $0.6 < m_T - m_0 < 2.5$~GeV/$c^2$.  The solid and dashed lines
show the local slope of the cocktail for the corresponding mass
ranges.}
\end{figure}

Figure~\ref{fig:meanmt} shows the 
inverse slopes calculated in two ranges, namely $0 < m_T <
0.6$~GeV/$c^2$ and $0.6 < m_T < 2.5$~GeV/$c^2$.  
For $m_0<0.4$~GeV/$c^2$ the spectra are
truncated due to the acceptance; therefore, we do not quote any slope
here.  Also for $m_0<0.1$~GeV/$c^2$ the slope in the range $0.6 < m_T
< 2.5$~GeV/$c^2$ is not shown because the cocktail subtraction has too
large systematic uncertainty in this region.  The solid and dashed
lines show the inverse slope of the cocktail, calculated
in the same way as the data, for the same mass ranges.  The inverse
slope parameters obtained from the two-exponential fit as well as the
exponential + direct photon fit in the same mass ranges are also shown.

Figure~\ref{fig:meanmt} suggests qualitatively different behavior for
low and high $m_T$.  In the large $m_T-m_0$ range ($0.6<m_T-m_0<2.5$
~GeV/$c^2$) the inverse slope $\langle T\rangle$ is approximately 300~MeV and
similar to that of the cocktail (shown in the dashed-dotted line).  In
the low $m_T-m_0$ range the inverse slope is approximately 100~MeV,
similar to that obtained with the two fit methods.  This latter one
(shown at $m_T-m_0 \sim 0.45$~GeV/$c^2$) is lower than that of the
cocktail in similar kinematic range (shown in the solid line).

The effective temperature of the lower inverse slope component $T
\simeq 100 $~MeV, obtained from the two fit methods as well as the
numerical calculation, is much lower than the inverse slope of hadrons
with similar masses (kaons) measured by PHENIX~\cite{AUpikp}.  The
slope of the kaon spectrum is larger than 200~MeV.  The hadron slopes
rise linearly with mass, consistent with the expectations from radial
expansion of the hadronic source.  If arising from thermal radiation
of the fireball, dominated by pion-pion annihilation $\pi^+ \pi^-
\rightarrow \rho \rightarrow e^+e^-$, the excess yield in the LMR would show similar
temperatures and a similar linear rise, reminiscent of radial flow of
a hadronic source~\cite{NA60_therm}.
The value of the low-$p_T$ inverse slope is lower than or similar to the
freeze-out temperature.  Also the inverse slope of dileptons with an
average mass smaller than 0.5~GeV/$c^2$ is more than a factor 2
smaller than that of kaons.  
The simplistic expectation, that 0.5~GeV/$c^2$ dilepton emission is
created similarly to kaons (from an equilibrated flowing source), is
not supported by the data.

\section{THEORY COMPARISON} \label{sec:discussion}

\begin{figure*}[t]
\includegraphics[width=0.45\linewidth]{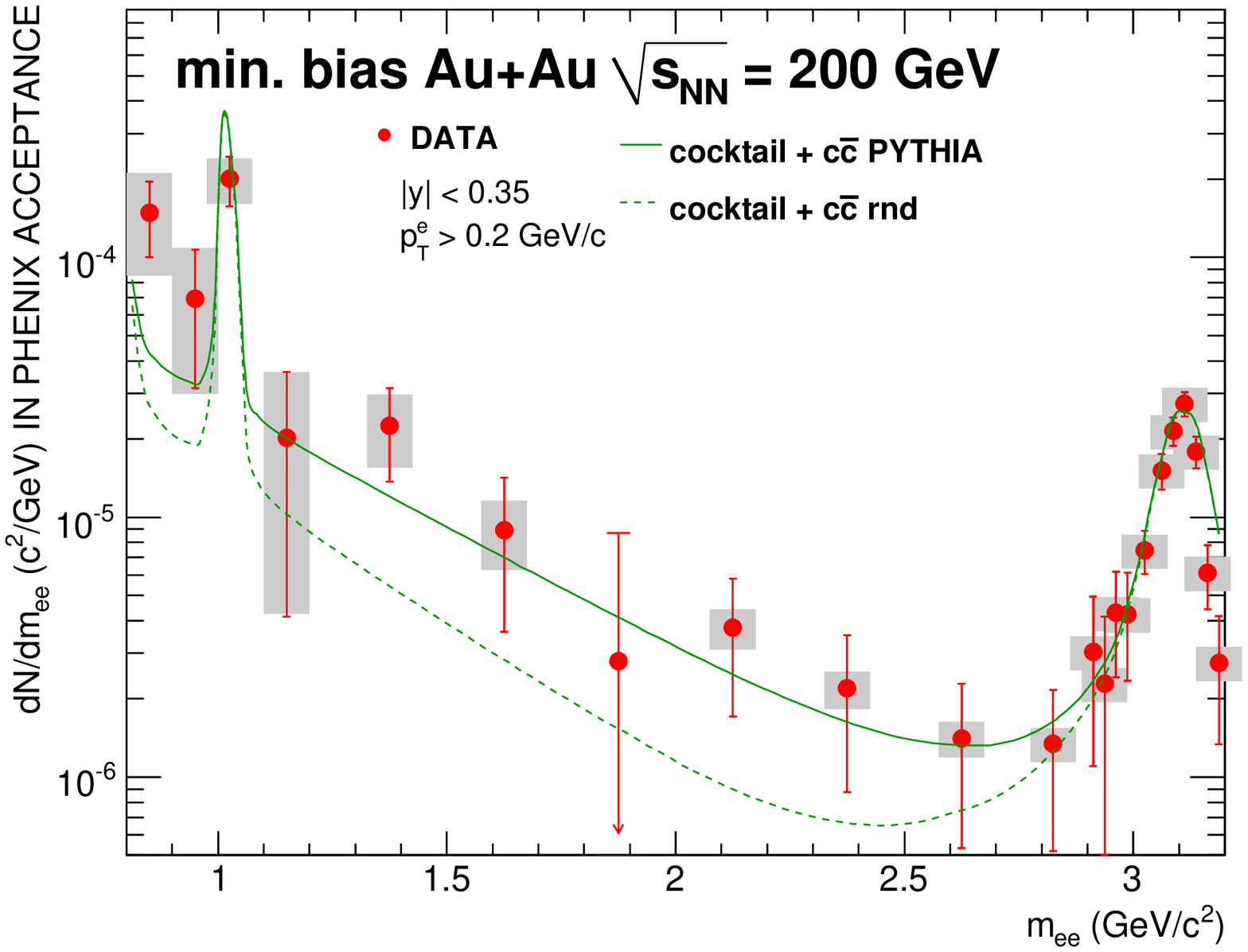} 
\includegraphics[width=0.45\linewidth]{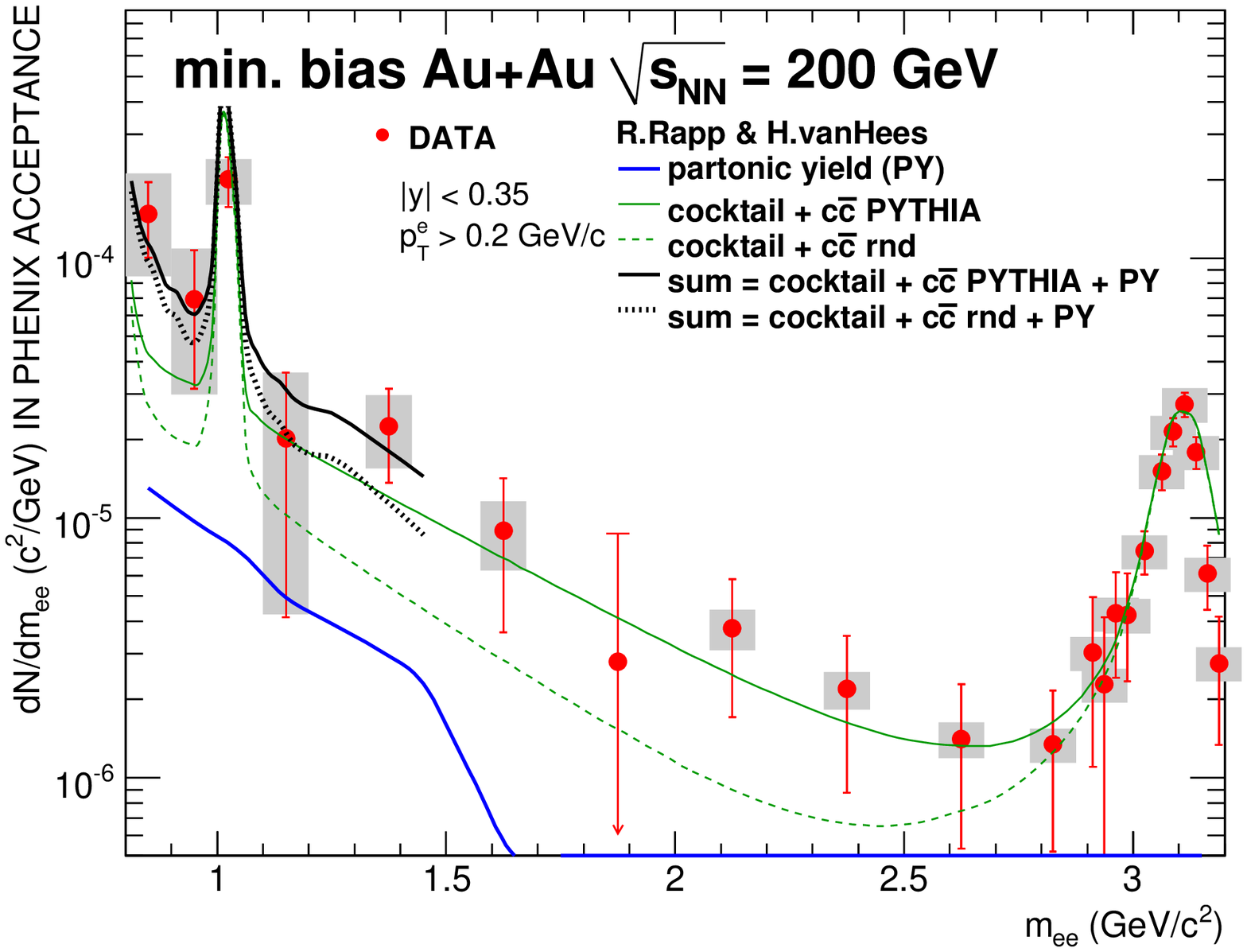} 
\includegraphics[width=0.45\linewidth]{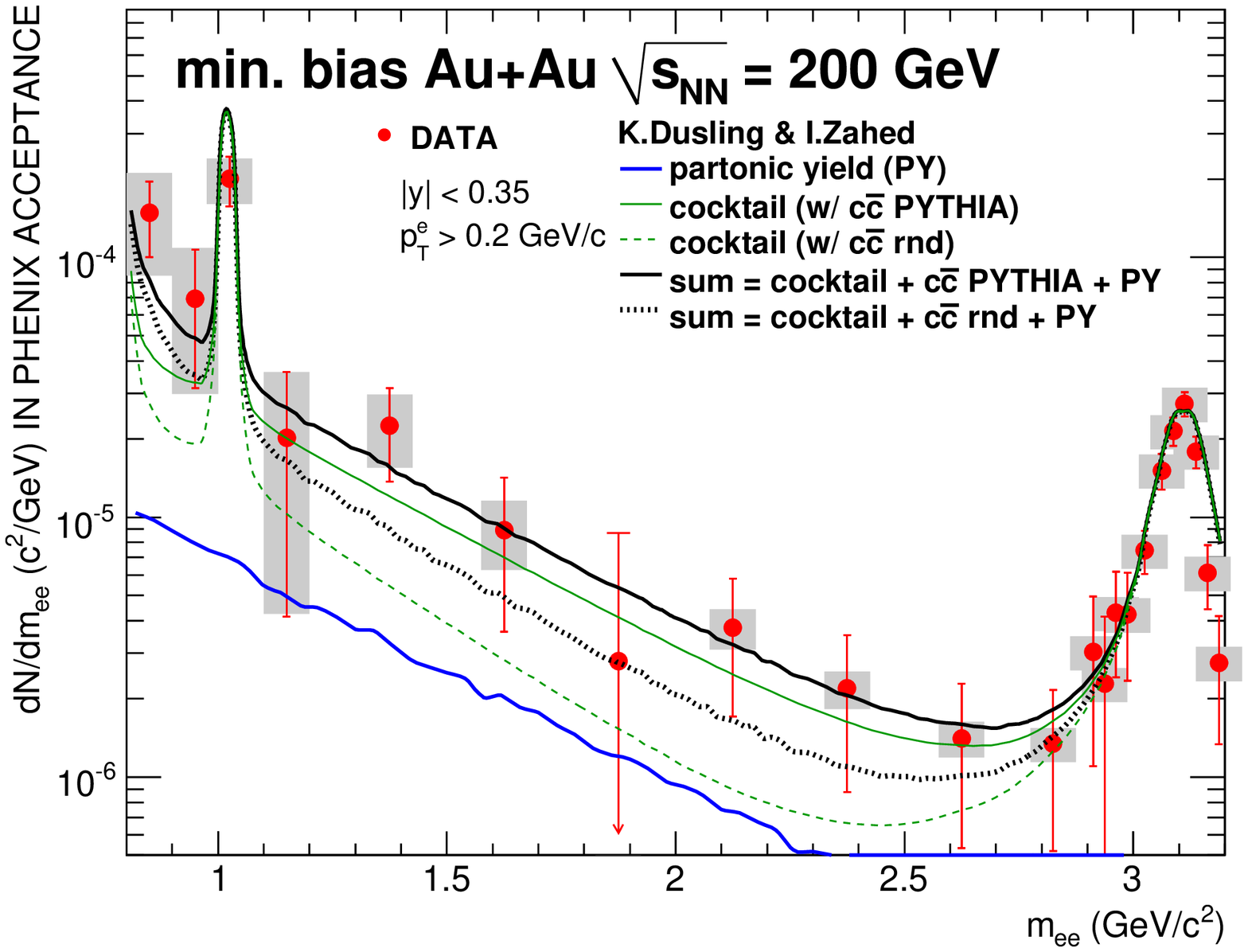} 
\caption{
\label{fig:mass_theo_IMR} (Color online)
Invariant mass spectra of $e^+e^-$ pairs in
Min. Bias Au~+~Au collisions in the IMR.  (top left) The data are 
compared to the sum of cocktail+charm.  The data are also compared 
to the sum of cocktail+charm and partonic contributions from different 
models.  The calculations are from 
(center) Rapp and van Hees~\protect\cite{Rapp1,rapp_RHIC,RalfRapp} 
and (right) Dusling and 
Zahed~\protect\cite{dusling_RHIC,dusling_mod,dusling_mod_RHIC}.
The partonic yields (PY) have been added to the two
scenarios for charmed mesons decays, i.e.  (i) {\sc pythia} and (ii) random
$c\bar{c}$ correlation.}
\end{figure*}

\begin{figure*}[t]
\includegraphics[width=0.45\linewidth]{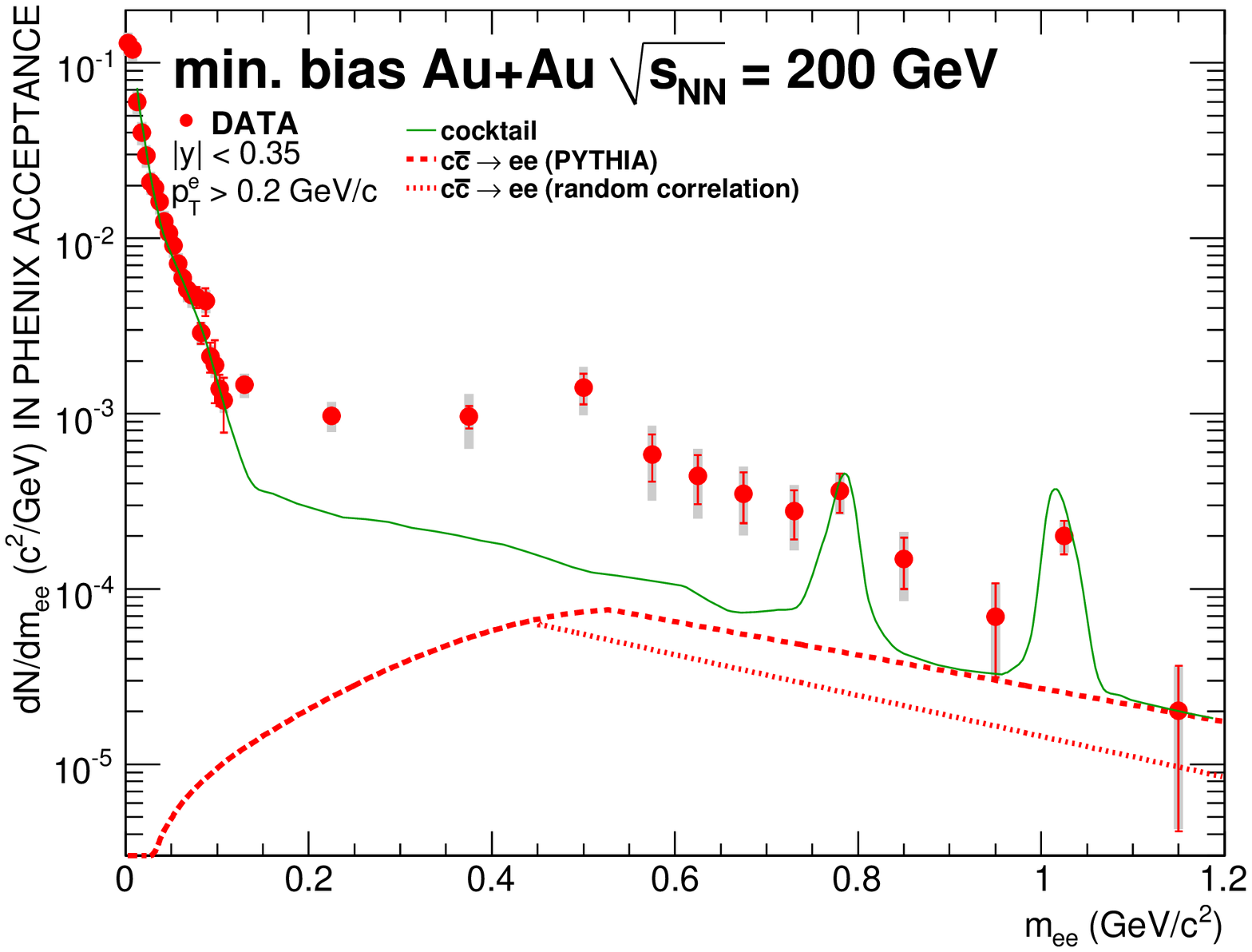} 
\includegraphics[width=0.45\linewidth]{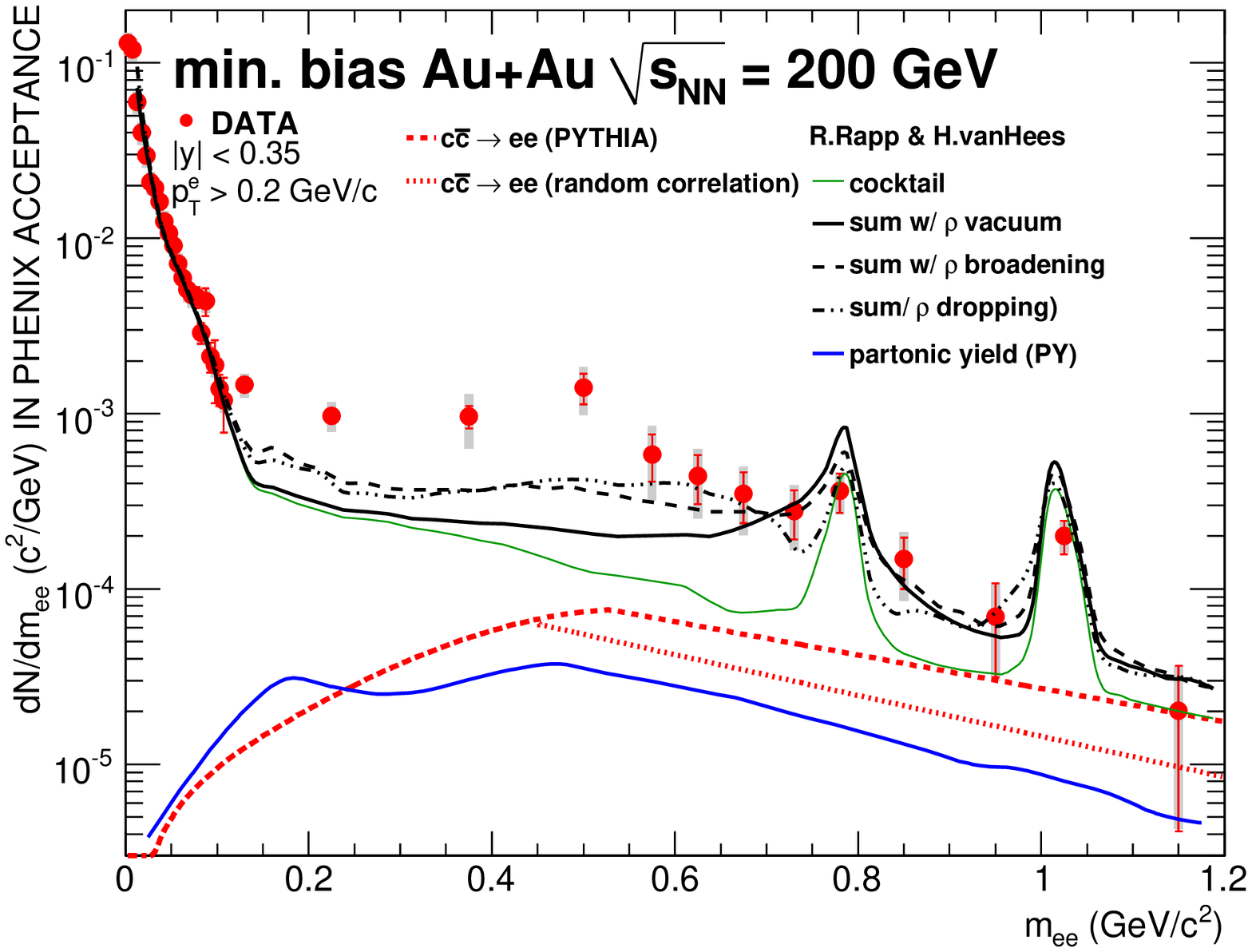} 
\includegraphics[width=0.45\linewidth]{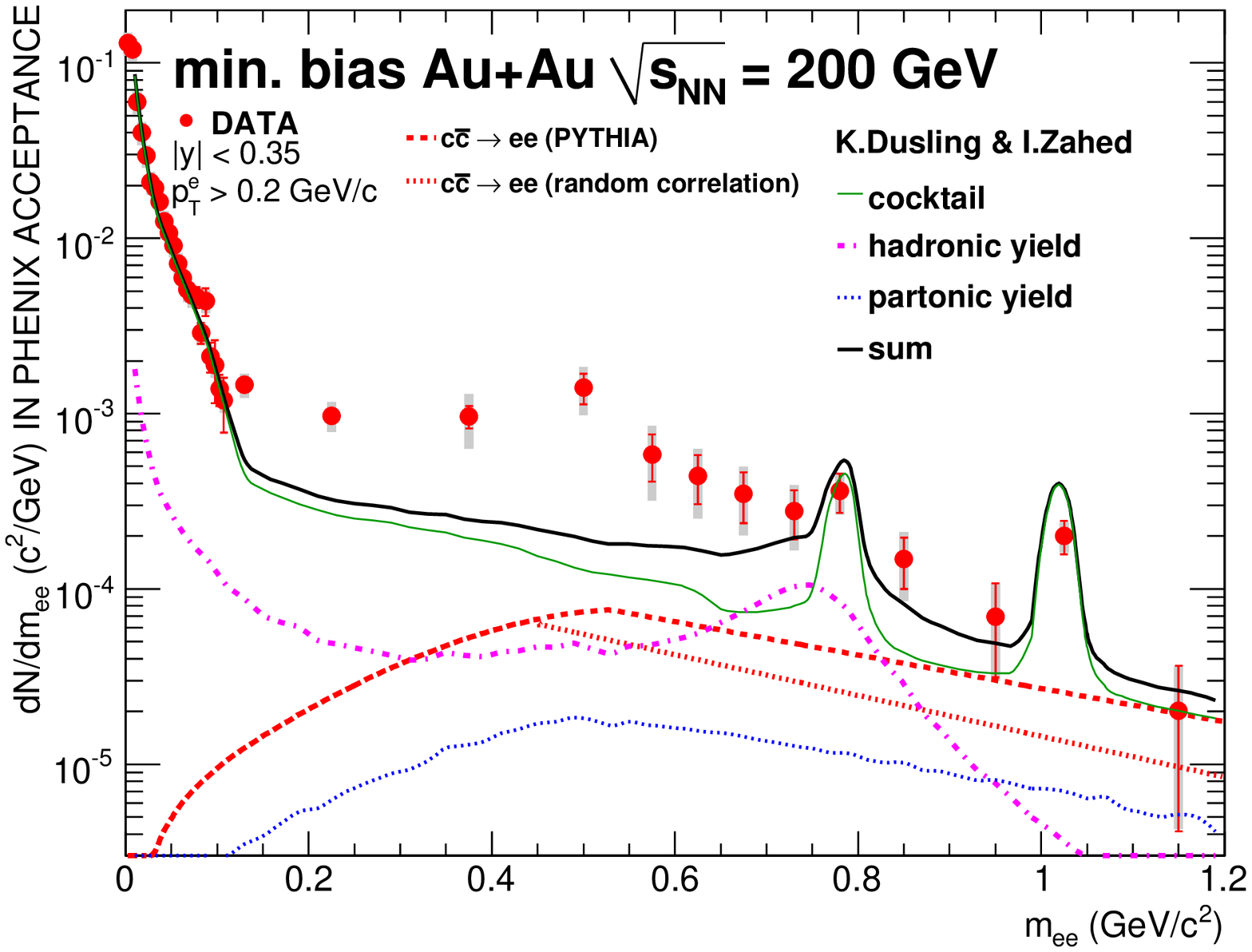} 
\includegraphics[width=0.45\linewidth]{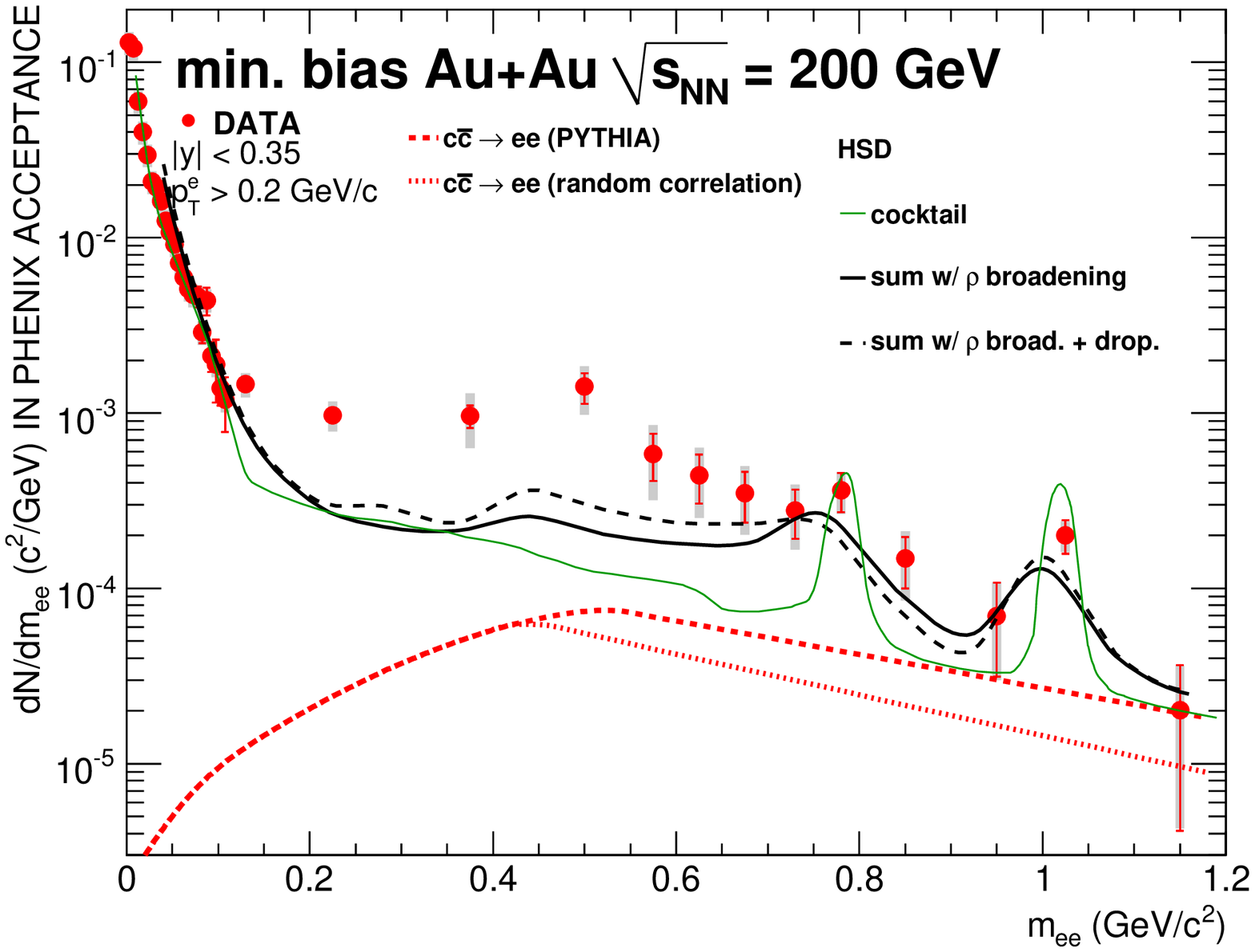} 
\caption{
\label{fig:mass_theo_LMR} (Color online)
Invariant mass spectra of $e^+e^-$ pairs in
Au~+~Au collisions in the LMR.  The data are compared to the sum of cocktail+charm
(top left).  The data are also compared to the sum of
cocktail+charm and hadronic+partonic contributions from different models.  The
calculations are from 
The calculations are from
(top right) Rapp and van Hees~\protect\cite{Rapp1,rapp_RHIC,RalfRapp},
(bottom right) Dusling and
Zahed~\protect\cite{dusling_RHIC,dusling_mod,dusling_mod_RHIC},
and Cassing and Bratkovskaya~\protect\cite{cassing_RHIC,cassing0,cassing_HSD,WolfgangCassing}.
}
\end{figure*}
The Au~+~Au $e^+e^-$ spectra are now compared to different models of
$e^+e^-$ production in the LMR and in the IMR.
These models, employed at SPS energies, identified the pion
annihilation process as the main source of thermal dileptons in the
hadronic phase of the fireball.  However, this process,
mediated by the intermediate $\rho$ meson, failed to describe
the observed enhancement in the LMR~\cite{CERES} at the SPS energy 
when vacuum properties of the $\rho$ are used.  
This suggested that in-medium modifications of the $\rho$
spectral function could be responsible for the enhancement 
of dilepton yield below the $\rho$ mass.
The proposed modifications mostly followed two different approaches:
\begin{itemize}
\item The \emph{dropping mass} scenario followed the scaling conjecture of
G.E.~Brown and M.~Rho~\cite{brown0}, which postulates that the mass of
vector mesons decreases in dense matter as the
quark condensate $\langle\bar{q}q\rangle$ decreases due to
partial restoration of chiral symmetry.
This leads to a decrease of the mass of $\rho$ meson from its vacuum value
(0.77~GeV/$c^2$) and causes enhancement of the dilepton yield below the $\rho$ mass.

\item The \emph{broadening mass} scenario explains the LMR dilepton
enhancement by hadronic many-body interactions~\cite{rapp0}.  The
spectral function in a hot and strongly interacting hadron
resonance gas is calculated.  The many-body interactions cause
the broadening of the $\rho$ resonance, leading to enhancement of
dilepton yield below $\rho$ mass.  Other hadronic many-body interactions
contribute to low mass enhancement.

\end{itemize}

There seems now to be consensus that the dropping mass scenario alone
cannot adequately reproduce the SPS data.
These two scenarios, sometimes in combination,
are used by different groups to calculate the
dilepton yield at the top RHIC energy ($\sqrt{s_{NN}}=200$~GeV).
Here we compare calculations by the following three
groups to the PHENIX data.  

\begin{itemize}
\item
Rapp and van~Hees~\cite{Rapp1,rapp_RHIC,RalfRapp} calculate the rate of
dilepton emission from a hadronic gas in thermal equilibrium.
In the calculation, the electromagnetic spectral function in the vacuum
is constrained by the data of $e^+e^-$ annihilation into hadrons.
The spectral function in the medium is modified by hadronic many-body
interactions.  The spectral function is characterized by the
light vector resonances $\rho$(770), $\omega$(782) and $\phi$(1020) at
low-mass according to the vector dominance model (VDM), and a perturbative
quark-antiquark continuum at higher masses.  Dilepton production
from a partonic phase outshines the hadronic gas radiation
for $m_{ee}>$~1.5~GeV/$c^2$ at RHIC energy due to
high initial temperatures.

\item
Dusling and Zahed~\cite{dusling_RHIC,dusling_mod,dusling_mod_RHIC} 
use a chiral reduction formalism
to calculate the electromagnetic current-current
correlation function in the medium.
The experimental data of $e^+e^-$ annihilation, $\tau$-decay, two-photon
fusion reactions, and pion radiative decays are used to constrain the
correlation function.  The dilepton emission
rates from hadronic gas at finite temperature and baryon density are then
computed from a hydrodynamic evolution model of Au~+~Au collisions.  The
partonic contribution, which does not become dominant below
the $\phi$ mass, is computed using the Born $q\bar{q}$
annihilation term.

\item
Cassing and 
Bratkovskaya~\cite{cassing_RHIC,cassing0,cassing_HSD,WolfgangCassing} 
use a microscopic relativistic
transport model (HSD) that incorporates the relevant off-shell
dynamics of the vector mesons.  This model is well established to
describe the yields, the rapidity distributions and the transverse
momentum spectra of hadrons in $p+A$ and $A+A$ collisions from SIS
to RHIC energies.  The model reproduces well the dilepton mass
spectrum in $p+p$ collisions.  In Au~+~Au the model includes a modified
$\rho$ spectral function according to a collisional broadening
scenario as well as a tunable dropping $\rho$ mass scenario.  No
yield from the partonic QGP phase is available at the moment from
the HSD model.
\end{itemize}

We received numerical values of these model calculations from the
authors.  The $e^+e^-$ rates from these calculations are filtered into the
PHENIX acceptance and compared to the data.

The theory calculation by Rapp and
van~Hees~\cite{Rapp1,rapp_RHIC,RalfRapp} is done for a fixed impact
parameter $b=8$~fm.  In this calculation the number of charged tracks
$N_{ch} = 230$, which is consistent with the measured $N_{ch}$ in
Au~+~Au MB data~\cite{ppg019}.  The theory calculation by Dusling and
Zahed~\cite{dusling_RHIC,dusling_mod,dusling_mod_RHIC}
is done for a fixed impact parameter for $b=0$~fm or $N_{\rm part} = 378$.

Since the calculations were provided for central collisions, or for
collisions with a fixed impact parameter, in the comparison to our Min.
Bias data we normalize the theory calculations by
$N_{\rm part}^{\rm model} / N_{\rm part}^{\rm Min. Bias}$,
i.e.  assuming that the dielectron yield scales with $N_{ch}$.  However,
this scaling procedure may have introduced some bias in the comparison, as
the data show (Fig.~\ref{fig:ratio}) that the dielectron yield increases
faster than $N_{\rm part}$ (and $N_{ch}$ is proportional to
$N_{\rm part}$).  A more detailed comparison requires knowledge of
centrality dependence of the dielectron yield both in the data and in the
theoretical model.

For comparison to the data, we add these calculations to the
hadronic cocktail and the charm decays.  The contribution from the
freeze-out $\rho$ meson is subtracted from the cocktail to avoid
double-counting.
Because the calculation with the HSD transport model by Cassing
and Bratkovskaya~\cite{cassing_RHIC,cassing0,cassing_HSD,WolfgangCassing}
can sample any impact parameter, the final Min. Bias
cross section is obtained by performing the integration over impact
parameter $b$ with a proper geometrical weight.  In this case the
model also calculates the hadronic contributions, which are in a good
agreement with the cocktail.  Only the charm contribution is taken from
the {\sc PYTHIA} calculation~\cite{pythia} or from our calculation with random 
correlation.

In Section~\ref{sub:discussion_photon} the Au~+~Au photon
spectrum is compared to several theoretical predictions.
These employ hydrodynamical models to calculate thermal
photon emission from the thermalized partonic and hadronic phases of
the reaction, added to NLO pQCD calculations that describe prompt
photon emission from perturbative parton-parton scatterings in the
first tenths of fm/c of the collision process.  

The experimental conditions reached at mid-rapidity in central heavy
ion collisions at RHIC of nearly zero net baryon density and
longitudinally boost-invariance in the initial conditions facilitate
the applicability of hydrodynamics to describe the reaction evolution.
In addition, the thermalization times usually assumed in the
hydrodynamical models ($\tau_{therm} \lesssim$~1fm/$c$) are, for the
first time at RHIC, above the lower limit imposed by the transit time
of the two colliding nuclei ($\tau_0 = 2R/\gamma \approx 0.15$~fm/$c$
for Au~+~Au at 200 GeV).

Hydrodynamical approaches describe, under the assumption of local
conservation of energy and momentum, the evolution of the system using
the equations of motion of perfect relativistic hydrodynamics
complemented with a set of initial conditions (e.g.  initial
temperature $T_{\rm init}$ at thermalization time $\tau_0$), the
equation-of-state of the system, and the freeze-out conditions.  These
models have been very successful in describing quantitatively most of
the differential observables of bulk hadronic production (in particular
those sensitive to early-times pressure gradients).

The same hydrodynamical models, with initial conditions chosen so as
to reproduce the bulk hadron data, are now employed to carry out the
description of thermal photon production over the whole space-time
evolution of the system.

\subsection{Comparison in the IMR and constraint on possible QGP radiation} \label{sub:discussion_IMR}
In this Section we compare the model calculations with the data in the
IMR and investigate whether the experimental data can constrain the
QGP radiation in the IMR.  We also study whether we can constrain
contributions from charm and QGP radiation in the LMR.

All theoretical models predict that there is large contribution from
QGP radiation in the IMR.  The QGP radiation competes with the
dileptons from correlated charm, which contribute much more than any
other cocktail contribution in the IMR.  There is a large uncertainty
in the charm contribution in Au~+~Au, since charm quarks are known to
suffer energy loss in the medium.  Therefore the $e^+e^-$ mass shape
from semi-leptonic decays of charmed quarks may be modified.
The shape calculated from {\sc pythia}~\cite{pythia} provides an upper limit for
the expected $e^+e^-$ yield.  The shape with random $c\bar{c}$
correlation
is softer for $m_{ee}>0.5$~GeV/$c^2$, and this provides
the lower limit.
Below 0.5~GeV/$c^2$ the two shapes are almost identical.  

These two scenarios for the open charm contribution are added to
the predictions for the QGP radiation provided by the models described
above and are compared to the experimental data in
Fig.~\ref{fig:mass_theo_IMR}.  In all the models the yield in the QGP
phase arises entirely from the $q\bar{q}\rightarrow e^+e^-$
annihilation process.  The magnitude of the yield is closely linked to
the thermalization time $\tau_0$.  A larger $\tau_0$ translates into
a reduction of the initial temperature and thus of the total QGP radiation.
In Fig.~\ref{fig:mass_theo_IMR} the differences between the models
are attributed to different initial conditions used for the
hydrodynamic evolution of the QGP phase ($\tau_0$=0.2~fm/$c$ for
Dusling and 
Zahed~\cite{dusling_RHIC,dusling_mod,dusling_mod_RHIC}, 
=0.6~fm/$c$ or Rapp and van~Hees~\cite{Rapp1,rapp_RHIC,RalfRapp}).

In the IMR the data have large statistical errors and systematic
uncertainties.  Therefore, they do not allow discrimination either 
between the two proposed scenarios for charm production 
({\sc pythia}~\cite{pythia} or random $c\bar{c}$ correlation) 
or among the three theoretical models.

\subsection{Inclusive Low Mass Excess} \label{sub:discussion_LMR}
The data in the IMR do allow setting an upper limit on the
contribution arising from charm or from $q\bar{q}\rightarrow e^+e^-$
going to the LMR.  Indeed we can saturate the IMR yield either with
charm or with the partonic yield calculated by the theorists and see
what their contribution would be in the LMR.  Thus, 
neither the charm nor the contribution from $q\bar{q}\rightarrow
e^+e^-$ can be solely responsible of the LMR enhancement.

In the LMR the shape and the yields of the $e^+e^-$ mass spectra from
charm calculated by {\sc pythia}~\cite{pythia} are not very different from those given by
a calculation which assumes random $c\bar{c}$ correlation, because the
shape is mostly determined by the geometrical acceptance.  Thus the
dilepton yield measured in the IMR gives strong constraint on the
charm contribution in the LMR.  Since the calculated charm contribution
is consistent with the data and is less than the hadronic cocktail
below $m_{ee} < 0.5$~GeV/$c^2$, we conclude that charm contribution
alone cannot explain the large enhancement observed in the LMR.  A
similar consideration can be given for the QGP radiation.  The
contribution from $q\bar{q}\rightarrow e^+e^-$ process is negligible
in LMR~I.

\begin{figure}[!ht]
\includegraphics[width=1.0\linewidth]{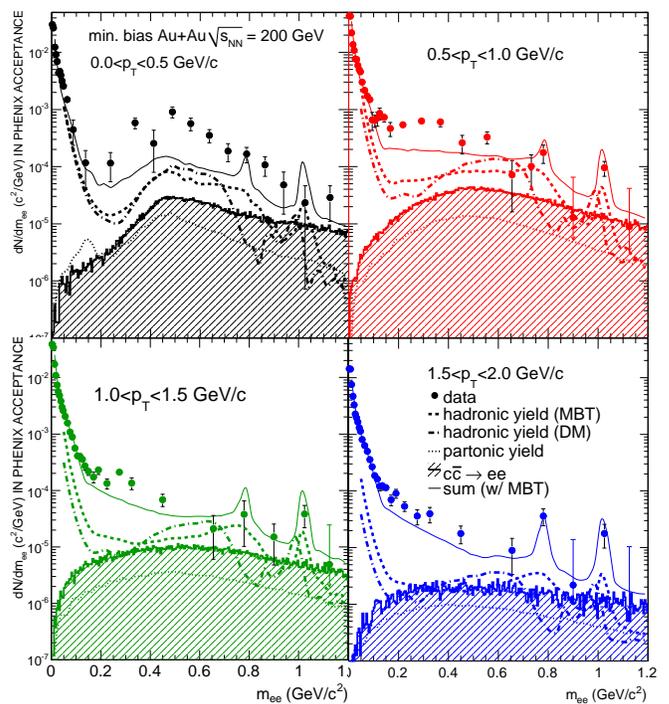}
\caption{
\label{fig:mass_pt_rapp} (Color online)
Invariant mass spectra of $e^+e^-$ pairs
in Min. Bias Au~+~Au collisions for different $p_T$ windows compared to
the expectations from the calculations of Rapp and 
van~Hees~\protect\cite{Rapp1,rapp_RHIC,RalfRapp},
separately showing the partonic and the hadronic yields and the
different scenarios for the $\rho$ spectral function, namely ``Hadron
Many Body Theory'' (HMBT) and ``Dropping Mass'' (DM).  The calculations
have been added to the cocktail of hadronic decays (where the
contribution of the freeze-out $\rho$ meson is subtracted) and charmed
meson decays products.
}
\end{figure}

\begin{figure}[!ht]
\includegraphics[width=1.0\linewidth]{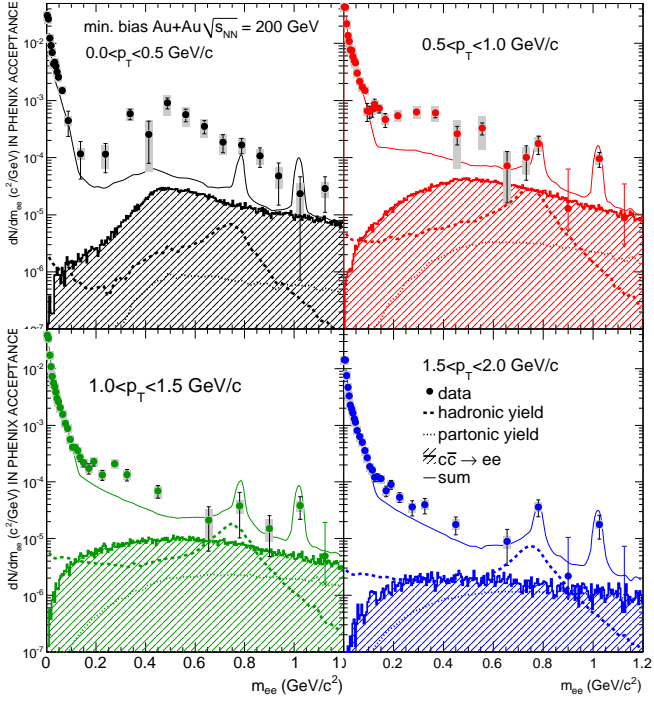} 
\caption{ \label{fig:mass_pt_dusling} (Color online)
Invariant mass spectra of $e^+e^-$ pairs
in Min. Bias Au~+~Au collisions for different $p_T$ windows compared to
the expectations from the calculations of 
Dusling and 
Zahed~\protect\cite{dusling_RHIC,dusling_mod,dusling_mod_RHIC},
separately showing the partonic and the hadronic yields.  The
calculations have been added to the cocktail of hadronic decays (where
the contribution of the freeze-out $\rho$ meson is subtracted) and
charmed meson decays products.
}
\end{figure}

\begin{figure}[!ht]
\includegraphics[width=1.0\linewidth]{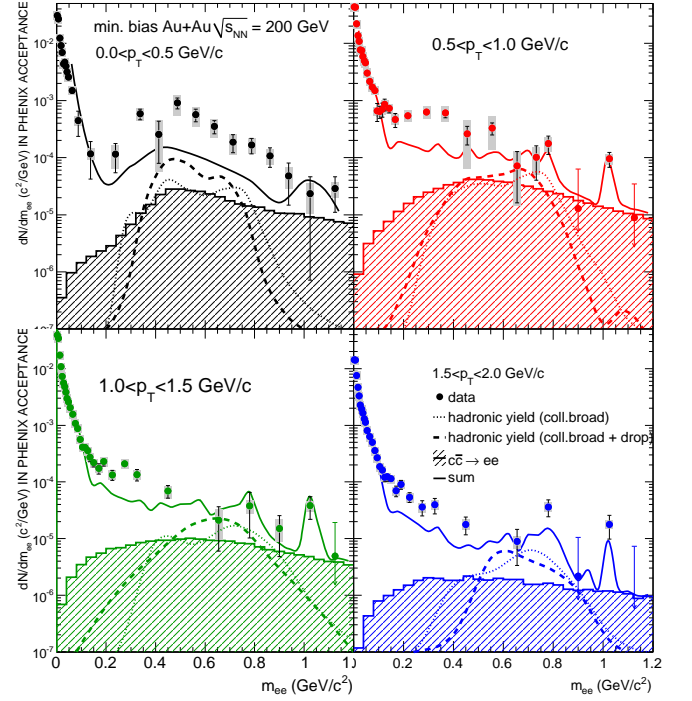} 
\caption{ \label{fig:mass_pt_bratkovskaya} (Color online)
Invariant mass spectra of $e^+e^-$
pairs in Min. Bias Au~+~Au collisions for different $p_T$ windows
collisions compared to the expectations from the calculations of
Cassing and 
Bratkovskaya~\protect\cite{cassing_RHIC,cassing0,cassing_HSD,WolfgangCassing}, 
separately showing the partonic and
the hadronic yields calculated with different implementations of the
$\rho$ spectral function, namely according to collisional broadening,
with or without a dropping mass scenario.  The calculations which
include the dropping mass scenario have been added to the cocktail of
hadronic decays (which is calculated by the HSD model itself) and
charmed meson decays products.  
}
\end{figure}


Figure~\ref{fig:mass_theo_LMR} compares the inclusive mass
spectrum in the LMR with the cocktail+charm only and with
cocktail+charm and the calculations by three groups: 
Rapp and van~Hees~\cite{Rapp1,rapp_RHIC,RalfRapp},
Dusling and 
Zahed~\protect\cite{dusling_RHIC,dusling_mod,dusling_mod_RHIC}, 
and Cassing and 
Bratkovskaya~\cite{cassing_RHIC,cassing0,cassing_HSD,WolfgangCassing}.
In all three models the $e^+e^-$ yield in the LMR 
arises mostly from the hadronic phase.
Rapp and van~Hees~\cite{Rapp1,rapp_RHIC,RalfRapp} propose
three different scenarios of vector mesons spectral functions: 
(i) no medium effects, (ii) dropping $\rho$ mass, 
and (iii) broadening $\rho$ mass.  Dusling and 
Zahed~\cite{dusling_RHIC,dusling_mod,dusling_mod_RHIC}
use a broadening $\rho$ mass scenario in the hadronic phase and 
a pion chemical potential of $\mu_{\pi}$=50~MeV.  Cassing and 
Bratkovskaya~\cite{cassing_RHIC,cassing0,cassing_HSD,WolfgangCassing} 
propose two scenarios:  (i) broadening $\rho$ mass, 
and (ii) dropping \emph{and} broadening mass.

The common characteristic of the in-medium effects in these models is
a slight suppression of the yield in
the $\rho$-$\omega$ region compared with the unmodified $\rho$ scenario
and an enhancement in  the region $0.4<m_{ee}<0.7$~GeV/$c^2$.  
The $\phi$ survives as a pronounced resonance, although its width is
broadened.  These features become less
distinct once the cocktail contribution 
and the smooth yield from the QGP (which
constitutes 15-20\%) are added.  
The differences in the yields of $e^+e^-$ pairs in the
various models are attributed to differences in the medium effects
on the spectral function, different durations of the
lifetime of the fireball in the hadronic phase, and different
evolutions of the temperature as a function of time.

While the calculations proposed by Rapp and 
van~Hees~\cite{Rapp1,rapp_RHIC,RalfRapp} agree
with the data for $m_{ee}>0.5$~GeV/$c^2$, the ones of Cassing and
Bratkovskaya~\cite{cassing_RHIC,cassing0,cassing_HSD,WolfgangCassing}
touch the lower end of the systematic uncertainty in
the same mass region.  The yields calculated by Dusling and 
Zahed~\protect\cite{dusling_RHIC,dusling_mod,dusling_mod_RHIC}
appear everywhere too low to add significant contribution
in the LMR, where the data are enhanced with respect to the hadronic
cocktail.

All of the models under predict the data for $0.2<m_{ee}<0.5$~GeV/$c^2$
by at least a factor of two.
It should be noted that the contributions in the region $m_{ee}<0.4$
~GeV/$c^2$ are very different in the three models compared.  In Rapp
and van~Hees~\cite{Rapp1,rapp_RHIC,RalfRapp}
this contribution arises from processes like
$a1\rightarrow \pi\gamma^*
\rightarrow \pi e^+e^-$ or $N\rightarrow N\gamma^* \rightarrow N
e^+e^-$.  Those
processes in the HSD model are suppressed by a few orders of magnitude
and are not seen at all compared to the major Dalitz decays.  In
Dusling and 
Zahed~\cite{dusling_RHIC,dusling_mod,dusling_mod_RHIC}
the main contribution below the two-pion 
threshold comes from $\Pi_A$, the axial-vector contribution in medium.
However the absolute yield of this process is too low because it is
concentrated at very low-$p_T$ which is suppressed by
our acceptance cut ($p_{T}^{\rm single}>$~0.2~GeV/$c$).

\subsection{$p_T$ Dependence of Low Mass Excess} \label{sub:discussion_LMR_pt}

Figs.~\ref{fig:mass_pt_rapp},~\ref{fig:mass_pt_dusling} and
~\ref{fig:mass_pt_bratkovskaya} show the $e^+e^-$ invariant mass
spectra in different $p_T$ windows from data compared to the sum of
all these contributions for the predictions of 
Rapp and van~Hees~\cite{Rapp1,rapp_RHIC,RalfRapp},
Dusling and Zahed~\cite{dusling_RHIC,dusling_mod,dusling_mod_RHIC}, 
and Cassing and 
Bratkovskaya~\cite{cassing_RHIC,cassing0,cassing_HSD,WolfgangCassing},
respectively.  The contribution from the hadronic and the partonic
medium and the charm expectations from {\sc pythia}~\cite{pythia}  are shown separately.
The charm distribution from {\sc pythia} is somewhat harder than the
calculations which assume random correlation of the $c\bar{c}$ pair
but not very different in the LMR, where the shape of the distribution
is essentially determined by the detector acceptance.

From the comparison we learn that in general the yield from these
theoretical predictions is insufficient to explain the observed
enhanced dilepton production, both at low-and high-momenta.  At
low-$p_T$, where the enhancement reaches approximately a factor of five, the
shape of the enhancement shown by the data is quite different from any of the
theoretical models.  

At high-$p_T$ ($p_T>1.0$~GeV/$c$) the enhancement
is about a factor two over the cocktail and its shape is quite similar
to that of the cocktail.  In the previous Section we showed that
this enhancement can be attributed to internal
conversion of virtual direct photons.  In the theoretical calculations of
direct photon emission at RHIC energies, the contribution from the
QGP phase, e.g.  quark-gluon Compton scattering, is the dominant
source of real thermal photons for $p_T>1$~GeV/$c$.  The process that
produces real photons in the QGP should also contribute low mass
$e^+e^-$ pairs at high $p_T$.  However, none of the three models
includes such processes
(e.g.  $q+g \rightarrow q+\gamma^* \rightarrow q + e^+e^-$).
The QGP radiation in these models only include $q+\bar{q}$ annihilation.
This could explain the discrepancy between the models and the data for 
$p_T > 1.0$~GeV/$c$.

\begin{figure*}[t]
\includegraphics[width=0.3\linewidth]{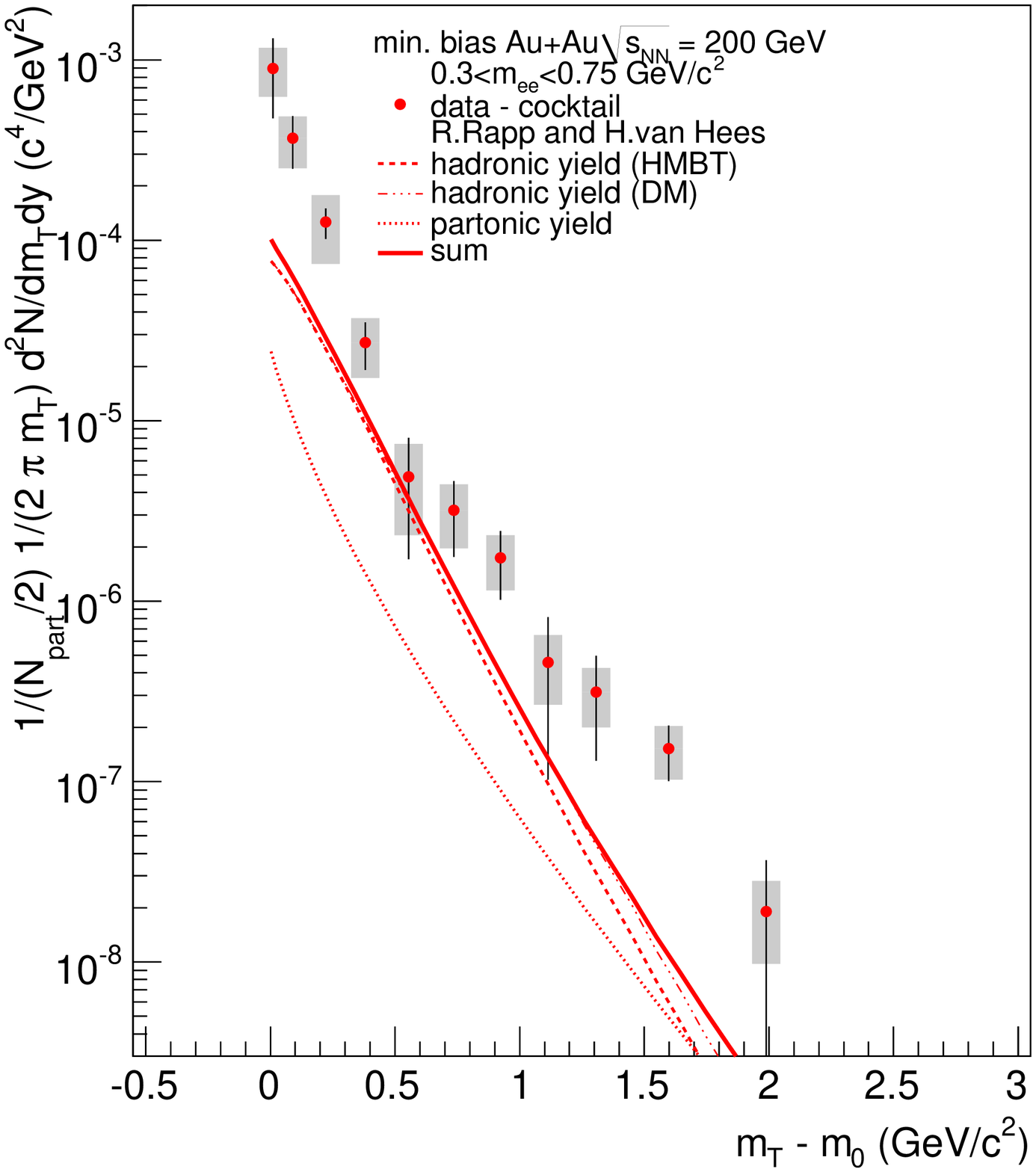}
\includegraphics[width=0.3\linewidth]{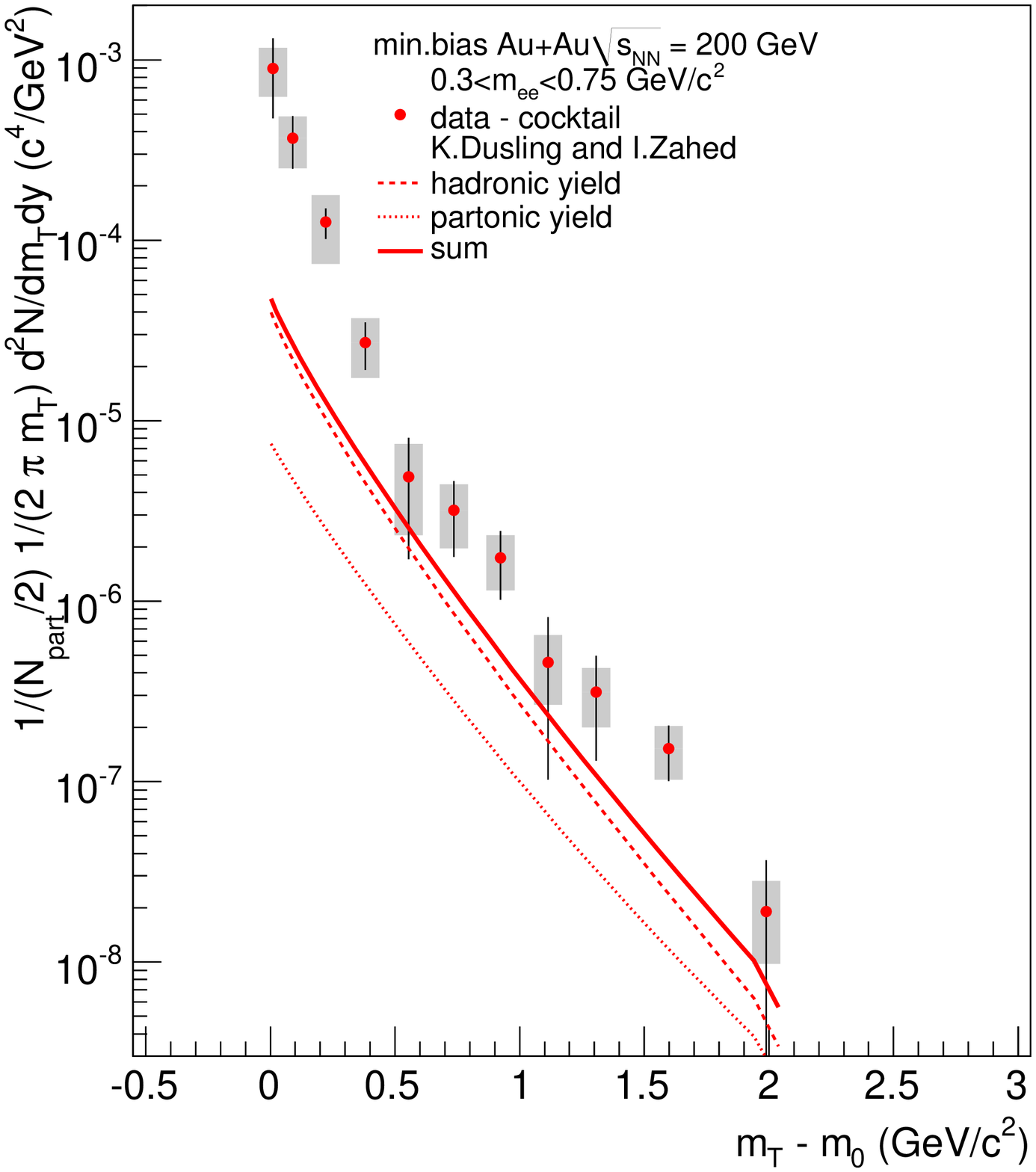}
\includegraphics[width=0.3\linewidth]{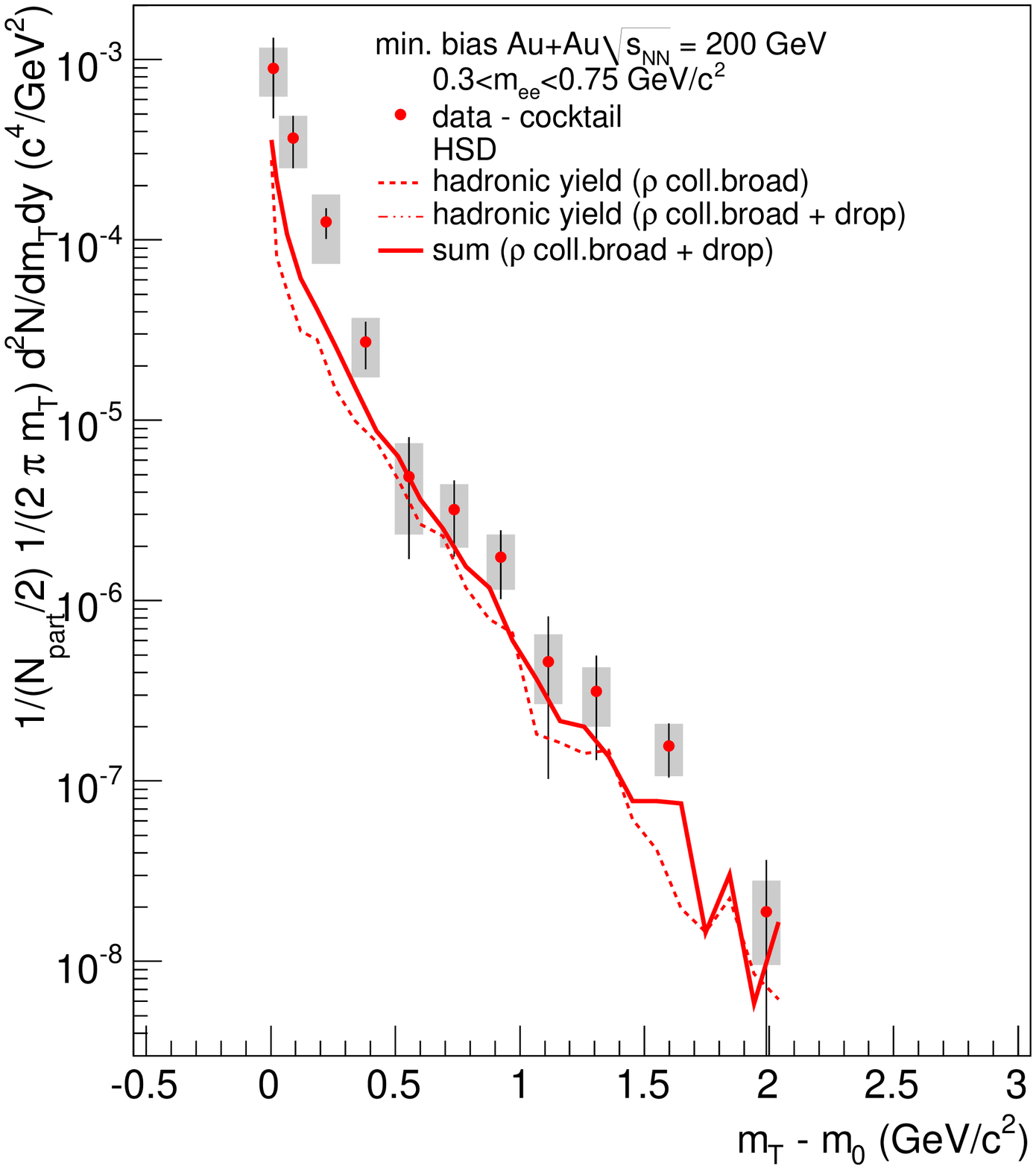}
\caption{\label{fig:pt_theo} (Color online)
$p_T$ spectra of $e^+e^-$ pairs for
$0.3<m_{ee}<0.75$~GeV/$c^2$ in Min. Bias Au~+~Au collisions compared to
the expectations from the calculations of respectively 
R.~Rapp and van~Hees~\cite{Rapp1,rapp_RHIC,RalfRapp}, 
Dusling and 
Zahed~\cite{dusling_RHIC,dusling_mod,dusling_mod_RHIC}, 
Cassing and 
Bratkovskaya~\protect\cite{cassing_RHIC,cassing0,cassing_HSD,WolfgangCassing}.  
The spectra are fully acceptance corrected.  
The curves show separately partonic and hadronic
yields.  For the curves of Rapp and van~Hees~\cite{Rapp1,rapp_RHIC,RalfRapp} 
the two scenarios:
Hadron Many Body Theory (HMBT) and Dropping Mass (DM) are shown.  The
sum is calculated with HMBT.  The calculations are compared to the
data from which the contributions of the cocktail of hadronic decays
and charmed meson decays have been subtracted.}
\end{figure*}

\subsection{$p_T$ Spectra in the Low Mass Region} \label{sub:discussion_pt_spectra}
The $p_T$ spectra of the excess (i.e.  after subtracting the
hadron cocktail and the charm from the dilepton spectra) can be also compared
to the theoretical models.  We already noted that the yields from the
partonic medium in the theoretical models are produced only via the
$q\bar{q}\rightarrow e^+e^-$ annihilation process.
Processes like $q+g\rightarrow qe^+e^-$ are not included.

Figure~\ref{fig:pt_theo} shows the $p_T$ spectrum in the mass
window $0.3<m_{ee}<0.75$~GeV/$c^2$ after subtracting the contribution
from the cocktail and the charm.  The spectrum is compared to the
theoretical calculation from 
Rapp and van~Hees~\cite{Rapp1,rapp_RHIC,RalfRapp}, 
Dusling and 
Zahed~\cite{dusling_RHIC,dusling_mod,dusling_mod_RHIC}, 
and Cassing and 
Bratkovskaya~\cite{cassing_RHIC,cassing0,cassing_HSD,WolfgangCassing}, 
respectively.  The figure
shows separately the $e^+e^-$ yields from the partonic phase and the
hadronic phase (with two possible implementations of the $\rho$
spectral function) and their sum is compared to the data.  In all the
models the sum of the cocktail contribution and the $e^+e^-$ yield
from medium-effects is insufficient to explain the experimental data,
and divergences are observed both at low-and high-$p_T$.
While for Rapp and van~Hees~\cite{Rapp1,rapp_RHIC,RalfRapp} 
and for Dusling and 
Zahed~\cite{dusling_RHIC,dusling_mod,dusling_mod_RHIC} 
the disagreement with the data is strong at low-$p_T$, for
Cassing and 
Bratkovskaya~\cite{cassing_RHIC,cassing0,cassing_HSD,WolfgangCassing} 
a better agreement is achieved over the full $p_T$ range.  However, 
the data seem still higher than the theoretical calculations.

\subsection{Theoretical Comparison to Direct Photon Measurement} \label{sub:discussion_photon}

\begin{figure}[!ht]
\includegraphics[width=1.0\linewidth]{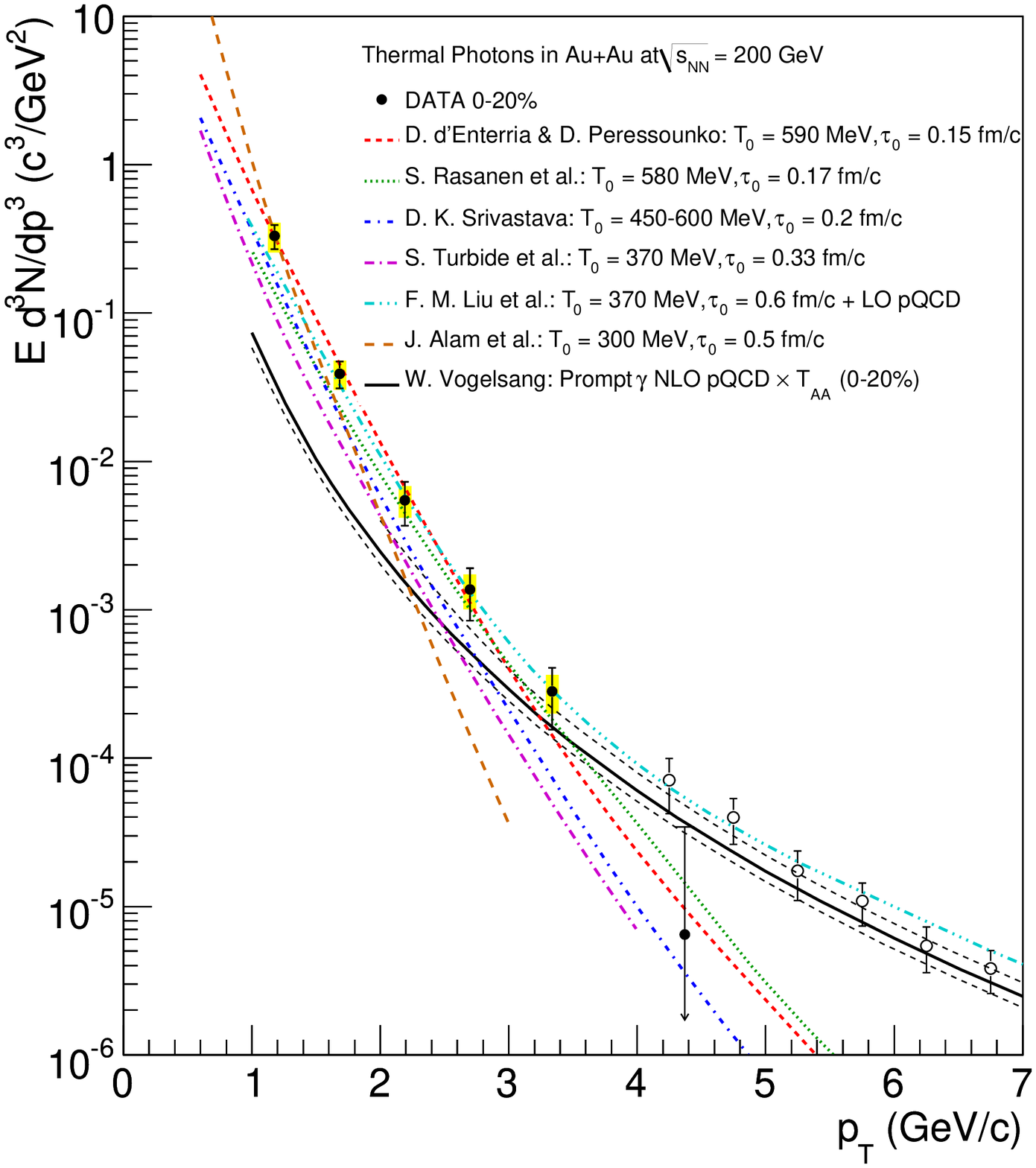}
\caption{\label{fig:compare_thermal} (Color online)
Theoretical calculations of thermal photon
emission~\cite{Turbide:2003si,d'Enterria:2005vz,huovinen02,srivastava01,alam01,liu08}
are compared with the direct photon data in central 0-20\% Au~+~Au
collisions shown separately and added to pQCD calculations.  In
contrast to the others, the curve by \cite{liu08} includes pQCD
contributions.  The black solid curve show the pQCD calculation, scaled
by $T_{\rm AA}$.  The QCD scale $\mu$ is set to $p_T$ for this
calculation.  The two black dashed curves around the black solid curve
show the scale uncertainty, with the upper curve and the lower curve
corresponds to $\mu = 1/2 \cdot p_T$ and $\mu = 2 \cdot p_T$, respectively.}
\end{figure}

In section~\ref{sub:photon} we have extracted the direct photon
yield from the analysis of LMR~I.
The obtained direct photon spectrum in central Au~+~Au, shown in
Figure~\ref{fig:photon_spectra}, shows excess over $T_{\rm AA}$ scaled
$p+p$ data, and the shape of the excess is well described by a
pure exponential with inverse slope $T \simeq 220$ MeV.
If the direct photons in Au~+~Au collisions are of thermal origin, the
inverse slope $T$ is related to the initial temperature $T_{\rm init}$ of
the dense matter.  In hydrodynamical models, $T_{\rm init}$ is 1.5 to 3 times $T$
due to the space-time evolution~\cite{d'Enterria:2005vz}.

\begin{figure}[!ht]
\includegraphics[width=1.0\linewidth]{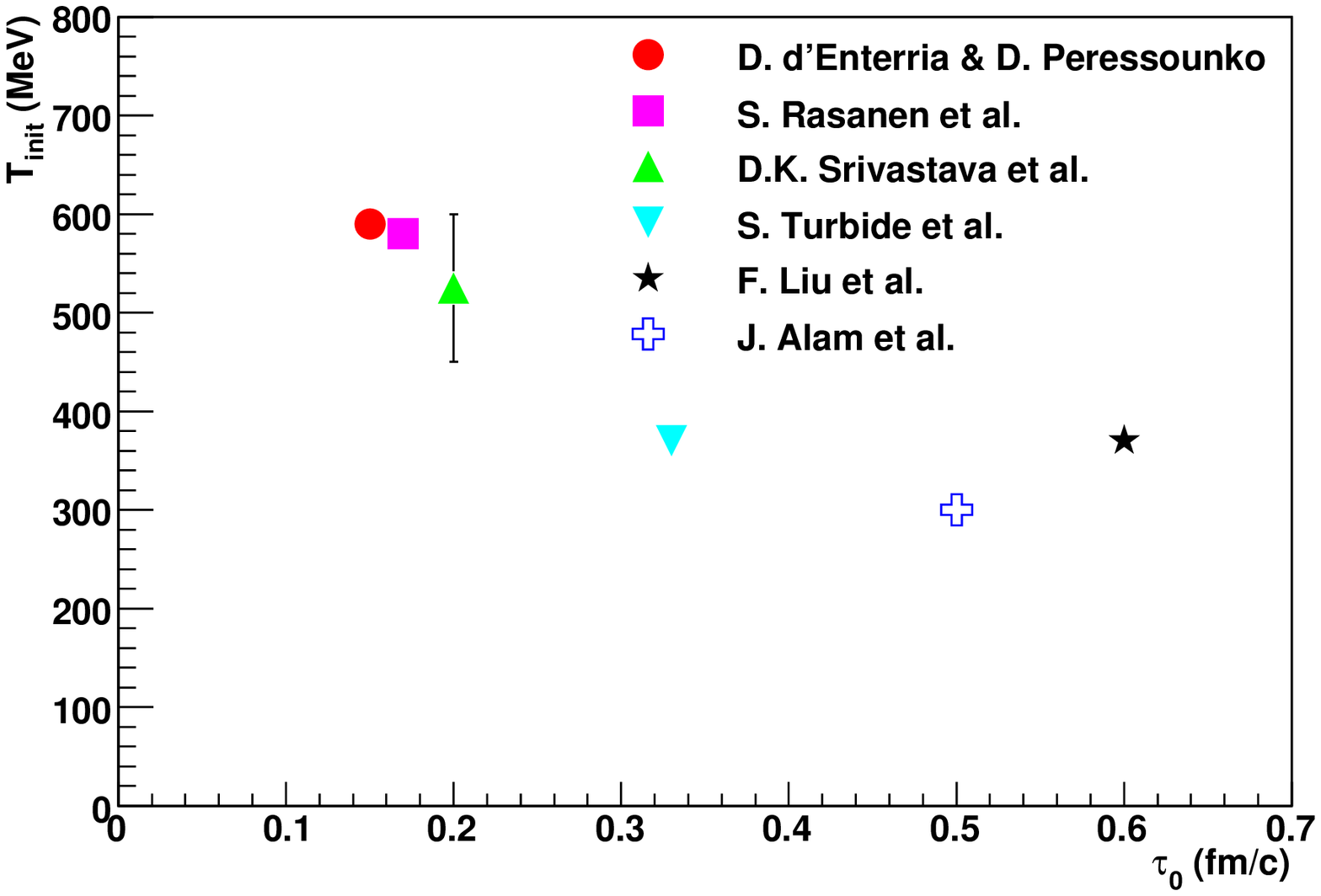}
\caption{\label{fig:T_vs_tau} (Color online)
$T_{\rm init}$ vs.  $\tau_0$ for various theoretical calculations
shown in Fig.~\ref{fig:compare_thermal}.}
\end{figure}

Figure~\ref{fig:compare_thermal} compares the direct photon data in
central 0-20\% Au~+~Au collisions with several theoretical calculations
of thermal photon emission added to the pQCD calculations
~\cite{vogelsang}.  Note that the curve by \cite{liu08} includes pQCD
contributions, while the others do not.  For $p_T<3$~GeV/$c$, the
thermal contribution dominates over pQCD.  These hydrodynamical models
can reproduce the high $p_T$ central Au~+~Au data within a factor of
two.  These models assume formation of a hot QGP with initial
temperature ranging from $T_{\rm init} = 300$~MeV at thermalization time
$\tau_0 = 0.6$ fm/$c$ to $T_{\rm init} = 600$~MeV at $\tau_0=0.15$
fm/$c$~\cite{Turbide:2003si,d'Enterria:2005vz,huovinen02,srivastava01,alam01,liu08}.
Figure~\ref{fig:T_vs_tau} summarizes the $T_{\rm init}$ and $\tau_0$ for
theoretical calculations shown in Fig.~\ref{fig:compare_thermal}.
There is a clear anti-correlation between $T_{\rm init}$ and $\tau_0$.
Lattice QCD predicts a phase transition from hadronic phase to quark
gluon plasma at $\simeq$ 170 MeV.

\section{SUMMARY AND CONCLUSIONS} \label{sec:summary}
PHENIX has measured dilepton production in Au~+~Au and $p+p$ collisions at
$\sqrt{s_{NN}}$~=~200~GeV.
The measured $e^+e^-$ yield is compared with the cocktail of
known sources of light hadron decays.
Cocktail sources are mostly measured by PHENIX in
the same experimental run via hadronic decay channels.  Extrapolations
to low-$p_T$, where experimental data are not always available
are obtained using an $m_T$-scaling procedure.

In the $p+p$ data the $e^+e^-$ invariant mass spectrum has been
measured in the mass range from 0 to 8~GeV/$c^2$ for all pair-$p_T$.
The intermediate-mass-region (IMR) is dominated by semi-leptonic
decays of heavy flavor mesons for which the extracted production cross
sections is consistent with fixed-order next-to-leading-log (FONLL)
predictions and with the PHENIX measurement of single
electrons~\cite{ppg065}.

The low-mass-region (LMR) can be described by known contributions from
light meson decays, and virtual direct photons for which the extracted
cross section is consistent with NLO pQCD calculations and with PHENIX
measurements of real photons.

In Au~+~Au collisions, the data are consistent with the expectations
from correlated $c\bar{c}$ production for $m_{ee}>0.5$~GeV/$c^2$.
However, this interpretation is ambiguous, due to the interplay
between possible two different medium effects: energy loss of charm quarks
in the medium which would deplete the yield in the IMR, and QGP
radiation, which would increase the yield in the IMR.

In the low mass region the Au~+~Au Min. Bias inclusive mass spectrum shows an enhancement
by a factor of $4.7\pm0.4^{\rm stat}\pm1.5^{\rm syst}\pm0.9^{\rm model}$
compared to the expectation from the hadronic cocktail.
The enhancement is concentrated at low $p_T$ ($p_T<1$~GeV/$c$).
The integrated yield increases faster with the centrality of the collisions
than the number of participating nucleons.


At low mass ($m_{ee} <$ 0.3~GeV/$c^2$) and high $p_T$ (1
$<p_T<$5~GeV/$c$) an enhanced $e^+e^-$ pair yield is observed both in
$p+p$ and Au~+~Au collisions.  The mass dependence of the excess is
consistent with that expected for virtual direct photon production.
This excess is used to infer the yield of real direct photons by
extrapolating to $m_{ee} = 0$.  A perturbative QCD calculation is
consistent with the real direct photon cross section in $p+p$
extracted by this method, while in central Au~+~Au collisions much
larger yields compared with the $p+p$ cross section scaled with
$T_{AA}$ are observed.  In central Au~+~Au collisions, the excess over
the $p+p$ cross section scaled by $T_{\rm AA}$ is exponential in
$p_T$, with inverse slope 
$T = 221 \pm 19^{\rm stat} \pm 19^{\rm syst}$~MeV.


In Au~+~Au collisions at very low $p_T$ there is a further, very
significant, enhancement that increases strongly with centrality.  The
$p_T$ spectrum of dileptons in this region has been analyzed with two
fit methods and a numerical calculation.  An inverse slope of $T \simeq
100$ MeV has been extracted for $m_T < 0.6$~GeV/$c$, lower than for
hadrons with similar mass and similar to the freeze-out temperature.

The Au~+~Au data are compared to different models which provide
additional $e^+e^-$ yield in both the LMR and the IMR.  
In the IMR the data have too large uncertainty to discriminate
different possible scenarios of charm production and QGP radiation.
In the LMR no quantitative agreement has been found yet with the models.

The yield of direct photons in Au~+~Au collisions is compared with 
several hydrodynamical models of thermal photon emission at RHIC 
energies.  The models assuming the formation of a hot system with 
initial temperature ranging $T_{{\rm init}}\simeq 300--600$~MeV at 
times $\tau_0 \simeq 0.6--0.15$~fm/$c$ are in qualitative agreement 
with the data.  Lattice QCD predicts a phase transition from hadronic 
phase to quark gluon plasma at $T~\simeq$~170 MeV.

In conclusion, we presented measurements of the $e^+e^-$ continuum in
$p+p$ and Au~+~Au at $\sqrt{s_{NN}}=200$~GeV in a wide range of mass and
transverse momenta.  In Au~+~Au collisions, a large enhancement of the
yield of $e^+e^-$ pairs is observed at low mass and low $p_T$ in
Au~+~Au.  The yield of direct photons is deduced from low mass, high
$p_T$ $e^+e^-$ pairs.  Future measurements with an upgraded PHENIX
detector with higher statistics, together with further advance in
theory, will allow more detailed study of the properties of the hot
dense matter formed in heavy ion collisions at RHIC.

\section*{ACKNOWLEDGEMENTS}

We thank the staff of the Collider-Accelerator and Physics
Departments at Brookhaven National Laboratory and the staff of
the other PHENIX participating institutions for their vital
contributions.  We acknowledge support from the 
Office of Nuclear Physics in the
Office of Science of the Department of Energy, the
National Science Foundation, Abilene Christian University
Research Council, Research Foundation of SUNY, and Dean of the
College of Arts and Sciences, Vanderbilt University (USA),
Ministry of Education, Culture, Sports, Science, and Technology
and the Japan Society for the Promotion of Science (Japan),
Conselho Nacional de Desenvolvimento Cient\'{\i}fico e
Tecnol{\'o}gico and Funda\c c{\~a}o de Amparo {\`a} Pesquisa do
Estado de S{\~a}o Paulo (Brazil),
Natural Science Foundation of China (People's Republic of China),
Ministry of Education, Youth and Sports (Czech Republic),
Centre National de la Recherche Scientifique, Commissariat
{\`a} l'{\'E}nergie Atomique, and Institut National de Physique
Nucl{\'e}aire et de Physique des Particules (France),
Ministry of Industry, Science and Tekhnologies,
Bundesministerium f\"ur Bildung und Forschung, Deutscher
Akademischer Austausch Dienst, and Alexander von Humboldt Stiftung (Germany),
Hungarian National Science Fund, OTKA (Hungary), 
Department of Atomic Energy (India), 
Israel Science Foundation (Israel), 
National Research Foundation (Korea),
Ministry of Education and Science, Russia Academy of Sciences,
Federal Agency of Atomic Energy (Russia),
VR and the Wallenberg Foundation (Sweden), 
the U.S. Civilian Research and Development Foundation for the
Independent States of the Former Soviet Union, the US-Hungarian
NSF-OTKA-MTA, and the US-Israel Binational Science Foundation.

\appendix
\section{Background Normalization} \label{app:2sqrt}
\subsection{Pairing of Electrons and Positrons} \label{sec:pairing}

In the following we assume that, as dictated by the charge
conservation law, $e^-$ and $e^+$ are always produced in pairs and
that most of these pairs are produced statistically independent of
each other.  Let us say $N$ pairs are produced in a particular event
and $N$ is given by a probability distribution $P(N)$.  Of the $N$
pairs only a fraction $\varepsilon_p$ is fully reconstructed, and then
the number of reconstructed pairs $n_p$ is given by a binomial
distribution $B$ sampling out of $N$ ``events'' with a probability
$\varepsilon_p$.

\begin{itemize}
\item Probability to get $n_p$ pairs from $N$ true pairs: $\omega(n_p)
= B(n_p, N, \varepsilon_p)$
\item with an average: $\langle n_p\rangle = \varepsilon_p N$
\item and variance: $\sigma_p^2 = \varepsilon_p N (1-\varepsilon_p)$
\end{itemize}

Of the remaining pairs one track is reconstructed with a probability
$\varepsilon_+$ or $\varepsilon_-$.  For a given $N$ and $n_p$ the
number of additional single positive tracks $n_+$ and negative tracks
$n_-$ follow a multinomial distribution $M$ with three possible
outcomes for each of the $N-n_p$ unreconstructed pairs: no track, one
$+$ track or one $-$ track.

The probability to get $n_+$ and $n_-$ single tracks from $N$ true
pairs with $n_p$ reconstructed pairs, i.e., from $(N-n_p)$ not
fully reconstructed pairs is: 
\begin{eqnarray}
\omega(n_+, n_-) = M(n_+, n_-; N-n_p,\varepsilon_+, \varepsilon_-)
\nonumber \\
\omega(n_+) = \sum_{n_-=1}^{N-n_p} M(n_+, n_-; N-n_p,\varepsilon_+, \varepsilon_-)
\nonumber \\
\omega(n_-) = \sum_{n_+=1}^{N-n_p} M(n_+, n_-; N-n_p,\varepsilon_+, \varepsilon_-)
\end{eqnarray}

\begin{itemize}
\item with average: $\langle n_{\pm}\rangle = \varepsilon_{\pm} (N-n_p)$
\item variance: $\sigma_{\pm}^2 = \varepsilon_{\pm} (N-n_p) (1-\varepsilon_{\pm})$
\item and covariance: $\rm{cov}(n_+, n_-) = -(N-n_p) \varepsilon_+ \varepsilon_-$
\end{itemize}

In this case the number of unlike-sign pairs for a given $N$ and
$n_p$ is:
\begin{eqnarray}
\langle n_{+-} \rangle &=& n_p^2 + n_p \sum_{n_+=1}^{N-n_p} n_+ \omega(n_+) + n_p
\sum_{n_-=1}^{N-n_p} n_- \omega(n_-) \nonumber \\&~& + \sum_{n_+=1}^{N-n_p} \sum_{n_-=1}^{N-n_p} n_+ n_-
\omega(n_+, n_-)\nonumber\\
&=& n_p^2 + n_p \varepsilon_+ (N-n_p) + n_p \varepsilon_- (N-n_p) +
\langle n_+ n_- \rangle\nonumber\\
&=& n_p^2 + n_p \varepsilon_+ (N-n_p) + n_p \varepsilon_- (N-n_p) +
\nonumber \\&~& 
\varepsilon_+ \varepsilon_- (N-n_p)^2 - \varepsilon_+
\varepsilon_- (N-n_p)\nonumber\\
&=& n_p^2 + \varepsilon_+ N n_p - \varepsilon_+ n_p^2 + \varepsilon_-
N n_p - \varepsilon_- n_p^2 \nonumber \\&~& + \varepsilon_+ \varepsilon_- N^2 - 2
\varepsilon_+ \varepsilon_- N n_p + \varepsilon_+ \varepsilon_-
n_p^2  \nonumber \\&~& -\varepsilon_+ \varepsilon_- N + \varepsilon_+ \varepsilon_-
n_p\nonumber\\
&=& \left(n_p + \varepsilon_+ \left(N-n_p\right)\right)\left(n_p +
\varepsilon_- \left(N-n_p\right)\right) -
\nonumber \\&~&
\varepsilon_+ \varepsilon_- (N-n_p)
\end{eqnarray}

Similarly we can calculate the number of like-sign pairs:
\begin{eqnarray}
2\langle n_{++} \rangle &=& \sum_{n_+=1}^{N-n_p} (n_p + n_+)(n_p + n_+ -1)
\omega(n_+)\nonumber\\
&=& n_p^2 - n_p + \langle n_+^2 \rangle - \langle n_+ \rangle + 2 n_p
\langle n_+ \rangle\nonumber\\
&=& n_p^2 - n_p + \varepsilon_+^2 (N-n_p)^2 +
\varepsilon_+(1-\varepsilon_+)(N-n_p) \nonumber\\&~&
- \varepsilon_+ (N-n_p) + 2 \varepsilon_+ n_p (N-n_p)\nonumber\\
&=& n_p^2 - n_p + \varepsilon_+^2 (N-n_p)^2 - \varepsilon_+^2(N-n_p) \nonumber\\&~&
+ 2 \varepsilon_+ n_p (N-n_p)
\end{eqnarray}
and
\begin{eqnarray}
2\langle n_{--} \rangle &=& n_p^2 - n_p + \varepsilon_-^2 (N-n_p)^2
\nonumber \\&~& - \varepsilon_-^2(N-n_p)
+ 2 \varepsilon_- n_p (N-n_p)
\end{eqnarray}

To obtain the expected number of like- and unlike- sign pairs for a
fixed number of real pairs $N$ we need to average over all possible
reconstructed pairs $n_p$:
\begin{eqnarray}
\langle N_{+-} \rangle &=&  \sum_{n_p} \langle n_{+-} \rangle B(n_p, N, \varepsilon_p) \nonumber\\
&=& (1 - \varepsilon_+ - \varepsilon_- + \varepsilon_+ \varepsilon_-)
\langle n_p^2 \rangle \nonumber \\ &~&
+ (\varepsilon_+ N + \varepsilon_- N - 2
\varepsilon_+ \varepsilon_- N  + \varepsilon_+ \varepsilon_-) \langle
n_p \rangle  \nonumber\\ &~&
+ \varepsilon_+ \varepsilon_- N^2 - \varepsilon_+
\varepsilon_- N \nonumber\\
&=& (1 - \varepsilon_+ - \varepsilon_- + \varepsilon_+
\varepsilon_-)(\varepsilon_p^2N^2 + \varepsilon_p(1-\varepsilon_p)N) \nonumber\\
&~& + (\varepsilon_+ N + \varepsilon_- N - 2 \varepsilon_+ \varepsilon_- N
+ \varepsilon_+ \varepsilon_-) \varepsilon_p N\nonumber\\
&~& + \varepsilon_+ \varepsilon_- N^2 - \varepsilon_+ \varepsilon_-
N \nonumber\\
&=& (\varepsilon_p^2 - \varepsilon_p^2 \varepsilon_+ -
\varepsilon_p^2 \varepsilon_- + \varepsilon_p^2 \varepsilon_+
\varepsilon_- + \varepsilon_p \varepsilon_+ + \varepsilon_p
\varepsilon_- \nonumber\\&~&  - 2 \varepsilon_p \varepsilon_+ \varepsilon_- +
\varepsilon_+ \varepsilon_-)(N^2-N) + \varepsilon_p N\nonumber\\
&=& (\varepsilon_p + \varepsilon_+ (1-\varepsilon_p))(\varepsilon_p
+ \varepsilon_-(1-\varepsilon_p))  \nonumber\\&~&  (N^2-N) + \varepsilon_p N
\end{eqnarray}

Now we calculate the like-sign background:
\begin{eqnarray}
2 \langle N_{++} \rangle &=&  \sum_{n_p} 2 \langle n_{++} \rangle
B(n_p, N, \varepsilon_p) \nonumber\\
&=& \varepsilon_p^2 N^2 + \varepsilon_p(1-\varepsilon_p)N -
\varepsilon_p N + \varepsilon_p^2 \varepsilon_+^2 N^2 
\nonumber\\&~& 
+ \varepsilon_+^2 \varepsilon_p(1-\varepsilon_p) N 
- 2 \varepsilon_+^2 \varepsilon_p N^2 + \varepsilon_+^2 N^2
\nonumber\\&~& 
- \varepsilon_+^2 N + \varepsilon_+^2 \varepsilon_p N + 2
\varepsilon_+ \varepsilon_p N^2 \nonumber\\&~& - 2 \varepsilon_+ \varepsilon_p^2 N^2 - 2 \varepsilon_+ \varepsilon_p
(1-\varepsilon_p) N\nonumber\\
&=& \varepsilon_p^2 (N^2-N) + \varepsilon_+^2 \varepsilon_p^2
(N^2-N) + \varepsilon_+^2 \varepsilon_p N \nonumber\\&~&
- 2 \varepsilon_+^2
\varepsilon_p N^2 + \varepsilon_+^2(N^2-N) + \varepsilon_+^2
\varepsilon_p N\nonumber\\
&~& + 2\varepsilon_+ \varepsilon_p N^2 - 2 \varepsilon_+
\varepsilon_P^2 N^2 - 2 \varepsilon_+ \varepsilon_p N + 2
\varepsilon_+ \varepsilon_p^2 N\nonumber\\
&=& (\varepsilon_p^2 + \varepsilon_+^2 +
\varepsilon_+^2\varepsilon_p^2)(N^2-N) - 2 \varepsilon_+^2
\varepsilon_p (N^2-N) \nonumber\\&~& + 2 \varepsilon_+ \varepsilon_p (N^2-N) - 2
\varepsilon_+ \varepsilon_p^2 (N^2-N)\nonumber\\
\langle N_{++} \rangle &=& \frac{1}{2} (\varepsilon_p + \varepsilon_+ (1-\varepsilon_p))^2(N^2-N)
\end{eqnarray}
and
\begin{eqnarray}
\langle N_{--} \rangle &=& \frac{1}{2} (\varepsilon_p + \varepsilon_- (1-\varepsilon_p))^2(N^2-N)
\end{eqnarray}

Finally we need to average over all $N$ to get the foreground unlike-sign pairs:

\begin{eqnarray}
\langle FG_{+-} \rangle &=& 
\sum_{N} \langle N_{+-} \rangle P(N)\nonumber\\
&=& 
(\varepsilon_p + \varepsilon_+(1-\varepsilon_p))(\varepsilon_p +
\varepsilon_-(1-\varepsilon_p)) \nonumber\\&~&	
\cdot (\langle N^2 \rangle - \langle N
\rangle) + \varepsilon_p \langle N \rangle\nonumber\\
&\equiv& \langle BG_{+-} \rangle + \langle S \rangle
\end{eqnarray}

The unlike-sign foreground $FG_{+-}$ consists of the sum of the unlike-sign background $BG_{+-}$ and the signal $S=\varepsilon_p \langle N \rangle$.  
Similarly the like-sign foreground is calculated as: 

\begin{eqnarray}
\langle FG_{++} \rangle &=& \sum_{N} \langle N_{++} \rangle
P(N)\nonumber\\
&=& \frac{1}{2}(\varepsilon_p + \varepsilon_+(1-\varepsilon_p))^2(\langle N^2 \rangle - \langle N
\rangle)\nonumber\\
&\equiv& \langle BG_{++} \rangle\\
\langle FG_{--} \rangle &=& \sum_{N} \langle N_{--} \rangle
P(N)\nonumber\\
&=& \frac{1}{2}(\varepsilon_p +
\varepsilon_-(1-\varepsilon_p))^2(\langle N^2 \rangle - \langle N
\rangle)\nonumber\\
&\equiv& \langle BG_{--} \rangle
\end{eqnarray}

The like-sign foreground contains no signal.

So due to the fact that electrons and positrons are always created in
pairs, the unlike-sign background is the geometric mean of the like-sign backgrounds, independent of the primary multiplicity distribution.
\begin{equation}
\langle BG_{+-} \rangle = 2 \sqrt{\langle BG_{++} \rangle \langle BG_{--} \rangle}
\end{equation}

Let us compare the background to the product of the average track
multiplicities.  For a fixed $n_p$:
\begin{eqnarray}
\langle n_+ \rangle &=& \sum_{n_+=1}^{N-n_p} (n_p + n_+)\omega(n_+)\nonumber\\
&=& n_p + \langle n_+ \rangle\nonumber\\
&=& n_p + \varepsilon_+ (N-n_p)
\end{eqnarray}
averaged over all possible $n_p$:
\begin{eqnarray}
\langle N_+ \rangle &=& \sum_{n_p=0}^{N} \langle n_+ \rangle\omega(n_p)\nonumber\\
&=& \varepsilon_p N + \varepsilon_+ N - \varepsilon_+ \varepsilon_p N\nonumber\\
&=& (\varepsilon_p + \varepsilon_+ (1-\varepsilon_p)) N
\end{eqnarray}
or averaged over all possible $N$:
\begin{eqnarray}
\langle FG_+ \rangle &=& \sum_N \langle N_+ \rangle P(N)\nonumber\\
&=& (\varepsilon_p + \varepsilon_+ (1-\varepsilon_p)) \langle N \rangle
\end{eqnarray}
and thus:
\begin{equation}
\langle FG_+ \rangle \langle FG_- \rangle = (\varepsilon_p +
\varepsilon_+ (1-\varepsilon_p))(\varepsilon_p +
\varepsilon_- (1-\varepsilon_p))\langle N\rangle^2
\end{equation}
or
\begin{equation}
\frac{\langle BG_{+-}\rangle}{\langle FG_+ \rangle \langle FG_- \rangle}
= 1 + \frac{\sigma^2 - \langle N \rangle}{\langle N \rangle^2}.
\end{equation}

So in general $\langle BG_{+-} \rangle \neq \langle FG_+ \rangle \langle
FG_- \rangle$, except for the special case that $P(N)$ is a Poisson
distribution.  Note this is the opposite conclusion one derives in the
case that the sources of $+$ and $-$ tracks are independent,  i.e., $+$
and $-$ tracks are produced as singles and not as pairs as in the case
of muons.  In that case $\langle FG_+ \rangle \langle FG_- \rangle$ is
the correct background normalization.

\section{Relation between real photons, virtual photons, and electron
pairs}\label{app:ICA_theory}
\subsection{Introduction}

\begin{figure}[htb]
\includegraphics[width=1.0\linewidth]{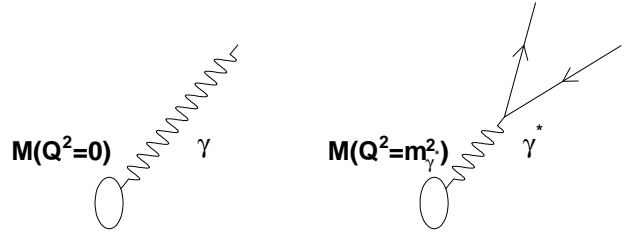}
\caption{\label{fig:ICA} 
Diagram for real photon production (left) and 
its associated process producing an $e^+e^-$ pair (right).}
\end{figure}

Figure~\ref{fig:ICA} illustrates that, in general, any source of high 
energy real photons can also emit virtual photons which materialize into 
electron pairs.   On the left side a real photon is emitted
by a source labeled as $M(Q^2=0)$.  
On the right side is an analogous diagram, where a virtual
photon with mass $m_{\gamma^*}$ is emitted.
The virtual photon can then convert to an $e^+e-$ pair if 
$m_{\gamma^*} > 2 m_e$.
This $e^+e^-$ pair production process is a QED correction to the
real photon production process and is often called internal conversion.

In the energy region, where electroweak effects are negligible,
an electron pair can only be produced
through a virtual photon.  (Here we don't include $e^+e^-$ pairs from 
correlated weak decays such as
$c \bar{c} \rightarrow e^+e^-$.)
Thus any electron pair production
process can be described as production of a virtual photon and its subsequent
decay into an $e^+e^-$pair.  

The yields of
virtual photons $dN_{\gamma^*}$ and electron pairs $dN_{ee}$ are related:
\begin{eqnarray}
\frac{d^2N_{ee}}{dM^2} &=& \frac{\alpha}{3\pi}\frac{L(M)}{M^2}
dN_{\gamma^*}, \label{eq:b1} \\ 
L(M)&=&\sqrt{1-\frac{4m_e^2}{M^2}}\left(1+\frac{2m_e^2}{M^2}\right),
\end{eqnarray}
where $M$ is the mass of the virtual photon or the electron pair
($M = m_{\gamma^*} = m_{ee}$) and
$\alpha$ is the fine structure constant ($\alpha\simeq 1/137$).
The factor, $\frac{\alpha}{3\pi}\frac{L(M)}{M^2}$, is a universal factor
describing the decay of the virtual photon into an $e^+e^-$ pair.
This relation is exact to first order in the electromagnetic
coupling $\alpha$.

Equation~\ref{eq:b1} can be written as
\begin{equation}
\frac{d^2N_{ee}}{dM^2} = \frac{\alpha}{3\pi}\frac{L(M)}{M^2} S(M,q)
dN_\gamma.  \label{eq:ICA}
\end{equation}
Here we have introduced $S(M,q) = dN_{\gamma^*}(M)/dN_\gamma$ to factor out the difference
between real photon emission and virtual photon emission.
The factor $S(M,q)$ is process dependent and accounts for effects such as form factors,
phase space, and spectral functions.  $S(M,q)$ approaches 1 for small $M$,
$S(M,q) \rightarrow 1$ for $M\rightarrow 0$.
Additionally, since $L(M) \simeq 1 - 6 m_e^4/M^4$ for $m_e \ll M$,
$L(M)=1$ is a very good approximation.
Thus the relationship between the electron pair yield and the direct
photon yield simplifies to
\begin{eqnarray}
\frac{d^2N_{ee}}{dM^2} \simeq \frac{\alpha}{3\pi}\frac{1}{M^2}dN_{\gamma},\\
\frac{d^2N_{ee}}{dM} \simeq \frac{2\alpha}{3\pi}\frac{1}{M} dN_{\gamma}.\label{eq:master}
\end{eqnarray}
The relation between real photon production and electron pair production shown by
Equation~\ref{eq:master} is valid if $M \ll E_\gamma$, i.e.  if the virtual photon
is quasi-real.
In this region, the yield of electron pairs in the mass range
$m_1 < M < m_2$ is related to the photon yield as
\begin{eqnarray}
dN_{ee}(m_1<M<m_2) &\simeq& \int^{m_2}_{m_1} \frac{2\alpha}{3\pi}\frac{1}{M} dN_{\gamma} \\
   &\simeq& \frac{2\alpha}{3\pi} \log{\frac{m_2}{m_1}} dN_\gamma.\label{eq:g_ee_ratio}
\end{eqnarray}

In the following, we discus two examples of internal conversions: Dalitz decays and
high $p_T$ Drell-Yan production.  We then discuss the relationship between 
direct photons and electron pairs from thermal sources.  At the end, we illustrate
the relationship between electron pairs and virtual photons using a theoretical model
calculation.

\subsection{Dalitz decays}
Dalitz decays of pseudo-scalar and vector mesons are prime examples of
internal conversion.  In these processes, a virtual photon, instead of a real photon,
is emitted in the decay of a hadron and subsequently decays into
an $e^+e^-$ pair.  The relation between
$A\rightarrow Be^+e^-$ and $A\rightarrow B \gamma$ is given by~\cite{landsberg}
\begin{equation} \label{eq:Dalitz1}
\frac{d\Gamma(A\rightarrow Be^+e^-)}{dM^2} = 
\frac{\alpha}{3\pi} \frac{L(M)}{M^2} S_{\rm AB}(M) \Gamma(A\rightarrow B\gamma),
\end{equation}
\begin{eqnarray}
\lefteqn{S_{\rm AB}(M) = |F_{\rm AB}(M^2)|^2 \times }   \nonumber \\
&& \Bigl[(1+\frac{M^2}{m_A^2-m_B^2})^2-\frac{4m_A^2M^2}{(m_A^2-m_B^2)^2} \Bigr]^{3/2},
\end{eqnarray}
where $m_A$ and $m_B$ are the mass of hadrons $A$ and $B$,
and $F_{\rm AB}$ is the electromagnetic transition form factor.
$S_{AB}(M)$ here is an example of $S(M,q)$ in Equation~\ref{eq:ICA}.
For decays of pseudo-scalar mesons
($P=\pi^0, \eta, \eta'$) the relationship between the photonic decay
($P \rightarrow \gamma \gamma$) and the corresponding Dalitz decay
($P \rightarrow e^+e^-\gamma$)
is given by the well known Kroll-Wada formula~\cite{kroll-wada,landsberg}.
\begin{eqnarray}
\frac{d\Gamma(P\rightarrow e^+e^-\gamma)}{dM^2} &=&
\frac{2\alpha}{3\pi}\frac{L(M)}{M^2} S_\mathrm{KW}(M) \Gamma(P\rightarrow 
\gamma\gamma), \nonumber \\
S_\mathrm{KW}(M)&=&|F_{P}(M^2)|^2 \Bigl(1-\frac{M^2}{m_P^2}\Bigr)^3, 
\label{eq:DalitzPS}
\end{eqnarray}
where $m_P$ is
the meson mass and $F_P(M^2)$ is the electromagnetic form factor.  
Note that the factor 2 in $\frac{2\alpha}{3\pi}$ accounts for the fact that each of the two
decay photons can convert to an electron pair.
The form factor is usually parameterized as $F_P(Q^2) =
1/(1-Q^2/\Lambda_P^2)$.  Experimental measurements of the transition form
factor by Lepton-G~\cite{lepton-g}
and Cello~\cite{CELLO} show $\Lambda_P \simeq M_{\rho}$, consistent with
the vector meson dominance model (VDM).  

\subsection{High $p_T$ Drell-Yan process}
In $p+p$ collisions, the cross section for Drell-Yan 
electron pair
production can be expressed in terms of the cross section for virtual photon production as follows~\cite{VWJQ}:
\begin{eqnarray}
\frac{d^3\sigma_{pp\rightarrow l^+l^- X}}{dM^2dp_T^2dy}=\frac{\alpha}{3\pi}\frac{L(M)}{M^2}\frac{d^2\sigma_{pp\rightarrow \gamma^* X}}{dp_T^2dy}.\label{eq:DY_exact}
\end{eqnarray}
Where $M$, $p_T$, and $y$ are the mass, the transverse momentum, and the rapidity of
the virtual photon.  
For $p_T \gg M$, the virtual photon cross section becomes equal to
the real photon cross section ($d\sigma_{\gamma^*} \rightarrow d\sigma_{\gamma}$ as
$M/p_T \rightarrow 0$).  Thus the electron pair cross section can be described as
\begin{eqnarray}
\frac{d\sigma_{pp\rightarrow e^+e^- X}}{dM^2dp_T^2dy} \simeq \frac{\alpha}{3\pi}\frac{L(M)}{M^2}\frac{d\sigma_{pp\rightarrow \gamma X}}{dp_T^2dy}.\label{eq:DY_approx}
\end{eqnarray}

Direct photon production via gluon-Compton scattering
($q+g \rightarrow q+\gamma$) has an associated electron pair production
process ($q+g \rightarrow q+\gamma^* \rightarrow q + e^+e^-$).
For the lowest order pQCD calculation, the following relation between
the two processes is obtained by an explicit calculation.
\begin{eqnarray}
\frac{d^2\sigma_{ee}}{dM^2dt} = 
\frac{\alpha}{3\pi}\frac{L(M)}{M^2}\frac{d\sigma_{\gamma}}{dt}\times
S_{qg}(u,t,s),
\end{eqnarray}
where
\begin{eqnarray}
\lefteqn{S_{qg}(u,t,s) = (1+\frac{2u}{t^2+s^2}M^2)} \nonumber \\
&&= 1 - \frac{2x} {(x+\sqrt{1+x^2})(3x^2+1+2x
\sqrt{1+x^2})}.\label{eq:gCompton}
\end{eqnarray}
Here $x=p_T/M$ and $s, t, u$ are the Mandelstam variables
defined as $s = (p + k)^2$, $t = (p - k')^2$, $u = (p - p')^2$;
$p, k, k',p'$ are 4-momenta of the incoming quark, the incoming
gluon, the outgoing (virtual or real) photon, and the outgoing quark,
respectively.  
The factor $S_{qg}(u,t,s)$ accounts for the difference between the
virtual photon cross section and the real photon cross section, and
$S_{qg}(u,t,s)$ becomes unity for small $M$.  It is an example
of $S(M,q)$ in Equation~\ref{eq:ICA}.
For $90^{\circ}$ scattering in the c.m.s.,
$S_{qg} \simeq (1-p_T^2/5M^2)$.  Thus $S_{qg} \rightarrow 1$ as
$p_T^2/M^2 \rightarrow 0$.  This means that the approximation $S_{qg} \simeq 1$
is valid as long as $p_T^2/M^2 \ll 1$ even if $M$ is relatively large.

In a kinematic region, where $p_T/M$ is not very large,
contributions from parton fragmentation into direct photon (real and virtual)
become significant.  Unfortunately
hadronic effects in parton fragmentation into photons are poorly understood.
Kang, Qui and Vogelsang discussed~\cite{VWJQ} the theoretical uncertainties
in this process.  However, the uncertainty due to this effect is relatively small,
except for very low $p_T$ ($p_T \lesssim 1$~GeV/$c$)~\cite{VWJQ}.

\subsection{Thermal radiation}\label{app:ICA_thermal}
In heavy ion collisions, thermal radiation from the hot and dense matter formed in the collision may
contribute to both real direct photon production and electron pair production.
The emission rate of electron pairs per space-time volume from a thermal source can be
described in terms of the electromagnetic (EM) spectral function as~\cite{rapp0,vHvRapp}
\begin{eqnarray} \label{eq:spectral_fc}
\frac{dR_{ee}}{d^4q} = -\frac{\alpha^2}{3\pi^3}\frac{L(M)}{M^2}{\rm Im}\Pi^{\mu}_{em,\mu}(M,q;T)f^B(q_0,T) \\
     = -\frac{\alpha^2}{3\pi^3}\frac{L(M)}{M^2}{\rm Im}(2\Pi^{T}_{em}+\Pi^{L}_{em})f^B(q_0,T).
\end{eqnarray}
Here $\Pi^{T}_{em}$ and $\Pi^{L}_{em}$ are the transverse and the longitudinal components of the
in-medium photon self-energy tensor $\Pi^{\mu}_{em,\nu}$, and $f^B(q_0,T)=1/(e^{q_0/T}-1)$ is the Boltzmann factor.

Using the same notation, the emission rate of virtual photons is
described as~\cite{rapp0,Turbide2008}
\begin{eqnarray}
q_0\frac{dR_{\gamma^*}}{d^3q} = -\frac{\alpha}{2\pi^2}{\rm Im}\Pi^{\mu}_{em,\mu}(M,q;T)f^B(q_0,T).
\end{eqnarray}
The real photon emission rate is obtained in the limit of $M\rightarrow 0$.
The longitudinal polarization contribution $\Pi^L_{em}$ vanishes for real photons:
$\Pi^{T}_{em}(M,q;T)\rightarrow \Pi^{T}_{em}(0,q;T), \Pi^{L}_{em}(M,q;T)\rightarrow 0$ for $M \rightarrow 0$.
The real photon emission rate is given by
\begin{eqnarray}
q_0\frac{dR_{\gamma}}{d^3q} = -\frac{\alpha}{\pi^2}{\rm Im}\Pi^{T}_{em}(M=0,q;T)f^B(q_0,T).
\end{eqnarray}
The virtual photon and electron pair emission rates are thus related by
\begin{eqnarray}
q_0\frac{dR_{ee}}{dM^2d^3q} &=& \frac{1}{2} \frac{dR_{ee}}{d^4q} \nonumber \\
&=& \frac{\alpha}{3\pi}\frac{L(M)}{M^2}q_0\frac{dR_{\gamma^*}}{d^3q}, \label{eq:ee_vph_exact} \\ 
q_0\frac{dR_{ee}}{dM^2d^3q} &\simeq& \frac{\alpha}{3\pi}\frac{L(M)}{M^2}q_0\frac{dR_\gamma}{d^3q}.  \label{eq:ee_p_approx}
\end{eqnarray}
Note that Equation~\ref{eq:ee_vph_exact}, describing the relationship
between the virtual photon and electron pair emission, is exact to
order $\alpha$ in QED and is exact to all orders of the strong
coupling constant.  Note also that this relationship between electron
pairs and photons is the same as that shown in Equation~\ref{eq:DY_exact}.
This reflects the fact that an $e^+e^-$ pair can only be produced
through virtual photon and that the conversion rate of a virtual
photon into an $e^+e^-$ pair is described by a universal factor,
$\frac{\alpha}{3\pi}\frac{L(M)}{M^2}$.  Equation~\ref{eq:ee_p_approx} is
equivalent to Equation~\ref{eq:DY_approx} and is an approximation for small
$M$, where $dR_\gamma^* \simeq dR_\gamma$.

The relations above are for the emission rates per space-time volume.  The yields $dN_{ee}$ and
$dN_{\gamma}$ are obtained from space-time integral over the rates.
\begin{eqnarray}\label{eq:virtual}
\frac{d^2N_{ee}}{dM^2} &=& \frac{\alpha}{3\pi}\frac{L(M)}{M^2} dN_{\gamma^*},\\
\frac{d^2N_{ee}}{dM} &=& \frac{2\alpha}{3\pi}\frac{L(M)}{M} dN_{\gamma^*} \\
   &=& \frac{2\alpha}{3\pi}\frac{L(M)}{M} S(M,q) dN_{\gamma}.
\end{eqnarray}
These are the same equations as Equation~\ref{eq:b1} -- Equation~\ref{eq:ICA}.
Here, following Equation~\ref{eq:ICA}, the difference between real and
virtual photons is factored out in $S(M,q)$.  The $S(M,q)$ factor for
radiation from thermal sources can be written as
\begin{eqnarray}
S(M,q) = \frac{\langle{\rm
Im}(2\Pi^T_{em}(M,q)+\Pi^L_{em}(M,q))f^B(q_0)\rangle} {\langle{\rm
Im}2\Pi^{T}_{em}(0,q)f^B(q_0;M=0)\rangle} \label{eq:S}
\end{eqnarray}
Here $\langle \rangle$ indicates the space-time average.

Deviation of $S(M,q)$ from unity can arise from non-zero
$\Pi^{L}_{em}(M,q)$ in the medium and a change of $\Pi^{T}_{em}(M,q)$
from $\Pi^{T}_{em}(0,q)$.  The behavior of $\Pi^{T}_{em}$ and
$\Pi^{L}_{em}$ is model-dependent.  However, on very general grounds
we can conclude that the in-medium spectral function $\Pi_{em}$ is a
smooth function of $M$ for the low-mass region ($M <$ a few 100
MeV/$c^2$).

Hadronic interactions yield and almost flat behavior in $S(M)$ as we see in a
model calculation by Rapp later.  $q+g$ scattering gives an
almost flat contribution in $S(M)$; see $S_{qg}$ of gluon Compton scattering.
As we see later, the contribution from
$q\bar{q}$ annihilation is not constant and is $\propto M^2$ (see Equation~\ref{eq:qq}).
This means that it is
strongly suppressed in the low-mass region relative to hadronic and $qg$
scattering contributions.  Thus $S(M,q)$ is a smooth and almost constant function of $M$ 
at low masses.  Furthermore, the short time ($\le 1$ fm/$c$)
between rescattering among hadrons and partons in the medium should
smear any feature in $S(M)$ smaller than a mass scale of a few 100 MeV/$c^2$.

\begin{figure*}[t]
\includegraphics[width=0.4\linewidth]{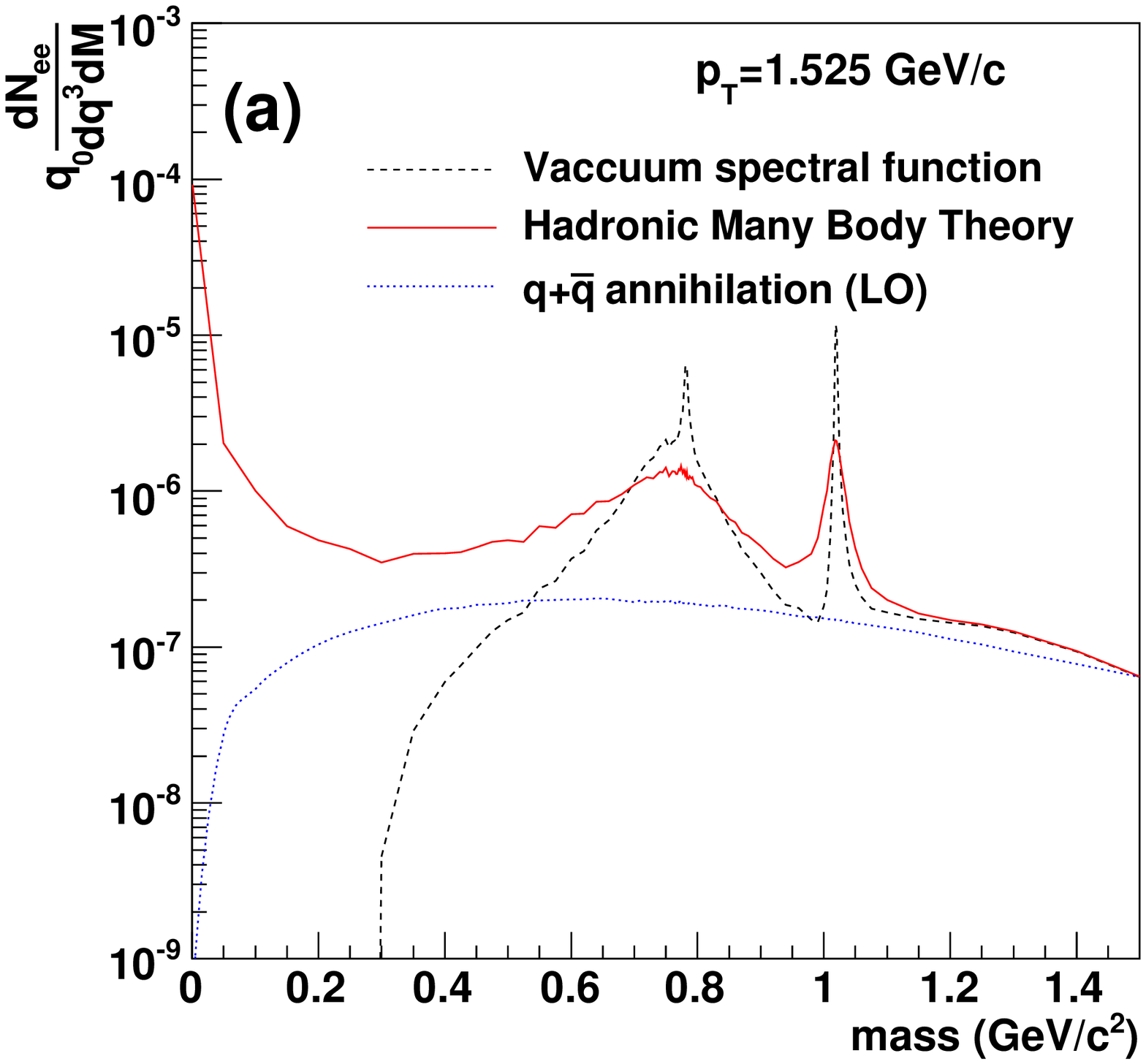}
\includegraphics[width=0.4\linewidth]{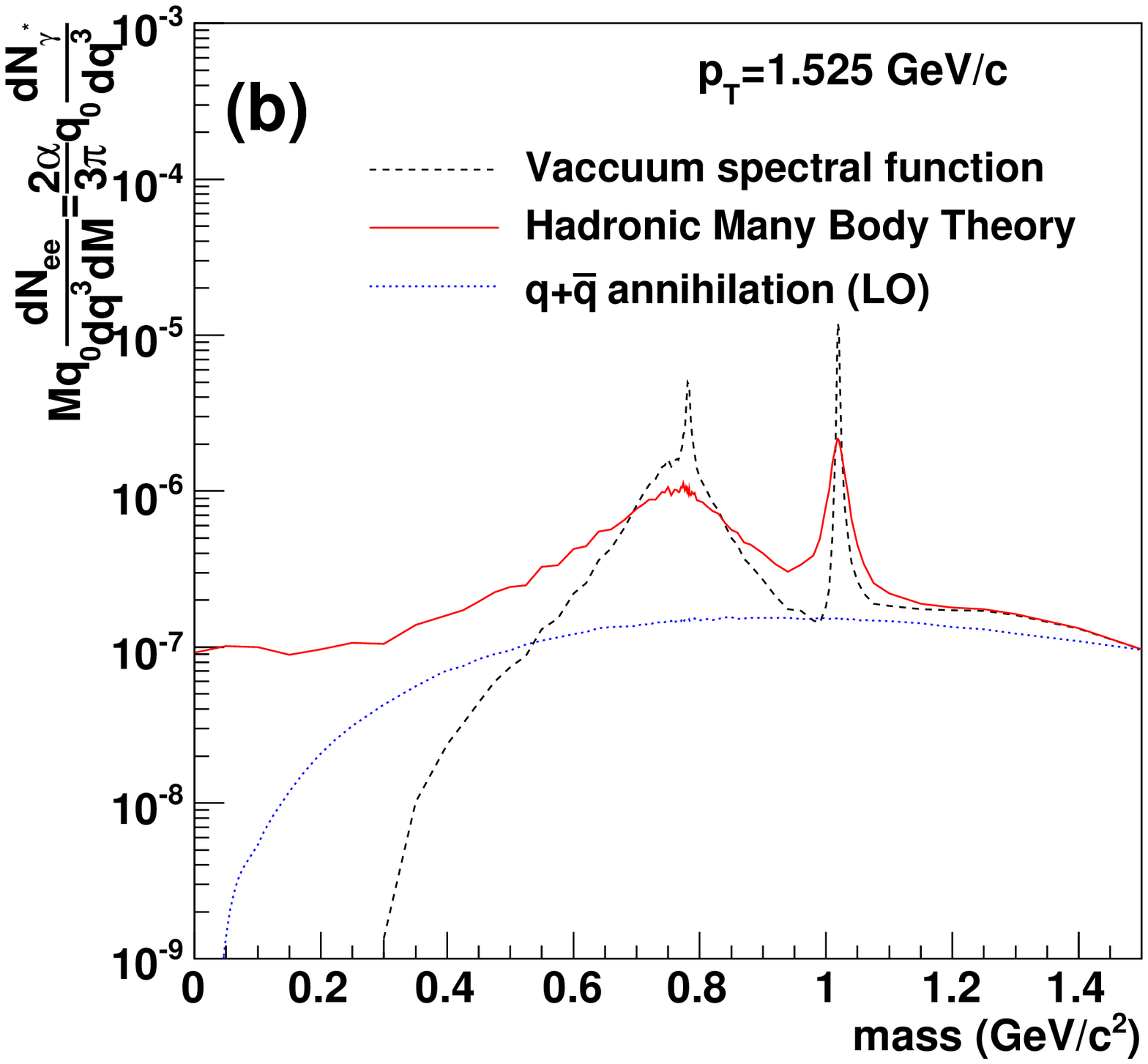}
\caption{\label{fig:RalfRapp} (Color online)
(a) Electron pair emission rate and (b) virtual photon emission rate calculations at a fixed pair $p_T$~=~1.525~GeV/$c$~\cite{RalfRapp}.  
The solid curve shows the hadronic many-body
theory in the medium.  The dashed curve shows the calculation when the EM
spectral function in the vacuum is used.  The dotted curve shows the
$q\bar{q}$ annihilation contribution.}
\end{figure*}

\subsection{Conversion of $e^+e^-$ pairs to virtual photons }
Equation (\ref{eq:virtual}) can be used to determine the virtual photon yield
from the electron pair yield.  The equation can be rewritten as
\begin{eqnarray}
q_0\frac{dN_{\gamma^*}}{d^3q} &=& S(M,q) q_0\frac{dN_\gamma}{d^3q} \\
&\simeq& \frac{3\pi}{2\alpha} M q_0\frac{dN_{ee}}{d^3qdM}.\label{eq:lepton_photon}
\end{eqnarray}
Using this relationship, the virtual photon spectrum as a function of mass $M$ and $p_T$
can be determined from the double differential electron pair spectrum.
In particular, the mass dependence for a given $p_T$ bin can be measured.
This is a direct measurement of the shape of the $S(M,q)$ function and
the shape of $\langle{\rm Im}\Pi^{em}(M,T)f^B(T)\rangle$ as a function of $M$.
The real photon yield is then obtained by extrapolation of
$dN_{\gamma^*}(M)$ to the $M \rightarrow 0$ limit.

\subsection{Electron pairs and virtual photons in a theoretical model calculation}
Figure~\ref{fig:RalfRapp} illustrates the relationship between the electron pair mass spectrum 
and the virtual photon cross section.
Figure~\ref{fig:RalfRapp} (a) shows the double differential electron pair spectrum,
$(1/p_T)dN_{ee}/dMdp_Tdy$ at $p_T=1.525$~GeV/$c$ from a model calculation of electron pair
production by Rapp~\cite{RalfRapp}.  The dashed and solid curve show electron pairs from the hadronic
gas.  
The calculation shown in the dashed curve uses
the spectral function $\Pi_{em}$ that is unchanged from its vacuum value so the line shapes
of vector mesons ($\rho, \omega, \phi$) are unmodified.  The calculation shown in the solid curve,
uses a spectral function calculated by a model Lagrangian of hadronic many-body interactions, and
the line shapes of vector mesons are broadened due to the interactions.
It also includes the contributions like
$a_1$(1260)$\rightarrow \pi+e^+e^-$,
$\rho \rightarrow \pi + e^+e^-$,
and $N+\pi \rightarrow N^* \rightarrow Ne^+e^-$ that fill out the low-mass
regions below two-pion threshold.  In the low-mass region, the mass spectrum steeply increases
with decreasing $M$.  This steep behavior is due to the $1/M$ factor in 
$\gamma^* \rightarrow e^+e^-$.
The dotted curve shows the contribution from the leading order (LO) $q\bar{q}$
annihilation which is augmented with a $q=0$ Hard Thermal Loop (HTL)
correction factor~\cite{Braaten}.  

Figure~\ref{fig:RalfRapp}(b) shows the same calculations presented as
the differential yield of virtual photons as a function of mass.  The electron
pair yield shown in Fig~\ref{fig:RalfRapp}(a) is converted to the
virtual photon yield using Equation~\ref{eq:lepton_photon}.  In this plot,
the solid curve becomes almost constant for $M<0.3$~GeV.  The steep $1/M$
behavior of the electron pair spectrum is removed, and much more
smooth behavior of the virtual photon spectrum is revealed.  The plot
shows that the virtual photon yield is almost constant as function of
$M$.  The value of the solid curve at $M=0$ corresponds to the real
photon yield.  Thus in this model calculation, the yield of virtual
photons for $0.1<M<0.3$~GeV is almost the same as that of real photons.
The solid curve illustrates that in a consistent theory calculation the
yield of virtual photons is a smooth function of $M$ and it becomes the
real photon yield in the limit of $M=0$, $dN_\gamma^*(M,p_T)
\rightarrow dN_\gamma(p_T)$ as $M \rightarrow 0$.

The flat behavior of the solid curve in this low-mass region comes from the fact
that hadronic scattering processes such as
$a_1$(1260)$\rightarrow \pi+e^+e^-$,
$\rho \rightarrow \pi + e^+e^-$, and
$N+\pi \rightarrow N^* \rightarrow Ne^+e^-$
dominate this low-mass region.
These processes are
internal conversion of the corresponding real photon
production processes, i.e., $a_1 \rightarrow \pi + \gamma$,
$\rho \rightarrow \pi + \gamma$, $N+\pi \rightarrow N+\gamma$.
Thus production of virtual photons from these processes should be
very close to that of real photons at low mass.
In the language of the spectral function $\Pi_{em}$,
these processes contribute to $\Pi_{em}$ at $M=0$ as well as
$M>0$.  Their contribution to $\Pi_{em}$ is a smooth, almost a constant
function of virtual photon mass $M$.

In Fig.~\ref{fig:RalfRapp}(b), the contribution from $q\bar{q}$ annihilation is shown by
the dotted curve.
In perturbation theory,
the $q\bar{q}$ contribution is given by
\begin{eqnarray}
{\rm Im}\Pi_{em}(M) = \frac{M^2}{12\pi}(1+\frac{\alpha_s}{\pi}+...)N_c\Sigma(e_q)^2.  \label{eq:qq}
\end{eqnarray}
The quark annihilation contribution behaves as $\propto M^2$ in the
low-mass region.  Thus it is strongly suppressed and has little
contribution in the low-mass region.  In the high-mass region, the
$M^2$ behavior of the quark annihilation is suppressed by the
Boltzmann factor.

There is a large uncertainty in the approximation used in the
dotted curve, and it is shown here just to illustrate that the $q\bar{q}$
annihilation contribution is sub-leading contribution in the low-mass
region ($M <$ a few 100 MeV/$c^2$).  Many effects can alter the shape
and magnitude of the $q\bar{q}$ contribution.  The calculation shown
in Fig.~\ref{fig:RalfRapp} uses zero quark mass.
The effective quark mass in
the medium is uncertain, but in most theoretical calculations it is
the order of the temperature of the medium.
Quarks in the medium should also have a large width (more than a few
100 MeV/$c^2$) due to the short time between interactions.  A non-zero
effective quark mass would further suppress the quark annihilation
contribution in the very low-mass region while a finite quark width would
smear the $M^2$ behavior and can cause non-zero $q\bar{q}$
contribution at $M=0$.

It should be noted that the dotted curve does not include processes like
$q+g \rightarrow q+\gamma^*$ that are associated with real direct
photon production.  This is because HTL calculation of
thermal radiation from QGP is only available in the real photon
case.  Contributions from processes associated with real photon
production can be much larger than those from the LO $q\bar{q}$ annihilation
shown in Fig.~\ref{fig:RalfRapp} in the low-mass region.  
Turbide, Gale, and Rapp~\cite{Turbide:2003si} calculated real photon
production in a hadronic gas using the same model and compared
it with real  photon production in the QGP phase using the complete
leading order HTL analysis.  They found that real photons from
the QGP outshine those from a hadronic gas
for $p_T>1.5$~GeV/$c$ in Au~+~Au collisions at RHIC.  Since the
virtual photon yield should be continuous from $M=0$ (i.e.  real
photons) to $M>0$, this implies that the contribution from the QGP,
including processes like $q+g \rightarrow q+\gamma^*$, is as large as or
even greater than that shown in the solid curve in Fig.~\ref{fig:RalfRapp}(b)
in the low-mass and high-$p_T$ region
(i.e.  $M<$ a few 100 MeV/$c^2$ and $p_T>1.5$~GeV/$c^2$) at RHIC energies.

\def\IJMPA{{Int. J. Mod. Phys.}~{\bf A}}
\def\JPG{{J. Phys}~{\bf G}}
\def\NCA{Nuovo Cimento}
\def\NIM{Nucl. Instrum. Meth.}
\def\NIMA{{Nucl. Instrum. Meth.}~{\bf A}}
\def\NPA{{Nucl. Phys.}~{\bf A}}
\def\NPB{{Nucl. Phys.}~{\bf B}}
\def\PLB{{Phys. Lett.}~{\bf B}}
\def\PLC{Phys. Repts.\ }
\def\PRL{Phys. Rev. Lett.\ }
\def\PRD{Phys. Rev. D~}
\def\PRC{Phys. Rev. C~}
\def\ZPC{{Z.~Phys.}~{\bf C}}
\def\EPJ{Eur.~Phy.~J.~{\bf C}}
\def\etal{{\it et al.}}



\begin{thebibliography}{99}
\bibitem{whitepaper} K.~Adcox \etal (PHENIX Collaboration), \NPA {\bf 757}, 184 (2005).
\bibitem{ppg019} S.~S.~Adler \etal (PHENIX Collaboration), \PRC {\bf 71}, 034908 (2005).
\bibitem{ppg003} K.~Adcox \etal (PHENIX Collaboration), \PRL {\bf 88}, 022301 (2002). 
\bibitem{ppg051} S.~S.~Adler \etal (PHENIX Collaboration), \PRL {\bf 96}, 202301 (2006). 
\bibitem{ppg056} S.~S.~Adler \etal (PHENIX Collaboration), \PRL {\bf 96}, 032301 (2006). 
\bibitem{ppg066} A.~Adare \etal (PHENIX Collaboration), \PRL {\bf 98}, 172301 (2007).
\bibitem{ppg022} S.~S.~Adler \etal (PHENIX Collaboration), \PRL {\bf 91}, 182301 (2003).
\bibitem{ppg062} A.~Adare \etal (PHENIX Collaboration), \PRL {\bf 98}, 162301 (2007).
\bibitem{ppg073} S.~Afanasiev \etal (PHENIX Collaboration), \PRL {\bf 99}, 052301 (2007).
\bibitem{Stankus} P.~Stankus, Ann. Rev. Nucl. Part. Sci. {\bf 55}, 517 (2005).
\bibitem{rapp_shuryak} R.~Rapp and E.~Shuryak, \PLB {\bf 473}, 13 (2000).
\bibitem{masui_satz} T.~Matsui and H.~Satz, \PLB {\bf 178}, 416 (1986).
\bibitem{pbm}P.~Braun Munzinger and J.~Stachel, \PLB {\bf 490}, 196
(2000) and P.~Braun-Munzinger, K.~Redlich, and J.~Stachel,
``Quark Gluon Plasma. Vol 3'', World Scientific, Singapore, nucl-th/0304013 (2003).
\bibitem{coalescence} R.~L.~Thews, M.~Schroedter, J.~Rafelski, \PRC
{\bf 63}, 054905 (2001).
\bibitem{Rapp1} R.~Rapp, \PRC {\bf 63}, 054907 (2001).
\bibitem{Kaempfer} K.~Gallmeister, B.~Kampfer, and O.~P.~Pavlenko \PRC {\bf 57}, 3276 (1998) and 
B.~Kaempfer, O.~P.~Pavlenko, and K.~Gallmeister, \PLB {\bf 419}, 412 (1998). 
\bibitem{Shuryak} E.~V.~Shuryak \PRC {\bf 55}, 961 (1997); 
C.~M.~Hung, E.~V.~Shuryak, \PRC {\bf 56}, 453 (1997). 
\bibitem{rapp_RHIC} 
R.~Rapp, nucl-th/0204003; 
W.~Liu, R.~Rapp, \NPA {\bf 796}, 101 (2007). 
\bibitem{dusling_RHIC} K.~Dusling and I.~Zahed, \NPA {\bf 825}, 212 (2009); 
K.~Dusling PhD thesis, Stony Brook University (2008).
\bibitem{cassing_RHIC} E.~L.~Bratkovskaya, W.~Cassing, O.~Linnyk, \PLB {\bf 670}, 428 (2009).
\bibitem{brown0} G.~E.~Brown and M.~Rho, \PRL {\bf 66}, 2720 (1991).
\bibitem{Cobb:1978gj} J.~H.~Cobb \etal, \PLB {\bf 78}, 519 (1978).
\bibitem{Albajar:1988iq} C.~Albajar \etal (UA1 Collaboration), \PLB {\bf 209}, 397 (1988).
\bibitem{HEL} M.~Masera, (HELIOS/3 Collaboration), \NPA {\bf 590}, 103c (1995).
\bibitem{CER1} G.~Agakichiev \etal (CERES Collaboration), \PRL {\bf 75}, 1272 (1995).
\bibitem{rapp0} R.~Rapp, J.~Wambach, Adv. Nucl. Phy. {\bf 25}, 1 (2000) 
and references therein.
\bibitem{cassing0} W.~Cassing and E.~Bratkovskaya, Phys. Rep. {\bf 308}, 65 (1999) 
and references therein.
\bibitem{NA60_rho} R.~Arnaldi \etal (NA60 Collaboration), \PRL {\bf 96}, 162302 (2006).
\bibitem{CER2} G.~Agakichiev \etal (CERES Collaboration), \EPJ {\bf 41} 475 (2005).
\bibitem{HADES} G.~Agakichiev \etal (HADES Collaboration), \PRL {\bf 98}, 052302 (2007).
\bibitem{NA60_pt} R.~Arnaldi \etal (NA60 Collaboration), \PRL {\bf 100}, 022302 (2008).
\bibitem{HELIOS3} A.~L.~S.~Angelis \etal (HELIOS/3 Collaboration), \EPJ {\bf 13}, 433 (2000).
\bibitem{NA38} M.~C.~Abreu \etal (NA38 Collaboration), \PLB {\bf 368} 230(1996).
\bibitem{NA50} M.~C.~Abreu \etal (NA50 Collaboration), \EPJ {\bf 14}, 443 (2000).
\bibitem{NA60_therm} R.~Arnaldi \etal (NA60 Collaboration), \EPJ {61}, 711 (2009).
\bibitem{ppg075} S.~Afanasiev  \etal, arXiv:nucl-ex/07063034; to be published.
\bibitem{ppg085} A.~Adare \etal (PHENIX Collaboration), \PLB {\bf 670}, 313 (2009).
\bibitem{ppg086} A.~Adare \etal (PHENIX Collaboration), 
arXiv:0804.4168 [nucl-ex]; to be published.
\bibitem{Adcox:2003zm} K.~Adcox \etal (PHENIX Collaboration), \NIMA {\bf 499}, 469 (2003).
\bibitem{Adler:2001a} C.~Adler \etal (STAR Collaboration), \NIMA {\bf 470}, 488 (2003).
\bibitem{Aronson:2003a} S.~H.~Aronson \etal (PHENIX Collaboration), 
\NIMA {\bf 499}, 480 (2003).
\bibitem{Adcox:2003a} K.~Adcox \etal (PHENIX Collaboration), \NIMA {\bf 499}, 489 (2003).
\bibitem{Aizawa:2003a} M.~Aizawa \etal (PHENIX Collaboration), \NIMA {\bf 499}, 508 (2003).
\bibitem{Aphecetche:2003a} L.~Aphecetche \etal (PHENIX Collaboration), 
\NIMA {\bf 499}, 521 (2003).
\bibitem{Allen:2003a} M.~Allen \etal (PHENIX Collaboration), \NIMA {\bf 499}, 549 (2003).
\bibitem{ppg065} A.~Adare \etal (PHENIX Collaboration), \PRL {\bf 97}, 252002 (2006).
\bibitem{ppg001} K.~Adcox \etal (PHENIX Collaboration), \PRL {\bf 86}, 3500 (2001). 
\bibitem{ppg002} K.~Adcox \etal (PHENIX Collaboration), \PRL {\bf 87}, 052301 (2001).
\bibitem{pythiamb} T.~Sj{\"o}strand, \etal, Computer Phys. Commun. 135, 238 (2001);
we used {\sc pythia} 6.319 with MSEL=0 and the following processes switched on: 
MSUB 11,12,13,28,53,68, PARP(91)=1.5 ($\langle k_t \rangle$), 
MSTP(32)=4 (Q$^2$ scale), and CKIN(3)=2.0 (min. parton $p_T$).
\bibitem{ppg083} A.~Adare \etal (PHENIX Collaboration), \PRC {\bf 78}, 014901 (2008).
\bibitem{ppg077} PHENIX in preparation.
\bibitem{geant} GEANT User's Guide, 3.15, CERN Program Library.
\bibitem{PDG} W.~M.~Yao \etal (Particle Data Group), Journal of Physics, 
{\bf G33}, 1 (2006).
\bibitem{pi0} A.~Adare \etal (PHENIX Collaboration), \PRD {\bf 76}, 051106 (2007).
\bibitem{pikp} S.~S.~Adler \etal (PHENIX Collaboration), \PRC {\bf 74}, 024904 (2006).
\bibitem{eta} S.~S.~Adler \etal (PHENIX Collaboration), \PRC {\bf 75}, 024909 (2007).
\bibitem{omega} S.~S.~Adler \etal (PHENIX Collaboration), \PRC {\bf 75}, 051902 (2007).
\bibitem{phi} Y.~Riabov \etal (PHENIX Collaboration), \JPG {\bf 34}, No.8, S925 (2007).
\bibitem{jpsi} A.~Adare \etal (PHENIX Collaboration), \PRL {\bf 98}, 232002 (2007).
\bibitem{AUpi0} S.~S.~Adler  \etal, \PRL {\bf 91}, 072301 (2003).
\bibitem{AUpikp} S.~S.~Adler \etal (PHENIX Collaboration), \PRC {\bf 69}, 034909 (2004).
\bibitem{AUphi} S.~S.~Adler \etal (PHENIX Collaboration), \PRC {\bf 72}, 014903 (2005).
\bibitem{AUjpsi} A.~Adare  \etal, \PRL {\bf 98}, 232301 (2007).
\bibitem{etaprime} V.~Ryabov, \NPA {\bf 827}, 1-4, 395c (2009).
\bibitem{AUomega} A.~Milov, 
\bibitem{XY} A.~Author, \emph{Proc.~of 15th Int.~Workshop on Deep-Inelastic 
Scattering and Related Subjects, Munich}, 731 (2007)
\verb$http://dx.doi.org/10.3360/dis.2007.127$; 
arXiv:0707.1258 [nucl-ex].  
Y.~Nakamiya,  \JPG Nucl.~Part.~Phys. {\bf 35}, 104158 (2008). 
\bibitem{psiprimePP} C.~da~Silva, arXiv:0907.4696 [nucl-ex].
\bibitem{psiprime} R.~Gavai \etal, Int. J. Mod. Phys. {\bf A10}, 3043 (1995).
\bibitem{milov} A.~Milov, \EPJ {\bf 61}, 721 (2009). 
\bibitem{naglis} M.~Naglis, \EPJ {\bf 61}, 835 (2009). 
\bibitem{bec} J.~Manninen, F.~Becattini, \PRC {\bf 78} 054901 (2008). 
\bibitem{kroll-wada} N.~M.~Kroll and W.~Wada, Phys. Rev. {\bf 98}, 1355 (1955).
\bibitem{lepton-g} R.~I.~Dzhelyadin \etal, \PLB {\bf 102}, 296 (1981).
\bibitem{landsberg} L.~G.~Landsberg, Phys. Rep. {\bf 128}, 301 (1985).
\bibitem{gounaris-sakurai} G.~J.~Gounaris and J.~J.~Sakurai, 
\PRL {\bf 21}, 244 (1968).
\bibitem{cacciari} M.~Cacciari, P.~Nason, and R.~Vogt, 
\PRL {\bf 95} 122001 (2005).
\bibitem{pythia} We used {\sc pythia} 6.152 with parameters 
as in K.~Adcox \etal (PHENIX Collaboration), 
\PRL {\bf 88}, 192303 (2002) and 
CTEQ5L PDF as in H.L.~Lai {\it et al.}  \EPJ {\bf 12}, 375 (2000). 
\bibitem{white} K.~Adcox \etal (PHENIX Collaboration), 
\NPA {\bf 757}, 184 (2005).
\bibitem{ppg042} S.~S.~Adler \etal (PHENIX Collaboration), 
\PRL {\bf 94}, 232301 (2005).
\bibitem{ppg060} S.~S.~Adler \etal (PHENIX Collaboration), 
\PRL {\bf 98}, 012002 (2007).
\bibitem{Adler:2003pb} S.~S.~Adler \etal (PHENIX Collaboration), 
\PRL {\bf 91}, 241803 (2003).
\bibitem{CERES} G.~Agakichiev \etal (CERES Collaboration), \EPJ {\bf 4}, 231 (1998).
\bibitem{RalfRapp} R.~Rapp, private communication. 
The numerical table of the double differential yield 
of lepton pair based on the same theoretical model of~\cite{vHvRapp} 
is provided by Rapp.
\bibitem{dusling_mod} K.~Dusling, D.~Teaney, and I.~Zahed, \PRC {\bf 75}, 024908 (2007).
\bibitem{dusling_mod_RHIC} K.~Dusling, private communication.  The
numerical table of the double differential yield of lepton pair based
on the same theoretical model of~\protect\cite{dusling_RHIC} 
is provided by Dusling; for more recent calculations see
K.~Dusling and I.~Zahed, arXiv:0911.2426 [nucl-th].
\bibitem{cassing_HSD} E.~L.~Bratkovskaya, W.~Cassing, \NPA {\bf 807} 214 (2008). 
\bibitem{WolfgangCassing} E.~L.~Bratkovskaya, private communication. The
numerical table of the double differential yield of lepton pair based
on the same theoretical model of~\cite{cassing_RHIC} is provided by Bratkovskaya.
\bibitem{d'Enterria:2005vz} D.~d'Enterria and D.~Peressounko, \EPJ {\bf 46}, 451 (2006).
\bibitem{vogelsang} L.~E.~Gordon and W.~Vogelsang, \PRD {\bf 48}, 3136 (1993); 
W.~Vogelsang, private communication (2008).
\bibitem{Turbide:2003si} S.~Turbide, R.~Rapp, and C.~Gale, \PRC {\bf 69}, 014903 (2004).
\bibitem{huovinen02} P.~Huovinen, P.~V.~Ruuskanen, and S.~S.~Rasanen,
\PLB {\bf 535}, 109 (2002).
\bibitem{srivastava01} D.~K.~Srivastava and B.~Sinha, \PRC {\bf 64}, 034902 (2001).
\bibitem{alam01} Jan-e~Alam, S.~Sarkar, T.~Hatsuda, T.K.~Nayak, and B.~Sinha \PRC 
{\bf 63}, 021901(R) (2001).
\bibitem{liu08} F.~M.~Liu, T.~Hirano, K.~Werner, and Y.~Zhu, \PRC {\bf 79}, 014905 (2009). 
\bibitem{CELLO} H.~J.~Behrend \etal (CELLO Collaboration), Z. Phys. {\bf C 49}, 401 (1991).
\bibitem{VWJQ} Z-B.~Kang, J-W. Qiu, and W. Vogelsang, \PRD {\bf 79}, 054007 (2009).
\bibitem{vHvRapp} H.~van Hees and R.~Rapp, \NPA {\bf 806}, 339 (2008).
\bibitem{Turbide2008} S.~Turbide, C.~Gale, E.~Frodermann, 
and U.~Heinz, \PRC {\bf 77}, 024909 (2008).
\bibitem{Braaten} E.~Braaten, R.~D.~Pisarski, T.-C.~Yuan, \PRL {\bf
64}, 2242 (1990).

\end{thebibliography}
\end{document}